\documentclass[submission, Phys]{SciPost}
\hypersetup{
    colorlinks,
    linkcolor={red!50!black},
    citecolor={blue!50!black},
    urlcolor={blue!80!black}
}

\usepackage{graphicx} % Required for inserting images

\usepackage{graphicx}
\usepackage[curve,matrix,arrow,color]{xy}

\usepackage{subcaption}
\usepackage{pstricks}
\usepackage{pstricks-add}
\usepackage{tikz}
\usetikzlibrary{arrows.meta}
\usepackage{verbatim}
\usetikzlibrary{calc}
\usetikzlibrary{decorations.pathreplacing}
\usetikzlibrary{decorations.markings}

\usepackage{mathtools}
\usepackage{upgreek}
\usepackage{amsmath}
\usepackage{amsfonts}
\usepackage[title]{appendix}
\usepackage{amsmath, amssymb, amsthm}
\usepackage{color,soul}
\usepackage{float}
\usepackage{placeins}
\usepackage{hyperref}
\usetikzlibrary{intersections} 
\usetikzlibrary{er,positioning}
\usetikzlibrary{arrows,shapes}
\usetikzlibrary{decorations.markings}
\usepackage{blkarray}
\usepackage{cite} % To regroup citations: [1-3] instead of [1,2,3]
\usepackage{booktabs}
\usepackage{pgfplots}
\pgfplotsset{compat=1.18}

\newcommand{\beq}{\begin{equation}}
\newcommand{\eeq}{\end{equation}}

\newcommand{\N}{{\cal{N}}}

\begin{document}

\title{Nonlinear sigma models, antiperiodic boundary conditions,\\ spin chains, and 't Hooft anomalies}

\author{Nicholas Read\textsuperscript{1,2} and Hubert Saleur\textsuperscript{3,4}}

\maketitle
\begin{center}
{\bf 1} Department of Physics, Yale University, P.O. Box 208120, New Haven,\\ 
CT 06520-8120, USA\\
{\bf 2} Department of Applied Physics, Yale University, P.O. Box 208284, New Haven,\\ 
CT 06520-8284, USA\\
{\bf 3} Institut de Physique Th\'eorique, Universit\'e Paris Saclay, CEA, CNRS, F-91191 Gif-sur-Yvette, France
\\
{\bf 4} Department of Physics and Astronomy, University of Southern California, Los Angeles, CA 90089-0484, USA
\\
%%%%%%%%%% END TODO: AFFILIATIONS
%%%%%%%%%% TODO: EMAIL
% Provide email address of corresponding author(s)
%\\[\baselineskip]
%$\star$ \href{mailto:yhang@usc.edu}{\small email1}
%%%%%%%%%% END TODO: EMAIL
\end{center}

\begin{abstract}
We consider two sets of related models: initially, these are $SU(2)$ antiferromagnetic spin chains with $N$ sites
of spin $S$, and the $O(3)$ nonlinear sigma model in two dimensions with topological coefficient $\Theta$ a multiple of $\pi$
(and later, the extensions of these with any semisimple Lie group symmetry). In a continuum description at large $N$ and large $S$, 
it is known that the low-energy behavior of the spin chain is given by the sigma model
with $\Theta=2\pi S$. We study these models with $N$ odd and with 
antiperiodic (A) boundary condition (b.c.), respectively [in place of the usual $N$ even and periodic b.c.], which correspond.
The A b.c.\ in the sigma model involves the $\mathbb{Z}_2$ inversion symmetry $\vec{n}\to-\vec{n}$, and amounts to a flux 
of a $\mathbb{Z}_2$ gauge field through a spacetime torus; summing over the two b.c.s for each direction would amount to 
gauging the $\mathbb{Z}_2$ inversion symmetry. We show directly that, if and only if $(-1)^{\Theta/\pi}=-1$, the gauging cannot be carried out, 
meaning there is an 't Hooft anomaly. The partition function 
for the A b.c.\ exists, but is not gauge invariant. Consequently, the sum over b.c.s cannot be made modular invariant;
the modular anomaly for $\Theta=\pi$ (mod $2\pi$) was found in the low-energy conformal field theory long ago.
The gauged model would be a sigma model with target space $\mathbb{R}\mathbb{P}^2\cong \mathbb{S}^2/\mathbb{Z}_2$, 
and hence this model does not exist for $\Theta=\pi$ (mod $2\pi$). A related result is that, using semiclassical quantization, 
in the spin chain we obtain the known values of the ground-state crystal momentum, which at leading order 
depend only on $N$ modulo $4$ and $2S$ modulo $2$. For $\Theta=\pi$ (mod $2\pi$) and the A b.c., inversion is in fact 
lifted to an element of a double cover of $O(3)$ called $Pin_+(3)$. 
For a large class of spin chains and associated sigma models we find similar results, but now $(-1)^{\Theta/\pi}$
is replaced by the value $\pm 1$ of the square of the time-reversal operator acting on a single spin, 
which is still determined by the coefficients of the topological terms, in a way that depends on the symmetry group.

\end{abstract}

\bigskip
\today
\bigskip

\newpage 

\tableofcontents
\section{Introduction}

The twin topics of antiferromagnetic $SU(2)$ quantum spin chains and the $O(3)$ ``nonlinear'' sigma model in one-dimensional space 
(or $1+1$-dimensional 
spacetime) have both been studied for many years. Briefly, the Hamiltonian of the nearest-neighbor antiferromagnetic spin $S$ chain is 
\begin{equation}
H=J_0\sum_{\ell=0}^{N-1}\vec{S}_\ell\cdot\vec{S}_{\ell+1}
\label{spinchHam}
\end{equation}
where $\vec{S}_\ell$ are a set of $N$ $SU(2)$ spin operators, indexed by $\ell=0$, $1$, \ldots, $N-1$ (with $\ell$ and $\ell+N$ 
viewed as equivalent), acting naturally in the 
tensor product of $N$ copies of the same irreducible spin-$S$ representation of $SU(2)$, that is $\vec{S}_\ell^2=S(S+1)$ 
for all $\ell$, and $J_0>0$ is a parameter.
On the other hand, the $O(3)$ sigma model is defined by the action
\beq
{\cal S}=\frac{1}{2g^2}\int dxdt\, \left[v(\partial_x \vec{n})^2+\frac{1}{v}(\partial_t \vec{n})^2\right] 
+\frac{i\Theta}{4\pi}\int dxdt\;\vec{n}\cdot(\partial_x\vec{n}\times \partial_t\vec{n}),
\label{O3action}
\eeq
where $\vec{n}(x,t)$ is a continuous unit three-component vector function of position $x$ and time $t$ (note that in this paper $t$ is real, 
but the metric on spacetime, which is used in the first term, is the standard Euclidean one 
when written in terms of the coordinates $x$, $vt$). The partition function $Z$ of the quantum or statistical field theory 
can then be defined as the path integral over $\vec{n}(x,t)$ with the ``paths'' [i.e.\ configurations of $\vec{n}(x,t)$] 
weighted by $e^{-{\cal S}}$. The velocity $v>0$ may be needed to match the choice of 
coefficient $J_0$, $v\propto J_0$, when the two models are related. (We could set $J_0$ or $v$ to $1$, but we usually retain them.) 
$g>0$ is a coupling constant. 
The expression
\begin{equation}
{\cal N}=\frac{1}{4\pi}\int dxdt\;\vec{n}\cdot(\partial_x\vec{n}\times \partial_t\vec{n})\label{Ndef}
\end{equation}
is normalized so that it is an integer when spacetime is a torus $\mathbb{T}^2$, or more generally whenever spacetime is an 
orientable manifold without boundary, such as a $2$-sphere $\mathbb{S}^2$ [we emphasize that $\vec{n}(x,t)$ is assumed continuous over 
the manifold]; it does not involve a metric on the manifold, but does involve a choice of an orientation [there is also an orientation
on the ``target space'' $\mathbb{S}^2$ in which $\vec{n}(x,t)$ lives; it is implicit in the scalar triple product]. 
The term $\Theta {\cal N}$ in $\cal S$ (with the $i$ removed) is the so-called topological term (we call the remainder 
of the action ``non-topological''),
while $\cal N$ is frequently called the instanton number [$\vec{n}(x,t)$ configurations with ${\cal N}=1$ and minimal non-topological 
action are called instantons]. 
Then in $e^{-{\cal S}}$, $\Theta$ and $\Theta+2\pi$ are equivalent, and so in the quantum theory $\Theta$ is viewed as defined 
only modulo $2\pi$. (For the sake of pedagogy, in the main text we consider only these cases, in which most points already appear. 
In a final Appendix, we explain the generalization to any semisimple Lie group, and the relation of our results with time-reversal symmetry. 
Readers are encouraged to work through most of the main text before attempting the Appendix.)

The close connection between these two models was unearthed by Haldane in the early 1980s \cite{HALDANE1983464,PhysRevLett.50.1153} 
(originally, without the $\Theta{\cal N}$ term; see also Ref.\ \cite{AflHal87}), who found that in the semiclassical 
(large $S$) limit, the long-wavelength description of the spin chain is the sigma model, 
in which the field $\vec{n}$ represents the local N\'eel order parameter of the spin chain, with $g^2\sim 1/S$ and $\Theta=2\pi S$. 
[The N\'eel order parameter would describe the ground state if the system were classical: for $N$ even, the lowest energy would be 
for $\vec{S}_{\ell+1}=-\vec{S}_\ell$ for all $\ell$, and $\vec{n}$ could be defined as 
$\vec{n}(x)=(-1)^\ell\vec{S}_\ell/S$ (or the same times $-1$), where $x=\ell a$ and $a$ is the lattice constant, so $L=Na$. 
One expects equal-time 
correlations of nearby spins to resemble this picture quantum mechanically at low or zero temperature, at least for $S$ large.] For $S$ 
an integer (i.e.\ $2S$ even), $\Theta=0$ (mod $2\pi$), and the quantum field theory (QFT) of the sigma model (after renormalization) is 
expected to be gapped (massive) in the large distance [or infrared (IR)] limit, so Haldane predicted that the spin chain is gapped. 
We will have only a little to say about this case. 
For $2S$ odd (we will refer to such numbers $S$ as ``half integer''), the Lieb-Schultz-Mattis (LSM) theorem \cite{Lieb1961} 
says that such a spin chain has a low-lying excited state with energy no larger than $\sim J_0/N$ above the ground state energy, 
and so is either gapless or has a broken discrete symmetry in the $N\to\infty$ limit. For these cases, $\Theta=\pi$ (mod $2\pi$), 
and it is believed \cite{AflHal87} that the (renormalized) theory in the IR is the $SU(2)$ level $1$ conformal field theory (CFT), 
the same as occurs in the IR limit of the antiferromagnetic nearest-neighbor spin $S=1/2$ Heisenberg model of a spin chain, 
which can be solved exactly following Bethe's original ansatz for that model. (The CFT of the spin-$1/2$ chain was already partially 
understood in the 1970s \cite{PhysRevB.12.3908}, though not in relation to Kac-Moody current algebras.)

All of this has been well studied for the cases (which correspond) of an even-length chain ($N$ an even integer), and the sigma model 
with the boundary condition (b.c.) that $\vec{n}$ be periodic in $x$ under $x\to x+L$ for length $L$, as used above. 
Further, because the $\vec{S}$ 
or $\vec{n}$ variables are bosonic (i.e.\ in the Heisenberg picture for the quantum theory in terms of operators, 
those at distinct $j$ or $x$ commute at equal times), general time-ordered correlation functions of spins or fields
that may be at various different times are periodic functions in each time $t$ variable, with period $\beta$ ($\beta$ is inverse 
temperature). Likewise, the ordinary partition function 
\begin{equation}
Z={\rm Tr}\,e^{-\beta H},
\label{partfn}
\end{equation}
(where the Hamiltonian and the trace are those of the spin chain or of the operator formulation
of the sigma model), when expressed as a path integral (as above for the sigma model), 
involves a periodic temporal b.c.\ (i.e.\ b.c.\ in $t$) with period $\beta$ on the spins or fields, for both the spin chain 
and the sigma model. Thus we mainly consider spacetime that is a torus $\mathbb{T}^2$.

In this paper we consider what occurs when these b.c.s on the torus are generalized. To do so, we first examine the 
global internal symmetries of the sigma model. For periodic b.c.s in both directions, which we will denote as PP b.c.s 
(the first P being space, 
the second time), the sigma model with $\Theta=0$ is invariant under the global inversion symmetry 
$\vec{n}(x,t)\to -\vec{n}(x,t)$ for all $x$, $t$, as well as under all global $SO(3)$ rotations of $\vec{n}$. Together, these symmetries 
form the orthogonal group $O(3)$ of symmetries highlighted in the very name we use for the model. For $\Theta\neq0$, 
the topological term is not invariant under 
inversion if $\Theta$ is viewed as a well-defined real number, however, when $\Theta$ is viewed as defined only modulo $2\pi$, one 
observes that the term is invariant modulo $2\pi$ for both $\Theta=0$ and $\Theta=\pi$ (modulo $2\pi$), 
and hence the path integral is invariant under inversion in both these cases.
In this paper, forthwith we consider the sigma model for only these two values of $\Theta$, which happen to be those relevant 
for the spin chains, and then the inversion symmetry (in $\vec{n}$ space) makes sense. [In these cases, in addition 
to translation symmetry (translating $\ell$ or $x$), the sigma model and spin chain are also invariant under 
both spatial reflection (or parity) symmetry, say $x\to -x$, and time-reversal symmetry, 
$t\to -t$.]

To obtain a more general spatial b.c.\ in the spin chain (but one still invariant under translations), 
we can examine the chain of odd length $N$. For the N\'eel order, this would lead to an antiperiodic spatial b.c.\ 
(i.e.\ in $x$), so that 
$\vec{n}(x+L,t)=-\vec{n} (x,t)$ for all $x$, $t$, and we can place a single branch cut, so that $\vec{n}$ reverses sign on crossing it, 
arbitrarily at $x=-a/2$. We already know that the inversion symmetry makes sense in the sigma model when the b.c.\ is periodic 
in both directions. Now we can also consider twists in the b.c.s in the sigma model by the inversion operation in one or both of the 
$x$, $t$ directions; these will be similarly denoted by A, so we have three more cases PA, AP, and AA. We note that such twists of a 
b.c.\ by a symmetry can be viewed as introducing a background gauge potential for the group involved, with nontrivial flux through 
one or both of the basic cycles on the torus $\mathbb{T}^2$ defined by the $(x,t)$ plane with $x$ defined modulo $L$ and $t$ 
defined modulo $\beta$. The background potential can be viewed (by a choice of gauge) as imposing a jump $\vec{n}\to -\vec{n}$ on crossing some branch 
cut on the torus, and the location of the cut is a question only of the gauge choice, so can be changed by a gauge transformation; 
here, a gauge transformation means a change of $\vec{n}(x,t)$ by multiplication by a $\pm 1$-valued 
function of $x$, $t$ that is single valued (i.e.\ periodic) on the torus. [Incidentally, to avoid any possible confusion, 
we emphasize that here the non-topological part of the action is viewed as gauge invariant by virtue of being ``minimally coupled''
to the gauge potential (via covariant derivatives), in other words as not ``seeing'' a branch cut or an A b.c., 
other than through the global effect on the, otherwise continuous, $\vec{n}$ configurations; there is no non-topological 
``defect action'' associated with a branch cut.] Finally, one can consider summing the partition function 
over the distinct choices of the flux for both cycles, and this corresponds to ``gauging'' the symmetry, that is, 
making the gauge field dynamical, as we now explain. 

If, at a given $(x,t)$, $\vec{n}(x,t)$ and $-\vec{n}(x,t)$ are viewed as equivalent (i.e.\ modulo the action 
of the $\mathbb{Z}_2$ group generated by inversion), then the equivalence classes form the quotient space of 
$\mathbb{S}^2$ by $\mathbb{Z}_2$, which is the two-dimensional real projective space 
$\mathbb{R}\mathbb{P}^2\cong \mathbb{S}^2/\mathbb{Z}_2$. (The inverse statement is that the simply-connected covering space
of $\mathbb{R}\mathbb{P}^2$ is $\mathbb{S}^2$.) For this reason, we may consider whether gauging the sigma model
by the $\mathbb{Z}_2$ symmetry is related to a different sigma model in which the ``target'' space
is now $\mathbb{R}\mathbb{P}^2$. Because our definition of a sigma model involves imposing continuity
on the field configurations, this is in fact the case. Continuity of the $\mathbb{R}\mathbb{P}^2$ valued field means, 
in particular, that in a sufficiently small neighborhood of any point, it can be viewed (by choice of gauge) as coming 
from a continuous $\vec{n}$-field configuration. The fact that the target space is $\mathbb{R}\mathbb{P}^2$ rather than
$\mathbb{S}^2$ only makes a difference when the global topology of configurations on the torus $\mathbb{T}^2$
is considered. While the fundamental (i.e.\ first homotopy) group $\pi_1(\mathbb{S}^2)$ of $\mathbb{S}^2$ is the trivial group, 
that of $\mathbb{R}\mathbb{P}^2$ is $\pi_1(\mathbb{R}\mathbb{P}^2)=\mathbb{Z}_2$ (a direct result of the former, and of the 
quotient by $\mathbb{Z}_2$). Consequently, on passing around one of the nontrivial cycles on the torus, the values of the field in 
$\mathbb{R}\mathbb{P}^2$ either form a contractible loop in $\mathbb{R}\mathbb{P}^2$, or a nontrivial one (corresponding to
the $-1$ element of $\mathbb{Z}_2$, written multiplicatively), which means that $\vec{n}$ returns to minus its original value. 
Thus this topology gives rise to the four independent b.c.\ choices (or fluxes) on the torus, when expressed in terms 
of the $\vec{n}$ field. It means that, for continuous $\vec{n}$ fields, summing over the four b.c.s, or equivalently over the four
choices for two gauge fluxes, is sufficient to treat the gauge potential as dynamical and so
quantize the $\mathbb{Z}_2$ gauge theory by summing over gauge potentials, modulo gauge transformations,
provided the theory is gauge invariant for each b.c. or background gauge field. 
[The $\mathbb{R}\mathbb{P}^2$ sigma model would also admit ``twist fields'' or ``$\mathbb{Z}_2$ vortices''; 
a twist field at a point $(x',t')$ means that the configurations $\vec{n}(x,t)$ have a topological defect imposed 
at $(x',t')$, at which $\vec{n}$ is now not continuous but, 
related to the nontrivial $\pi_1(\mathbb{R}\mathbb{P}^2)=\mathbb{Z}_2$, $\vec{n}$ reverses sign as 
$(x,t)$ goes around $(x',t')$ \cite{Mermin}. We will not consider these directly at present, though they are related to the partition 
functions with A b.c.s which appear below, but discuss them further, together with gauging, in Sec.\ \ref{Discussion}.]
In another language, in CFT this is called (up to a different normalization) an ``orbifold'' of the 
model by the discrete symmetry. 

For $\Theta=0$ (mod $2\pi$), all these boundary condition choices for the $\vec{n}$ field make sense for 
the path integral of the $O(3)$ sigma model, and gauge invariance holds, so we do obtain the $\mathbb{R}\mathbb{P}^2$ sigma model. 
For $\Theta=\pi$ (mod $2\pi$), the topological term in its original form is no longer a topological invariant when a branch cut 
is present, and has to be reconsidered; the remaining discussion in this Introduction is entirely for this case. 
In section \ref{subsec:pathintmodtrans} below, we will show that, by adding a term to the action, we can recover, 
not a well-defined integer
multiple of $i\pi$ like $i\pi\cal N$ as in the original case, but only a quantity well defined modulo $2\pi i$, so after 
exponentiation we obtain a factor $\cal I$ in the integrand of the path integral. $\cal I$ depends on the choice of branch cuts, 
takes only the values $\pm 1$ for any combination of P or A b.c.s, and agrees with $e^{-i\pi\cal N}$ for the PP case; we adopt this as our 
definition for these b.c.s in general. $\cal I$ is topologically invariant, by which we mean here that it is unchanged under continuous changes 
of the $\vec{n}$ field (homotopies) that respect the b.c.s, with the branch cuts held fixed. It turns out that while $\cal I$ is, 
of course, well (i.e.\ uniquely) defined as before for PP, for AP, PA, and AA it is only defined up to an overall sign, because
there are topologically-inequivalent arrangements of branch cuts that give opposite signs. For the latter three combinations
of b.c.s, it also fails to be fully gauge invariant; instead it is invariant only under a restricted class of gauge transformations
(we can continue the discussion by using a fixed choice of gauge, or of branch cuts, for each b.c.). 
[With one or both b.c.s A, there are still distinct homotopy classes of $\vec{n}$ configurations as for PP. For a fixed choice of 
branch cut(s), the relative signs of distinct configurations in distinct homotopy classes are well defined in all cases.] 
Then while the path integral is unambiguous for PP, and the contribution of the $\vec{n}=$ constant configuration, 
which has lowest non-topological action, is positive, for the other three b.c.s it is ambiguous by a sign---there are 
two partition functions for each case (at least if they are 
nonzero). For the case AP (or PA), there is one natural ``obvious'' or ``standard'' choice of branch cut, a single straight line 
running in the time (resp., space) direction, and for this our invariant gives $+1$ for the configuration $\vec{n}(x,t)$ 
with the lowest value of the non-topological term in the action of the $O(3)$ sigma model. For the AA case, the obvious choice 
is to have both these cuts present at once, and our construction requires in effect that the crossing be resolved so that they 
do not cross; there are two equally valid ways to do so, which give opposite signs for any given configuration $\vec{n}$.
(For AP and PA b.c.s, the opposite sign occurs for less obvious choices of branch cuts.)

As an aside from the main discussion, we point out here that it does not require that the action in the sigma model 
be invariant under the full $O(3)$ symmetry, but only requires that the non-topological part be invariant under 
the $\mathbb{Z}_2$ inversion symmetry. More generally, we could modify the non-topological part to keep it invariant 
under a subgroup of $O(3)$ that contains inversion, but not
under the full $O(3)$. For example, the derivative-squared terms in the non-topological term involves a metric on the 
$2$-sphere $\mathbb{S}^2$ defined by 
$\vec{n}^2=1$ (as well as one on spacetime), for which we used the standard $O(3)$-invariant one. Keeping 
this condition on $\vec{n}=(n_\mu)_{\mu=1,2,3}$, we could use a different metric on $\mathbb{S}^2$, 
but restrict ourselves to those invariant 
under inversion (see for example Ref.\ \cite{FATEEV1993521}). (The topological term does not involve the metric, and does not 
need to be modified.) We could also add to the action the integral of a non-derivative term, such as $\int dxdt\;n_3^2$, 
that is invariant under inversion (as in Ref.\ \cite{PhysRevLett.50.1153}), and so on. (Either type of modification is sometimes called 
``anisotropy''.) Everything we said up to this point about the partition functions 
for different boundary conditions still holds for these cases. We will continue to emphasize the $O(3)$ invariant model, as we find it 
the most interesting, and it can also guide the interpretation in cases in which the symmetry is broken down to a subgroup containing 
inversion. We note the evident fact that the Hamiltonian of the spin chains could likewise also be deformed so that the internal 
symmetry is explicitly broken to a subgroup of $SO(3)$ or $SU(2)$, for example as in the well-known antiferromagnetic $XXZ$ model for 
the $S=1/2$ case (or likewise for any $S$), or with a quadrupolar anisotropy term $\sum_\ell S_{\ell 3}^2$ (for $S>1/2$) 
added to the original Hamiltonian, 
and the basic correspondence with the sigma model with that same symmetry is still expected to hold {\it mutatis mutandis} 
(the $XXZ$ type of deformation might give rise to a change in metric \cite{FATEEV1993521}, while quadrupolar anisotropy 
gives rise to the 
$\int dxdt\;n_3^2$ term \cite{PhysRevLett.50.1153}). Indeed, we could break the internal symmetry in the spin chain completely, leaving
only translation, reflection in space (with a site fixed), and time-reversal symmetries. However, an important point here is that, 
to arrive at the deformed sigma model, one must consider the spin chain Hamiltonian with full $SO(3)$ [or $SU(2)$] symmetry, 
plus the desired symmetry-breaking terms,
and then the coefficients of the latter must go to zero as $\sim a$, so that, for example, a domain wall has a width 
determined by the sigma model parameters, not of order $a\to0$. For the spin chains also, we will usually concentrate on the unbroken case. 

For the inversion symmetry in the spin chain, at first we find ourselves in a quandary. Initially, and classically, it has a 
global $SO(3)$ internal symmetry that acts on the spins. In quantum mechanics, $S$ is quantized to be integer or half integer, 
and for $2S$ odd it is 
well known that $SO(3)$ acts ``projectively'', because rotation by $2\pi$ now acts by $-1$ on each spin. Equivalently, and more 
usefully, the symmetry group that acts linearly (unitarily) is the simply-connected double cover of $SO(3)$, known as $Spin(3)$, 
which is isomorphic to the better-known group $SU(2)$. What about the inversion in $O(3)$? It cannot be 
$\vec{S}_\ell\to - \vec{S}_\ell$, 
because that does not preserve the commutation relations of the $\vec{S}_\ell$. Notice that the relation of $Spin(3)$ and $SO(3)$ 
is that there is a ``projection'' (a group homomorphism) of the former to the latter, which has non-trivial kernel, the subgroup 
$\mathbb{Z}_2=\{1,-1\}\subseteq Spin(3)$ [it is the center of $Spin(3)$], that maps to $1$ (the identity) in the latter; then each 
element of $SO(3)$ has two ``lifts'' to elements of (i.e.\ inverse images under the homomorphism in) $Spin(3)$. Analogously, one can 
look for a double cover (with two connected components) of $O(3)$ such that the lifts of elements of $SO(3)$ form $Spin(3)$ [recall 
that $SO(3)$ is the subset of proper rotations, with determinant $+1$]. For the general case of $O(n)$, there exist, up to isomorphism, 
only two such groups for each $n$ (we discuss them in detail in Appendix \ref{PinG}), known in general as 
$Pin_\pm(n)$ groups, so we must consider $Pin_\pm(3)$. We might think that we should view the quantum spins in the chain as irreducible 
unitary representations of one or other of these larger groups, as well as of the subgroup $SU(2)$. Then, because for both $Pin_+(3)$ 
and $Pin_-(3)$ the two lifts of the inversion commute with all elements (see Appendix \ref{PinG}), they act by scalar multiplication 
on each spin (with the scalar squaring to either $+1$ or $-1$). It follows that the action on the spin chain is also just 
multiplication by a scalar, which cannot give rise to inversion of $\vec{n}$ and does not seem likely to be useful. 

Consequently, the non-trivial inversion symmetry, if present and linear, can only be a symmetry that emerges in the long-wavelength 
or low-energy limit, either semi-classically in the sigma model or in the IR; it does not exist, at least not in
the form of an inversion symmetry, in the spin chain. In that case it will be necessary to work 
out what symmetry group emerges, how it acts on the states (presumably, there will be a single 
symmetry group under which all states transform), and how it relates to the spin chain. This is most interesting when $N$ and $2S$ 
are odd, in which case the total spin of any state is half integer and a double cover 
of $O(3)$ is required. Note that then, for both cases $Pin_\pm(3)$, the two lifts of inversion differ simply 
by a minus sign, because inversion squares to the identity in $O(3)$, so either both lifts of the inversion square to $+1$, 
or they both square to $-1$. Then for half-integer spin, there are altogether four possible eigenvalues for the lifts of inversion,
which are $1$, $-1$ [for $Pin_-(3)$ only], and $i$, $-i$, [for $Pin_+(3)$ only]. (Some texts on quantum mechanics discuss this in the 
context of space inversion or parity acting on half-integer spins, but not all do so correctly; one that is better is Ref.\ 
\cite{Schiff}.) If the global internal symmetry group of the spin chain is not the full $SU(2)$, but a subgroup, 
then a similar discussion 
applies. In particular (up to isomorphism) there are only two distinct double covers of the $\mathbb{Z}_2$ group generated by 
inversion, which are the two groups of order $4$, namely $\mathbb{Z}_2\times\mathbb{Z}_2$ and $\mathbb{Z}_4$, and these give 
rise to the respective possible eigenvalues.

Armed with this information about the possible groups, we will be able to draw a conclusion for the sigma model.
If we view the path integral as an expression for the partition function, as in eq.\ (\ref{partfn}), 
then the choice of P or A spatial b.c.\ gives rise to two sectors in the Hilbert space of the model,
to which the trace must be restricted. For either case, for the temporal P b.c.\ the partition function 
could be taken to be that written there; that is the natural choice of sign 
(for spatial A, in the path integral it corresponds to the earlier standard choice of branch cut,
that runs parallel to the $t$ axis).
For an A temporal boundary condition, an additional operator must be inserted in the trace,
which is the unitary operator that implements the global symmetry operation acting in that sector. 
For spatial P, the use of the standard branch cut corresponds semiclassically to an operator 
that has eigenvalue $+1$ on the ground state. But for AA, we already know that the path integral
is ambiguous by a sign, and either choice appears equally valid. This suggests that there are two 
equally good choices for the operator that implements inversion, and they differ by a sign. 
However, for the spacetime geometry we considered so far, which is a ``rectangular'' torus,
we will find that the AA partition function $Z_{AA}$ is zero, due to another minus sign, 
a relative sign between configurations that for this geometry have the same non-topological action by symmetry,
giving complete cancellation. In order to proceed with the argument, we must first generalize the spacetime geometry, as follows. 

If we represent the $(x,t)$ plane as the complex plane with 
points $z=x+ivt$ and with the standard Euclidean metric, then the domain (topologically, a torus) on which we placed the b.c.s was, 
up to now, taken to be a rectangle with corners $0$, $L$, $L+iv\beta$, and $iv\beta$. More generally, we can consider a parallelogram 
with corners at $0$, $L$, $(\uptau+1)L$, $\uptau L$, where $\uptau$ is a complex number with ${\rm Im}\,\uptau>0$.
$\uptau$ can be thought of as the (complex) aspect ratio of the domain or of the torus that is obtained by identifying the 
boundaries of the domain in an obvious way (i.e.\ under $z\to z+L$ and $z\to z+\uptau L$; these two translations generate 
a discrete group of translations (a lattice) isomorphic to $\mathbb{Z}^2$, and the spacetime torus is $\mathbb{C}/\mathbb{Z}^2$ 
for that choice of generating translations).
For the partition functions for general $\uptau$, the definition in terms of operators becomes
\begin{equation}
Z(\uptau)={\rm Tr}\, e^{-\uptau_2 L H/v +i\uptau_1 L P},
\end{equation}
where $\uptau_1={\rm Re}\,\uptau$, $\uptau_2={\rm Im}\,\uptau$. This definition holds for either P or A spatial b.c.\
(with the restriction of the trace to the corresponding sector of Hilbert space, as before), 
and for the temporal P b.c.; for either spatial b.c., the partition function is a generating function (of $\uptau$) from which
the multiplicities for each value of the pair $(H,P)$ can be recovered. For the temporal A b.c., again an operator that implements 
(possibly, a lift of) the inversion must be inserted inside the trace. 

For general $\uptau$ values, the partition functions [including $Z_{AA}(\uptau)$] do not vanish. 
Now the ambiguity in sign for $Z_{AA}(\uptau)$ already suggests that the states in the spatial A sector 
all have half-integer spin, and that the AA partition function is ambiguous in sign because there are two lifts of inversion 
for this sector, which differ by $-1$, which is in the center of the double cover (whichever one it is) 
of $O(3)$ (or of a subgroup of $O(3)$ that contains inversion). 
Of course, as we expect the spatial A b.c.\ and $\Theta=\pi$ to arise for spin chains with
$N$ and $2S$ odd, the total spin values must all be half integer in this sector, and so the ambiguity 
in the sign of the AA partition function must occur. We will derive semiclassically the fact that the 
total spin is half integer directly in the sigma model as well. [For the PA partition function,
there is also an ambiguity in sign. Existence of two inversion operators, acting on the quantum states
in the spatial P sector, that differ by overall sign makes sense, as either choice of eigenvalue (parity) is valid,
however this does not correspond to any interesting double cover of $O(3)$, and we may as well
pick one of the two. One of the two does appear more natural, namely that corresponding to the standard branch cut
and lowest value of the non-topological term, as already chosen above.]

We can obtain more detailed information on the eigenvalues of the lifts of inversion in the spatial A sector
of the sigma model from our partition functions with a little more effort. For this, we recall that another aspect of trying 
to define things for all four choices of boundary condition concerns their relations under modular transformations. 
These are commonly considered in CFT, but are equally meaningful in two-dimensional QFT generally. These are 
not a true symmetry of the system, but rather merely a change in parametrization of the spacetime geometry, and as such the full 
partition function of the theory must be invariant under all 
modular transformations (and similarly for correlation functions). To explain this, again consider the spacetime geometry parametrized 
by $\uptau$. An equivalent (i.e.\ isometric) description of the same domain or torus is obtained if $\uptau$ is 
replaced by its image under either of the two transformations $S:\uptau\to-1/\uptau$ and $T:\uptau\to\uptau+1$. 
Indeed, these two transformations generate a group, the modular group $\overline{\Gamma}\cong PSL(2,\mathbb{Z})
=SL(2,\mathbb{Z})/\{\pm 1\}$, under elements of which the structure of the torus is invariant. We saw already that the torus can be 
viewed as $\mathbb{C}/\mathbb{Z}^2$, the plane modulo a lattice which can be viewed as $\mathbb{Z}^2\subset\mathbb{C}$ by choosing 
generators $L$, $\uptau L$; a modular transformation
simply corresponds to choosing different generators, but leaving the lattice unchanged, so that the torus remains the same 
as a metric space, that is, up to an orientation-preserving isometry. The non-topological actions we consider are 
assumed to be invariant under such a transformation (``reparametrization''), as in the examples given earlier.

Boundary conditions on $\vec{n}$ on the torus with parameter $\uptau$ were defined as the behavior, P or A,  of $\vec{n}$
under each of $z\to z+L$, $z\to z+\uptau L$. It is easy to see that the effect of modular transformations on the b.c.s is that 
$S$ leaves $PP$ and $AA$ invariant but exchanges $PA$ and $AP$, while $T$ leaves $PP$ and $PA$ invariant, but exchanges $AP$ and $AA$.
Thus the modular group acts on the set of b.c.s, and there are two 
orbits under that action: one consists solely of $PP$, while $PA$, $AP$, and $AA$ form the other. Notice also that, on the set of b.c.s, 
$S^2$ and $T^2$ both act as the identity permutation.

To calculate the modular transformations of the partition functions precisely, we will first observe that under a continuous change
in $\uptau$ and corresponding change in $\vec{n}(x,t)$, the four partition functions are continuous functions
of $\uptau$. In section \ref{subsec:pathintmodtrans}, we carry out a study of how, invoking continuity, modular transformations 
change the $\vec{n}$ field 
configuration, the branch cuts, and our invariant $\cal I$, and hence how they change the partition functions. 
We find, starting from the standard choices (as above) for the definitions of the PP, AP, PA partition 
functions at one value of $\uptau$ and for given branch cuts there 
\begin{eqnarray}
Z_{PP}(-1/\uptau)&=&Z_{PP}(\uptau),\nonumber\\
Z_{AP}(-1/\uptau)&=&Z_{PA}(\uptau),\nonumber\\
Z_{PA}(-1/\uptau)&=&Z_{AP}(\uptau),\nonumber\\
Z_{AA}(-1/\uptau)&=&-Z_{AA}(\tau),
\label{StransZ}
\end{eqnarray}
while similarly
\begin{eqnarray}
Z_{PP}(\uptau+1)&=&Z_{PP}(\uptau),\nonumber\\
Z_{PA}(\uptau+1)&=&Z_{PA}(\uptau),\nonumber\\
Z_{AP}(\uptau+1)&=&\pm Z_{AA}(\uptau),\nonumber\\
Z_{AA}(\uptau+1)&=&\mp Z_{AP}(\uptau),
\label{TtransZ}
\end{eqnarray}
where we indicated that both signs in the last two lines must be reversed if we make the opposite choice of sign of $Z_{AA}$.
[As $Z_{AA}(\uptau)=0$ when $\uptau$ is pure imaginary, vanishing of some other partition functions at particular $\uptau$ values 
follows from these transformations.]
Thus in words, we find that for the sigma model, $S$ maps the PP, PA, and AP partition functions 
to one another linearly exactly as for the b.c.s, that is with coefficients $+1$, but maps either of the two AA 
partition function to the other, or in effect reverses the sign. 
$T$ maps PP and PA to themselves with coefficient $1$, but maps the AP partition function to one of the AA partition functions,
while $T^{-1}$ maps it to the other, and inversely the action of $T$ and $T^{-1}$ on the two AA partition functions follows from these.
It follows that $S^2$ acts as the identity (by $+1$ in all cases), while $T^2$ acts as $-1$ on each of AP and AA, 
and as $+1$ on the other two. [Note that these statements for $S^2$ and $T^2$ do not depend on the (fixed) 
choice of branch cuts, and are gauge invariant.] This behavior prevents us from obtaining 
modular invariance by forming a sum over the orbits if the b.c.s other than PP are to appear: the coefficients
of the AP and AA partition functions must be zero because $T^2=-1$ on these (and also that of AA must be zero because $S$ acts 
as $-1$ on it), while the action of $S$ also implies that the coefficients of AP and PA must be equal (for our choices
of branch cuts). Note that the opposite sign
of the AP or PA partition function can be produced by beginning with one of the standard partition functions and using some 
composite elements of the modular group, the simplest being $T^2$ or $ST^2S$, respectively (or the same with $T^2$ replaced by 
$T^{-2}$).

From the operator view of $T$ as involving translation by $L$ inside the trace, it follows from these results
that the momenta of the $P$ eigenstates have the form $2\pi n/L$ ($n$ an integer) in the spatial P sector (because the PP 
and PA partition functions are invariant under $T$), but $2\pi(n\pm 1/4)/L$ in the spatial A sector, because
$T^2=-1$. As the temporal A b.c.\ means the lift of an inversion is inserted in the trace, that means that such a lift must have 
eigenvalues $i$ or $-i$ on all states in the A sector. These values are not possible for inversion [in $O(3)$].
They occur in the double cover of $O(3)$ that we denote $Pin_+(3)$ (again, see Appendix \ref{PinG}). 
This group has center $\mathbb{Z}_4$, and the lifts 
of the inversion in $O(3)$ are the two elements of order $4$ in the center, which square to $-1$, which is in the center of $SU(2)$;
conversely, these properties determine that the group is (isomorphic to) $Pin_+(3)$.  
It is remarkable that this was obtained 
without even first showing that the spatial A states have half-integer spin (but that will emerge 
in a moment). A further observation is that the fact that $Z_{AA}(\uptau)=0$ for $\uptau$ imaginary implies that,
for each state with eigenvalue $i$ under inversion, there is another with eigenvalue $-i$ and the same energy eigenvalue.
Again, this whole discussion still holds [though with the groups $O(3)$ and $Pin_+(3)$ replaced by subgroups]
even if the non-topological action of the sigma model is invariant only under a smaller 
group of internal symmetries, as long as it contains inversion, and also is invariant under spatial reflections.

Returning to the symmetries of the spin chain, a candidate for the lifts of inversion in the long-wavelength regime of 
the spin chain presents itself, suggested by our discussion of the N\'eel order parameter: translation of the chain by one site sends 
$\vec{n}$ to $-\vec{n}$ when $N$ is even. (This view of the origin of inversion also appears in Ref.\ \cite{Sulejmanpasic},
but otherwise they consider a different problem.) In this case, for $N$ even, 
one can think of $\vec{n}$ as the direction of $\vec{S}_\ell$
for $\ell$ even, and of $-\vec{S}_\ell$ for $\ell$ odd. This defines sublattices, say $\rm A$ and $\rm B$, that are well 
defined when $N$ is even. For $N$ odd, if we assign sites to sublattice $\rm A$ starting from say $\ell=0$ and moving to the right, 
on traversing the system we find that $\rm B$ sites are reassigned as $\rm A$, and {\it vice versa}. This can be fixed if we introduce a branch 
cut while making the definition. This then becomes the same branch cut used for the antiperiodic b.c.\ on $\vec{n}$. We will see 
in Sec.\ \ref{sec:spinch} how to deal with 
translations in this case. If $\vec{S}_\ell$ is slowly varying (using some semiclassical description), and has N\'eel order 
$\vec{n}(x)$ locally at $x=\ell a$, then for large $N$ translation by one site is almost the same as sending $\vec{n}\to-\vec{n}$. 
For this reason we will pay attention to the (crystal) momenta of low-lying states of the spin chain, including the ground state.  
For particular cases ($N$ even or $S=1/2$) these values are known from rigorous or exact arguments, and for other cases there is a 
consensus based on numerical results. The values depend on whether $2S$ is even or odd. 
We show in Section \ref{sec:spinch} that semiclassical calculations are in agreement with all these results.

If a continuum description as a QFT, such as the sigma model, exists, then the lattice constant $a$ must tend to zero
and drop out everywhere (after renormalization). The translations by fixed fractions of $L$ can, in the limit, 
form a group of continuous translations, while the translation by one or a few sites can remain a symmetry and become 
a separate discrete group of operators that commute with translations. We will show how this works in Sec. \ref{sec:spchsig}. 

For the spin chain, translation by $N$ sites leaves the chain invariant, by construction. For the sigma model with
spatial b.c.\ A, translation by $L$ reverses the temporal b.c.. The temporal A b.c.\ 
should result (in operator language) by the insertion of an operator implementing (one of the lifts of) the inversion 
inside the trace. Thus it can be undone by a translation. If we define inversion to be translation by $+1$ site, and define
translations in the sigma model as translations by {\em multiples of four} sites in the spin chain, then for $N$ odd
the additional translation by $\pm 1$ is required in order to obtain the identity. (This is not exactly the way in which we will
define inversion in the end, but is perhaps simpler to explain here.) Then everything is consistent
between the spin chain and the sigma model for odd $N$. The implication is again that the double cover of $O(3)$ is isomorphic 
to the group $Pin_+(3)$, with center $\mathbb{Z}_4$, and the latter occurs also when less internal symmetry is present. 
[These results are again all for $\Theta=\pi$ (mod $2\pi$). For $\Theta=0$ (mod $2\pi$), again translation by $L$ is
$e^{-iPL}=1$ for the spatial P b.c., but $e^{-iPL}$ is the inversion operator with eigenvalues $\pm 1$ for the spatial A b.c.,
and there is a similar relation with the spin $S$ spin chain with $2S$ even.]

All these results will be obtained by semiclassical methods, in section \ref{sec:O3} for the sigma model 
with $\Theta=\pi$ (mod $2\pi$), and in section \ref{sec:spinch} for the spin chain with $2S$ odd; see Appendix \ref{PetWeyl} also. 
In the IR limit,
we can consider the partition functions using results for the $SU(2)$ level $1$ CFT. There are right- and left-moving (chiral)
isomorphic copies of $SU(2)$ level 1 current (or Kac-Moody) algebras, and there are just two distinct unitary representations 
of either of these chiral algebras, distinguished by the $SU(2)$ spin $j=0$ or $1/2$ of the corresponding primary fields. 
The characters of these are defined for the right-movers as ${\rm Tr}\, q^{L_0-c/24}$ [$c$ is the central charge; 
$c=1$ for $SU(2)$ level $1$], taken in one of the two representations, where $L_0$ is one of the generators of the Virasoro algebra, 
and $q=e^{2\pi i\uptau}$. For the left movers, $\overline{L}_0$ replaces $L_0$, and $\overline{q}$
replaces $q$ (the overline on $\overline{q}=e^{-2\pi i\overline{\uptau}}$ or other complex numbers or functions denotes 
complex conjugation, whereas that on $\overline{L}_0$ does not, but is conventional). [$c$ is the same for right and left movers.] 
The Hamiltonian and momentum operators in any CFT are $H/v=2\pi(L_0+\overline{L}_0 -c/12)/L$ and 
$P=2\pi(L_0-\overline{L}_0)/L$, respectively. 

The characters are given by 
 \begin{eqnarray}
 \chi_0(\uptau)&=&\frac{1}{\eta(\uptau)}\sum_{m\in\mathbb{Z}}q^{m^2},\nonumber\\
 \chi_{1/2}(\uptau)&=&\frac{1}{\eta(\uptau)}\sum_{m\in\mathbb{Z}}q^{(m+1/2)^2},
 \end{eqnarray}
where $\eta(\uptau)=q^{1/24}\prod_{n=1}^\infty(1-q^n)$ is the Dedekind function. We note that the conformal weights 
$h_0=0$, $h_{1/2}=1/4$ of the $j=0$, $1/2$ primary fields appear here as the lowest $L_0$ eigenvalues in each character.
Then in principle the full Hilbert space of the left times right theory may have four sectors, labeled $(j_R,j_L)$, 
where each $j$ is either $0$ or $1/2$. Then for the spatial P and A sectors of the sigma model or spin chain,
we restrict the Hilbert space to either integer or half-integer total spins, respectively, and for the P temporal b.c.\ 
nothing more is inserted in the trace. Hence the P sector consists of the primary fields with $(h_R,h_L)$ conformal weights
$(0,0)$ or $(1/4,1/4)$, and their descendants, while the spatial A sector includes the ``twist fields'' of the sigma model,
as mentioned earlier, and the corresponding primary fields have conformal weights $(0,1/4)$ or $(1/4,0)$. Then the 
PP and AP partition functions in the IR can only be
\begin{eqnarray}
Z_{PP}(\uptau)&=&|\chi_0(\uptau)|^2+|\chi_{1/2}(\uptau)|^2,\nonumber\\
Z_{AP}(\uptau)&=&\chi_0(\uptau)\overline{\chi}_{1/2}(\uptau)+\chi_{1/2}(\uptau)\overline{\chi}_0(\uptau)\label{ZPPZAP}
\end{eqnarray}
(up to a common factor $e^{-e_0{\rm Im}\, \uptau L^2/v}$, where $e_0$ is a constant free energy per unit length in $x$; 
this factor is always dropped).
For the PA partition function, we need to modify the PP partition function by inserting the unitary operator that represents inversion
inside the trace. The $SU(2)$ currents are invariant under inversion (to preserve the commutation relations), so the only possible 
effect is a factor $\pm1$ in front of either term in $Z_{PP}$. As the $\vec{n}$ field should correspond to some components of the 
$(1/2,1/2)$ primary field, the two terms should acquire opposite signs. Then the natural choice is 
\begin{equation}
Z_{PA}(\uptau)=|\chi_0(\uptau)|^2-|\chi_{1/2}(\uptau)|^2\label{ZPA},
\end{equation}
at least up to a sign. Note this is not simply assigning a factor $(-1)^{j_{\rm tot}}$ depending on the total (i.e.\ diagonal,
right plus left) $SU(2)$ spin $j_{\rm tot}$ of the states: there are some states of spin zero that are odd, and some states 
of spin one that are even, and so on. (These two particular cases are sometimes termed pseudoscalar and pseudovector, and such 
possibilities are well known). For the same reason, the sign of the whole expression might be reversed. Finally, for
the AA partition function, including $\pm i$ for the lift of inversion, we should have 
\begin{equation}
Z_{AA}(\uptau)=\pm[i\chi_0(\uptau)\overline{\chi}_{1/2}(\uptau)-i\chi_{1/2}(\uptau)\overline{\chi}_0(\uptau)]\label{ZAA},
\end{equation}
where the overall sign is undetermined. (We can see that $Z_{AA}(\uptau)=0$ when $\uptau$ is imaginary.)
The relative minus between the two terms must be present, similar to the PA case, 
because fusion of the $(1/2,1/2)$ primary with $(0,1/2)$ gives $(1/2,0)$ and {\it vice versa}. Note that all four partition 
functions are real, as expected from the sigma model or the spin chain.

In the basis $\chi_0,\chi_{1/2}$, the modular transformations of the characters are obtained by analytic continuation
in $\uptau$, and can be represented as the matrices 
\begin{equation}
S\mapsto\left(\begin{array}{cc}
\frac{1}{\sqrt{2}}& \frac{1}{\sqrt{2}}\\
\frac{1}{\sqrt{2}}& -\frac{1}{\sqrt{2}}
\end{array}\right),~T\mapsto e^{-i\pi/12}\left(\begin{array}{cc}
1&0\\
0& i
\end{array}\right)
\end{equation}
where we used the conformal weights $h_0=0$, $h_{1/2}=1/4$ of the primary fields. Using these, the four partition functions transform
under modular transformations in exactly the same way that we saw above. In particular, we have 
\beq
SZ_{PA}(\uptau)=\chi_0\overline{\chi}_{1/2}+\chi_{1/2}\overline{\chi}_0=Z_{AP}
\eeq
and therefore
\begin{eqnarray}
TSZ_{PA}&=&-i\left(\chi_0\overline{\chi}_{1/2}-\chi_{1/2}\overline{\chi}_0\right)\nonumber\\
T^2SZ_{PA}&=&-\left(\chi_0\overline{\chi}_{1/2}+\chi_{1/2}\overline{\chi}_0\right)\nonumber\\
&=&-SZ_{PA}
%T^3SZ_{PA}&=&i\left(\chi_0\overline{\chi}_{1/2}-\chi_{1/2}\overline{\chi}_0\right)\nonumber\\
%T^4SZ_{PA}&=&SZ_{PA}
\end{eqnarray}
These are identical to the transformations obtained from the sigma model partition functions above.
Note in particular that the fractional conformal spins (i.e.\ the fractional part of $L_0-\overline{L}_0$) are $\pm 1/4$
in the $(1/2,0)$ and $(0,1/2)$ sectors, which agree with the semiclassical results, and are responsible for the fact that $T^2=-1$
in the spatial A sectors. Remarkably, these values already emerge in the semiclassical regime of the sigma model or of the spin chain.

It is known that any modular-invariant sequilinear combination of the $SU(2)$ level $1$ characters must be a multiple of $Z_{PP}(\tau)$
\cite{CIZ}, which is invariant because $S$ and $T$ act as unitary matrices on the characters. As we said above, this is clear 
in terms of our four partition functions. 

We note that Gepner and Witten \cite{gw86} gave a closely related discussion of modular non-invariance in the setting of
the $SU(2)$ level $k=1$ WZW model. The WZW model is a field theory of a field $g(x,t)$ (not to be confused with the coupling constant 
in the sigma model) that takes values in $SU(2)$; its definition involves a positive number $k$ which is quantized {\it a priori}
to integer values, analogously to quantization of $SU(2)$ spin \cite{Witten1984}. 
The theory is conformally invariant and can be identified with the $SU(2)$ level $k$ CFT. For $k=1$ and PP b.c.s, its partition 
function is hence the same as $Z_{PP}$ here. Gepner and Witten considered orbifolding by the $g\to -g$ global internal symmetry of the 
model, which leaves the currents invariant. 
For the corresponding temporal b.c., because $g$ corresponds to the $(1/2,1/2)$ primary 
field, the $Z_{PA}$ partition 
function is the same as here, and they used the action of modular transformations on the b.c.s and on the characters, as above, to 
obtain the others, concluding that the orbifold partition function is not modular invariant. While their point of view may be 
slightly different, in terms of $SU(2)$ level $1$ our conclusion is (of course) essentially the same as theirs. [They interpret it in 
terms of the fact that the WZW model with $g$ taking values in $SO(3)\cong SU(2)/\mathbb{Z}_2$ has level $k/2$ (in terms of $k$ for the 
$SU(2)$ model), which is not integer for $k$ odd, so in particular the corresponding orbifold theory of $SU(2)$ level $1$ does 
not exist.] We discuss the relation of these theories in further detail in section \ref{sec:relmod}.

To conclude this lengthy summary of our results, we can say that the $O(3)$ sigma model with $\Theta=\pi$ (mod $2\pi$)
gives rise to a consistent theory when the spatial b.c.\ is A (as well as when it is P), as long as we consider fixed branch cuts. 
However, while usually in a QFT translation by $L$ 
leaves all states invariant, that is not the case here. Moreover, the inversion operation on the $\vec{n}$ field acts 
projectively (i.e.\ with an ambiguity in sign) on the states, with eigenvalues $\pm i$, which are related
to those of translation by $\pm L$. Either of these statements could be called an ``anomaly''. The second implies that
the coupling of the theory to a nontrivial background $\mathbb{Z}_2$ gauge potential (the temporal A b.c.) 
is ambiguous by a sign, which can be called an 't Hooft anomaly \cite{Hooft1980}. That is, there is a breakdown of $\mathbb{Z}_2$
gauge invariance, even at fixed b.c., and even without consideration of modular transformations. 
These results imply that quantizing (summing over b.c.s), or orbifolding by, the gauge symmetry results in a partition function 
that is neither gauge nor modular invariant unless three of its terms have coefficient $0$, and the resulting partition function 
is the same as that of the ungauged theory, up to a factor. (Indeed, with the orbifold normalization \cite{gw86}, the resulting 
partition function is not normalized as a trace in some Hilbert space as it should be, implying that the gauged or orbifold theory 
does not exist.) Gepner and Witten call this lack of modular invariance a global anomaly in reparametrization invariance 
(in the CFT at least). More generally, these (closely related) anomalies mean that the QFT of the $\mathbb{R}\mathbb{P}^2$ 
sigma model with $\Theta=\pi$ 
does not exist, and this is still the case when the $SO(3)$ symmetry is broken to a subgroup. None of this requires passing 
to the IR limit in which a CFT is obtained: it holds at higher momenta or shorter wavelengths in the continuum theory as well.
Indeed, the properties found there arguably survive in the IR, and this ``explains'' the conformal weights of the CFT
as well as the anomalies.

A further aspect of our results is that we consider, as well as translations in space, also the spatial reflection
and time reversal symmetries of the models. For $\Theta=\pi$ (mod $2\pi$) and the spatial A b.c., we find that
the group generated by translations and reflections, which in other cases is isomorphic to $O(2)$ [we call it $O(2)_{\rm sp}$], 
is replaced by a double cover $Pin_-(2)_{\rm sp}$ (see Appendix \ref{PinG}). The two lifts of inversion also belong 
to this group. Then, while we have mentioned the double cover $Pin_+(3)$ of $O(3)$, which has the structure 
$Pin_+(3)\cong [Spin(3)\times \mathbb{Z}_4]/\mathbb{Z}_2$, a more complete view is that the group containing all 
internal and spatial symmetries becomes $[Spin(3)\times Pin_-(2)_{\rm sp}]/\mathbb{Z}_2$.

The form of parts of our results is the same as parts of the discussion by Cheng and Seiberg (CS) \cite{ChengSeiberg}. They give 
a general discussion of 't Hooft anomalies and twisted b.c.s, including the effect of translations on the b.c.s on the torus, 
of which ours are a special case. They relate the behavior of the partition functions under the modular $T$ operation to
the presence of a nontrivial projective representation of $\mathbb{Z}_2$; the projective representation is equivalent to a linear 
representation of the double-cover group $\mathbb{Z}_4$. The mixing of internal symmetries and translations as arising from 
translations on the spin chain also appears there, as do examples with invariance under only a subgroup of $O(3)$, such as the 
antiferromagnetic $XXZ$ spin chain. The relation of translations in the spin chain and the continuum translations and 
the $\mathbb{Z}_2$ symmetry in the continuum limit appears there as well, and then $\mathbb{Z}_2$ symmetry in the continuum 
is termed an ``emanent'' symmetry.
Our discussion differs in that we consider the $O(3)$ sigma model and its inversion symmetry, and give a direct definition 
within that model of the topological invariant $\cal I$ which exhibits the sign ambiguity and gauge non-invariance 
of the antiperiodic partition functions, and also plays a key role in the modular transformation properties. CS studied the anomalies 
in a model CFT in the IR, and made a hypothesis that the anomalies arise in the spin chain of $N$ sites to obtain 
the ground-state momenta of the latter, whereas we use semiclassical methods to calculate the ground-state momenta directly, 
finding agreement with the mixing of symmetries in the continuum. 
We use the language of double covers rather than that of projective representations. Finally, we consider the full modular group, 
and not only the $T$ transformation. We expect that the relations between these works may help illuminate both.

We also emphasize that, not only is there no modular-invariant partition function involving all four b.c.\ choices
when $\Theta=\pi$ (mod $2\pi$), but the partition function is not fully gauge invariant for every possible b.c.; 
instead there are three cases that have a residual dependence on the choice of branch cuts. On the other hand,
they are all gauge invariant for $\Theta=0$ (mod $2\pi$). In this case, we can view the gauged sigma model as the 
$\mathbb{R}\mathbb{P}^2$ sigma model. We give some further discussion of what this means in terms of gauge theories 
and of the massive QFT in the IR in section \ref{sec:relmod}. 

The QFT literature also contains earlier work in which a ``pure'' $\mathbb{Z}_2$ anomaly, preventing gauging 
of inversion symmetry exactly as we discuss, was found in two spacetime dimensions in sigma models and elsewhere 
\cite{10.21468/SciPostPhys.6.1.003}, as well as a ``mixed'' anomaly involving 
both $\mathbb{Z}_2$ and $SO(3)$ symmetries 
\cite{Seiberg1703} (see also Ref. \cite{ChengSeiberg}); 
the proof of the LSM theorem has also been interpreted as involving 
the mixed anomaly \cite{10.21468/SciPostPhys.16.4.098}. Our approach to the pure $\mathbb{Z}_2$ anomaly 
in the $O(3)$ sigma model seems more direct than those. We should mention that the recent Ref.\ 
\cite{Sulejmanpasic} considers anomalies in the $O(3)$ sigma model at $\Theta=\pi$ (mod $2\pi$) that involve inversion (there 
called ``charge conjugation'' $\cal C$), but the anomalies occur on attempting to gauge spacetime reflections (parity 
and time reversal) by summing over topologies for spacetime, including non-orientable ones; these are different from 
gauging the inversion symmetry on the fixed spacetime, a torus. 

In addition to the spin chain with continuous time, we can also consider the case of spacetime that 
is discrete in both directions, based on the six-vertex lattice model (in a limit in which time is continuous, it becomes
the $S=1/2$ spin chain). In this model, we consider the length to be an odd integer in one or both directions, 
corresponding to the antiperiodic b.c.\ cases, and again find an ambiguity in the sign of the partition function. 
As the methods used for this case appear different, we include it in a separate Appendix \ref{sixver}.

Finally, it is natural to consider how our results extend to chains with continuous internal symmetry groups other than $SO(3)$ or $SU(2)$.
In Appendix \ref{OtherSp}, we show that our results extend to a large class
of models with more general symmetry groups, that extends the spin-$S$ $SU(2)$ spin chain and the related $O(3)$ sigma model 
with $\Theta$ a multiple of $\pi$; every sigma model in this class possesses an inversion symmetry of the target space. 
This extension may be useful because it in no way depends on the existence of a CFT in the IR limit
when $\Theta=\pi$ (mod $2\pi$). Actually, in these models there is in general more than one parameter $\Theta$; there may also be more than
one velocity replacing $v$, so Lorentz invariance is broken in the continuum theories.
In these spin chains there is still a version of the LSM theorem, found by Affleck and Lieb \cite{AffleckLieb}, which states that 
if the center of the internal symmetry group acts nontrivially on the spin in the spin chain, then there is a low-lying excited state 
when $N$ is large, implying either that there are gapless excitations or that there is breaking of some discrete symmetry. Nontrivial action
of the center implies that some $\Theta$ parameter must be $\pi$ (mod $2\pi)$.
In many of these cases the IR limit is believed not to be a CFT or another gapless theory, 
but instead is gapped (massive), with spontaneous breaking of translation symmetry in the $N\to\infty$ limit. We determine the 
subset of these generalized models in which there is a pure $\mathbb{Z}_2$ anomaly as discussed in this Introduction and in the 
remainder of the paper. The anomaly occurs precisely when time reversal 
on a single spin in the chain (the nature of which can also be determined from the sigma model and its $\Theta$ parameters) 
squares to $-1$ rather than to $+1$; the former case can occur only when the center acts nontrivially, but such an action 
is not a sufficient condition. The result agrees with the special case of the spin-$S$ chains and $O(3)$ sigma models, 
but the crucial role of time-reversal symmetry, rather than just the center of the symmetry group, only comes fully into view 
when the generalizations are considered.

%%%%%%%%%%%%%%%%%%%%%%%%%%%%%%%%%%%%%%%%%%%%%%%%%%%%%%%%%%%%%%%%%%%%
\section{Background on spin chains and the sigma model}
\label{SpinCh}

In this short section, we review some parts of the relation of $SU(2)$ spin chains with the $O(3)$ sigma model
that will be essential to the remainder of the paper. 

To make a connection between quantum spins $\vec{S}$ and a path integral formulation, we first require
an action that, when $e^{-{\cal S}}$ is integrated over paths in a path integral, gives rise to the partition function of
a single quantum spin (with zero Hamiltonian at present). Such an action is 
\beq
{\cal S}=2iS\int_0^\beta dt\,\vec{\cal A}(\vec{n})\cdot\partial_t\vec{n},
\eeq
where $\vec{n}=\vec{n}(t)$ is again viewed as a unit vector, $\vec{n}\in\mathbb{S}^2$, and $\vec{\cal A}$ is a vector potential 
on the sphere
with $SO(3)$-invariant field strength. Of course, such a vector potential is not defined as a vector-valued function on the
whole sphere if the total flux of the field strength $\overrightarrow{\cal B}=\nabla\times\vec{\cal A}$ is nonzero, but this can 
be dealt with via either 
the old ``Dirac string'' point of view, or the more modern language of $\vec{\cal A}$ as a connection on a line bundle over the sphere. 
The line bundle is the setting for quantum mechanics. Either point of view requires that the total flux 
$\Phi=S\int \overrightarrow{d^2n}\cdot\overrightarrow{\cal B}$ (where $\overrightarrow{d^2n}$ is the $SO(3)$-invariant measure on the 
sphere, viewed here as a vector normal to its surface) emerging through the sphere be quantized, and it will be helpful to see this 
in more detail. 

An explicit formula for the action can be obtained by a form of Stokes's theorem, using part of the expression $\cal N$,
already seen in the introduction. If $\vec{n}(t)$, $t\in [0,\beta]$, describes a closed curve in $\mathbb{S}^2$, 
with an orientation given by the direction of increasing $t$, then we can find a region ${\cal D}\subset \mathbb{S}^2$ 
with boundary $\partial{\cal D}$ the given curve, consistent with its orientation. We view $\cal D$ as the image of a region
of a spacetime, with the topology of a disk, on which we again use coordinates $x$, $t$ [note that the map from $(x,t)$ into 
$\mathbb{S}^2$ is not required to be one to one]. Then we can define
\begin{equation}
\int_{\partial \cal D} dt\,\vec{\cal A}(\vec{n})\cdot\partial_t\vec{n}=\frac{1}{2}\int_{\cal D} dxdt\;\vec{n}\cdot(\partial_x\vec{n}\times 
\partial_t\vec{n}).\label{intermact}
\end{equation}
(Both sides are invariant under reparameterization of the coordinates $t$ or $x$, $t$, respectively.) The right hand side depends on the choice of the
map from the disk into $\mathbb{S}^2$. But two distinct choices with the same boundary values $\vec{n}(t)$ differ 
by an integral over a spacetime sphere [formed by joining two disks at their boundary, and mapped
into $\mathbb{S}^2$] that has the same form as $\cal N$ in eq.\ (\ref{Ndef}), but multiplied by $2\pi$. Hence different choices give
rise to expressions for the action $\cal S$ that differ by a multiple of $2\pi i$, which makes no difference to the path integral,
provided $2S$ is an integer, or $S=0$, $1/2$, $2$, \ldots. (The cases $S$ and $-S$ give isomorphic quantum systems, so we may choose 
$S\geq0$ as usual.) 

Quantization of the system described by the first-order action $\cal S$ involves quantization of a system with second-class constraints,
or alternatively one can introduce a kinetic energy term $\int_0^\beta dt\,(\partial_t\vec{n})^2/(2m)$ into the action, and send the mass $m$ to 
zero at the end. We omit these technical details; the end result is that, due to the constraints, the vector $\vec{S}=S\vec{n}$ 
is an operator that has components that do not commute (as they might have been expected to do), and in fact their commutations 
relations are
those of a quantum spin, with $\vec{S}^2=S(S+1)$. Classically, the first-order action implies that the two local coordinates
near any point on the sphere are canonically conjugate, and $\mathbb{S}^2$ becomes a phase space, with
finite volume (surface area) $4\pi S$ in canonical units. In Bohr-Sommerfeld quantization, the volume of such a phase space must 
be an integer multiple of $h$, Planck's constant, so that the integer is the number of quantum states (dimension of a Hilbert space).
We have set $\hbar$ equal to $1$, or $h=2\pi$, so the number of states would be $2S$, but the more precise quantization,
mentioned before, of course gives $2S+1$, which agrees at large $S$, up to a correction of relative order $1/S$.
A further point of view is
that the original first-order action can be viewed as the holonomy or adiabatic (or Berry) phase for a quantum 
spin in a coherent state that is forced to describe the closed curve $\partial \cal D$ as a function of time.
The path integral is sometimes derived from this point of view, and called a ``coherent-state path integral''
(for spin).

A spin chain can now be represented by the action
\beq
{\cal S}= \int_0^\beta dt\,H + 2i\sum_\ell\int_0^\beta dt\,\vec{\cal A}(\vec{S}_\ell)\cdot\partial_t\vec{S}_\ell,
\eeq
where $H=H((\vec{S}_\ell)_\ell)$ is the spin-chain Hamiltonian, and we use the notation $\vec{S}$ rather than $S\vec{n}$ 
[we can define $\vec{\cal A}(\vec{S})=\vec{\cal A}(\vec{n})$ for each spin].
For the antiferromagnetic $SO(3)$-invariant spin chain, as in eq.\ (\ref{spinchHam}), with an even number $N$ of sites,
the classical model has the N\'eel ground state, as we have discussed. To obtain an effective field theory,
with a view to a semiclassical (or large $S$) treatment, we can consider the so-called staggered magnetization
as the N\'eel order parameter $\vec{n}(x)$, and the remaining degrees of freedom as conjugate momenta $\vec{j}(x)$.
That is, we can express $\vec{S}_\ell=S(-1)^\ell\vec{n}(x)+\vec{j}(x)$, where $x=\ell a$, $\vec{n}^2=1$, and 
$\vec{n}(x)\cdot\vec{j}(x)=0$; $\vec{j}(x)$ is the
angular momentum (or ``spin'') density. With this change of variable, and treating the fields $\vec{n}(x,t)$ and $\vec{j}(x,t)$
as slowly varying in spacetime (we will not concern ourselves with the overcounting of degrees of freedom if $x=\ell a$ 
is viewed as discrete), we then aim to integrate out $\vec{j}(x,t)$ (to leading order in $1/S$) in the path integral, 
to leave the second-order action of the continuum $O(3)$ sigma model, as in eq.\ (\ref{O3action}). Again, we will skip most 
of the details.

The part of the derivation that we do need is that giving rise to the topological term in the sigma model action, 
which is the remnant of the sum of Berry phase terms. For these, it is sufficient to evaluate the sum
with $\vec{S}_\ell=S(-1)^\ell\vec{n}_\ell$ and $\vec{n}(x,t)$ slowly varying in both $x$ and $t$.
For a pair of neighboring $\ell$s, the $\vec{S}$s are nearly opposite, and if they were exactly opposite for all $t$
the Berry phases would cancel, modulo $2\pi$. As they are not exactly equal, there remains the flux through 
a thin closed strip (cylinder) on $\mathbb{S}^2$. If we divide the sites into pairs $(1,2)$, $(3,4)$, and so on
then, for any slowly-varying collection of $\vec{n}_\ell(t)$, the sum of the contributions of these strips
will be $2S$ times {\em half} of $2\pi{\cal N}$ for the corresponding continuous $\vec{n}(x,t)$ (because only 
alternate strips defined by $\vec{n}_\ell(t)$ were retained). That is,
we obtain the topological term in eq.\ (\ref{O3action}), with $\Theta=2\pi S$ as claimed.

%%%%%%%%%%%%%%%%%%%%%%%%%%%%%%%%%%%%%%%%%%%%%%%%%%

\section{\texorpdfstring{$O(3)$}{O(3)} sigma model semiclassically}
\label{sec:O3}

In this section we consider the $O(3)$ sigma model semiclassically, with any of the four boundary conditions on the torus. By semiclassically, 
we mean that we work initially with the action only, but will use the fact that it is exponentiated in the path integral, 
while later we carry out semiclassical quantization of the collective coordinates that describe the lowest-energy 
texture $\vec{n}(x,t)$ at given $t$. Both types of calculation should be valid at weak coupling $g$ in the sigma model, which means 
the semiclassical limit. 

\subsection{Path integral, gauge non-invariance, and modular transformations}
\label{subsec:pathintmodtrans}

\subsubsection{Invariants \texorpdfstring{${\cal N}'$ and $\cal I$}{N' and I}}

The first thing to do, perhaps the most important step for this section and the next, is to generalize the topological
term in eq.\ (\ref{O3action}) to the case when one or both b.c.s on the torus is antiperiodic (A). As before, we need only 
consider $\Theta=\pi$ (mod $2\pi$), and it suffices to consider $\Theta=\pi$, though any odd multiple of $\pi$ would be similar. 
This reveals the anomaly that is central to the paper.

We begin by recalling that, setting $\Theta=\pi$, $\Theta {\cal N}$ is given by the integral
\begin{eqnarray}
\pi{\cal N}[\vec{n}]&=&\frac{1}{4}\int dxdt\;\vec{n}\cdot(\partial_x\vec{n}\times \partial_t\vec{n})
\label{integraldef}\\
&\in&\pi\mathbb{Z}
\end{eqnarray}
which is $\pi$ times an integer [provided that $\vec{n}(x,t)$ is continuous 
(and differentiable) everywhere], and so a topological invariant.
Here $x$, $t$ can represent any arbitrary coordinates on any manifold, not only the torus $\mathbb{T}^2$ (with PP b.c.s on 
$\vec{n}$), but also for example the sphere $\mathbb{S}^2$ (for some cases, the integral must be defined using an atlas of coordinate 
charts, each covering part of the manifold, but we will not need such details). The integrand can be viewed as the pullback 
to the manifold of a $2$-form or magnetic field strength that is constant on $\vec{n}\in \mathbb{S}^2$, and could be described as
the field of a magnetic monopole at the center of the sphere. The fact that, when divided by $\pi$, it is an integer 
played a role in the quantization of spin to values $S$ with $2S$ integer, as discussed in the preceding section. 

To explain how we extend the definition of $e^{-i\pi{\cal N}}$ to allow A b.c.s, it will be useful first
to consider how to modify the definition for the present cases so as to make it invariant under at least some of the gauge
transformations by $\pm1$, which were already discussed in the introduction. For simplicity, consider first 
a gauge transformation by multiplication of $\vec{n}$ by a scalar that is $-1$ in a region 
(the ``transformation region'') on $\mathbb{T}^2$, and $+1$ otherwise, and here assume the transformation region is simply connected
(and hence contractible) in $\mathbb{T}^2$. Thus, after transformation, the field $\vec{n}(x,t)$ is 
continuous except on the boundary of the transformation region. The boundary in $\mathbb{T}^2$ of the transformation region 
consists of a single closed curve, which we can view as a branch cut. 

The contribution to $\pi\cal N$ of the transformation region changes sign after the gauge transformation. Thus we must 
add {\em twice} the original contribution of that region to the value after transformation to recover the original value,
which we know is a topological invariant. We associate this added term with the values of $\vec{n}$ just outside the 
cut. The $\vec{n}$ field before transformation gives us a smooth continuation of the values just outside
to a ``cap'' or disk that fills the hole left by removing the region from the integral. Given only the values
just outside, we can take {\em any} smooth continuation to the disk with boundary values as given. The contribution  
of the disk to $\pi\cal N$ may not be a multiple of $\pi$, but different continuations to the disk give the same
value modulo $\pi$. That is not sufficient for our purposes, but if we now multiply by $2$, different
continuations give the same value modulo $2\pi$, as we used before. 

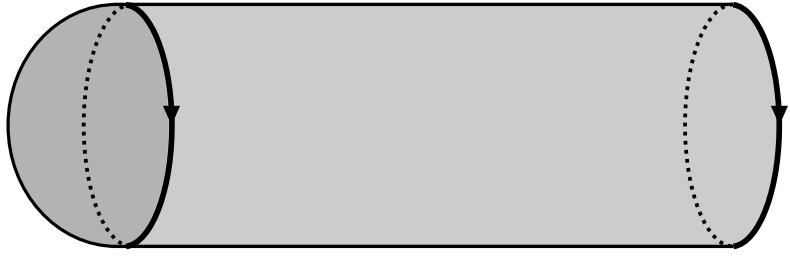
\begin{figure}
\begin{center}
\begin{tikzpicture}[scale=0.8]
    
    \def\a{2}  
    \def\b{0.8} % 
    \def\L{5}   
    \fill[gray!40] (-\L,\a) rectangle (\L,-\a);

    \fill[gray!40] (\L,0) ellipse ({\b} and {\a});

   \draw[black, line width=.75mm] (\L+.04, -2) arc[start angle=-87, end angle=87, x radius={\b}, y radius={\a}];

         \draw[black, line width=.5mm, dotted] (\L+.04, 2) arc[start angle=90, end angle=270, x radius={\b}, y radius={\a}];
  \draw[black, very thick, arrows={-Stealth[inset=0pt, angle=45:8pt]}] (\L+.8,.1) -- (\L+.8,0);

    \draw[very thick] (-\L-.2,\a) -- (\L+.04,\a);
    \draw[very thick] (-\L-.2,-\a) -- (\L+.04,-\a);

    \fill[gray!60] 
        (-\L-.15, 2) arc[start angle=90, end angle=270, x radius=1.8, y radius={\a}] 
        -- (-\L-.1, -2) arc[start angle=-90, end angle=90, x radius={\b}, y radius={\a}] 
        -- cycle;

   \draw[black, very thick] (-\L-.15, 2) arc[start angle=90, end angle=270, x radius=1.8, y radius={\a}];

  \draw[black, line width=.75mm] (-\L, -2) arc[start angle=-87, end angle=87, x radius={\b}, y radius={\a}];
 
          \draw[black, line width=.5mm,dotted] (-\L+.1, 2) arc[start angle=90, end angle=270, x radius={\b}, y radius={\a}];
   
          \draw[very thick] (-\L-.2,\a) -- (\L,\a);
    \draw[very thick] (-\L-.2,-\a) -- (\L,-\a);
   
        \draw[black, very thick, arrows={-Stealth[inset=0pt, angle=45:8pt]}] (-\L+.752,.1) -- (-\L+.752,0);
\end{tikzpicture}
\end{center}
\caption{ A cap (shaded dark) is attached to the spacetime manifold (here, a torus with a branch cut that is a closed curve 
along the time axis $x=0$) at one of the boundary components obtained after cutting along the curve, smoothly extending 
both the local coordinates and the orientation on spacetime over the cap, and also smoothly extending the local $\vec{n}$ field
to the cap. The arbitrary choice of a component at which to attach is shown by the (consistent) orientations of the arrows 
on the two boundary components, such that spacetime is continued from the part to the {\em left} of the cut when 
moving along the arrow. The interior of the rectangle in spacetime could contain addition cuts, leading to additional similar disks 
attached. } 
\label{firstFig}
\end{figure}

A similar prescription works for any branch cut, not only those resulting from a gauge transformation as in the simple example.
For any cut (a closed non-self-intersecting closed curve on $\mathbb{T}^2$, not intersecting any other cut either),
we first choose one side of the cut (we show in a moment that it does not matter which side).
The spacetime manifold can be viewed as actually cut along the branch cut, so that it acquires a boundary with two components
(each a single closed curve), one of them singled out as ``chosen''. Then we can attach a disk to the spacetime manifold so that the
boundary of the disk is identified with the chosen boundary component (and no disk is attached to the other). 
We extend both the local $(x,t)$ coordinates and the orientation on spacetime from the chosen side of the cut over the attached
disk and also, using the values of $\vec{n}(x,t)$ near the cut on the chosen side, smoothly extend $\vec{n}$ to the disk. 
Then the contribution we will add for that cut is again {\em twice} the contribution to $\pi\cal N$ of the added disk 
(using the continuation of the orientation) and its $\vec{n}$ field. For an example, see Fig.\ \ref{firstFig}. 

If instead we choose the other side of the cut, then the values of $\vec{n}$ just off the cut (i.e.\ within the original spacetime)
are minus those on the original side. We can attach a disk to the spacetime on that side instead, and continue both $\vec{n}$ and 
the spacetime orientation over it, and evaluate twice its contribution to $\pi\cal N$. If we subtract this contribution from 
that for attaching the disk to original chosen side of the cut, then the minus sign from subtracting the second contribution 
cancels that from the reversed sign of $\vec{n}$, while the spacetime orientations agree, and the two disks identified at their 
boundaries form a sphere $\mathbb{S}^2$. 
Then the difference can be viewed as $2\pi{\cal N}$ evaluated over that spherical spacetime for a smooth $\vec{n}$ field, and hence it 
is a multiple of $2\pi$. Thus we can use the same prescription for 
either choice of a side of the cut, and get the same result modulo $2\pi$. 

Inspired by Stokes's theorem, we can represent the choice of a side of a cut in the original spacetime by attaching an arrow 
to the cut (in the same direction all along the closed curve; we term this ``proper''), such that the values 
of $\vec{n}$ chosen are always those to the {\em left} on moving along the cut in the direction (or orientation) specified by the 
arrow (this too is shown in Fig.\ \ref{firstFig}). 
We emphasize that while the choice of orientation for each cut makes no difference to the corresponding boundary terms 
(mod $2\pi$), they will be useful in manipulating the expressions later. 

Thus finally our invariant is given by ${\cal I}[\vec{n}]=e^{-i\pi{\cal N}'[\vec{n}]}$,  where
\begin{equation}
\pi{\cal N}'[\vec{n}]=\pi{\cal N}[\vec{n}]+\sum_l \frac{1}{2}\int_{D_l} dxdt\;\vec{n}_l\cdot(\partial_x\vec{n}_l\times 
\partial_t\vec{n}_l),
\end{equation}
in which $\pi\cal N$ is the same integral as before, but now over the complement in spacetime of the union of the cuts,
$l$ labels the finite set of branch cuts (which never meet either themselves or each other),
$D_l$ is a spacetime disk attached to the chosen side of the $l$th cut, with coordinates $(x,t)$ on $D_l$ 
continued from the local coordinates near the $l$th cut on the chosen side (and similarly for the spacetime
orientation), and $\vec{n}_l(x,t)$ is the smooth continuation over $D_l$ of the values of $\vec{n}(x,t)$ on the chosen side of the 
$l$th cut (see figure \ref{firstFig} for an example). Note that $\cal I$ is a functional of $\vec{n}(x,t)$, which includes 
the branch cuts, though the choice of an orientation (i.e.\ of a ``side'') for each cut does not affect its value. 
In $\pi{\cal N}'$, we will often refer to the term $\pi\cal N$ as the ``bulk'' part, and to the terms from the cuts as ``boundary'' 
parts, one for each cut. [Further, for general $\Theta$ or $S=\Theta/(2\pi)$, we would define ${\cal I}[\vec{n}]=e^{-i\Theta{\cal N}'[\vec{n}]}$.
For now, we continue to assume $\Theta=\pi$.]

As the choice of side (i.e.\ orientation) of each cut in $\pi{\cal N}'$ does not matter (mod $2\pi$), we could assign half of each 
added term (for one choice of orientation and a choice of the continuation to the attached disk) in $\pi{\cal N}'$ to the disk attached 
at the side of the cut originally chosen, and half to a similar disk attached to the other. More precisely, that means that, for each branch cut,
cutting the spacetime along the cut, attaching a disk to both boundary components, with $\vec{n}$ and the spacetime orientation
continued smoothly onto both the added disks, and evaluating $\pi{\cal N}$ over the resulting spacetime without boundary
(and without the factor of $2$ for the disks this time), and carrying this out for all branch cuts, we get the same quantity 
$\pi{\cal N}'$, provided that, for each cut, the 
continuation of $\vec{n}(x,t)$ over the disk for the second choice of side for the cut is minus that on the disk for the first choice
of side at each $(x,t)$ {\em throughout the disk}. Note that under this identification of corresponding spacetime points on the disks, 
the spacetime orientations on the disks are opposite, and the contribution to $\pi \cal N$ from each disk is then the same,
due to the sign change of both $\vec{n}$ and the orientation. An example is shown in Figure \ref{secondFig}.
[This construction may look simpler than what we did before, but the use of arrows (orientation) to show a choice of side
is still useful in calculations. One can try to generalize this prescription by relaxing the rule
for the two continuations for each cut, but this would have to be done with care to ensure
that $\pi{\cal N}'$ is unchanged.] As this expresses $\pi{\cal N}'$ as the same $\pi\cal N$ for a continuous $\vec{n}$ field
over a manifold without boundary (except that the manifold is different from the original spacetime), this justifies denoting it 
$\pi{\cal N}'$. We also see that $\pi{\cal N}'$ is automatically a multiple of $\pi$, 
and a topological invariant (i.e.\ unchanged under homotopy of $\vec{n}(x,t)$ with the branch cuts held fixed). 
It is also clearly the case that $\pi{\cal N}'$ is unchanged under continuous deformations of the set of cuts, provided 
that such a deformation is an {\em isotopy}, that is, the cuts must not intersect or touch each other or themselves 
at any stage during the
deformation. [Isotopies of submanifolds appear, for example, in knot theory, as the correct definition of topological 
equivalence of knots.] Let us point out that a cut that is contractible
to a point by an isotopy (i.e.\ is contractible and does not surround any other cut) can thus be deformed to 
a point and then deleted (this was the case in the gauge 
transformation considered earlier). Also, if isotopy 
brings together portions of branch cut(s) that have opposite orientation on that portion, then those portions
can be canceled and the branch cut(s) merged in the fashion familiar from line integrals (which is what we have 
for each cut in the first prescription for the boundary terms). These facts mean that in general we need only be concerned with cuts 
that wrap the spacetime torus. Of course, $\pi{\cal N}'$ is also invariant under homotopy of $\vec{n}$ and isotopy 
of cuts performed simultaneously. 

\subsubsection{Invariant \texorpdfstring{$\cal C$}{C} and gauge non-invariance of \texorpdfstring{$\cal I$}{I}}

Next we check whether $\cal I$ is gauge invariant; in fact it is gauge invariant only under a {\em restricted} 
class of gauge transformations. First, we consider only gauge transformations such that the boundary of the transformation region
consists of a collection of closed curves that do not meet anywhere, and we note that we can assign an arrow (a proper orientation) 
arbitrarily to each such closed curve. After the gauge transformation, the new set of branch cuts is the
union of the original set of cuts and the boundary components of the transformation region. Then, by the previous discussion, 
gauge invariance should already be clear for a gauge transformation such that the boundary of the transformation region does 
not meet existing branch cuts in the $\vec{n}$ field. This result can be combined with the facts that $\cal I$ 
is invariant under isotopies of the cuts, and that portions of oriented cuts can be merged and canceled,
provided their orientations are opposite in the portions that merge. 
Further, one can build up a gauge transformation in stages, using these observations. Then, for example, a gauge transformation
in a connected contractible region (so with a single contractible closed curve as boundary), through which one or more 
of the branch cuts already present pass, leaves the invariant unchanged. [The transformation can be done in stages using regions
that do not intersect the cuts(s), and limits in which portions of cuts merge and cancel can be taken using isotopy.] 
After the transformation, 
the resulting configuration of cuts can be viewed with any crossing of cuts avoided (or ``resolved''), which 
for the gauge transformations considered here can always be done such that the resulting orientations are proper, 
and unchanged outside the transformation region. When resolving a crossing, we view such a crossing 
as four lines emerging from a disk at four points, resolving the crossing means that either the top right two points, and the bottom 
left two points, are connected inside the disk, or else the top left two, and the bottom right two, are connected. 
Given such a resolution, if the cuts that cross consist of two portions after the crossing is resolved, then there are four ways to orient 
the two portions of cut; if note, there are two. 
When a gauge transformation is built up using transformations in contractible 
connected regions that do not meet the cuts already present, and then the resulting cuts moved around using the rules,
proper orientations are always obtained. We will return shortly to demonstrating that invariance does not
hold for all gauge transformations.

Now we apply this discussion to the case of an A b.c.\ on the torus, which we can take initially to be described by $\uptau=i$,
with $x+L$ identified with $x$, and $t+\beta$ ($v\beta=L$) identified with $t$, and begin with the AP 
case with the natural choice of branch cut, running in the $t$ direction, at say $x=0$ (mod $L$). To fix
a gauge for the time being, we view $\vec{n}(x,t)$ as defined by the values in the fundamental domain 
$(x,t)\in[0,L]\times [0,\beta]$. Thus $\vec{n}(L^-,t)=-\vec{n}(0^+,t)$ for all $t$, and is otherwise continuous in the fundamental 
domain, with $\vec{n}(x,\beta)=\vec{n}(x,0)$. Thus, the configuration of $\vec{n}(x,t)$ can be viewed as the image in $\mathbb{S}^2$ 
of a cylinder, parametrized by the fundamental domain with the given boundary conditions. The integral, 
taken over $x$ in the open set $x\neq L/2$, is now not generally an integer multiple of $\pi$, and depends on the boundary values of 
$\vec{n}$ on either side of the branch cut, the two circles that bound the cylinder, which each map to a closed curve or ``loop''
on $\mathbb{S}^2$, parametrized by $t$. Either loop is mapped to the other by inversion, because of the boundary condition. 

\begin{figure}[h]
    \centering
   
    \begin{minipage}{0.48\textwidth}
        \centering
        \begin{tikzpicture}[scale=1.2] 
            \fill[gray,opacity=0.3] (0,0) rectangle (4,2); % 
           \draw[black,very thick] (0,0)--(4,0);
           \draw[black, very thick](4,0)--(4,2);
          \draw[black, very thick] (4,2)--(0,2); 
            \draw[black, very thick,dash pattern=on 2pt off .5pt] (0,0) -- (0,2);
            
            \draw[blue, very thick] (.08,0) -- (.08,1);
            \draw[blue, very thick, arrows={-Stealth[inset=0pt, angle=45:6pt]}] (.08,1) -- (.08,1.1);
            \draw[blue, very thick] (.08,1) -- (.08,2);
            
            \draw[red, very thick] (3.92,0) -- (3.92,1);
           \draw[red,  very thick, arrows={-Stealth[inset=0pt, angle=45:6pt]}] (3.92,1) -- (3.92,1.1);
            \draw[red, very thick] (3.92,1) -- (3.92,2);
            
            \draw[green,,very thick] (2,0) -- (2,1);
            \draw[green, very thick, arrows={-Stealth[inset=0pt, angle=45:6pt]}] (2,1) -- (2,1.1);
            \draw[green, very thick] (2,1) -- (2,2);
        \end{tikzpicture}
    \end{minipage}%
    \hfill
    \begin{minipage}{0.48\textwidth}
        \centering
        \begin{tikzpicture}[scale=1.2] 
            \def\radius{2}

            \def\lat{1.2}

            \pgfmathsetmacro\xrad{sqrt(\radius*\radius - \lat*\lat)}

            \shade[ball color=blue!10!gray, opacity=0.3] (0,0) circle (\radius);

            \draw[ very thick, blue] (-\xrad,\lat) arc[start angle=180, end angle=360, x radius=\xrad, y radius={\xrad/4}];

            \draw[dotted, thick, blue] (\xrad,\lat) arc[start angle=0, end angle=180, x radius=\xrad, y radius={\xrad/4}];

            \draw[very thick, red] (-\xrad,-\lat) arc[start angle=180, end angle=360, x radius=\xrad, y radius={\xrad/4}];

            \draw[dotted, thick, red] (\xrad,-\lat) arc[start angle=0, end angle=180, x radius=\xrad, y radius={\xrad/4}];

            \draw[very thick, green] (-\radius,0) arc[start angle=180, end angle=360, x radius=\radius, y radius={\radius/4}];

            \draw[dotted, thick, green] (\radius,0) arc[start angle=0, end angle=180, x radius=\radius, y radius={\radius/4}];

            \draw[blue, very  thick, arrows={-Stealth[inset=0pt, angle=45:6pt]}] ({0.09*\xrad}, {0.801}) -- ({0.09*\xrad + 0.1}, {0.805});

            \draw[red, very  thick, arrows={-Stealth[inset=0pt, angle=45:6pt]}]  ({0.09*\xrad}, {-1.598}) -- ({0.09*\xrad + 0.1}, {-1.599});

            \draw[blue, thick, ->](0,0) -- ({0.6*\xrad}, {0.86});

            \draw[red, thick, dotted, ->](0,0) -- (-{0.6*\xrad}, -{0.86});

            \draw[green,  very thick, arrows={-Stealth[inset=0pt, angle=45:6pt]}] ({0.*\radius}, -0.5) -- ({0.*\radius + 0.1}, -0.5);
        \end{tikzpicture}
    \end{minipage}
    \caption{The left part shows a rectangle with AP boundary conditions on $\vec{n}(x,t)$. As time increases along the blue, green and red curves on the rectangle, the tip of the vectors $\vec{n}(x,t)$ moves along the corresponding curves on the sphere, indicated at right). In the second formulation of the invariant, $\pi{\cal N}'$ is evaluated
    by calculating $\pi{\cal N}$ for the spacetime extended by adding caps at both boundary components for each branch cut, 
    and then in this example the image in $\mathbb{S}^2$ covers the sphere. In that formulation, the arrows are not involved in the formula, and serve only to guide the eye.} \label{secondFig}
\end{figure}
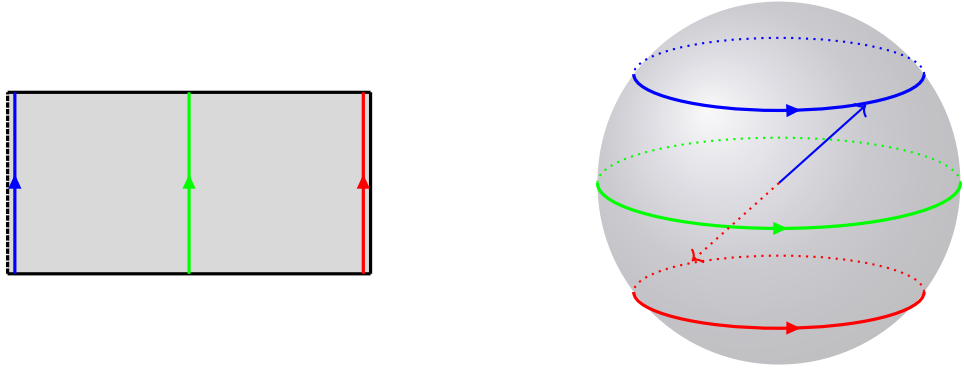

It is straightforward to picture the addition of a disk to each end of the cylinder, and continue $\vec{n}$ over
each disk so that they map to opposite points on $\mathbb{S}^2$, as required in our second form of the
construction. Fig.\ \ref{secondFig} is an example, in which we assumed that the values on one end of the cylinder
go once around a circle of latitude on the sphere as $t$ goes from $0$ to $\beta$, and those at the other end satisfy 
the b.c.\ on $\vec{n}(x,t)$ on the cylinder (or the original torus), that is, the value $(L^-,t)$ is at the antipode
of the value at $(0^+,t)$. (For $x$ between these values, $\vec{n}(x,t)$ interpolates them continuously on the sphere,
at each $t$.) This shows how $\pi{\cal N}'$ is $\pi$ in this case. 

Examining this case, we notice that the additional boundary contribution for the single branch cut (in either formulation), 
when multiplied by $i$, is the action for a single spin $1/2$, as discussed in the previous section. 
This is not a coincidence. When we discuss the odd-$N$ spin chain, say for $S=1/2$, such a contribution will arise directly 
because there is one additional spin compared with the even $N$ case, as well as the A b.c.\ instead of P. We want to emphasize 
that right now we are showing that this term can be understood directly in the continuum sigma model, without reference to the spin 
chain.

We mentioned how, using invariance of $\cal I$ under isotopies of the branch cuts, we need only consider branch cuts that ``wrap'', 
or wind around a cycle of, the torus. [Formally, the winding number for an oriented curve on the torus can be identified with 
an element of the fundamental, or first homotopy, group $\pi_1(\mathbb{T}^2)$. The sign of the winding number will not matter for our 
purposes because we can reverse orientation of a cut, and also we must consider only non-intersecting curves.] Further, 
when a pair of cuts is present such that one can be deformed (without intersections occurring) to coincide with the other, 
by choosing the two orientations we see that 
the coincident parts of those cuts can be canceled, whether they wrap the torus or not. Then we need only consider the case 
of a single non-self-intersecting 
branch cut on the torus that winds around one or both directions, as other cases of sets of branch cuts can all be reduced to this. 
Relative to a description of the torus with a fixed choice of $\uptau$, such a cut can either wind just once in one of the two  
directions and not the other, or it can wind once in one direction and any nonzero number of times in the other. In either case, cutting
spacetime along the cut produces a cylinder, as we had before. Some basic examples
are our standard choices of cuts (for $\uptau=i$ for simplicity), namely a straight line in the $t$ direction for the AP b.c., a 
straight line in the $x$ direction for the PA b.c., and two straight lines that intersect for the AA b.c.. The last one will have 
to be resolved, and then can be viewed as a straight line running diagonally, but there are two choices (up to isotopy) 
for its form. None of these are (restricted) gauge transforms of, or isotopic to, one another.

\begin{figure}[h]
\begin{center}
\begin{tikzpicture}[yscale=-1][scale=1.2]
    
    \fill[gray,opacity=0.3] (0,0) rectangle (4,2);
    \draw[black,thick] (0,0) rectangle (4,2);
    
     \draw[black,line width=.75mm] (0,.8) -- (4,1.2);
     \draw[black, line width=.75mm] (2,0) -- (2,.61);
     
         \draw[black,line width=.75mm] (2.091,.7) arc[start angle=95, end angle=180, radius=.1cm]; 
         \draw[black,line width=.75mm] (2.09,.7) -- (4,.8);
         
         \draw[black,line width=.75mm] (2,1.5) -- (2,2);
         \draw[black, line width=.75mm] (0,1.2) -- (1.92,1.4);
         
  \draw[black,line width=.75mm] (1.9,1.4) arc[start angle=-90, end angle=0, radius=.1cm]; 

 \draw[black, line width=.75mm,-{Latex[scale=.5]}] (2., 1) -- (2.1, 1.01);
 \draw[black, line width=.75mm,-{Latex[scale=.5]}] (2, .5) -- (2, .6);
 \draw[black, line width=.75mm,-{Latex[scale=.5]}] (2, 1.8) -- (2, 1.9);

  \end{tikzpicture}

  \end{center}
  \caption{The case of a branch cut that winds twice in the $x$ direction and once in the $t$ 
direction; the double winding can be removed at the cost of reversing the sign of $\cal I$.}\label{thirdFig}
\end{figure}
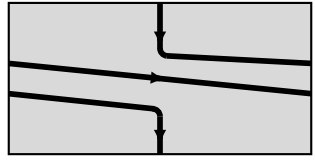

It will be useful now to examine the case of a branch cut that winds twice in the $x$ direction and once in the $t$ 
direction (see Fig. \ref{thirdFig}).
Suppose we orient it with an arrow that goes to the right along horizontal portions and down
on the vertical portions, passing out to the right twice and out at the bottom once (re-entering on the opposite side at the 
appropriate point). If we deform it so that the horizontal portions coincide and form one ``double'' closed curve on the torus 
that wraps once around the space direction, the $\vec{n}$ values on the two sides of this ``double'' branch cut
are equal and it might appear we can cancel (i.e.\ delete) the two coinciding curves. But because the arrows on the two curves
are in the same direction, not opposite, we cannot simply cancel them here. Instead, we examine the boundary terms for
the branch cut in this limit where the horizontal portions coincide and circle the torus, using the first formulation of the boundary 
term, which involves a single disk. To the left (relative to its orientation) of the vertical portion of the cut 
(that is, to its right in the Figure, because the arrow runs downward), the $\vec{n}$ values are the same just above 
and just below the double curve. It follows that, in the limit, there are two points on the boundary of the disk with the same 
$\vec{n}$ value. We can deform the $\vec{n}$ values over the disk so that they remain at that value on a curve that crosses the disk 
and connects those two points. Then we can 
contract the disk to form two disks joined at a single point on the boundary, and the $\vec{n}$ field configuration
over the disk still obeys the requirements. Thus the single boundary term can now be viewed as the sum of two such terms, 
one for the double curve, the other for the vertical curve. For the double curve, it comes with a parametrization
by $x$ running from $0$ to $2L$ and, say, $t=t_0$, and because it is double it has the property that 
$\vec{n}(x,t_0)$ (evaluated just above the single cut {\em before} the limit was taken in which it becomes double)
has the property that $\vec{n}(x+L,t_0)=-\vec{n}(x,t_0)$ for all $x\in [0,2L]$ (taken modulo $2L$). 
It is a fact that the boundary contribution to $\pi{\cal N}'$ of a cut with such $\vec{n}$ values is $\pi$ (modulo $2\pi$). 
[To see this, first consider a loop that is a great circle on $S^2$, which clearly has the desired property. It gives $\pi$ 
(mod $2\pi$), and the result is unchanged under deformations of the loop that preserve the required property.
Another method of calculation, which is direct and uses time-reversal symmetry, is given in Appendix \ref{OtherSp}.] This shows that, 
when we have a cut that winds in this way, the double winding can be canceled at the cost of a change of sign by $(-1)^{\Theta/\pi}$. 
(If a cut does not 
wind in the time direction, double winding in the space direction means the cut has two nonintersecting disjoint components, 
each a curve that winds once, and these can always be canceled,
because we can choose the arrows in opposite directions, and then they cancel without leaving a sign, as we have seen.)
Finally, for more general windings, we can use this rule for double winding to reduce every case to that of a single cut 
winding at most once in each direction, and obtain the correct sign of $\cal I$.

We pause here to emphasize that the short calculation just given for the Berry phase factor associated to a closed loop
$\vec{n}(x,t_0)$ with the property $\vec{n}(x+L,t_0)=-\vec{n}(x,t_0)$ for all $x\in [0,2L]$ is in fact the central point
concerning the anomaly. We call this Berry phase factor $\cal C$. In Appendix \ref{subsec:thm1}, we show that this phase factor
is given in full generality simply by the square of the time reversal operation we call ${\cal T}'$, acting on the
irreducible representation associated with a single spin, as ${\cal C}={\cal T}'^2=\pm 1$, which can also be determined 
using the topological terms in the sigma model. In terms of the parameters in the $O(3)$ model, or for a spin of magnitude $S$ 
for $SU(2)$, it is well known that ${\cal T}'^2=(-1)^{2S}$, which is the same as 
\begin{equation}
{\cal C}=(-1)^{\Theta/\pi},
\end{equation}
as found here.

Returning to the issue of gauge invariance of $\cal I$ in the $O(3)$ model, we first make a side remark. The cut we considered can be 
viewed as consisting of one vertical line and two horizontal lines, with two crossings avoided in a particular way. If we begin
with one cut that is a single vertical line, and then make a gauge transformation in a region that is a cylinder wrapping
the torus, with the two horizontal lines as boundary, then we get a configuration of cuts with two crossings.
However, the invariant is not unchanged under this gauge transformation if the crossings are resolved in
the manner shown in Fig.\ \ref{thirdFig}. The point here is that the gauge transformation as stated 
does not have a contractible transformation region. We can use a similar contractible region, and take a limit
in which the region touches itself to wrap the cylinder. But performing that transformation before taking the
limit, as required for our restricted transformations, in the end gives a configuration of a single cut with avoided
crossings different from (isotopically inequivalent to) Fig.\ \ref{thirdFig}, and in fact the single cut winds in the 
time direction once, with no net winding in the space direction. (This means the orientation is different away 
from the avoided crossings also; there are two horizontal portions, one with arrow in the $+x$ direction, and another in the 
$-x$ direction.) This is consistent with our discussion of restricted gauge transformations under which $\cal I$ is invariant, 
but shows that there are branch cuts and gauge transformations under which $\cal I$ is not invariant.

Now we can use the preceding result to address the two ways of resolving the crossing of two cuts for AA b.c.s. Suppose first that
the non-intersecting cuts connect left to bottom and right to top on the rectangle, which is one of our two
standard forms. We now add an additional double horizontal winding on the vertical part near the top, which reverses the sign
of $\cal I$. 
By choosing it to wind in the correct sense (up and to the right), we can deform it to cancel part of the
original avoided crossing just using the rules for restricted gauge transformation, which do not involve a change of sign.
This produces the other standard form of branch cut for AA b.c.s, and so we have shown that these have opposite sign
for the invariant $\cal I$ for the same $\vec{n}$ field configuration.
This is the origin of the ambiguity of sign in the AA partition function, discussed in the Introduction. 

This result can also be viewed as gauge non-invariance. If we consider only branch cuts without orientation, which
we sometimes (and later) discuss in terms of a $\mathbb{Z}_2$ gauge potential that is nonzero on the cut, then we
might imagine that the two avoided crossings can be related by a gauge transformation in a small contractible region
at the avoided crossing. If we were able to cancel {\em unoriented} portions of cuts (as we would be if full gauge invariance held),
then it would transform one avoided crossing to the other. However, we have seen that $\cal I$ is 
in fact not invariant under such a gauge transformation (because it requires use of an orientation of each cut, or some 
equivalent prescription). The same gauge non-invariance of $\cal I$ occurs for PA and AP b.c.s, but may be less obvious. As they can be mapped to
the case of AA b.c.s simply by a different choice of the cycles on the torus (corresponding to the effect of a modular transformation,
for example), the result follows.

At this stage we can also prove the result about the effect of reflection of spacetime (parity or time-reversal transformations).
For $\uptau$ pure imaginary (and similarly for $\uptau$ a modular transform of such $\uptau$), the spacetime maps to itself 
under reflection through any line parallel to the $x$, or to the $t$ direction, and the non-topological action is invariant 
under such a transformation
(we will assume this remains true for actions that break global internal $O(3)$ symmetry down to a subgroup containing inversion).
For $\Theta=\pi$ (mod $2\pi$), we mentioned in the introduction for PP b.c.s the fact that, while the reflection reverses the 
orientation of the spacetime, and so reverses the sign of $\cal N$ (and more generally of ${\cal N}'$ for the extended spacetime
when branch cuts are present), when the action is exponentiated the contribution to $\cal I$ is invariant. For general b.c.s, this 
holds true for PP, AP, and PA b.c.s, as the standard branch cut (if any) in each of these cases maps back to itself under 
reflection up to a restricted gauge transformation. But for AA, the choice of avoided crossing is reversed by reflection, so restoring 
the original choice changes the sign of $\cal I$. That is, configurations of $\vec{n}$ with AA b.c.s
that are related by reflection, which occur in the same geometry when $\uptau$ is imaginary, have opposite values of $\cal I$
when the latter is calculated with a fixed choice of branch cut. As their non-topological action is the same, the contributions of the 
two to the path integral cancel. This is why $Z_{AA}(\uptau)=0$ for $\uptau$ imaginary, and also for modular transformations 
of such $\uptau$.

\subsubsection{Modular non-invariance}

We can also obtain the results stated in the Introduction for the effect of $S$ transformation on the partition functions. 
We consider the standard choices of branch cuts
(i.e.\ for PP, none). Then implementing $S$ as a continuous change of $\uptau$ to $-1/\uptau$ and of $\vec{n}$, 
we return to the original set of b.c.s or branch cuts, except that AP and PA are interchanged, and the two ways to avoid crossing
for AA are exchanged. For AA, restoring the original avoided crossing after $S$ transformation gives a minus sign. 
Then the contributions to the path integral for the partition function are evaluated exactly as without the $S$ transformation, 
except for the interchange of AP and PA and the sign change in the AA case, and that gives the results in eq.\ (\ref{StransZ}).  

The arguments that $T^2$ acting on $Z_{AP}$ gives $-Z_{AP}$, and on $Z_{AA}$ gives $-Z_{AA}$, are similar. 
As $Z_{AA}$ can be obtained from $Z_{AP}$ by $T$, it is sufficient to consider $Z_{AP}$.
First, for any configuration of $\vec{n}$ with the standard branch cut for AP b.c., changing $\uptau$ continuously from $\uptau$ 
to $\uptau+2$ changes the cut in the time direction into that with an additional two windings in the space direction, 
discussed above. The standard branch cut can be restored at the cost of a minus sign in $\cal I$. This proves the result.

These results for $S$ and $T^2$ have the following implication for how the corresponding continuous deformations, 
$\uptau\mapsto -1/\uptau$ and $\uptau\mapsto \uptau+2$, act on the homotopy classes of $\vec{n}$ configurations 
with fixed branch cuts for each b.c.\ AA and AP respectively. ${\cal I}[\vec{n}]$ is unchanged 
under such a continuous change in $\uptau$, homotopy of $\vec{n}$, and isotopy of the branch cut, but then changing back 
to the original choice of branch cut reverses the sign of $\cal I$. Then it must be that, while the b.c.s and branch cuts
have been restored to their original forms, $\vec{n}$ (with the given branch cut fixed) has been changed such that $\cal I$ 
has acquired another minus sign. This means that the $\vec{n}$ configuration after the continuous change $\uptau\to -1/\uptau$
or $\uptau\to\uptau+2$ must be in a different homotopy class (for the fixed b.c.s and branch cuts) from the one in which it began. 
Clearly this gives a bijective map of homotopy classes, and classes that correspond under this bijection have opposite signs 
of the invariant.

Though it is not strictly necessary, we will go deeper and obtain the form of the bijection itself for $T^2$ on the AP b.c.\
(some readers may prefer to skip a few paragraphs, to that beginning ``the action of $T$ itself \ldots''; 
the case of $S$ on AA is similar, but more difficult, and we omit it). 
We will usually discuss the homotopy classes by thinking of the torus topologically as a square with opposite edges identified, 
and placing the branch cuts (if any), which we take to be the standard one(s), on zero, two, or four edges of the square. 
(We sometimes use complex plane notation $z$ for position in this square. The square arises automatically if $\uptau=i$, 
and other cases are similar.) First we point out that, because $\mathbb{S}^2$ is simply 
connected, any loop in it can be deformed to any other, and we can do so by a homotopy of $\vec{n}$ with the given b.c.s. 
Then the values just inside an edge of the square can be deformed to some standard values, the precise choice of which will
not make a difference. For example, for PP we can require that 
$\vec{n}$ be some constant value on the boundary of the square. For AP (or PA), we can take $\vec{n}=\vec{n}_0$, a constant, say 
$\vec{n}_0$ on the equator, and independent of $t$ (resp., $x$) on the east and west (resp., north and south) edges of the square, 
and require that it winds along the equator in the positive direction to reach the opposite point $-\vec{n}_0$ as $x$ 
(resp., $t$) goes from $0$ to $L$ (resp., $0$ to $\beta=L/v$). For AA, we can take $\vec{n}(0,0)=\vec{n}_0$, with winding in the 
positive sense along the equator to $-\vec{n}_0$ as $z$ goes from $0$ to $L$ and as it goes from $0$ to $iv\beta$ (using the 
complex plane notation here). Alternatively, we could take the same form on the bottom edge of the square, but wind in the reverse 
direction as $t$ goes from $0$ to $\beta$. 

If we impose one of these boundary behaviors of $\vec{n}$ on the boundary of the square, then we can consider
the homotopies of $\vec{n}$ configurations with the fixed boundary values. For PP, the boundary value is $\vec{n}_0$, 
a constant, and the classes of homotopy-equivalent $\vec{n}$ configurations are the exactly the elements of the homotopy group 
$\pi_2(\mathbb{S}^2)$, with all the points on the boundary of the square identified (so it too becomes $\mathbb{S}^2$) 
as the basepoint in the domain, and $\vec{n}_0$ as the basepoint in the codomain $\mathbb{S}^2$. In this case the configuration 
$\vec{n}(x,t)=\vec{n}_0$ for all $x$, $t$ is a representative of the identity element in the homotopy group, and the correspondence 
of homotopy classes with integers is canonical up to a choice of sign. For the cases AP (and PA), a natural choice for a standard
reference configuration is that in which $\vec{n}(x,t)$ is independent of $t$ (resp., $x$) throughout the square. Similarly,
for AA, there is a standard choice for each of the two standard boundary values described above. In the first, $\vec{n}(x,t)$
is constant on each straight line that runs parallel to the straight line from the top left corner to the bottom right corner,
while for the second it is constant on diagonal lines parallel to that between the other two corners. Note that the bulk
contribution to ${\cal N}'$ is zero for all these standard configurations, because the partial derivative along one
direction (a coordinate direction for PP, AP, PA, and a diagonal direction for AA) is zero everywhere. If we now replace 
the part of these $\vec{n}$ configurations lying inside a simply connected region (a disk), that lies strictly
inside the boundary of the square, with another general configuration, then we can reach any homotopy class. By deforming $\vec{n}$ by 
homotopy in the vicinity of the boundary of this region so that it equals $\vec{n}_1$ everywhere on its boundary (which again can 
always be achieved), we can classify the different homotopy classes by again using the $\pi_2$ homotopy group of $\mathbb{S}^2$, 
this time with $\vec{n}_1$ as the basepoint in the codomain $\mathbb{S}^2$. For all four b.c.s, the different classes can be 
labeled with an integer, the value of $\cal N$ for the configuration of $\vec{n}$ in this disk, with value fixed at $\vec{n}_1$ 
on the boundary of the disk. 

By these arguments, for all A b.c.s, we have separated the homotopy classification into two parts, one concerned with the boundary 
values on the square, the other with the homotopy classes for a sphere or equivalently for PP. However, for the three cases with 
an A b.c., it is not clear that this labeling is canonical, even up to sign; a change in what was used as the standard boundary 
values on the boundary of the square may affect the value of the integer. For AP (or PA) b.c.s, we can take the standard configuration
to be any one in which $\vec{n}(x,t)$ is independent of $t$ (resp., $x$) throughout the square, and any of these can be deformed 
by a homotopy to any other. Any such choice gives the same
value for the integer $\cal N$ associated to the disk, and so appears reasonably canonical. For AA, we had two choices of
standard configuration, which were not obviously homotopic for given boundary values of $\vec{n}$ on the square. 

However, it turns out that, for b.c.s other than PP, the integer is not the whole story. To understand this, suppose that 
$\vec{n}$ is constant over 
a small disk away from the branch cut. If we replace the configuration in that region by a ``bubble'', consisting of a map of a 
$2$-sphere into $\mathbb{S}^2$, with the points on the boundary of the original disk mapping to a single point, where the map
is in the homotopy class assigned the integer $+1$ in $\pi_1(\mathbb{S}^2)$ (this modification of the configuration is related
to the concept of ``connected sum'' of spaces in topology), then the integer we assigned to the homotopy
classes of maps with AP b.c.s changes by $+1$. But now we can perform a homotopy on the $\vec{n}$ field, such that the location of
the bubble moves along a loop that crosses a cut exactly once, and eventually returns to its starting point. It is not difficult 
to see that the bubble then becomes an element in the class $-1$ in $\pi_1(\mathbb{S}^2)$ (for example, by computing $\cal N$ 
for it, and using our initial remarks on the effect of inversion in the Introduction), so now it shifts our homotopy class by $-1$. 
(For a bubble with a general winding number, say $m$, in $\pi_2(\mathbb{S}^2)$, which shifts the integer by $m$, the homotopy maps
it to a shift by $-m$, a difference of $2m$.) Then for any given starting $\vec{n}$ configuration, by creating bubbles
of opposite winding number so that the integer homotopy class is unchanged, then moving one bubble around a loop that crosses a cut
while the other remains fixed, we can shift the integer by any multiple of $2$. 
Hence we find that the homotopy classes of $\vec{n}$ configurations with at least one A b.c.\ 
must be classified (relative to the reference configuration) by integers modulo $2$; there are only two homotopy classes
in each of these three cases. [The number of classes cannot be further reduced to one, as is shown by the existence of
configurations with both values ${\cal I}=\pm 1$ of our invariant when $\Theta=\pi$.] Note that these considerations 
do not affect the invariant 
$\pi{\cal N}'$, which was only defined modulo $2\pi$ anyway, and that $\cal I$ is unaffected by the homotopy that moves the 
bubble around a loop. 

Now we can consider the effect of $T^2$ on ${\cal I}[\vec{n}]$. It will be sufficient to consider its effect
for AP b.c.s, and we can begin with the standard configuration just described. For this, both the bulk and boundary contributions
to ${\cal N}'$ are zero. After applying $T^2$ by continuous change of
$\uptau$ and homotopy of $\vec{n}$, we obtain a configuration in which $\vec{n}$ is still on the equator for all $(x,t)$, 
and starting from the origin it moves continuously along the equator in the positive direction, ending at the opposite point, as $x$ 
increases at $t=0$. It moves continuously along the equator in the negative sense as $t$ increases at $x=0$, ending at the same 
point it started at. (We see that the AP b.c.s are satisfied.) The bulk contribution to $\pi{\cal N}'$ is still zero. A picture
of the map of the cylinder to the sphere again has the form of Figure 2, although increasing $t$ maps to motion in the negative sense 
on the sphere, and the two circles of constant latitude are degenerated, as both lie on the equator. Then we can evaluate
the invariant, and we see that $\pi{\cal N}'=-\pi$ for this configuration (it may help to first deform the circles of latitude to
slightly away from the equator). By deforming the circles of latitude to the north and south 
poles (respectively), we can recover the original (constant in $t$) boundary values on the sphere, but as $x$ increases at fixed $t$ 
$\vec{n}$ moves from one pole to the other, instead of along the equator. By a final global rotation we can recover the original 
configuration at $t=0$, and we see that, as $t$ increases, the curve from some $\vec{n}_0$ to $-\vec{n}_0$ rotates, with its endpoints
fixed; thus it sweeps out the full sphere.
Hence we have found that the effect of $T^2$ is to map the standard $\vec{n}$ configuration to one with $\pi{\cal N}'=-\pi$. 
We can do a similar calculation with a disk inserted inside the square containing an $\vec{n}$ configuration 
of nontrivial $\cal N$ (taking the value $\vec{n}_1$ on the boundary of that disk), and it works out in the same way:
$T^2$ adds $-\pi$ to $\pi{\cal N}'$. Because the homotopy classes are labeled by integers modulo $2$, this shows that the two classes
are interchanged by $T^2$ (like a ``spectral flow''). This concludes the analysis of $T^2$. 

The action of $T$ itself on the various b.c.s was given already. For PP and PA b.c.s, $T$ maps each homotopy class of $\vec{n}$ 
configurations (as defined above) into itself. $T$ applied to AP maps the standard branch cut to the AA standard form running
from bottom left to top right, while $T^{-1}$ maps it to the form bottom right to top left, and these differ by an avoided crossing,
so give AA partition functions of opposite sign, as discussed in the Introduction. Putting all these results together, we have obtained 
all the formulas in eq.\ (\ref{TtransZ}). In the operator point of view, it also follows (as explained in the Introduction) that 
either lift of inversion has eigenvalues $i$, $-i$ in the spatial A sector. The corresponding eigenstates must come in pairs with 
the opposite eigenvalues $i$, $-i$ of each lift, and with the same energy eigenvalue (i.e.\ of the Hamiltonian $H$), so that both 
AA partition functions are zero when $\uptau$ is imaginary. These are some of the main results of this paper, and apply
in the $\Theta=\pi$ (mod $2\pi$) case. 

\subsubsection{Discussion}

In concluding this subsection, we emphasize that we have shown that, for $\Theta=\pi$ (mod $2\pi$),
when one or more b.c.\ is A, the invariant $\cal I$, and hence the corresponding partition function, is
not gauge invariant under all gauge transformations, but only under a restricted class of them. 
This is a very direct demonstration of a 't Hooft anomaly, that is, an obstruction to gauging the inversion symmetry.
No such anomaly occurs for $\Theta=0$ (mod $2\pi$).

To sharpen this point further, we can discuss our results in terms of the $\mathbb{R}\mathbb{P}^2$ sigma model.
The possibility of a topological term (of the type under consideration) in a sigma model, and hence of a topological invariant 
like $\cal I$ that can be inserted as a factor into the path integral, is usually phrased first in terms of terms of the second homotopy
group $\pi_2$. If, as in our case, we have $\pi_2(\mathbb{R}\mathbb{P}^2)=\mathbb{Z}$, with the classes labeled by, say, 
an integer ${\cal M}$, then we might expect that we can insert the factor $e^{-i\Theta{\cal M}}$ into the path integral, 
where $\Theta$ is well defined as an angle, modulo $2\pi$. In principle, that is correct (but see below), but we saw at 
the beginning of the Introduction that there is a difficulty 
already for the PP b.c.\ case, because the integral there for $\cal N$, which we would hope can stand in for $\cal M$, 
is not invariant under the global inversion operation
on $\vec{n}$, but reverses sign. [As explained in this section, for PP b.c.s 
the homotopy classes of $\mathbb{R}\mathbb{P}^2$ configurations, or of $\vec{n}$ configurations with fixed branch cuts, 
reduce to the homotopy classes of such configurations for spacetime a sphere $\mathbb{S}^2$, which are exactly those 
classified by $\pi_2$. Moreover, as we have seen, in these cases an $\mathbb{R}\mathbb{P}^2$ configuration can be ``lifted'' 
to obtain a continuous $\vec{n}\in\mathbb{S}^2$ configuration (without branch cuts), and this is why the homotopy classification 
is the same.] Thus $\cal N$ is not well defined for the $\mathbb{R}\mathbb{P}^2$ sigma model because it is not 
invariant under global inversion, a particular case of a gauge transformation (equivalently, $\mathbb{R}\mathbb{P}^2$ 
is not orientable). That led us to restrict ourselves to $\Theta=0$ or $\pi$ 
(mod $2\pi$), for which $e^{-i\Theta{\cal N}}=\pm 1$ is well defined. In this section, we also analyzed the three cases with at least 
one A b.c.\ on $\vec{n}$, which corresponds to the $\mathbb{R}\mathbb{P}^2$ field that moves around a nontrivial cycle in 
$\mathbb{R}\mathbb{P}^2$ on moving around the corresponding cycle on the torus (in these cases, the $\mathbb{R}\mathbb{P}^2$ 
configurations cannot be lifted to $\mathbb{S}^2$, but necessarily involve a branch cut in the $\vec{n}$ field). In these three 
sectors also, the homotopy classes of configurations in $\mathbb{R}\mathbb{P}^2$
can be labeled (relative to an arbitrary reference configuration) by an element of $\mathbb{Z}_2$. But we have seen that the attempt 
to obtain a similar expression involving a bulk integral of the same form as $\cal N$ still failed to be gauge invariant 
for these b.c.s also, which means that there is no such gauge-invariant expression, 
or invariant $\cal I$, for these homotopy classes, other than the trivial constant one, ${\cal I}=1$. This emphasizes that,
contrary to a widespread belief, the classification of homotopy classes cannot always be captured (even modulo $2$)
by an explicit expression involving an integral over the domain of the functions being classified. But from the point of 
view of local QFT, only an expression as an integral in the action, without either nonlocal global restrictions [e.g.\
restricting to PP b.c.s, which anyway just corresponds to the $O(3)$ model] or a choice of gauge for the $\vec{n}$ field, 
is acceptable, so we conclude that the $\mathbb{R}\mathbb{P}^2$ sigma model with $\Theta=\pi$ (mod $2\pi$) does not exist
as a local QFT.

Based on this section, we can also state a more general topological result. In general, to discuss modular transformations, 
we seek not only a topological invariant for a fixed torus $\mathbb{T}^2$ as domain of the field 
configuration in $\mathbb{R}\mathbb{P}^2$, but also it should be 
invariant under continuous deformations of the domain, that is, of $\uptau$. The set of all $\uptau$ modulo modular transformations 
forms a moduli space, which is not simply connected; its fundamental group is the group $\overline{\Gamma}$
of modular transformations, which have to be implemented as orientation-preserving homeomorphisms of the torus,
and the equivalence classes of the latter form $\overline{\Gamma}$ (called the ``mapping class group'' in this context). 
We showed that, when there is at least one A b.c., say AP, in other words, when there is a cycle on the torus that maps to a noncontractible loop
in $\mathbb{R}\mathbb{P}^2$, the continuous deformation of $\uptau$ along a path in $\uptau$ space
such that the b.c.\ is restored to the original choice [e.g.\ applying $T^{2}\in\overline{\Gamma}$; the modular transformations preserving 
one (not all three) such b.c.(s) form a subgroup $\cong \overline{\Gamma}_0(2)\subset\overline{\Gamma}$ of index $3$] has the effect, 
for certain such loops, of exchanging the two homotopy classes of the $\mathbb{R}\mathbb{P}^2$ configurations with that b.c.. This shows that this 
set-up is a fibre bundle with a two-element fibre (the two homotopy classes), analogous to the ``two'' edges (i.e.\ locally) of the M\"{o}bius strip, 
but over the space of all $\uptau$ (${\rm Im}\,\uptau>0$) modulo $\overline{\Gamma}_0(2)$, in place of $\mathbb{S}^1$ for the strip. 
That is, like the single edge (globally) of the M\"{o}bius strip, the so-called total space of the fibre bundle consists of 
a single connected component. A continuous function that is locally a constant
on this space can be viewed as a topological invariant of the $\vec{n}$ field configuration, and is gauge invariant by construction.
Because the total space has a single connected component, such a function must be globally constant, corresponding to the $\Theta=0$ (mod $2\pi$) case. 
Hence any attempt to construct a non-constant topological invariant, such as our $\cal I$, corresponding to $\Theta=\pi$ (mod $2\pi$),
must either fail to be gauge invariant, or fail to be continuous on the total space of the fibre bundle.  
Our invariant $\cal I$ changes continuously with $\uptau$ (when the branch cuts change isotopically also) but, for these b.c.s and 
for certain loops, it changes sign when the branch cuts are restored to the original choice. We see that the spectral-flow property 
of the homotopy classes and the gauge non-invariance of $\cal I$ [for $\Theta=\pi$ (mod $2\pi$)] are essentially equivalent: 
for consistency, each is a necessary consequence of the other. (We have seen that the spectral flow is connected with the failure of modular 
invariance.) This also shows that our invariant $\cal I$ is the only continuous one possible, apart from an overall constant factor.

Finally, we wish to emphasize the generality of these results, in the following sense. The basic anomaly, the gauge non-invariance of
$\cal I$ when $(-1)^{\Theta/\pi}=-1$, is a purely topological result. No invariance properties of the non-topological part of the action 
were involved, though when we wish to apply the results to the sigma model, inversion symmetry of that term would usually be assumed.
For the topological part, and the anomaly itself, we wish to emphasize the following. In the $O(3)$ sigma model as we formulated it,
the topological term is the integral of the Berry curvature $\vec{\cal B}=\nabla\times \vec{\cal A}$ (or $2$-form, or magnetic field) on $\mathbb{S}^2$, 
pulled back to some spacetime manifold. When we calculated the anomaly by reducing the problem to the line integral of $\vec{\cal A}$ 
along a closed curve (loop) with an equivariance property, we evaluated the line integral by using its equality for any such curve, 
and evaluating it for a great circle. All the steps (including also the equivalence of the two formulations of $\pi{\cal N}'$) 
invoked the fact that the Berry curvature is odd under inversion, which is a sufficient assumption. 
In fact, that assumption is not really needed. The second formulation of the invariant $\pi{\cal N}'$ showed that it is a homotopy invariant, 
and well defined modulo $2\pi$. $\pi{\cal N}'$ is also invariant under continuous deformations of the Berry curvature $\vec{\cal B}$ 
[i.e.\ it depends only on the cohomology class or Chern number (equal to $2S$) of $\vec{\cal B}$ on $\mathbb{S}^2$], 
of which it is an integral. Hence, any choice of the Berry curvature 
in the class determined by $\Theta/\pi$ can be deformed to reach the one with $SO(3)$ symmetry that we used, without changing 
the result for $\pi{\cal N}'$ or $\cal I$. That choice was convenient, as it simplified our calculation. 

For other results, some additional symmetries were assumed. The argument that $Z_{AA}(\uptau)=0$ when $\uptau$ is imaginary was one case,
which involved the assumption of either space or time reflection symmetry of the non-topological action.
The modular transformation arguments in the Introduction that inferred the eigenvalues of the lift of inversion involved 
translation symmetry as well. But no internal symmetry group larger than that containing inversion was required for these results. 
(In the modular transformation arguments, reparametrization invariance of the non-topological action under an orientation-preserving 
isometry of the spacetime was also involved implicitly, and of course holds for the integral expressions we gave as examples.)

%%%%%%%%%%%%%%%%%%%%%%%%%%%%%%%%%%%%%%%%%%%%
\subsection{Semiclassical quantization}
\label{Subsec:semicl}

In this subsection, we obtain various results, including those concerning the action of $T$ in the previous subsection, by a different, 
but still semiclassical, approach, in which instead of the Lagrangian (or action) and a path integral, we consider the Hamiltonian
point of view at weak coupling. We do some of this for $\Theta=0$ as well as for $\Theta=\pi$. 
As a starting point, the Hamiltonian of the $O(3)$ sigma model [for which we here assume the model has the full $O(3)$ symmetry group, 
and comment briefly on other cases later] takes the form
\beq
H=\int dx\,\left[\frac{v}{2g^2}(\partial_x \vec{n})^2 + \frac{g^2v}{2} \vec{j}(x)^2\right],
\eeq
where $\vec{j}(x)$ is the angular momentum (or ``spin'') density. 
Then classically (i.e.\ as $g^2\to0$), for the P spatial b.c., the lowest energy 
state is a constant, $\vec{n}(x)=\vec{n}_0$, say, for all $x$ for the spatial P b.c., while for the A spatial b.c.\ we have 
$\vec{n}(0^+)=\vec{n}_0$, $\vec{n}(L^-)=-\vec{n}_0$, and for $x$ in-between these values $\vec{n}(x)$ moves along 
a semicircular arc of a great circle between those values, with $(d\vec{n}/dx)^2$ constant. [Here we chose to put 
the branch cut for the A b.c.\ at $x=0$ (mod $L$).] 

For both b.c.s, the lowest-energy configurations are degenerate in energy, because in the sigma model such a configuration
always breaks the $O(3)$ symmetry. To obtain first results for the quantum-mechanical theory, we should consider the quantum mechanics
of a point, representing the system, as it moves on the space of degenerate lowest-energy configurations. [$x$-dependent fluctuations
in $\vec{n}(x)$ around the lowest-energy configurations appear as higher-energy excitations on top of these states, but can be 
neglected in the lowest semiclassical approximation.] For the P b.c., the space of solutions is clearly $\mathbb{S}^2$, as it is 
parametrized by the single vector $\vec{n}_0$, and we may as well call it $\vec{n}$ with the understanding that $\vec{n}(x)=\vec{n}$ 
for all $x$. For the A b.c., we have the choice of $\vec{n}_0$, and also the choice of semicircular 
arc leaving $\vec{n}_0$, which is labeled by a single angle. This three-dimensional space can also be described by fixing a standard
choice of $\vec{n}(x)$, and then the general configuration is obtained by an $SO(3)$ rotation of it. Hence the space of the so-called 
collective coordinates in this case is simply $SO(3)$, and we denote the collective coordinate by the matrix $O$.

\begin{figure}[h]
\begin{center}

\begin{tikzpicture}

\begin{scope}[shift={(-4.5,0)}, line cap=round, ->, gray]
  \draw[->] (0,0) -- (-1,0) node[anchor=south] {\small 1};  % x-axis
  \draw[->] (0,0) -- (0.7,-0.4) node[anchor=north] {\small 2}; % y-axis
  \draw[->] (0,0) -- (0,1) node[anchor=west] {\small 3};   % z-axis
\end{scope}

\def\radius{2}

\def\lat{1.2}

\pgfmathsetmacro\xrad{sqrt(\radius*\radius - \lat*\lat)}

\shade[ball color=blue!10!gray, opacity=0.3] (0,0) circle (\radius);

\draw[thick, green] (-\radius,0) arc[start angle=180, end angle=360, x radius=\radius, y radius={\radius/4}];

\draw[dotted, green] (\radius,0) arc[start angle=0, end angle=180, x radius=\radius, y radius={\radius/4}];

\foreach \angle in {180,195,...,360} {
    \draw[blue, arrows={-Stealth[inset=0pt, angle=90:2pt]}]
        (0,0) -- ({\radius*cos(\angle)}, {\radius/4*sin(\angle)});
}

\node[anchor=east] at ({-\radius}, 0) {$\vec{n}_0$};
\node[anchor=west] at ({\radius}, 0) {$-\vec{n}_0$};

\draw[green, thick, arrows={-Stealth[inset=0pt, angle=90:2pt]}] ({0.*\radius}, -0.5) -- ({0.*\radius + 0.1}, -0.5);

\end{tikzpicture}

\end{center}
\caption{An example of ground-state texture with  $\vec{n}_0$ in the 1 direction, and a semi-circular arc in the 12 plane.}
\end{figure}
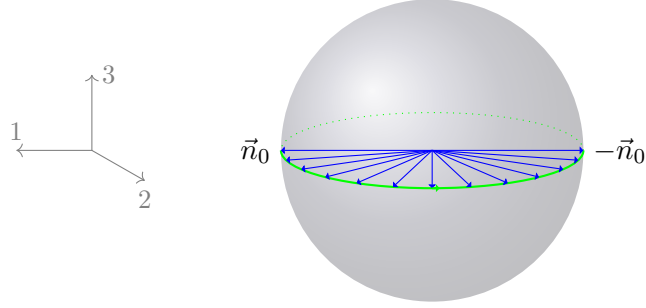

To complete the derivation of the ``effective'' theory of the motion on the collective coordinate space, we should consider
once again the action with which we defined the model. At small $g$, we can assume that, as a function of $x$ at each $t$, 
$\vec{n}(x,t)$ has one of the forms above, and the point in the collective coordinate space depends on $t$. Then the
non-topological term in the action will lead to an effective Hamiltonian for the collective coordinate, which by symmetry will be 
an angular-momentum squared term. For the P b.c., this is the rotor model with Hamiltonian (easily obtained from the action)
\beq
H_P=\frac{g^2v}{2L}\vec{J}^2,
\eeq
where $\vec{J}=\int dx\,\vec{j}(x)$, and $L/(gv)$ can be viewed as a moment of inertia.

For the A b.c., the collective coordinate space is $SO(3)$, and can be recognized
as what occurs in the motion of a rigid body with one point held fixed, or a ``top'', where the $SO(3)$ element $O$ describes
a rotation from some standard orientation of the body to its current orientation. Here there are two types of operations,
some of which form a symmetry group of the effective Hamiltonian. We can multiply $O$ on the left or on the right by
an element of $SO(3)$, and these operations form groups that we denote $SO(3)_L$ and $SO(3)_R$; the actions of these
two groups commute, so together we have a group $SO(3)_L\times SO(3)_R$ (we discuss all this in Appendix \ref{PetWeyl}). 
[Readers are cautioned that the left and right rotations, generators, and quantum numbers here do not correspond to
the right- and left- {\em moving} current algebras in the CFT discussed in the Introduction, in spite of some similar notation. 
However, there is some relation within the WZW model, in which the field has values in $SU(2)$, and the left and right 
rotation symmetries are left and right moving as well \cite{Witten1984}; that is not at all pertinent at present.] 
The corresponding generators, 
or angular momenta, will be denoted $\vec{J}_L=\vec{J}$ and $\vec{J}_R$, which each generate a copy of the $SO(3)$ Lie algebra
(they can be viewed as vectors of $3\times 3$ matrices). The left rotations
describe rotation of the rigid body in the ``laboratory'' frame, and when our sigma model has full $O(3)$ symmetry, as we assume here,
these left rotations form a symmetry group of the effective Hamiltonian (the texture is not invariant under this symmetry).
The right rotations describe rotation of the texture in the ``body'' frame, or before application of $O$,
and this is again not a symmetry of the texture (again, our conventions are set out in Appendix \ref{PetWeyl}). 
Thus we can consider the group $SO(3)_L\times SO(3)_R$,
and the texture breaks this but, at any given $O=O_0$ say, it leaves an unbroken subgroup $SO(3)_{O_0}$ that acts on $O_0$ as 
simultaneous multiplication by $O'^{-1}\in SO(3)$ on the right and by $O_0 O'O_0^{-1}$ from the left, under which the 
texture (or $O_0$) is invariant; again, see Appendix \ref{PetWeyl}.

The kinetic energy is then described using the moment of inertia tensor of the body, and that tensor is constant in the body axes, 
which are fixed in the body even as it rotates.
The moment of inertia tensor can be diagonalized in terms of eigenvalues ${\bf I}_\mu$, $\mu=1$, $2$, $3$, which 
are the principal moments of inertia. Then the kinetic energy of the rigid body can be expressed as 
$H=\sum_\mu J_{R\mu}^2/(2{\bf I}_\mu)$. 
For such a rigid body, in general the principal moments of inertia do not all have to be equal, and if not, then the 
$SO(3)_R$ symmetry in the body frame of reference is broken. But in our setting, if the semicircular arc of the 
standard $\vec{n}(x)$ lies in the 
$12$ plane, then we can take these to be the body frame, say with $\vec{n}_0$ in the positive $1$ direction, and then there is 
unbroken rotation symmetry around the $3$ axis, so ${\bf I}_1={\bf I}_2$, but it is likely that ${\bf I}_3$ has a different value.
For our case, because the texture has $(d\vec{n}/dx)^2$ constant, the moments of inertia for the collective coordinate are those 
of a rigid circle with uniform mass density 
(a ``hoop''), for which ${\bf I}_3=2{\bf I}_1=2{\bf I}_2$. Generally, in any case in which ${\bf I}_1={\bf I}_2\neq{\bf I}_3$,
there is a subgroup of $SO(3)_R$, which we can term $SO(2)_R$, of rotations about the $3$ axis, under which the Hamiltonian
is invariant. Then we find from the action
\beq
H_A = \frac{g^2v}{L}\vec{J}_R^2 - \frac{g^2v}{2L}J_{R3}^2,
\label{eq:HA}
\eeq
up to an additive constant of order $1/L$ that comes from the gradient-squared energy for the texture. 
Here the moments of inertia are ${\bf I}_1={\bf I}_2=L/(2g^2v)$, ${\bf I}_3=L/(g^2v)$.
This Hamiltonian $H_A$ commutes with the elements of (possibly, a double cover of) the symmetry group $SO(3)_L\times SO(2)_R$,
generated by $\vec{J}_L$ and $J_{R3}$. The existence of this symmetry group $SO(2)_R$ in our case is ultimately due to the
symmetry of the sigma model under spatial translations, which generate a copy of $SO(2)$, and are always assumed to 
commute with the (full) Hamiltonian $H$.

Finally, there is the role of the topological term in the action, which is first order in time derivatives, and imaginary. 
In general, such a term can be viewed as involving a vector potential (or ``connection'') on the space of field configurations, 
like the term $i\int dt\,A_\mu(\vec{x})dx^\mu/dt$ in the action for a particle moving in a magnetic field described by the 
vector potential $A_\mu$, with $\vec{x}(t)$ periodic in $t$. The exponential of this term is the holonomy of the connection
(it can also be viewed as a Berry phase factor).
Then for our case, using the P temporal b.c., when we reduce to the motion of the collective coordinate, we may pick up 
a connection on the collective coordinate space. (In Hamiltonian form, this usually appears in the formula for 
the kinetic momentum that appears in the Hamiltonian, $\pi_\mu=-i\partial_\mu-A_\mu$ in our example.) Its form can be inferred 
from the holonomy for a time-periodic path of the collective coordinate. For the spatial P b.c., $\vec{n}(x,t)$ is constant
in $x$ for all $t$, so the topological term is zero, and there is no holonomy and no vector potential in the Hamiltonian for
the collective coordinate (where it would appear in the expression for $\vec{J}$ in terms of partial derivatives).
On the other hand, for the spatial A b.c., if we consider a configuration in which $\vec{n}(x,t)$ is standard at each $t$,
the values at $x=0^+$, $L^-$ are fixed at $\vec{n}_0$, $-\vec{n}_0$, respectively, for all $t$, and it rotates about the axis 
through $\vec{n}_0$ by $2\pi$ as $t$ goes from $0$ to $\beta$, then we see that $\vec{n}$ covers the whole sphere, 
and for $\Theta=\pi$ the bulk 
contribution to $\pi{\cal N}'$ is $\pi$ while the boundary term is zero. (This is exactly as appeared near the end of the 
calculation of the effect of $T^2$ on the AP configurations in the preceding subsection, where Fig.\ \ref{secondFig}, 
with the circles of latitude at the poles, is again the relevant picture). For other time dependence, the holonomy is $-1$ to the power 
of the number of times the arc rotates by $2\pi$ about its endpoints.

This holonomy can easily be understood topologically. The collective-coordinate space $SO(3)$ is doubly connected;
its fundamental group is $\pi_1(SO(3))=\mathbb{Z}_2$. 
If the curvature (i.e.\ curl or magnetic field strength) of 
a connection (vector potential) on $SO(3)$ vanishes everywhere, then the holonomy around any closed loop in $SO(3)$ 
only depends on the homotopy class of that loop, and so gives rise to a unitary representation of the fundamental group 
$\cong\mathbb{Z}_2$, expressed multiplicatively as $\pm 1$. Hence there are only two possibilities for a connection on $SO(3)$ 
(up to gauge transformations): it is either trivial, or it is the unique nontrivial connection. We find the first case 
for $\Theta=0$ (mod $2\pi$), and the second for $\Theta=\pi$ (mod $2\pi$). 

Next we discuss the quantum mechanics of the collective coordinate spaces, first for the case with trivial connection
(the discussion of both cases was partly inspired by Ref.\  \cite{WITTEN1983433}).
In the case of the spatial P b.c., for $\Theta$ either $0$ or $\pi$ (mod $2\pi$), we have the quantum version of the 
rotor on $\mathbb{S}^2$, and no vector potential, and then of course the Hamiltonian has eigenvalues $g^2vj(j+1)/(2L)$, 
where $j=0$, $1$, \ldots, is the angular momentum quantum number, and the corresponding eigenfunctions are the spherical harmonics 
$Y_{jm}(\vec{n})$, which are single-valued functions on $\mathbb{S}^2$; there are $2j+1$ linearly-independent states for each $j$. 

In the case of the spatial A b.c.,
we have the quantum version of the rigid body above, and for $\Theta=0$ (mod $2\pi$), there is no connection, 
and it is fairly well known that, if all three principal moments of inertia are equal to $\bf I$ (the case of the ``isotropic top''), 
the energy eigenvalues are again given by $j(j+1)/(2{\bf I})$, but now the total degeneracy of each value is $(2j+1)^2$ (as stated for 
example in Ref.\ \cite{landau1981quantum}). (We discuss the corresponding wavefunctions in a moment, and postpone discussion of the 
splitting of the degeneracy when the top is anisotropic, as it is 
in our case.) This requires a little more explanation: First, the collective coordinate $O$ is again a $3\times 3$ orthogonal 
$\check{O}$ that represents $O$, and can be viewed as an exponential expression $\check{O}=e^{-i\vec{\theta}\cdot\vec{J}_L}$,
where the parameters $\vec{\theta}$ describe the matrix $O$. Then again we have left and right symmetry groups
$SO(3)_L$, $SO(3)_R$, with generators $\vec{J}_L$, $\vec{J}_R$ (see Appendix \ref{PetWeyl}). For the isotropic top, 
the Hamiltonian is invariant under the full group 
$SO(3)_L\times SO(3)_R$. Then the eigenstates can be labeled by their right and left angular momenta $(j_L,j_R)$, together with the 
corresponding azimuthal quantum numbers $(m_L,m_R)$. However, in fact we must have $j_L=j_R=j$, and the Casimirs are equal as 
operators, $\vec{J}_L^2=\vec{J}_R^2$ (see Appendix \ref{PetWeyl}, or for an incomplete traditional treatment see e.g.\ Ref.\ 
\cite{landau1981quantum}), as written for the energy. 

Another way to view the Hilbert spaces of states on the collective-coordinate space that arises here
(to which we will refer as the ``collective-coordinate Hilbert space''), and especially the fact 
that $j_L=j_R$, comes from harmonic analysis. 
If we consider a spin $j$ irreducible representation of $SO(3)$, for an integer $j$ at present, then
for each element of $SO(3)$ there is a $(2j+1)\times (2j+1)$ matrix, which is a continuous function of the element,
and represents that element acting in that representation (i.e.\ there is a group homomorphism
of $SO(3)$ into the group of unitary operators on $2j+1$-dimensional complex Hilbert space). In physics, these matrices
are known as ``Wigner $D$ matrices'', with elements $D^j_{m_L,m_R}(O)$. A similar statement holds for the 
irreducible representations of any compact group $G$. The Peter-Weyl theorem \cite{Dieck} states, in part, that the Hilbert space of 
all square-integrable functions (without branch cuts at present) on any compact group $G$ is spanned by the set of 
matrix elements of the representation matrices in all the finite-dimensional irreducible representations, from which a basis
for the Hilbert space can then be obtained. Note that, 
going back to $G=SO(3)$, the $(2j+1)^2$ functions that make up the spin-$j$ representation matrices transform among themselves 
under left or right multiplication of $O\in SO(3)$ by a rotation, and so they themselves span a representation of the simultaneous 
left and right actions, with quantum numbers $j_L=j_R=j$. It turns out that the wavefunctions that transform 
in the desired way are actually the complex conjugates of the $D$ matrix elements, $\overline{D^j_{m_Lm_R}(O)}$. 
All of this is discussed further in Appendix \ref{PetWeyl}.

To discuss wavefunctions on $SO(3)$ with nontrivial connection, we will use a gauge choice for the wavefunctions and the 
connection that involves vanishing vector
potential except on a branch cut (somewhat analogous to what we did for the A b.c.s on $\vec{n})$. 
(We could do this in terms of a line bundle with a connection, but we will refrain from the language
of bundles in this paper.) First, recall that rotations in $SO(3)$ can be described by a choice of a rotation axis,
given by a unit vector, say $\vec{m}$, and a rotation angle $\theta$ about that axis. $\theta$ is defined only modulo $2\pi$, so we
can restrict to $|\theta|\leq \pi$. Then a rotation in $SO(3)$ corresponds to a unique vector in the three-ball of vectors
$\theta\vec{m}$ of magnitude less than $\pi$, where each vector on the surface of the ball is identified with the opposite vector.
This also describes the real projective space $\mathbb{R}\mathbb{P}^3$, so we have $SO(3)\cong\mathbb{R}\mathbb{P}^3$
as manifolds. A loop of rotations that begins at the identity ($\theta=0$ modulo $2\pi$), leaves the boundary of the ball, 
reenters at the opposite side and returns to the origin is not contractible to a point, and it belongs to the unique nontrivial
homotopy class in $\pi_1(SO(3))\cong\mathbb{Z}_2$ (a loop that makes such an excursion twice, ending as a rotation by $4\pi$, 
can be contracted to a point, as demonstrated by a well-known party trick). If wavefunctions are required to be continuous up to 
boundary conditions, then the nontrivial connection means that the wavefunctions in the collective-coordinate Hilbert space 
in fact have a branch cut, 
which by choice of gauge we can place on the surface $\theta=\pi$ that we just described. Thus the functions must reverse sign on 
crossing the cut (similar to an A b.c.\ on the $\vec{n}$ field earlier). 

Now consider a rotation acting on $O$, say by left multiplication by an element of $SO(3)_L$. 
In the analogous case of a particle in a uniform magnetic field on $\mathbb{R}^3$, the naive action of a translation of $\mathbb{R}^3$
would change the vector potential, which cannot be invariant under all translations if the magnetic field is nonzero.
Then the naive translation must be followed by a gauge transformation to recover the original choice of gauge,
and the combination of the two is a so-called magnetic translation. In the case of our problem with a connection on $SO(3)$,
described using the branch cut, rotating the function naively would move the branch cut, and we must follow it with a gauge 
transformation to restore the cut to its original location. What this means is that under rotation, the cut stays fixed, 
and as points move across the cut the value of the wavefunction at the moving point changes to minus what it was when on 
the opposite side of the cut. Then after rotation by $2\pi$, this means that the wavefunction returns to itself but with opposite sign. 
Hence, for the case of non-trivial connection, all states in the collective-coordinate Hilbert space have half-integer spin. 
With this sole change, the earlier results for the trivial connection still hold for the nontrivial connection, 
giving similar conclusions, except that now $j_L=j_R=j$ is half integer.

For the wavefunctions of these eigenstates, consider the cover $SU(2)\cong Spin(3)$ of $SO(3)$, to which the Peter-Weyl 
theorem again applies, 
and now there are both integer and half integer values of $j$ (the defining representation is spin $1/2$ in this case). 
Imposing the additional condition that the functions be invariant under the center $\mathbb{Z}_2=\{\pm1 \}$ 
in $SU(2)$ (i.e.\ under the lifts of left or right $2\pi$ rotations, which are the same) restricts the space to the functions 
on $SO(3)$, with integer spins only. When we instead impose the condition that the functions on $SU(2)$ be odd under the action 
of the center, then these are spanned by (the complex conjugates of) the elements of representation matrices for half-integer 
spin representations. These 
functions, when viewed as functions on $SO(3)$ with a branch cut as described, then span the Hilbert space of states for the top
in the case with the nontrivial connection. (The whole picture generalizes from $SO(3)$ to any connected
compact Lie group with finite center; see Ref.\ \cite{Dieck} for related discussion.) 

It is important to point out here that, in this treatment, we have assumed that the sigma model is invariant under the 
full $O(3)$ symmetry [the $SO(3)$ subgroup of which corresponds on the collective-coordinate space to $SO(3)_L$], 
and not only under a proper subgroup, whereas in the preceding subsection we were able to include the latter more general case
(at least when the subgroup contains inversion). If instead the Hamiltonian or action of the sigma model is invariant only
under a subgroup of the global internal $O(3)$ group [again, this contains $SO(3)_L$; we are not referring here to
the breaking of $SO(3)_R$ by the unequal principal moments of inertia, which occurs even in the case of unbroken $SO(3)_L$], 
then the semiclassical treatment for the spatial $A$ b.c.\ would have to be modified. If we consider only the textures of lowest energy,
the collective coordinate space for these textures may be smaller than $SO(3)$. But even if the internal symmetry group
is reduced to only $\mathbb{Z}_2$ (generated by inversion), then while a texture will involve a domain wall, the collective-coordinate space 
will still be a circle, or several copies of a circle, due to the translation symmetry of the model, which is spontaneously broken by the texture,
and there will be an $SO(3)$ symmetry due to spatial translations that acts on the circle (discussed further in the following subsection). 
A path in this space in which a rotation by $2\pi$ (or translation by $2L$) occurs has invariant ${\cal I}=-1$ when $\Theta=\pi$ (mod $2\pi$), 
by a calculation similar to that for $T^2$ in the preceding subsection. Then there is still a nontrivial connection, and the $SO(2)$ 
``spins'' must again be half integer, as above in our main case of full $SO(3)$ symmetry. Consequently, the main results of the paper 
still hold in this case.

Returning to the case of the $O(3)$-invariant sigma model, we obtain the spectrum of the Hamiltonian on the 
collective-coordinate space $SO(3)$
for the case of an isotropic top, and for $\Theta=\pi$ as well as for $\Theta=0$. For an anisotropic case with 
${\bf I}_1={\bf I}_2\neq{\bf I}_3$ we obtain the spectrum similarly. Thus for generic $j$ and $m_R\neq 0$ 
and values of the constant, the degeneracy of the energy levels is $2(2j+1)$, 
and of course the $Spin(3)$ symmetry of the system in the laboratory frame remains. In particular, for our Hamiltonian $H_A$ 
as above, we have energy eigenvalues 
\beq
(g^2v/L)[j(j+1)-m_R^2/2].
\label{HAevals}
\eeq
Here $j$ is integer for $\Theta=0$ (mod $2\pi$), half integer for $\Theta=\pi$ (mod $2\pi$), and $m_R$, $m_L$ both run through the 
values $-j$, $-j+1$,\ldots, $j$, as usual. For each pair $(j,m_R)$, there is $2j+1$-fold degeneracy due to left (laboratory frame) 
rotation symmetry and the corresponding $m_L$ quantum number. In some cases, there is additional degeneracy involving different
pairs; for example, $j=4$, $m_R=\pm 4$ give the same energy as $j=3$, $m_R=0$.

We can now discuss the semiclassical ground state(s) in each case, beginning with their spin quantum numbers and degeneracies. 
For the spatial P b.c., we have the non-degenerate
ground state with $j=0$. For spatial A b.c.\ and $\Theta=0$ (mod $2\pi$), we again have a non-degenerate ground state with $j=0$.
But for spatial A b.c.\ and $\Theta=\pi$, we have fourfold-degenerate ground states with $j=1/2$. As our quantum rigid body is not 
in the isotropic case, the $(2j+1)^2$ degeneracy is split, but the lowest energy
is still $j=1/2$, and this degeneracy is not split; it remains $4$. Note that these states are always formed of two spin-$1/2$ 
multiplets under the left, or laboratory frame, rotation symmetry. Notice also that whether the spin values $j$ are integer or half 
integer persists in each case for the excited states over the ground state(s), including the excited states with fluctuations 
away from the standard textures that are described solely by the collective coordinate. These results for ground-state degeneracy
still hold when the internal global symmetry is reduced, so long as it includes inversion, as we will see further in the next subsection.

%%%%%%%%%%%%%%%%%%%%%%%%%%%%%%%%%%%%%%%%%%%%%%
\subsection{Translations and discrete symmetries}
\label{Sec:transdisc}

Next we consider the momentum quantum numbers of the semiclassical ground states in the collective-coordinate semiclassical 
quantization of the sigma model, the lifts of inversion, and later the discrete symmetries of spatial reflection and time reversal; 
finally, we extend all this to the symmetries of the QFT of the full sigma model as well. First we assume full $O(3)$ global internal symmetry,
and later the case of reduced symmetry.

In the collective-coordinate description,
for the spatial P b.c.\ sector, the $\vec{n}(x)$ field is constant, so all states in the subspace described by quantizing 
the collective coordinate have momentum zero. For the A b.c.\ sector, the standard texture is not invariant under translations, 
however a translation in $x$ can be undone by a rotation in $\vec{n}$ space. In terms of the standard textures with $\vec{n}$ forming
an arc in the $12$ plane starting from $\vec{n}_0$ at $x=0^+$, a translation in $x$ can be undone by rotation about the $3$ axis, 
which acts on $O$ by multiplication from the right. (In the case that only two of the three principal moments of inertia are equal,
these right rotations are still symmetries of the collective-coordinate space Hamiltonian as they act in the body frame.) 
Because there is a branch cut in the $\vec{n}$ configuration, such a transformation again involves some points crossing the cut (which 
remains fixed at $x=0$). 
After translation by $L$, the reference texture $\vec{n}(x)$ has reversed in sign at each $x$, which corresponds to right rotation 
by $\pi$. After translation by $2L$, the texture returns to itself. We can think of the translations in the continuum sigma model
as forming a symmetry group we will term $SO(2)_{\rm tr}$, acting on space by translations, not internally. Translation by $y$ 
is equivalent to internal right rotation by $\pi y/L$. [For the sign conventions, see Appendix \ref{PetWeyl}.] Hence we can also think 
of the collective coordinate space in yet another 
presentation as the homogeneous space $SO(3)\cong SO(3)\times SO(2)_{\rm tr}/SO(2)_{\rm diag}$, where $SO(2)_{\rm diag}$
is the simultaneous action of $SO(2)_{\rm tr}$ and the $SO(2)$ subgroup of $SO(3)_R$ of rotations about the $3$ axis, 
which imposes the equivalence discussed here. This form is used and extended in Appendix \ref{OtherSp}.] It is this point, 
that the texture breaks both the global $SO(3)$ symmetry and the translation symmetry, but leaves a subgroup $SO(2)_{\rm diag}$ 
of $SO(3)\times SO(2)_{\rm tr}$ unbroken, which produces the mixing of the two symmetry operations
that appear in our results.

For the case of spatial A b.c.\ and $\Theta=0$ (mod $2\pi$), the semiclassical states have $j$ an integer, and the 
momentum is related to the right azimuthal 
quantum number by $P=\pi m_R/L$, where $m_R$ is integer and $|m_R|\leq j$; in particular, 
the ground state has momentum zero. For the case $\Theta=\pi$ (mod $2\pi$), 
it is similar, but now $m_R$ is half integer; the ground states have momentum $P=\pm \pi/(2L)$. Alternatively, we can look at 
the eigenvalue of the unitary $e^{-iPL}$ for translation by $L$. Then for the spatial A b.c., we have $(-1)^{m_R}$ for 
$\Theta=0$ (mod $2\pi$),
so always $\pm 1$, and $e^{i\pi m_R}$ for $\Theta=\pi$ (mod $2\pi$) so always $\pm i$. Note that the eigenvalues are constant 
within a multiplet of left angular momentum (i.e.\ independent of $m_L$). 

We emphasize that, for the $O(3)$-invariant sigma model, we have recovered the eigenvalue spectrum of the operator 
that represents translation by $L$, which produced the $T$ transformations in the partition 
functions in the previous subsection. More generally, we can also determine the (lifts of the) inversion operation
in each case. For the spatial P b.c.\ sector, the inversion operation $\vec{n}\to-\vec{n}$ leads to the inversion operator
with eigenvalues $(-1)^j$ on the ($j$ an integer) semiclassical collective coordinate spaces, in which we can fix an overall sign
by declaring that the ground state has inversion eigenvalue $+1$ (spatial translation has no relation to these eigenvalues in this 
sector). For the spatial A b.c.\ sector, inversion $I$ can be directly identified as rotation by $\pi$ in the body frame of the
collective coordinate description, and so as an element of $SO(2)_R$. For $\Theta=0$ (mod $2\pi$), $I$ has eigenvalues 
$e^{-i\pi m_R}=(-1)^{m_R}$, and $I^2=1$. For $\Theta=\pi$, we have a lift $\widehat{I}$, which similarly has 
eigenvalues $e^{-i\pi m_R} = \pm i$, and $\widehat{I}^2=-1$. In all cases, the (lift of) inversion $I$ (or $\widehat{I}$)
commutes with $SO(3)_L$ [$Spin(3)_L$], and together they generate the group we can call $O(3)_L$ [resp., $Pin_+(3)_L$]. 

Similarly, we can consider the reflection symmetries in both space and time, known as parity and time-reversal symmetry.
These already played a role in the path integral formulation when we showed that $Z_{AA}(\uptau)=0$ for $\uptau$ imaginary.
Here we examine their action as operators on the quantum states in the semiclassical set-up. First, for reflections,
a reflection maps $x-x_0$ to $-(x-x_0)$ for any choice of $x_0\in[0,L]$, subject to the identification $x\equiv x+L$.
The continuum sigma model has (at least naively or classically) an $O(2)$ symmetry consisting of translations and reflections 
on the one-dimensional space (with coordinate $x$, periodic with period $L$). To be definite, let ${\cal R}$ be reflection through 
$x=0$.  The natural definition in the sigma model is that under reflection ${\cal R}$, $\vec{n}(x)\to \vec{n}(-x)$, 
without a change of sign of $\vec{n}$, and that is what we use here. In the semiclassical 
description and with the P spatial b.c., $\vec{n}(x)$ is a constant $=\vec{n}$, so reflection maps it to itself, and all the 
semiclassical states are invariant under reflection. (We can make one arbitrary choice of phase factor in ${\cal R}$,
and we define it so that ${\cal R}=1$ on the ground state.)

For the A spatial b.c., we have the standard configuration, with a branch cut which we place 
at $x=0$, and $\vec{n}(x)$ winds in the equator from the $1$ direction to minus that, in the positive sense, as $x$ increases from
$0$ to $L$. Then ${\cal R}$ leaves the branch cut unchanged, and other choices can be obtained by conjugating with a translation and 
applying a gauge transformation. With the branch cut 
fixed at $x=0$, reflection of $x$ through $x=0$ has the same effect on the standard configuration as a rotation by $\pi$ about the 
$1$ axis. In fact, a reflection, as well as a translation, can be undone by a corresponding internal right $O(2)$ transformation 
on the coordinate $O$, or hence is equivalent to an (inverse) such transformation. That is, here we are concerned with a subgroup 
$O(2)\subset SO(3)_R$ of the $SO(3)_R$ rotations, and we call this subgroup $O(2)_R$. 
Then for $\Theta=0$ (mod $2\pi$), all semiclassical states for the motion on the collective-coordinate space $SO(3)$ 
are invariant under rotations by $2\pi$, so reflection squares to ${\cal R}^2=1$. It reverses the sign of $m_R$, 
and on the energy eigenspaces, we have either $m_R=0$, in which case the states are invariant under reflection, or $m_R\neq 0$ and 
there is a pair with quantum numbers $m_R$, $-m_R$ with the same energy; reflection interchanges these two. Indeed, these respectively
non-degenerate and degenerate states form irreducible representations of $O(2)_R$ in each case, and this $O(2)_R$ 
symmetry commutes with the laboratory (i.e.\ left) rotation symmetry. 

For the other case $\Theta=\pi$ (mod $2\pi$), the right rotations are lifted to the double cover $Spin(3)_R\cong SU(2)_R$ 
of $SO(3)_R$. Then $O(2)_R$, as a subgroup, is also lifted to a double cover. Because all rotations by $2\pi$ lift to $-1$, 
either lift $\widehat{\cal R}$ of the reflection operation now squares to $-1$ also. This means that the double cover 
is in fact $Pin_-(2)_R\subset Spin(3)_R$ (some information on $Pin_-(2)$ is given in Appendix \ref{PinG}). The lifts of 
the reflection operation, which again interchange $m_R$, $-m_R$ (and which here cannot be equal), now have eigenvalues $\pm i$. 

Similarly, for time-reversal, the natural choice in the sigma model is that it leaves $\vec{n}$ 
invariant, though it reverses time and the signs of the components of the spin density $\vec{j}(x)$. In quantum mechanics, 
time reversal is of course represented by an anti-unitary operator ${\cal T}$. 
The case of collective-coordinate states in the spatial P b.c.\ sector present few problems (time reversal symmetry
of a quantum rotor follows from standard principles, and ${\cal T}^2=1$ for both values of $\Theta$), so we concentrate
on the A sector. The quantum states are obtained by a unitary map $\widehat{O}$ (which represents $O$) applied to the standard state
(or texture). If we use a basis for the Hilbert space of semiclassical states in which all matrix elements of $J_{L1}$, $J_{R1}$,
$J_{L3}$, and $J_{R3}$ are real and those of $J_{L2}$, $J_{R2}$ are imaginary (as is conventional \cite{Schiff}), 
then time-reversal ${\cal T}$ is represented by complex conjugation of the coefficients in that basis (or of wavefunctions, likewise), 
followed by rotation by $\pi$ (or $-\pi$) about the $2$ axis in {\em both} $Spin(3)_L$ and $Spin(3)_R$, 
that is by $e^{-i\pi (J_{L2}+J_{R2})}$ (up to a phase factor). This is necessary in order that time reversal commutes with
all left and right rotations, and gives ${\cal T}\vec{J}_L{\cal T}^{-1}=-\vec{J}_L$ (and similarly for $\vec{J}_R$), 
while preserving the commutation relations. It follows in particular that, in the spatial A b.c.\ sector, we again
have ${\cal T}^2=(-1)^2=+1$ when $\Theta=\pi$, as well as when $\Theta=0$ (mod $2\pi$). This may seem surprising, in view
of the half-integer spins for $\Theta=\pi$, but is an inevitable consequence of our discussion. 

We note that, in terms of the basis eigenfunctions, for the spatial P b.c.\ sector we have only integer $j$, and we may use the 
spherical harmonics, which obey 
\beq
\overline{Y_{jm}(\vec{n})}=(-1)^mY_{j,-m}(\vec{n}),
\eeq
an expression of time-reversal invariance; alternatively, there is a 
basis for the eigenspaces for all $j$ (for example, the basis consisting of cubic harmonics, for which see Appendix \ref{PetWeyl}), 
in which all basis functions are real and time-reversal invariant. Similarly, for the spatial A b.c.\ sector, the basis functions
are the complex conjugates (see Appendix \ref{PetWeyl}) of the Wigner $D$ 
matrices $D^j_{m_L,m_R}(O)$ [in which $O$ must be lifted to an element 
of $Spin(3)\cong SU(2)$ in general, and then the wavefunctions on $SO(3)$ have the branch cut], which span the Hilbert space, 
and obey 
\beq
\overline{D^j_{m_L,m_R}(O)}=(-1)^{m_L-m_R}D^j_{-m_L,-m_R}(O)
\eeq
for all $j=0$, $1/2$, \ldots, leading again to ${\cal T}^2=1$.
Here we can point out that, topologically, $SU(2)$ is the three-sphere $\mathbb{S}^3$, and the action of $SU(2)_L\times SU(2)_R$
on $SU(2)\cong\mathbb{S}^3$ is the same as the natural action of $SO(4)\cong[SU(2)\times SU(2)]/\mathbb{Z}_2$ on $\mathbb{S}^3$
[here the modding out by $\mathbb{Z}_2$, which is the diagonal subgroup of simultaneous action by the same element of the center 
in both $SU(2)_L$ and $SU(2)_R$, occurs because that subgroup leaves any point in $SU(2)\cong\mathbb{S}^3$ invariant].
Then the cubic harmonics on $\mathbb{S}^3$ (see Appendix \ref{PetWeyl}) furnish a basis of time-reversal-invariant 
real functions for the Hilbert space of functions on $SU(2)$, which is an alternative to the $D$ matrices, and manifestly 
gives ${\cal T}^2=1$.

We can also ask whether the various operators $I$ (or $\widehat{I}$), ${\cal R}$ (or $\widehat{\cal R}$) and 
${\cal T}$ commute. For the P b.c.\ sector this is trivial: they all commute. For the spatial A b.c.\ sector,
the operations $I$ and $\cal R$ (or $\widehat{I}$ and $\widehat{\cal R}$) are elements of $SO(3)_R$ [resp., $Spin(3)_R$], 
and so commute with $\cal T$, for both $\Theta=0$ and $\pi$ (mod $2\pi$). For $\Theta=0$, inversion $I$ is in the center of 
$O(2)_R\subset SO(3)_R$, so commutes with spatial reflection $\cal R$. However, for $\Theta=\pi$ (mod $2\pi$), 
we have instead the double cover $Pin_-(2)_R\subset Spin(3)_R$, and in this case a lift $\widehat{I}$ 
of inversion anticommutes with a lift $\widehat{\cal R}$ of a spatial reflection (see Appendix \ref{PinG}),
which implies that $\widehat{\cal R}\widehat{I}=-\widehat{I}\widehat{\cal R}$. [This is connected with the vanishing of
$Z_{AA}(\uptau)$ for $\uptau$ imaginary.]

This concludes the discussion of the collective-coordinate semiclassical ground states in the case with $O(3)$ internal symmetry. 
To consider the full sigma model,
still viewed as semiclassical (i.e.\ as having weak coupling), we point out that the (full) Hilbert space should
be spanned by the states generated by applying local operators (fields) to the ground state(s). The local operators canonically
correspond to the field $\vec{n}(x)$ and the spin density $\vec{j}(x)$ (the integral $\int_0^L dx\,\vec{j}(x)$ of which 
is the total spin angular momentum $\vec{J}$ that we have used), and to operator products of these. To complete the account, 
we therefore examine how they transform under various symmetry operations, though many cases have already been covered,
at least implicitly. Thus $\vec{n}(x)$ and $\vec{j}(x)$ transform internally as vectors under rotations in $SO(3)$, 
inversion maps $\vec{n}(x)\to-\vec{n}(x)$ and $\vec{j}(x)\to\vec{j}(x)$, and time reversal
leaves $\vec{n}(x)$ invariant, but sends $\vec{j}(x)$ to $-\vec{j}(x)$, and each pair of these operations commute; 
these local statements are independent of boundary conditions, and hold in all cases. For other results, we begin with the 
P spatial b.c.\ sector. Translation by $L$ gives the identity, 
so the momenta (which label the Fourier components) of $\vec{n}(x)$ and $\vec{j}(x)$ are of the form $2\pi m/L$ 
(throughout this paragraph, we let $m$ denote an arbitrary integer). Reflection through the origin sends $\vec{n}(x)$ to $\vec{n}(-x)$ 
and $\vec{j}(x)$ to $\vec{j}(-x)$.  Time reversal commutes with translation
by $x_0$, $e^{-iPx_0}$. 
For the A b.c.\ sector, the field $\vec{n}(x)$ is antiperiodic, so its Fourier components have momenta $2\pi(m+1/2)/L$,
while $\vec{j}(x)$ remains periodic, with momenta $2\pi m/L$. (More formally, one can define 
the system as periodic on a covering space of length $2L$, with the additional condition that $\vec{n}$ be antiperiodic under 
translation by $L$.) So far, everything we said in this paragraph holds for both $\Theta=0$ and $\Theta=\pi$ (mod $2\pi$).
But for $\Theta=0$, the semiclassical collective-coordinate Hilbert space states have momenta $\pi m/L$,
and now we see that likewise the states in the full Hilbert space have momenta $\pi m/L$. For $\Theta=\pi$, the semiclassical ground
states have momenta $\pm \pi/(2L)$. Hence in this case, all states in the Hilbert space have 
momenta $2\pi(m\pm 1/4)$, and $\vec{n}(x)$ maps the momenta between the two congruence classes of momenta modulo $2\pi/L$. 
Similarly, all states in the Hilbert space have integer spin for $\Theta=0$, but all have half-integer spin
for $\Theta=\pi$. 
Then the statements above about the squares of the symmetry operators, and their mutual (anti-)commutation properties,
also hold as operator statements in the full Hilbert space in each case, and other consequences involving the eigenvalues
of the unitary discrete symmetries can be easily obtained.

Finally, we summarize all our results for the discrete symmetries $\widehat{I}$, $\widehat{\cal R}$, and $\cal T$
in the sigma model. For the discrete symmetries, here and below we will make use of the group-theoretic commutator 
$[a,b]=aba^{-1}b^{-1}$. For the spatial A b.c,\ sector, we have 
the relations (recall that $\Theta$ is an integer multiple of $\pi$)
\beq
\begin{array}{ll}
\widehat{I}^2=(-1)^{\Theta/\pi},& [\widehat{I},\widehat{\cal R}]=(-1)^{\Theta/\pi},\\
\widehat{\cal R}^2=(-1)^{\Theta/\pi},& [\widehat{I},{\cal T}]=1,\\
{\cal T}^2 = 1,&[\widehat{\cal R},{\cal T}]=1.
\label{arr:discrelssigma}
\end{array}
\eeq
We may mention again that the relations involving only $\widehat{I}$ and $\widehat{\cal R}$, and with $\Theta=\pi$ (mod $2\pi$),
can be understood by viewing these as representing elements of $Pin_-(2)\subset Spin(3)$, acting by right multiplication on the 
semiclassical states. In fact, here $\widehat{I}$ and $\widehat{\cal R}$ together generate the (finite) quaternion group $Q_8$ of order $8$.
For the spatial P b.c.\ sector, and also whenever $\Theta=0$ (mod $2\pi$), $\widehat{I}$, $\widehat{\cal R}$ 
reduce to $I$, $\cal R$, and $(-1)^{\Theta/\pi}$ is replaced by $1$. 

The fact that for the A b.c.\ and $\Theta=\pi$ (mod $2\pi$) the lift of inversion does not commute with the lift of
reflection (though it does commute with translations) makes clear that it is necessary to view $\widehat{I}$,
$\widehat{R}$, and the generator $P$ of translations as part of the same Lie group, which is not a direct product. 
We have identified this group as the double cover $Pin_-(2)$ of $O(2)$, and we can add a subscript, viz.\ $Pin_-(2)_{\rm sp}$
and $O(2)_{\rm sp}$, to make clear that typical elements of this group are spatial symmetries (i.e.\ they change the $x$ coordinate,
except for at most two $x$ that are fixed), 
not internal ones (which leave $x$ invariant). This point of view will be important in Appendix \ref{OtherSp}.
In collective coordinate terms, this group can be identified with the subgroup $Pin_-(2)_R\subset Spin(3)_R$, as we have seen, 
but exists {\it a priori}, and more generally, because of the spatial symmetries of the sigma model.

We can argue that the results so far of this subsection continue to hold even if the internal symmetry is reduced,
as long as it includes inversion (we also assumed translation, spatial reflection, and time-reversal symmetries). 
We can use the same arguments in the collective coordinate picture, as discussed briefly
in the preceding subsection, provided we allow the collective coordinate to leave the space of lowest energy textures, and pass through
some textures of higher energy, in order to obtain the results for reflection ${\cal R}$. Essentially, deforming the model that has full 
$O(3)$ internal symmetry (or its double cover) leads to modifications to the semiclassical wavefunctions, without changing the
results for the translations and the discrete symmetries [which form $Pin_-(2)_{\rm sp}$, plus time reversal symmetry];
the relations in eq.\ (\ref{arr:discrelssigma}) remain unchanged. For the A b.c.\ and when $(-1)^{\Theta/\pi}=-1$, four-fold ground-state 
degeneracy is necessary in order to represent those relations on the ground states (a representation as matrices is given 
in Sec.\ \ref{Sec:sigcft} below).

Returning to the case with symmetry under a subgroup of $SO(3)$ that contains $SO(2)$, as an aside from the main discussion we can make contact 
with the mixed anomaly, which involves both inversion and $SO(3)$ symmetry, and with the LSM theorem \cite{Lieb1961} also; 
these were related in Ref.\ \cite{10.21468/SciPostPhys.16.4.098}. 
(The relation of this part with the pure $\mathbb{Z}_2$ anomaly discussed in most of the paper will
be clarified further in Appendix \ref{OtherSp}.) This involves the center of the symmetry group $SU(2)$, which can be viewed as another 
discrete symmetry [and likewise for a similar subgroup of the double cover of $SO(2)$]. Consider again the semiclassical quantization
on the collective-coordinate space for the spatial A b.c.. We saw that the resulting states all have half-integer spin if and only if
$\Theta=\pi$ (mod $2\pi$). The A b.c.\ corresponds to imposing a background $\mathbb{Z}_2$ gauge field, or twist, in the $x$ direction. 
We can similarly impose a background non-Abelian gauge field, or a twist in the $t$ direction, for the $SO(3)$ symmetry (say by rotation 
about the $3$ axis), chosen so that the net twist is by $2\pi$ (apart from this, the temporal b.c.\ is P). In the operator or quantum mechanics 
point of view, this is $SO(3)$-gauge equivalent to inserting an operator that implements a global rotation by $2\pi$ in the trace. 
This is an element of the center, and so from our earlier results
acts on the states in the spatial A sector by multiplication by $(-1)^{\Theta/\pi}$. We also argued that all states in the full, and not only
the collective-coordinate, Hilbert space have the same allowed values of the spin [i.e.\ always integer for P and, for A, half integer 
if and only if $\Theta=\pi$ (mod $2\pi$)]. 
If we view the inclusion of the twist in the $x$ direction as an operator that interchanges the P and A b.c.\ spaces, then we have shown that its
(again, group-theoretic) commutator with $e^{-2\pi iJ_3}$ is $(-1)^{\Theta/\pi}$. In the path-integral point of view (for $\uptau$ pure imaginary), 
we can now interchange space and time (by modular $S$). We can view this in terms of an operator that imposes a twist by $2\pi$ in the 
$x$ direction (here for the spatial P b.c.), as an $SO(3)$ gauge transformation; this operator can be taken to be $e^{-2\pi i \int_0^L dx\,xj_3(x)/L}$ 
[we will not enter here into the form of $\vec{j}(x)$ in canonical quantization of the sigma model]. In the same way, we see that the commutator 
of this operator with $I$ (inversion, or $\mathbb{Z}_2$ twist in the $t$ direction) is $(-1)^{\Theta/\pi}$, so the operator reverses the sign of the 
eigenvalue of $I$. If we recall that, in the spin chain, inversion corresponds 
to translation by one site, and also that $\Theta=2\pi S$, then this result closely resembles the main step of the LSM argument
(discussed further in the following Section \ref{sec:spinch}), 
which shows that an operator that twists the ground state in the $x$ direction by angle $2\pi$ changes the crystal momentum by $\pi/a$ (mod $2\pi/a$) 
precisely if $S$ is half integer. 
Our result here says there is a mixed anomaly when $\Theta=\pi$ 
(mod $2\pi$). The role of the center of the group is apparent in the argument; it did not involve 
the Berry phase factor $\cal C$ directly, but did use the $SU(2)$ quantum numbers of the semiclassical states.

%%%%%%%%%%%%%%%%%%%%%%%%%%%%%%%%%%%%%%%%%%%%%%%%
%%%%%%%%%%%%%%%%%%%%%%%%%%%%%%%%%%%%%%%%%%%%%%%%

\section{\texorpdfstring{$SU(2)$}{SU(2)} spin chains semiclassically}
\label{sec:spinch}

In this section, we derive results about the quantum numbers
and symmetries of low-lying states of the spin chains by semiclassical methods,
and relate them to those derived from the sigma model. As in much of the two preceding
subsections, we will only consider spin chains with the full $SO(3)$ [or $Spin(3)$]
symmetry, though some results do extend to more general cases.
Because of the different forms of semiclassical textures, dividing
our material first according to whether $N$ is even or odd seems the most natural choice.

%%%%%%%%%%%%%%%%%%%%%%%%%%%%%%%%%%%%%%%

\subsection{\texorpdfstring{$N$}{N} even}

For the case of $N$ an even integer, our results will be simple, but are included
for completeness.
In this case, all states in the Hilbert space have total spin $j$
that is an integer for any value of $S$. There are many exact results for systems
with some form of the Hamiltonian $H$ as in eq.\ (\ref{spinchHam}), among which 
here we will first refer only to the fact that the ground state
has spin zero for all values $S=1/2$, $1$, $3/2$, \ldots,
a proof of which goes back to Marshall (and Peierls) \cite{Marshall1955} and LSM \cite{Lieb1961}. 
The underlying idea is to apply the Perron-Frobenius theorem, which says that, for a matrix in which
all off-diagonal entries are non-positive, the eigen- (column-) vector
with the lowest eigenvalue has all entries real non-negative (or non-positive).
This theorem can be used after a suitable transformation from the standard basis, 
leading to a rule for the sign of the entries of the eigenvector in the 
original product basis of $(S_{\ell 3})_\ell$ eigenstates, known as the Marshall sign rule. (This idea will recur in
Appendix \ref{sixver}.)

We also consider the crystal momentum. We recall that, if $\tau$ is the operator
for translation of a state of the system on a one-dimensional lattice of $N$ sites
to the right by $1$ site, where the system is viewed as periodic, then $\tau^N=1$. 
The translation operator is useful when the Hamiltonian $H$ is translation invariant, as ours are,
meaning that $H$ commutes with $\tau$.
Then energy eigenstates can also be classified by the eigenvalues
of $\tau$, which are $N$th roots of unity. It is conventional
to express these, using the analogy with translations in continuous space,
in terms of the logarithm of the eigenvalue of $\tau$ divided by $-ia$
(recall that $a$ is the lattice spacing), as a ``crystal'' momentum
$P'$, which is defined only modulo $2\pi/a$. It is also conventional
to choose representatives for the crystal momentum that lie
in the so-called Brillouin zone, which here is, say, $P'\in (-\pi/a,\pi/a]$
(we will assume the use of this Brillouin zone).
For the system of $N$ sites, $P'$ is a multiple of $2\pi/L$, where $L=Na$.
When $a\to0$, say with $L$ fixed, an ordinary continuum momentum $P$ (i.e.\ one that is well-defined as a real number) 
is recovered, with the correct dimensions (we caution however that the limit must be taken with some care,
and that the definitions of some symmetry operators in the sigma
model and in the spin chain may not agree even in the limit, as we will see).

There is a rigorous proof of the values of the crystal momentum of the ground state 
for all cases with $N$ even, again based on the Marshall sign rule (see Refs.\  \cite{PhysRevLett.60.639,BARWINKEL2000227}). 
The result is that (for $N$ even), for $S$ integer, $P'=0$,
while for $S$ half integer, $P'=0$ for $N=0$ (mod $4$), $P'=\pi/a$ 
for $N=2$ (mod $4$); note that these are integer multiples of $2\pi/L$
in all cases. In fact, precisely the same proof goes through \cite{AffleckLieb} for the spin and momentum of the ground
state (when $N$ is even) if a translation-invariant sum of longer-range terms $\vec{S}_\ell\cdot\vec{S}_{\ell'}$ 
is added to the nearest-neighbor 
Hamiltonian in eq.\ (\ref{spinchHam}), provided that it (i) only connects sites $\ell$, $\ell'$ 
with $\ell-\ell'$ (mod $N$) odd, and (ii) is antiferromagnetic [one can further include the 
same sum with opposite sign, so that it is ferromagnetic, but with only terms where $\ell-\ell'$ 
(mod $N$) is even]. The point here is that the conditions ensure that the Marshall sign rule
still holds. We comment that the same values of the ground state quantum numbers, namely spin zero
and $P'=0$, are found in the AKLT states \cite{AKLT}, each of which is the ground state of an antiferromagnetic spin chain 
in which $S$ is integer (the Marshall sign rule is again obeyed in these states).

Here we wish to examine the properties of the spin chain semiclassically, and see if the known results
are recovered. We can begin with an analog of a classical N\'eel-ordered state of the spin chain, 
using coherent states. A coherent state for a single spin of magnitude $S$ [i.e.\ an irreducible representation of spin $S$
of $SU(2)$, with generators $\vec{S}$] can be defined as a state $|\vec{n}\rangle$ such that 
$\vec{n}\cdot\vec{S}|\vec{n}\rangle=+S|\vec{n}\rangle$, where $\vec{n}$ is a unit vector. Of course, 
the phase of $|\vec{n}\rangle$ has not been defined, and when we want to define coherent states for all $\vec{n}$ at once, 
we must be careful about the phase. Let us define $|3\rangle$ to be a coherent state for the choice $\vec{n}=(0,0,1)^T$,
(the $^T$ denotes transpose) and then define $|\vec{n}\rangle=\widehat{O}|3\rangle$ where $\widehat{O}$ is the rotation operator, 
in our spin-$S$
representation, of any rotation $O$ that maps $(0,0,1)^T$ to $\vec{n}$. This still does not uniquely define the phase
of $|\vec{n}\rangle$. If we apply a rotation about the $3$ axis, $e^{-i\theta S_3}$, to $|3\rangle$, the latter 
changes by the phase factor $e^{-i\theta S}$. Then such a rotation multiplied into $\widehat{O}$ from the right
changes the phase of our $|\vec{n}\rangle$, though it leaves $\vec{n}$ unchanged. (We will show that these phases
drop out.) Now define $|-3\rangle=e^{-i\pi S_2}|3\rangle$, which is a coherent state with 
$\vec{n}=(0,0,-1)^T$. We use such coherent states at each site of our chain of spins $S$:
define $|(-1)^\ell 3\rangle_\ell$ to be the coherent state of the spin at site $\ell$ that is $|(-1)^\ell 3\rangle_\ell=|3\rangle$ 
for $\ell$ even, $|-3\rangle$ for $\ell$ odd.
Now define
\beq
|\vec{n}^N\rangle = \widehat{O}\bigotimes_\ell |(-1)^\ell 3\rangle_\ell,
\eeq
where again $\widehat{O}$ is the operator representing any rotation $\in SO(3)$ that maps $(0,0,1)^T$ to the given $\vec{n}$,
$\vec{n}=O(0,0,1)^T$, but now acting in the tensor product space of the spins on $N$ sites, so 
$\widehat{O}=e^{-i\vec{\theta}\cdot\vec{J}}$ where $\theta$ is a vector of parameters, and we recall that 
$\vec{J}=\sum_\ell \vec{S}_\ell$. 
[Thus $\widehat{O}$ is itself a tensor product over $\ell$ of the representation of $O$ for the $\ell$th site.] In this case, 
right multiplication of $O$ by a rotation about the $3$ axis produces opposite phase factors from the even and odd sites, which cancel 
in the product. Hence the state
$|\vec{n}^N\rangle$ for $N$ sites ($N$ even) is a well-defined function of $O$ that is invariant under
right multiplication of $O$ by an element of the $SO(2)$ subgroup of rotations about the $3$ axis,
and so is determined by a point $\vec{n}$ in $SO(3)/SO(2)\cong \mathbb{S}^2$, the unit sphere.

If we now apply the operator $\tau$ that maps each spin one step to the right, we see that we obtain
a similar state, except that the coherent states $|3\rangle$, $|-3\rangle$ have been exchanged for each site.
This differs from the original state by application of $e^{\mp i\pi S_{\ell 2}}$ to the coherent state 
$|(-1)^\ell 3\rangle_\ell$ for the $\ell$th site, 
where in $e^{\mp i\pi S_{\ell 2}}$ the $-$ sign is for $\ell$ even, $+$ sign is for $\ell$ odd. This is then $(-1)^{(2S)N/2}=(-1)^{NS}$ 
(from $2\pi$ rotation on $N/2$ sites) times multiplication of $e^{-i\pi S_{\ell 2}}$ into $\widehat{O}$ 
from the right for each site $\ell$, which is simply the global symmetry rotation on the system. 
Finally, right multiplication of $O$ by a rotation by $\pi$ about the $2$ axis is the same as inverting the sign of the $\vec{n}$ 
vector. Thus we have
\beq
\tau|\vec{n}^N\rangle=(-1)^{NS}|(-\vec{n})^N\rangle,
\eeq
and $\tau$ does give inversion of $\vec{n}$.

Now we apply semiclassical quantization using $\vec{n}$ as the collective coordinate to this family of states, with
the approach used for the sigma model in the previous section. That is, we consider a quantum-mechanical particle
moving on $\mathbb{S}^2$ (with coordinate $\vec{n}$), and we can assume the standard Hamiltonian $\propto \vec{J}^2$,
where as before $\vec{J}=\sum_\ell \vec{S}_\ell$ is the total angular momentum. 
The total angular momentum $j$ is integer, and the energy eigenvalues are $\propto j(j+1)$, with degeneracy $2j+1$.
The energy eigenstates can be expressed as $|j,m\rangle =\int d^2n\,Y_{jm}(\vec{n})|\vec{n}\rangle$, where $d^2n$
denotes the $O(3)$ invariant probability measure on $\mathbb{S}^2$. Then the ground state is the singlet, $j=0$.
(The fact that the low-lying collective-coordinate states of this form in an antiferromagnet have collective kinetic
energies $\sim 1/N$ was pointed out long ago \cite{PhysRev.86.694} in an effort to reconcile the existence of N\'eel 
order in higher-dimensional antiferromagnets with the fact that the ground state must be a spin singlet. Of course, 
as a spin chain is one-dimensional,
there is no long-range N\'eel order in the IR.)
Further, $\tau$ acts on these states with eigenvalue $\tau=(-1)^{NS}(-1)^j$. The $(-1)^j$ arises
because for the spherical harmonics $Y_{jm}(-\vec{n})=(-1)^jY_{jm}(\vec{n})$, as is well known. Then the
$(-1)^{NS}$ tells us that the ground state ($j=0$) crystal momentum is $P'=\pi NS/a$ (mod $2\pi/a$), which agrees 
for all cases ($N$ even, $2S$ even or odd) with the known result mentioned above: if $NS$ is odd, then $N/2$ and $2S$ 
are both odd. More generally, the $|j,m\rangle$ eigenstates have crystal momentum $P'=\pi(NS+j)/a$ (mod $2\pi/a$). 

We can also consider the operations of spatial reflection (or parity) and of time reversal on the semiclassical states
constructed here. In the spin chain, we will denote them by ${\cal R}'$ and ${\cal T}'$,
to distinguish them from those in the sigma model for the time being. The group of translations and spatial reflections 
of the periodic chain of $N$ sites is (quite generally)
the dihedral group $D_N$ of order $2N$; its generators can be taken to be translation $\tau$ by one site, and reflection,
$\ell\mapsto N-\ell$, leaving one site, here the origin $\ell=0$, fixed (for $N$ even, $\ell=N/2$ is also fixed.) For the state 
$|\vec{n}^N\rangle$
constructed above, we see that spatial reflection fixing the origin leaves it invariant, so also the eigenstates 
$|j,m\rangle$ are invariant: ${\cal R}'|j,m\rangle=|j,m\rangle$. For time reversal ${\cal T}'$, we use the standard definition
of its action on a spin system, such that it commutes with the spin rotation operators such as $\widehat{O}$
and with translation $\tau$ and reflection ${\cal R}'$; hence it anticommutes with the generators $\vec{J}$ 
and reverses the sign of crystal momentum (modulo $2\pi/a$). 
As the total spin is integer, time reversal ${\cal T}'$ squares to $+1$. Its action on the coherent states,
however, is less usual; compare the preceding subsection. As it commutes with the operator $\widehat{O}$,
we only require its action on each $|(-1)^\ell3\rangle_\ell$, which maps to $e^{-i\pi S_{\ell 2}}|(-1)^\ell 3\rangle_\ell$. 
Then we find that
\beq
{\cal T}'|\vec{n}^N\rangle=|(-\vec{n})^N\rangle,
\eeq
similar to translation, though here the overall sign is absent (and anyway
would not be very meaningful). This action of time reversal on the $\vec{n}$ vector is well known in this context.

%%%%%%%%%%%%%%%%%%%%%%%%%%%%%%%%%%%%%%

\subsection{\texorpdfstring{$N$ odd}{N odd}}

For odd number $N$ of sites, there are fewer rigorous results
on the ground-state quantum numbers. However, for $S=1/2$
and nearest-neighbor interactions, there are exact results from
the Bethe ansatz, also discussed in Appendix \ref{sixver}. For
other cases, there are exact results for small $N$, and
numerical results. In addition, the AKLT states \cite{AKLT} 
have ground state properties that are known exactly for $N$ odd 
as well as for $N$ even, but these states exist only for $S$ integer.

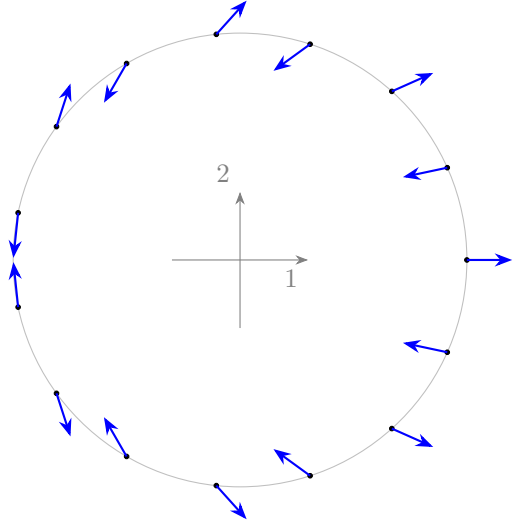
\begin{figure}[h]
\begin{center}

\begin{tikzpicture}[scale=3, >=Stealth]

\draw[gray!50] (0,0) circle(1);

\draw[->,  gray] (-.3,0) -- (.3,0) node[pos=1, below left] {\small 1}; % Horizontal axis
\draw[->,  gray] (0,-.3) -- (0,.3) node[pos=1, above left] {\small 2}; % Vertical axis

% Number of vectors
\def\N{15}

% Draw vectors
\foreach \j in {1,...,\N} {
 
  \pgfmathsetmacro{\phi}{360 * \j / \N}
  \pgfmathsetmacro{\bx}{cos(\phi)}
  \pgfmathsetmacro{\by}{sin(\phi)}

  \pgfmathsetmacro{\theta}{\j * 180 * (1 + 1/\N)}
  \pgfmathsetmacro{\dx}{0.2 * cos(\theta)}
  \pgfmathsetmacro{\dy}{0.2 * sin(\theta)}

  \filldraw[black] (\bx,\by) circle(0.01);

  \draw[->,thick,blue] (\bx,\by) -- ++(\dx,\dy);
}

\end{tikzpicture}
\end{center}\caption{A standard texture in the spin chain with $N$ odd, with both spins and lattice sites shown in the $12$ plane; 
here $N=15$. }
\label{nvecNodd}
\end{figure}

For $N$ odd, the N\'eel order is frustrated, and a low-energy
classical state must involve a texture in the $\vec{n}$ field, as we have already
discussed. We can construct a semiclassical texture from coherent states, similar to
what we did for $N$ even. Define the coherent state $|1\rangle$ for a single spin,
so that $\vec{n}\cdot\vec{S}|1\rangle=S|1\rangle$ where $\vec{n}=(1,0,0)^T$, and let $|1\rangle_\ell$ be this same coherent 
state for all $\ell$, so $S_{\ell 1}|1\rangle_\ell=+S|1\rangle_\ell$ for all $\ell$. Then define the following bijective 
function $\ell\mapsto\widetilde{\ell}(\ell)\equiv \widetilde{\ell}$
of $\mathbb{Z}_N$ to itself. It can be expressed through its inverse as 
$\ell(\widetilde{\ell})\equiv \ell=2\widetilde{\ell}$ (mod $N$), 
which is a bijection because $N$ is odd. Thus $\widetilde{\ell}=0$ when $\ell=0$, $\widetilde{\ell}=1$ when $\ell=2$, \ldots; 
$\widetilde{\ell}$ increases by $1$ each time $\ell$ increases by $2$, and continues to increase as $\ell$ passes through 
\ldots, $N-1$, $N+1\equiv 1$ (mod $N$), \ldots, finally reaching $\widetilde{\ell}=N\equiv 0$ when $\ell$ returns to 
$2N\equiv N\equiv 0$. The orientation of the coherent states at each site of the chain
will be described by the vectors $\vec{n}_{\ell}$. We choose them to be 
\beq
\vec{n}_\ell=(\cos \widetilde{\ell}(\ell)\theta_2, \sin\widetilde{\ell}(\ell)\theta_2,0)^T,
\eeq
where $\theta_2=2\pi/N$; see also Table \ref{Table} and Fig.\ \ref{nvecNodd}. This construction is periodic, 
and ensures that (if $N$ is large) the orientation $\vec{n}_\ell$ 
in terms of $\ell$ changes slightly when $\ell$ increases by $2$, and that for $\ell$ differing by $1$, 
$\vec{n}_{\ell+1}$ is nearly opposite $\vec{n}_\ell$. (Note that no branch cut is required for these $\vec{n}_\ell$
in the spin chain language.) The $\vec{n}_\ell$ vectors wind in the $12$ plane, just as in the sigma model
in the previous section. Correspondingly, define the coherent state for site $\ell$ by
\beq
|\vec{n}_{\ell'}\rangle_\ell=e^{-i\widetilde{\ell}(\ell')\theta_2 [S_{\ell 3}-S]}|1\rangle_\ell,
\label{cohstatell}
\eeq
where it will be useful to allow $\ell'$ to differ from $\ell$, and in the exponent a multiple of $\widetilde{\ell}$ was added,
chosen so that, as a function of $\widetilde{\ell}$, the coherent state is periodic with period $N$, and becomes a continuous path 
in $\mathbb{C}^2$ when $N\to\infty$. Then define a semiclassical state for a standard texture by 
\beq
|\vec{n}^N\rangle = \bigotimes_\ell |\vec{n}_\ell\rangle_\ell.
\label{StandText}
\eeq
(We can further apply a global rotation by $\widehat{O}$ on the left as we did for $N$ even, but let us not do 
so at the moment.) $|\vec{n}^N\rangle$ is manifestly invariant under translation by $N$ sites.

\begin{table}[h!]
\centering
\caption{Example with $N=7$. In a translation $\ell\to \ell+1$, each $|\vec{n}_l\rangle_l$ is rotated by $\pi+\frac{\pi}{7}$. 
}
\begin{tabular}{lccc}
\toprule
$\ell$ & $\tilde{\ell}$ &$ \tilde{\ell} \theta_2$ \\
\midrule
0 & 0 & 0  \\
1 & 4 & $\pi+\frac{\pi}{7}$  \\
2 & 1 & $\frac{2\pi}{7}$  \\
3 & 5 & $\pi+\frac{3\pi}{7}$  \\
4 & 2 & $\frac{4\pi}{7}$  \\
5 & 6 & $\pi+\frac{5\pi}{7}$  \\
6 & 3 & $\frac{6\pi}{7}$  \\
\bottomrule
\end{tabular}
\label{Table}
\end{table}

For an immediate simple calculation, we consider translation $\tau^2$ by two sites; this gives
\beq
\tau^2|\vec{n}^N\rangle =\bigotimes_{\widetilde{\ell}=0}^{N-1} 
|\vec{n}_{\ell(\widetilde{\ell}-1)}\rangle_{\ell(\widetilde{\ell})},
\eeq
where the subscript $\ell$ still labels the position of the spin in the chain.
The inner product of the translated state with the original is then
\beq
\langle \vec{n}^N|\tau^2|\vec{n}^N\rangle = \prod_{\widetilde{\ell}=0}^{N-1}\,_{\ell(\widetilde{\ell})}\langle 
\vec{n}_{\ell(\widetilde{\ell})}
|\vec{n}_{\ell(\widetilde{\ell}-1)}\rangle_{\ell(\widetilde{\ell})}
\eeq
The right-hand side is the same as the product of overlaps of coherent states for a single spin that 
moves along a discrete path on a great circle on $\mathbb{S}^2$ , $\prod_{\widetilde{\ell}=0}^{N-1}\langle \vec{n}_{\widetilde{\ell}}
|\vec{n}_{\widetilde{\ell}-1}\rangle$. (Note that, in such a product, which is an example of a so-called Bargmann invariant,
any changes in the phases of $|\vec{n}_\ell\rangle$ for each $\ell$ cancel.) It is well known that, when a discrete closed path of 
coherent states is smoothly varying so that it becomes a continuous closed curve when the limit $N\to\infty$ is taken, 
the result is the holonomy or Berry phase factor of the single spin of magnitude $S$ moving along the path, as discussed in Section \ref{SpinCh}.
In our case, the path is a great circle 
on $\mathbb{S}^2$, and so the net phase factor is $(-1)^{2S}$. 
We note that this result is robust: if the path, say $\vec{n}(t)$, where $t$ is a continuous version
of $\widetilde{\ell}$, running from $0$ to $\beta$, is not a great circle, but obeys $\vec{n}(t+\beta/2)=-\vec{n}(t)$,
as it does in our case to ensure neighboring spins are nearly opposite, then we still find $(-1)^{2S}$, 
exactly as we already discussed in Sec.\ \ref{subsec:pathintmodtrans} (see also Appendix \ref{subsec:thm1}).

This result implies, first, that when $N\to\infty$, the state translated by two sites is the original state up to a phase factor,
so essentially an eigenstate of $\tau^2$; if it were not, the inner product would have magnitude less than $1$. Second, by taking 
a square root it implies that in semiclassical quantization (to which we turn in a moment), all states have crystal momentum 
close to $0$ or $\pi/a$ if $S$ is integer, and close to $\pm \pi/(2a)$ if $S$ is half integer. (One reason they can only be 
close, not equal, to these values is that $\pi/a$, $\pm \pi/(2a)$ are not integer multiples of $2\pi/L$ if $N$ is odd.)

We comment here that this result still holds if the spin-chain Hamiltonian only has $SO(2)\subset SO(3)$ internal symmetry (or its double cover),
because the lowest-energy texture can wind in the same way, but if the symmetry is reduced further to a discrete subgroup, say by adding 
weak symmetry-breaking terms, then there would be a domain wall
in the texture, and the Berry phase in the $N\to\infty$ limit will have a correction to the above result of order $a$ divided by the inverse width 
of the domain wall, not order $1/N\to 0$. To obtain the sigma model with explicit breaking of the $O(3)$ symmetry, it is always necessary to scale the 
symmetry breaking terms to go to zero as $a\to0$ (with $L=Na$ fixed) in the continuum, and in that case the spin chain and sigma model results agree. 

Next we wish to examine the semiclassical quantization of the collective coordinate $O$ from the point of view
of the spin chain. But first, to complete the mapping from the spin chain to sigma model when $N$ is odd, we consider the
fate of the Berry phase terms in the spin chain action, and compare with our invariant in the sigma model
with antiperiodic spatial b.c.. In this case, the N\'eel vector $\vec{n}$ must obey an antiperiodic spatial b.c.,
and we will again use a branch cut to describe it, with the cut at $x=0^-$. If we repeat the derivation in Sec.\ \ref{SpinCh} 
when $N$ is odd, we notice that, in the grouping of the spins into pairs of neighbors, say $(1,2)$, $(3,4)$, \ldots, 
$(N-2,N-1)$ as before, there is now one left over, namely that at $\ell=0$. Then the Berry phase for that
spin must be added to the topological term evaluated with the branch cut (which on its own is not invariant under homotopy);
note that both contain the same coefficient $S$. Then the result is exactly the invariant $\Theta{\cal N}'$, where $\Theta=2\pi S$, 
that we discussed in detail in Section \ref{subsec:pathintmodtrans} for the case $S$ half integer, or $\Theta=\pi$ (mod $2\pi$). 
For $S$ integer or $\Theta=0$ (mod $2\pi$), the whole corresponding term can be dropped as $\Theta{\cal N}'$ is always a multiple of $2\pi$.

We may now proceed to the quantization of the collective-coordinate space consisting of all the 
states $|O\rangle=\widehat{O}|\vec{n}^N\rangle$ 
that we have constructed. Similarly to the spatial P sector, for the spin quantum numbers of the semiclassical states,
this is essentially the same as the semiclassical treatment of the sigma model. That can include the use of the non-trivial
connection on $SO(3)$ that results from the invariant $\cal I$, which gave rise to spins $j$ that were half integer
when $\Theta=\pi$ (mod $2\pi$). However, for the spin chain, this fact is obvious because it corresponds to the case
$N$ odd, $2S$ odd. We can move directly to the results. The collective-coordinate space energy eigenstates are given 
by (see Appendix \ref{PetWeyl})
\beq
|j,m_L,m_R\rangle = \int d^3O\, \overline{D^j_{m_L,m_R}(O)}\, |O\rangle,
\eeq
where $D^j_{m_L,m_R}$ are again the Wigner matrices,
and $d^3O$ denotes Haar measure on $SO(3)$ for $j$ integer, or on $Spin(3)\cong SU(2)$ in general. The energy eigenvalues 
are proportional to those for the sigma model, eq.\ \ref{HAevals}. Hence the ground states are non-degenerate 
($j=0$) for $S$ integer, 
and four-fold degenerate ($j=1/2$) for $S$ half integer.

We now turn to the remaining quantum numbers, beginning with translations. The following calculation, though simple, 
is among the most important in the paper. For the translation of the $\ell$th site, we find
$\tau|\vec{n}_\ell\rangle_\ell=e^{i\pi(N+1)[S_{\ell+1, 3}-S]/N}|\vec{n}_{\ell+1}\rangle_{\ell+1}$ for all $\ell$.
For the standard texture, eq.\ (\ref{StandText}), we have similarly
\beq
\tau|\vec{n}^N\rangle = (-1)^{(N+1)S} e^{i\pi (1+1/N)J_3}|\vec{n}^N\rangle.
\eeq
(As a check, we must have $\tau^N|\vec{n}^N\rangle=|\vec{n}^N\rangle$, which follows from this formula for all odd $N$ and all $S$.)
When we apply this to the states $|j,m_L,m_R\rangle$ in the semiclassical quantization above, $e^{i\pi(1+1/N)J_3}$
gives rise to a right rotation, which commutes with global (left) rotations. We then find for $S$ integer,
in which case $j=m_L=m_R=0$ for the ground state, that the crystal momentum is $P'=0$ (for $N$ odd). On the other hand, for 
$S$ half integer, we find for the ground states ($j=1/2$, $m_{L,R}=\pm 1/2$)
\beq
P'=\left\{\begin{array}{ll} {\pm \left(1-\frac{1}{N}\right)\frac{\pi}{2a}} &{\hbox{($N=1$ mod $4$)}},\\
                          {\pm \left(1+\frac{1}{N}\right)\frac{\pi}{2a}} &{\hbox{($N=3$ mod $4$)}};
                          \end{array}\right.
\eeq
note that these values are integer multiples of $2\pi/L$ in each case,
and can be characterized as being as close as possible to $\pm\pi/(2a)$, subject to that condition.
These values of the spin, degeneracy, and crystal momentum of the ground states for $N$ odd and for all $S$ agree exactly with those 
found numerically (see e.g.\ Refs.\ \cite{BARWINKEL2000227,ChengSeiberg})
and for $S=1/2$ in the Bethe ansatz. (For $N=1$, both calculations agree with the trivial result of a unique multiplet 
of $2S+1$ states and $P'=0$, the full Hilbert space.) We also comment that for $S$ half integer, comparing with $N$ even, 
the (approximate) values $\pm \pi/(2a)$ lie midway between the values $P'=0$ for $N=0$ (mod $4$) and $P'=\pi/a$ for $N=2$ (mod $4$),
while for $S$ integer, $P'=0$ in the ground state for all $N$. The last result, $P'=0$ in the ground state
(or AKLT state) is also found for all $N$ in the AKLT spin chains, in which $S$ is integer \cite{AKLT}. 

We comment here that there is some similarity with the method of proof of the LSM theorem
\cite{Lieb1961}, which was for $N$ even. That proof involved applying a twist, given by the operator
\beq
\prod_\ell e^{-i \theta_2 \ell S_{\ell 3}},
\eeq
to the ground state of the chain (notice it involves $\ell$, not $\widetilde{\ell}$). Under translation $\tau$ by $1$ site, 
 so $S_{\ell 3}\to S_{\ell+1,3}$ for all $\ell$, the twist operator maps to the same operator multiplied by both $(-1)^{2S}$ and 
a global rotation $e^{i \theta_2 J_3}$. 
As the ground state is a $J_3$ eigenstate with eigenvalue $0$, this implies that for $S$ half integer, the crystal momentum 
of the twisted state differs by $\pi/a$ (modulo $2\pi/a$) from that of the ground state. (In any case, the global rotation by $\theta_2$
tends to the identity as $N\to\infty$; cf.\ the end of the preceding Subsection \ref{Sec:transdisc}.) We can apply the same argument to
such a twist of our semiclassical collective-coordinate states for $N$ even, and obtain exactly the LSM result for the
change of crystal momentum (as $m_L=j=0$ in the ground state).
Our state $|\vec{n}^N\rangle$ for $N$ odd likewise already contains a slowly-varying twist, but differs in that it reaches only 
$\pi$ on going the length of the system (to satisfy the b.c.), and also in that neighboring sites differ by a rotation 
close to $\pi$, in order to retain low energy. Under translation of $|\vec{n}^N\rangle$ for $N$ odd by $2$ sites 
(i.e.\ applying $\tau^2$), we obtain a phase factor close to $(-1)^{2S}$ by an argument almost identical to that 
for the LSM twist of the $N$ even ground state. Thus our results for the ground-state crystal momentum for $N$ odd 
are a sort of square root of the LSM result, in particular, the eigenvalues of $\tau$ are close to $\pm i$ for $S$ half integer, 
the square roots of $-1$. 

We now calculate the effect ${\cal R}'$ of reflection through the origin for our standard texture
state $|\vec{n}^N\rangle$, eq.\ (\ref{StandText}). This sends $\widetilde{\ell}$ to $-\widetilde{\ell}$ (mod $N$),
so reverses the sign in front of $S_{\ell 3}-S$ in the exponent. That is the same as a rotation
of the operator $S_{\ell 3}$ by $\pi$ about the $1$ axis: $e^{-i\pi S_1}S_3e^{i\pi S_1}=-S_3$, plus an additional 
phase. After a short calculation we find
\beq
{\cal R}'|\vec{n}^N\rangle = e^{i\pi N S}e^{-i\pi J_1}|\vec{n}^N\rangle.
\eeq 
Clearly we must have ${\cal R}'^2=+1$, and we see that ${\cal R}'^2= e^{2\pi i NS -2\pi i J_1}=(-1)^{4NS}=1$.
${\cal R}'$ rotates the texture by $\pi$ about the $1$ axis, which agrees with the right-rotation picture in the collective 
coordinate language for the sigma model. 

For time reversal ${\cal T}'$, we have to work a little harder. We know that, in the basis of $S_3$ eigenstates for each spin, 
with the choice of phases such that $S_\pm$ have real positive matrix elements in that basis, time reversal
acts by complex conjugation of coefficients, followed by rotation by $\pi$ about the $2$ axis. To apply this definition,
we first write the standard texture in eq.\ (\ref{StandText}) as $|\vec{n}^N\rangle=\widehat{O_{123}}|\vec{n}'^N\rangle$, 
where here we use $|\vec{n}'^N\rangle =\bigotimes_\ell e^{-i\widetilde{\ell}\theta_2S_{\ell 2}}|3\rangle_\ell$ and $|3\rangle_\ell$ 
is the same for each $\ell$, and equal to the basis state such that $S_{\ell 3}|3\rangle_\ell=S|3\rangle_\ell$ as before. 
(The phase of the standard texture state $|\vec{n}^N\rangle$  differs from that used earlier, but is more convenient here.)
We also choose $\widehat{O_{123}}$
to be the representation of the rotation $O_{123}$ by $2\pi/3$ about the unit vector $\frac{1}{\sqrt{3}}(1,1,1)^T$, so
$\widehat{O_{123}}=e^{-2\pi i(J_1+J_2+J_3)/(3\sqrt{3})}$. This rotation cyclically permutes 
the axes, $1$ to $2$ to $3$ to $1$, so $\widehat{O_{123}}S_{\ell 1}\widehat{O_{123}}^{-1}=S_{\ell 2}$, and cyclic permutations
thereof. We can also define $|1\rangle_\ell=\widehat{O_{123}}|3\rangle_\ell$, because the phase of $|1\rangle_\ell$ was 
not specified previously. Then because ${\cal T}'$ commutes with any rotation, in particular $\widehat{O_{123}}$,
it is sufficient to act on $|\vec{n}'^N\rangle$; $e^{-i\widetilde{\ell}\theta_2S_{\ell 2}}|3\rangle_\ell$ is real for all $\ell$, 
so on the texture $|\vec{n}'^N\rangle$, ${\cal T}'$ reduces to $e^{-i\pi J_2}$, which can be moved back past $\widehat{O_{123}}$. 
The result is that, for the choice of phase used in this paragraph,
\beq
{\cal T}'|\vec{n}^N\rangle = e^{-i\pi J_3}|\vec{n}^N\rangle,
\eeq
which in terms of the $\vec{n}_\ell$ vectors has the effect of reversing the sign of each, as in the spatial P b.c.\ case.
Similarly, ${\cal T}'^2 = (-1)^{2NS}=(-1)^{2S}$ because $N$ is odd here. Thus for $S$ half integer, ${\cal T}'$ squares to $-1$,
as it must in the standard definitions for time reversal ${\cal T}'$ in the spin chain when $N$ is odd.

Finally, we consider commutation properties of the various discrete symmetries acting in the spin chain. 
Translations, reflections, and time reversal together generate a group. Then for its representation in the spin chain 
(not a unitary representation, because time reversal is an antiunitary operator, the others unitary), we have the set
of exact operator relations (valid for all positive integer values of $N$ and $2S$):
\beq
\begin{array}{ll}
\tau^N=1,& {\cal R}'\tau{\cal R}'^{-1}=\tau^{-1},\\
{\cal R}'^2=1,& [\tau,{\cal T}']=1,\\
{\cal T}'^2 = (-1)^{2NS},&[{\cal R}',{\cal T}']=1
\label{discrels}
\end{array}
\eeq
(where again we use group-theoretic commutators). Here the relations not involving ${\cal T}'$ give a standard presentation 
of the dihedral group $D_N$
in terms of generators and relations. We already mentioned that time reversal commutes with translations and reflections,
as can be seen using a product basis for the space of states of the spin chain. 
We continue this discussion in the first subsection of the following section.

%%%%%%%%%%%%%%%%%%%%%%%%%%%%%%%%%%%%%%%%%%%%%%%%%%%%%%%
\section{Relations among models}
\label{sec:relmod}

In this section we first discuss the final relations among the models and their IR limits,
first for the spin chains at large $S$ with the sigma model at weak coupling, then later
for the IR QFT of the sigma model in the two cases $\Theta=0$, $\pi$ (mod $2\pi$).
We also discuss the meaning of the gauged sigma model when it is not anomalous.
In this section, we concentrate on cases with full $SO(3)$ [or $SU(2)$] internal symmetry.

%%%%%%%%%%%%%%%%%%%%%%%%%%%%%%%%%%%%%%%%
\subsection{Spin chain and sigma model}
\label{sec:spchsig}

We now relate the symmetries in the spin chain to those in the sigma model, 
beginning with the discrete symmetries.
With the help of results we derived semiclassically, in eqs.\ (\ref{discrels}) at leading order 
we can replace $\tau^N=1$, and simplify another relation involving $\tau$ when $N$ is large, to replace the first line 
of display (\ref{discrels}) by
\beq
\tau^2\simeq (-1)^{2NS},\qquad [\tau,{\cal R}']\simeq (-1)^{2NS},
\eeq
which we will first treat as exact equalities, leaving the subleading part of $\tau$ for discussion afterwards.
These relations are valid at leading order for any (large) $N$, whether even or odd, and for any (large) $S$, 
whether integer or half integer. 

It is clear that, as $N\to\infty$, translation by one site should, at leading order, become (a lift of) inversion, 
$\tau\sim \widehat{I}$ as $N\to\infty$; this holds for $N$ either even or odd, and for $S$ either integer or half integer. 
Hence, we have $\widehat{I}^2=(-1)^{2NS}$. Next, we have seen that
time reversal in the spin chain has the effect of reversing the sign of $\vec{n}$, so we construct ${\cal T}=\widehat{I}{\cal T}'$
in which this effect is removed; $\cal T$ is again antiunitary, but now ${\cal T}^2=+1$. (Some literature relates the two models 
in the reverse way, so that time reversal, and possibly also spatial reflection, in the sigma model agree with ${\cal T}'$, ${\cal R}'$ 
in the spin chain; see e.g.\ Ref.\ \cite{Sulejmanpasic}.) Finally, reflection in the spin chain squares to $1$,
thanks to a phase factor in the semiclassical expression. We remove that factor by defining $\widehat{\cal R}=e^{-i\pi NS}{\cal R}'$; 
the phase factor is real if $N$ or $2S$ is even, imaginary if $N$ and $2S$ are both odd, so it
gives $\widehat{\cal R}^2=(-1)^{2NS}$, and  in the second case it affects 
the commutation relation with time reversal (for the case $N$ even, we can omit the phase factor and define 
$\widehat{\cal R}={\cal R}'$). Then in sum, in the spin chain treated semiclassically, we obtain the relations
\beq
\begin{array}{ll}
\widehat{I}^2=(-1)^{2NS},& [\widehat{I},\widehat{\cal R}]=(-1)^{2NS},\\
\widehat{\cal R}^2=(-1)^{2NS},& [\widehat{I},{\cal T}]=1,\\
{\cal T}^2 = 1,&[\widehat{\cal R},{\cal T}]=1,\end{array}
\eeq
which are valid for all large $N$ and $S$.
If we recall that the b.c.\ P corresponds to $N$ even, A to $N$ odd, and that in the correspondence of the 
spin chain and the sigma model $2S=\Theta/\pi$ (mod $2$), then we see that these relations are exactly those [eq.\ (\ref{arr:discrelssigma})]
we obtained in the sigma model in the Section \ref{Sec:transdisc}. With this in mind, it will be more convenient in the following 
to continue referring to the parity of $2NS$ [i.e.\ to the sign of $(-1)^{2NS}$], rather than to the corresponding cases of spatial
b.c.\ and value of $\Theta$ modulo $2\pi$. (When $2NS$ is even, the hats on $\widehat{I}$, $\widehat{\cal R}$
can be removed.) Moreover, the action of these operators on the semiclassical texture 
states is the same as in the sigma model. This justifies the use here of the same notation
for the corresponding symmetry operations. 

Now we return to the operator $\tau$, and consider its eigenvalues more precisely. The crystal momenta $P'$ 
of the low-lying energy eigenstates in the semiclassical approximation are multiples of $2\pi/L$ (because $\tau^N=1$), and lie
close to the values $0$, $\pi/a$ for $2NS$ even, and $\pm \pi/(2a)$ for $2NS$ odd. (Recall our convention
of using a Brillouin zone for crystal momentum, rather than repeating the phrase ``modulo $2\pi/a$''.) 
Then within this description we are justified in defining an operator $\widehat{I}$ that has eigenvalues $1$, $-1$
in the first case, $i$, $-i$ in the second, according to the values of crystal momentum as described.
That is, we define two orthogonal subspaces of Hilbert space, the direct sum of which is the full 
space, and assume this is well defined. (The operator $\widehat{I}$ so-defined can be viewed as a ``spectral flattening''
of $\tau$, as each eigenvalue has been mapped to one of two possible values, in a manner we assume is well defined.
For $N$ finite, we can of course force it to be well defined, by dividing the Brillouin zone into two parts, with cuts close
to $P'=\pm \pi/(2a)$ for $2NS$ even, and $P'=0$, $\pi/a$ for the $2NS$ odd. Then the question is whether the somewhat arbitrary 
choices of these cuts disappear in the continuum limit.) Then we define the momentum of the states in the continuum model
as the difference from the crystal momentum associated to $\widehat{I}$, and with the same assumption this is now well defined
as a real number; thus the definition implies that $\tau=\widehat{I}e^{-iP a}$. The resulting momenta $P$ 
are integer multiples of $2\pi/L$ for $2NS$ even, but of the form $\pm 2\pi(m+1/4)$, $m\in\mathbb{Z}$, for $2NS$ odd, 
with the sign depending on whether the eigenvalue of $\widehat I$ is $i$ or $-i$, such that the crystal momentum 
is always a multiple of $2\pi/L$.  From the exact relation
${\cal R}'\tau{\cal R}'^{-1}=\tau^{-1}$ (or ${\cal R}\tau{\cal R}^{-1}=\tau^{-1}$), we find that, 
for the continuum $P$, we have ${\cal R}P{\cal R}^{-1}=-P$, as in the continuum sigma model
or indeed in most QFTs.

If we consider translation by $N$ sites in these terms, then we know that $\tau^N=1$. This gives the relation 
\beq
e^{-iPL}=\widehat{I}^{-N}.
\eeq
In the Introduction, we defined the lift of inversion using translation by $L$ in the sigma model or in the CFT,
as on the left-hand side of the equation. For $2NS$ odd, $\widehat{I}^4=1$, and then (as $N$ must be odd)
the right-hand side is either the lift $\widehat{I}$ we have defined here 
or its inverse, according to whether $N=3$ or $1$ (mod $4$), respectively. For $2NS$ even, $\widehat{I}$ reduces to $I$, 
with $I^2=1$ (or $I^{-1}=I$). From this point of view it seems natural to say that the non-degenerate ground state has $I=-1$
when $N$ is $2$ modulo $4$ and $S$ is half integer; this results from the crystal momentum $\pi/a$
of the ground state in this case.

If the momenta of the low-lying states are given in the various cases, then what we have described is exactly the ``emanent'' symmetry
discussed by Cheng and Seiberg: the discrete translations of the chain become continuous translations
of the continuum, but leave a non-trivial discrete symmetry because not all states have crystal momentum close
to a single value, such as $0$. The latter symmetry separates from the continuous translations in the limit, but
there is mixing between them because of constraints coming from the lattice. In our treatment, the quantum numbers of
states in the spin chain were obtained semiclassically, independently from the IR limit, and we identified the (lift of the) inversion
symmetry with part of an $O(3)$ [or a $Pin_+(3)]$ symmetry in the continuum sigma model, even before passing to
the IR limit [which is either a massive QFT or a CFT, depending on whether $\Theta=0$ or $\pi$ (mod $2\pi$)]. 

A more complete point of view, as mentioned already, is that in general in the sigma model with $\Theta=\pi$ (mod $2\pi$)
and spatial A b.c., the lift $\widehat{I}$ of the inversion can be viewed as belonging with the spatial symmetries, 
and the full symmetry group is 
\beq
[Spin(3)\times Pin_-(2)_{\rm sp}]/\mathbb{Z}_2,
\eeq
and not only the internal symmetry group 
$Pin_+(3)\cong [Spin(3)\times \mathbb{Z}_4]/\mathbb{Z}_2$; here the lift $\widehat{I}$ of the inversion generates 
the group $\mathbb{Z}_4\subset Pin_-(2)_{\rm sp}$. For the $P$ b.c., we can refer to $I$ as belonging to the group of internal 
symmetries $O(3)$, with no relation with $O(2)_{\rm sp}$, but for the spatial A b.c.\ and $\Theta=0$ (mod $2\pi$) its action 
coincides with that of spatial translation by $L$, which is in the center of $O(2)_{\rm sp}$, and we can view the 
full symmetry group as $[O(3)\times O(2)_{\rm sp}]/\mathbb{Z}_2$.

%%%%%%%%%%%%%%%%%%%%%%%%%%%%%%%%%%%%%%%%%%%%%%%%%%%%%%%%%%%
\subsection{Sigma model at \texorpdfstring{$\Theta=\pi$}{Theta=pi} and CFT in the IR limit}
\label{Sec:sigcft}

Finally, we can consider the QFTs found in the IR limit. For this, we may as well discuss the IR limit of 
the sigma model, since our results for the spin chain at large $S$ already map to that model.
We begin this discussion in this subsection, with the $SU(2)$ level $1$ CFT for the $\Theta=\pi$ (mod $2\pi$) sigma model
[or the spin chain, which is expected to flow to the same IR limit for all half-integer $S$], as discussed in the Introduction, 
for both P and A spatial b.c.\ sectors. [The $\Theta=0$ (mod $2\pi$) case will be discussed in the following subsection.]
The momentum values have already been covered,
so we turn again to discrete symmetries. The operators of the chiral (and anti-chiral) algebras are the right- and left-moving
current algebras generated by the corresponding currents. We already defined $\vec{j}(x)$ to be the spin density; now we redefine
this as $\vec{j}_t(x)$. There is also the spin current $\vec{j}_x(x)$, and together these obey 
$\partial_t \vec{j}_t+\partial_x\vec{j}_x=0$ whenever there is $SO(3)$ [or $SU(2)$] symmetry. In a CFT we 
then define $\vec{j}=\vec{j}_t+i\vec{j}_x/v$, $\overline{\vec{j}}=\vec{j}_t-i\vec{j}_x/v$, which are the right- and 
left-moving currents, respectively; we will also refer to the ``zero modes'', or global symmetry generators, 
associated with these currents, $\vec{J}=\int dx\, \vec{j}(x)$, $\overline{\vec{J}}=\int dx\,\overline{\vec{j}}(x)$. 
The right- (left-) moving Virasoro generators can be expressed quadratically in terms 
of the right- (resp., left-)  moving currents. The currents are invariant under inversion $\widehat{I}$ (which reduces to 
$I$ for the P sector), and both time reversal and spatial reflection (parity) interchange right- and left-movers, 
${\cal T}\vec{j}(x){\cal T}^{-1}=-\overline{\vec{j}}(x)$, ${\widehat{\cal R}}\vec{j}(x)\widehat{\cal R}^{-1}=\overline{\vec{j}}(x)$. 
Hence it is sufficient to examine the action of the discrete symmetries on the ground states of each sector, in other words 
on the (left times right) primary fields, with left and right $SU(2)$ spins $0$ or $1/2$. For the P sector, we already discussed 
how $I$ acts as $+1$ on the $(0,0)$ primary field (the identity) or ground state, and presumably as $-1$ on the $(1/2,1/2)$ 
primary field multiplet, which in the WZW model is the $2\times 2$-matrix field $g(x,t)$, so we have the symmetry $g(x,t)\to-g(x,t)$.
We pointed out that the latter is truly a symmetry of the CFT because it preserves the fusion rules involving these fields.
We noted before that the $\vec{n}$ field should become part of this multiplet in the IR. More precisely, and in more detail, 
we have the following. In the IR, we can express the relation as an asymptotic series (although we focus mainly on the first term),
\beq
\vec{n}(x,t)\sim {\rm tr}\,\vec{\sigma} g(x,t) +\ldots,
\eeq
in which $\vec{\sigma}$ is again the vector of Pauli matrices, constant coefficients could be inserted 
on the right-hand side, but can be changed by defining the expansion at a different scale, and 
the terms are in order of increasing scaling dimension of the fields. Finally, the limit in which the whole 
expansion holds asymptotically is that in which correlation functions are considered in an infinite system, 
and all locations of fields in the $z=x+ivt$ plane (relative to some origin) are scaled up by the same factor, as that factor 
tends to infinity. The subleading terms omitted must be current-algebra descendants of the $g$ field
that transform as vectors, and odd under inversion, like the left-hand side, under global $O(3)$. 

Time reversal and reflection can act simply on the ground states, mapping them to themselves [if the usual basis of $J_3$, 
$\overline{J}_3$ eigenstates is used, a $2\pi$ rotation is needed for both right- and left-movers in the $(1/2,1/2)$ case]. 
For the A sector, 
because we know the Hilbert space and algebra structure of the CFT in detail, there are many ways in which 
to define discrete symmetries with the required properties. We will impose definite transformations based on those 
we found in the $O(3)$ sigma model. There are four ground states, which form 
two spin-$1/2$ multiplets under the global (right plus left) $SU(2)$. We pick one spin $1/2$ multiplet to be the $(1/2,0)$
ground states, the other the $(0,1/2)$ ground states. This is a direct sum of two subspaces each with global spin $1/2$, 
but we will express it as the tensor product of the global spin $1/2$ multiplet with another space of dimension $2$ 
(another ``spin $1/2$'', but it is not associated with an $SU(2)$ symmetry); the second tensor factor
is used to distinguish right- from left- movers. Using the standard choice for time reversal again, the basis states
are real, and time reversal $\cal T$ acts as complex conjugation in this basis, followed by multiplication
by $i\sigma_y\otimes i\sigma_y$ (where $\sigma_x$, $\sigma_y$, $\sigma_z$ are again the standard Pauli matrices;
the overall phase factor is unimportant), and then ${\cal T}^2=1$. Then, reduced to this subspace, 
we have $\widehat{I}={\bf 1}\otimes i\sigma_z$, and $\widehat{\cal R}={\bf 1}\otimes i\sigma_x$, which are precisely
the lifts of $\pi$ rotations about the axes, as we have discussed already. It should be clear how these expressions generalize 
as block matrices to the full Hilbert space. These three operators obey the relations
above.

We have already seen how there is now an anomaly that prevents the construction of a modular-invariant CFT
in which the $g(x,t)\to-g(x,t)$ symmetry is gauged (see Ref.\ \cite{gw86}), and a similar one already in the sigma model. 
Then although the different b.c.\ sectors,
and correlation functions in each sector that may contain the $(1/2,0)$ or $(0,1/2)$ ``twist'' primary fields or their descendants, 
are well defined (though not single-valued; the twist fields are endpoints for branch cuts as in our earlier discussion, 
and the correlation functions are double valued due to the conformal spin $\pm 1/4$ of these fields), 
no meaning (or at least, no nonzero values) can be given to, for example, correlation functions containing 
$g(x,t)$ or the twist fields in the gauged (or orbifold) CFT, nor for that matter in the $\Theta=\pi$ (mod $2\pi$) sigma model 
back in the ultraviolet (UV). [Similarly in the spin chains, where one could perform an average of correlation functions over 
a small range of sizes $N$ that includes both some even and some odd $N$ values with equal total weight for even and for odd.
We note that such sums also arise in other contexts, where they can lead to different field theories; see e.g.\ the relation 
between dense $U(n)$ and $O(n)$ loop models' CFTs discussed in Ref.\ \cite{rjrs24}.]

%%%%%%%%%%%%%%%%%%%%%%%%%%%%%%%%%%%%%%%%%%%%%%%%%%%
\subsection{\texorpdfstring{$\Theta=0$, $\mathbb{R}\mathbb{P}^2$}{Theta=0, RP2} sigma model, and non-anomalous gauge theory}
\label{Discussion}

For the IR limit of the $\Theta=0$ (mod $2\pi$) case, we have a massive QFT. In fact, the $O(3)$ sigma model in this case 
is integrable as a QFT; its $S$-matrices were found long ago \cite{ZAMOLODCHIKOV1979253}, and later a Bethe ansatz solution 
was found also \cite{WIEGMANN1985209}. So it is known that the particle spectrum consists of an $SO(3)$ vector
of massive scalar particles, nothing more, and further it has $\mathbb{Z}_2$ symmetry, acting as $-1$ to the power of the 
particle number. The field $\vec{n}$ asymptotically becomes the field operator for the massive particle. Unlike the 
$\Theta=\pi$ case, for this case there is no anomaly to prevent the gauging of the inversion symmetry. For a system of 
finite length $L$, we obtain the gauged or orbifold theory, which looks as follows. We have the usual spatial P 
b.c.\ sector, with states of total momentum a multiple of $2\pi/L$, and the ``twisted'' or spatial A b.c.\ sector, with 
a ground state of momentum $0$, but each particle has momentum a half-integer multiple of $2\pi/L$ because of the A b.c.\ 
(by reference to the Fourier series for the correlation functions of the $\vec{n}$ field). Using the normalization associated 
with the orbifold point of view (as in e.g.\ Ref.\ \cite{gw86}), the sum 
of partition functions over the four possible b.c.s must be divided by $2$, which means that a projection onto states 
with $I=+1$ is imposed. This simply means that in both spatial b.c.\ sectors, there is a restriction to even number of particles. 
This has the consequence that all the states that survive projection have total momentum an integer multiple of $2\pi/L$.
Similarly, the sum over spatial b.c.s means that the net particle flux across a cycle such as the $t$ axis is also zero modulo $2$
(this is not so easy to see in Hamiltonian language), and the same is then true for any closed curve. 
The properties under discrete symmetries are the same as those discussed already in the corresponding cases of the sigma model 
or the integer-$S$ spin chain. Moreover, in the absence of the topological term, we can also generalize from the 
$O(3)$ (or $\mathbb{S}^2$) and $\mathbb{R}\mathbb{P}^2$ sigma models to $O(n)$ (or $\mathbb{S}^{n-1}$) and 
$\mathbb{R}\mathbb{P}^{n-1}\cong \mathbb{S}^{n-1}/\mathbb{Z}_2$ sigma models, and the results have the same form for all $n\geq 3$
(including the factorized $S$-matrices \cite{ZAMOLODCHIKOV1979253}).
For simplicity, we continue with $n=3$ only, but the generalization to $n>2$ will be evident.

We will argue that, in gauge-theory language, what we have described is a ``Coulomb'' phase of the $\mathbb{Z}_2$ gauge
theory, and in fact corresponds to vanishing of the gauge theory coupling. We also wish to relate the $\mathbb{R}\mathbb{P}^2$ sigma 
model to the same point of view. Because $\Theta=0$ here, we can make use of an ordinary lattice model, which we have not used up to now
in this paper (such models are distinct from both quantum spin chains and vertex models, the latter of which are discussed in Appendix 
\ref{sixver}). We will write down a couple of such models. In the first, we have unit vectors $\vec{n}_{\bf k}$ defined at lattice 
sites ${\bf k}\in\mathbb{Z}^2$ (we will say the lattice spacing is $a$ in length units), and also $\mathbb{Z}_2$ gauge potentials 
$U_{{\bf k}{\bf k}'}=\pm 1=U_{{\bf k}'{\bf k}}$ defined only 
on the edges $\langle{\bf k},{\bf k}'\rangle=\langle{\bf k}',{\bf k}\rangle$ that connect nearest neighbors 
${\bf k}$, ${\bf k}'$ in $\mathbb{Z}^2$. Then the lattice action for the model is \cite{PhysRevD.53.3445}
\beq
{\cal S} = -\beta'\left[\sum_{\langle{\bf k},{\bf k}'\rangle} U_{{\bf k}{\bf k}'}\vec{n}_{\bf k}\cdot\vec{n}_{{\bf k}'} 
+\upmu \sum_{\bf k}F_{\bf k}[U]\right],
\eeq
where $\beta'>0$ and $\upmu$ (not to be confused with the index $\mu$) are extended-real parameters in the model 
(i.e.\ we allow the values $\pm\infty$; $\beta'$ is an inverse temperature of this classical system, and is
not to be confused with $\beta$ which was length in the time direction, or inverse temperature in the quantum system),
and the $\mathbb{Z}_2$ flux $F_{\bf k}[U]$ through a plaquette is defined as the product of $U$s on the four edges of a square,
\beq
F_{\bf k}[U]=U_{{\bf k},{\bf k}+(1,0)^T}U_{{\bf k}+(1,0)^T,{\bf k}+(1,1)^T}U_{{\bf k}+(1,1)^T,{\bf k}+(0,1)^T}
U_{{\bf k}+(0,1)^T,{\bf k}}.
\eeq
The sum $\sum_{\langle{\bf k},{\bf k}'\rangle}$ runs once through each distinct edge $\langle{\bf k},{\bf k}'\rangle$.
The partition function (without gauge fixing) is the integral of $e^{-{\cal S}}$ over $\vec{n}_{\bf k}$ for all sites 
$\bf k$ [with $O(3)$ invariant probability measure] and sum over $U_{{\bf k}{\bf k}'}=\pm 1$ for all edges; the lattice can be 
restricted to a rectangle or parallelogram with periodic b.c.s on $\vec{n}$ and $U$, but we will not require the details of this. 
This defines a $\mathbb{Z}_2$ gauge invariant model.

If all $U$s are dropped from the action and the summation over $U_{\bf k}=\pm 1$ omitted, we obtain a lattice version of the 
$O(3)$ sigma model, that is, the classical Heisenberg ferromagnet, with action ${\cal S} = -\beta'\sum_{\langle{\bf k},{\bf k}'\rangle} 
\vec{n}_{\bf k}\cdot\vec{n}_{{\bf k}'} $.
When $\beta'$ is large, the configurations of $\vec{n}$ are slowly varying, and the model can be approximated by the
continuum $O(3)$ sigma model. There is a continuum limit as $a\to0$, provided $\beta'$ goes to infinity in a certain way
$\sim\ln(1/a)$ as $a\to0$, because of asymptotic freedom in the UV in the $O(3)$ sigma model, where $\beta'\sim 1/g^2$.
Now restoring the $U$s in the model, the coefficient $\upmu$ can be viewed as a chemical potential for $\mathbb{Z}_2$ fluxes
($F_{\bf k}=-1$), or as an inverse coupling $\upmu\sim 1/e^2$ for the $\mathbb{Z}_2$ gauge field. The fluxes produce $\mathbb{Z}_2$ 
vortices in the $\vec{n}$ field. That is, at large $\beta'$, a low energy $\vec{n}$ configuration must have a vortex, or in other words 
a branch cut emanating from the vortex (flux). From continuum considerations, such a configuration has energy that diverges
logarithmically as $a\to0$ [in fact faster, as $\sim(\ln (1/a))^2$, because $\beta'$ also diverges when the continuum limit is taken]. 
If $\upmu>0$, then fluxes or vortices are further suppressed by the weight $e^{-\beta' \upmu}$. 
Thus for $\upmu=+\infty$, the lattice model reduces to a lattice version of the gauged continuum $O(3)$ sigma model
(without vortices), and the continuum limit appears likely to be the gauged continuum $O(3)$ sigma model also for
large but finite positive $\upmu$. 

For $\upmu=0$, it is possible to sum over all $U$ in the partition function, and obtain the effective action
\beq
{\cal S} = -\sum_{\langle{\bf k},{\bf k}'\rangle} \ln \cosh(\beta'\vec{n}_{\bf k}\cdot\vec{n}_{{\bf k}'}),
\eeq
which as $\beta'\to\infty$ reduces to the sum of $\sim \beta'|\vec{n}_{\bf k}\cdot\vec{n}_{{\bf k}'}|$ over all edges.
For either form, we have a function of $(\vec{n}_{\bf k}\cdot\vec{n}_{{\bf k}'})^2$, which is manifestly gauge invariant
under inversion of $\vec{n}_{\bf k}$ for any $\bf k$. Hence this describes a model of an $\mathbb{R}\mathbb{P}^2$-valued field
at each lattice site. The field is quadrupolar (in contrast to the vector $\vec{n}$), and can be expressed using a symmetric matrix $\widetilde{Q}$
with elements $\widetilde{Q}_{\mu,\mu'}=n_\mu n_{\mu'}$ (recall that $\mu=1$, $2$, $3$); from $\widetilde{Q}^2=\widetilde{Q}$ (as matrices), 
$\widetilde{Q}$ can also be viewed 
as a rank-one projection operator (projection onto $\vec{n}$), in a three-dimensional 
vector space. As ${\rm tr}\, \widetilde{Q}=1$ is a constant, we can discard it, so the remaining traceless part $Q$ with components
$Q_{\mu,\mu'}=n_\mu n_{\mu'}-\delta_{\mu,\mu'}/3$ transforms as a quadrupole, that is, as spin $2$ under $SO(3)$.
Such degrees of freedom, and models for them, arose in liquid-crystal physics, and one simple model on the lattice has the action 
\cite{Maier1959,Maier1960,PhysRevA.6.426} 
(here for two dimensions, as usual in this paper) 
\beq
{\cal S} = -\beta'\sum_{\langle{\bf k},{\bf k}'\rangle} {\rm tr}\, Q_{\bf k} Q_{{\bf k}'}
\eeq
[note that ${\rm tr}\, \widetilde{Q}_{\bf k} \widetilde{Q}_{{\bf k}'}=(\vec{n}_{\bf k}\cdot\vec{n}_{{\bf k}'})^2$].
Either of these models for the $Q$ field action would be expected to possess a continuum limit,
and an action for a continuum $\mathbb{R}\mathbb{P}^2$ sigma model would have the form
\beq
{\cal S}=\frac{1}{2g^2}\int dxdt\; {\rm tr}\left[v(\partial_x Q)^2+\frac{1}{v}(\partial_t Q)^2\right], 
\eeq
similar to that of the $O(3)$ sigma model [the perturbative renormalization group (RG)
in the coupling $g^2$ gives the same results for the two]. In this continuum limit, vortices in $Q$ are again absent.
[We note that this continuum $\mathbb{R}\mathbb{P}^2$ sigma model should be classically integrable, because the Lax pair 
construction depends only on the equations of motion (which are local on $\mathbb{R}\mathbb{P}^2$), though it has 
only half as many integrals of motion as for the $O(3)$ sigma model.
Nothing seems to be known about the quantum integrability.]
If the $U$ fields are summed out at $\upmu\neq0$, the resulting effective action is not as local as for $\upmu=0$,
but the physics may be similar when $|\upmu|$ is small. 

Some readers may wish to discuss the $\mathbb{R}\mathbb{P}^2$ sigma model for a lattice (not continuum)
model, and treat it similarly to the Kosterlitz-Thouless (KT) theory of the $O(2)$ sigma model (or ``classical $XY$ model''), 
here with $\mathbb{Z}_2$ vortices
(instead of $\mathbb{Z}$); note however that in the present case the sigma model coupling has an RG
flow even in the absence of vortices, unlike in KT. We are saying that, 
when the lattice-scale coupling $g^2\sim \beta'^{-1}$ is very small, the system is
in the low-temperature phase in which the vortices are strongly bound, and play no role in the IR.

From these arguments (see also Ref.\ \cite{PhysRevD.53.3445}), it does not seem surprising that some numerical simulations 
\cite{PhysRevD.58.074510} of the gauged $O(3)$ model (as above) at $\upmu>0$ 
show a continuum limit that is the same for all $\upmu\geq 0$. Then it is expected that the gauged continuum $O(3)$ sigma 
model will have the same IR behavior as the ungauged model (though only correlation functions
of gauge invariant observables survive gauging, and there are constraints from the sum over b.c.s, as discussed already), 
just as we have described.
Recent simulations on more general lattice actions for the $Q$ variables \cite{PhysRevE.106.014104,PhysRevE.107.014117}, starting
from the $\upmu=0$ gauged model, also seem to confirm our conclusions about the continuum limit (but there is still some 
controversy concerning an exotic transition in $\mathbb{R}\mathbb{P}^2$ lattice models \cite{PhysRevE.106.014104,PhysRevE.107.014117}).
The situation that there are excitations
(particles, in the view as a one-space-dimensional quantum system) that transform as $O(3)$ vectors,
including transforming as odd under inversion, and are not confined, is usual in a Coulomb phase of a gauge theory,
consistent with the behavior at $e^2=0^+$ {($\upmu=\infty$). In this phase, $\mathbb{Z}_2$ fluxes (vortices) are absent. 
The field $Q$ is gauge invariant, and in the IR would create or destroy a pair of particle excitations (i.e.\ it has 
a particle worldline passing through its location).

One may consider the possibility of another phase, in which fluxes or vortices proliferate and produce confinement of the 
$O(3)$-vector particles. In this case, the remaining particle excitations would have even integer spin, such as spin $2$, 
like the $Q$ field. [One might have expected this to occur in the continuum $\mathbb{R}\mathbb{P}^2$ sigma model,
but we are arguing that there the correct description is a deconfined phase of a $\mathbb{Z}_2$ gauge theory.] 
However, it is difficult to see how such a phase can arise in these models with a simultaneous description 
as a continuum QFT in the UV. Ref.\ \cite{PhysRevD.58.074510} does find
a phase transition in the lattice model at negative $\upmu$, but it is not clear if there is any associated continuum
description. We note that an $O(3)$ vector of free massive scalar fields, coupled to a $\mathbb{Z}_2$ gauge theory,
might have both a continuum description and a confining phase, but differs in the UV from the continuum sigma models.

%%%%%%%%%%%%%%%%%%%%%%%%%%%%%%%%%%%%%%%%%
\section{Conclusion}

In this paper, we considered $SU(2)$-invariant (and some other), translation-invariant, antiferromagnetic Heisenberg spin chains 
with a periodic boundary condition (b.c.), for lengths $N$ that can be odd as well as even. For these models, odd length frustrates 
the classical N\'eel order, so a classical ground state must have a spatially-varying order parameter. The order parameter 
takes values $\vec{n}$ in the unit sphere $\mathbb{S}^2$, and the continuum field theory that arises semiclassically 
at large spin $S$ is the $O(3)$ nonlinear sigma model \cite{HALDANE1983464,PhysRevLett.50.1153}, and contains a topological term 
with coefficient $\Theta=2\pi S$. For a spin chain of odd length $N$, the sigma model field acquires the b.c.\ that 
$\vec{n}(x+L)=-\vec{n}(x)$ (where $L=Na$, and $a$ is the lattice spacing), an antiperiodic b.c.\ that involves the 
inversion symmetry $\vec{n}\to-\vec{n}$ present in 
the sigma model. The latter symmetry does not appear directly in the spin chain language, however translation by 
plus or minus one site in the spin chain has the effect of inversion of $\vec{n}$, and can be used as a proxy to recover inversion
in the continuum limit.

This motivates the study of the antiperiodic b.c.\ in the sigma model. An expression for the partition function
at a temperature $1/\beta$ as a path integral can be constructed for the sigma model with periodic b.c.s, 
which is an integral over $\vec{n}$-field configurations in two-dimensional Euclidian spacetime.
But for antiperiodic b.c.\ and $\Theta=\pi$ (mod $2\pi$) we found that, though a promising extension to antiperiodic 
b.c.s could be obtained, it was not fully independent of the definitions involved in the construction. To describe 
the antiperiodic b.c., we chose to use branch cuts, on passing through which $\vec{n}$ reverses sign. The position 
of the cut can be changed, and that is equivalent to making a gauge transformation, which reverses the sign of $\vec{n}$ 
on some subset of spacetime. In these terms, a branch cut is a line on which there is a $\delta$-function of $\mathbb{Z}_2$
gauge (``vector'') potential (or ``connection''), and this potential is changed by a gauge transformation. (Here of course, 
the $\mathbb{Z}_2$ group is that generated by the inversion symmetry.) Then our extension of the topological term
to a more general topological invariant turned out not to be invariant under all gauge transformations 
when an antiperiodic b.c.\ was present. This gauge-non-invariance is what is termed an 't Hooft anomaly.
The consequence is that, if we attempt to quantize the gauge theory by summing over the distinct b.c.s,
or equivalently by summing over all gauge equivalence classes of gauge potentials, this ``gauged'' $O(3)$ sigma model
does not exist as a QFT. The gauging is equivalent to passing to a sigma model with the target space $\mathbb{S}^2$
replaced by $\mathbb{R}\mathbb{P}^2\cong \mathbb{S}^2/\mathbb{Z}_2$, the real projective space. Our result means
that this model does not exist with topological coefficient $\Theta=\pi$ (mod $2\pi$), and hence not for 
any $\Theta\neq0$ (mod $2\pi$). If such a model existed, it would have a modular-invariant partition function.
Our result led at the same time to an anomaly in modular invariance in the gauged model, that is, no modular invariant
with the appropriate normalization could be constructed.

It is by now well understood that the spin chains with $2S$ odd, and the sigma model with $\Theta=\pi$ (mod $2\pi$),
flow at large distances (the infrared, or IR) to a conformal field theory based on $SU(2)$ current algebra at level $1$.
In that context, anomalies in the symmetry action and modular invariance have been discussed in the past
\cite{gw86,PhysRevLett.118.021601,ChengSeiberg} (in the CFT of Ref.\ \cite{gw86} a distinct, though similar and arguably related, 
inversion symmetry was involved).
Our results show that the anomalies already appear in the continuum $O(3)$ sigma model, before the IR limit
is taken. This type of effect has been discussed before as well \cite{Seiberg1703, 10.21468/SciPostPhys.6.1.003}, 
though our approach seems more direct for the sigma model.

We then considered the symmetries and quantum numbers of low-energy states in both the continuum sigma model
and the spin chain, in order to gain insight into the effects mentioned. For an antiperiodic b.c.\ and $\Theta=\pi$ (mod $2\pi$),
the inversion symmetry must be lifted to one of two elements in a group $Pin_+(3)$, a double cover of $O(3)$. 
We obtained this explicitly from semiclassical quantization of the sigma model, using the quantum mechanics
of a rigid body, which arises because of the spatially varying $\vec{n}$-field texture. The spin chain can be handled
in a similar way, and we found complete agreement between the two, including for the discrete symmetries
of spacetime reflections in space (parity) and time (time reversal). The spin-chain results included
the ground state crystal momentum, which agrees with rigorous and numerical results for different values
of $N$ and $S$, with periodicity in $N$ with period $4$ for $2S$ odd. As $N\to\infty$, translation symmetry gives
rise to both the lifts of inversion and the continuous translation symmetry in the sigma model.
These points were also discussed in Ref.\ \cite{ChengSeiberg}, with reference to a CFT in the IR as well as spin chains,
but there the spin chain results were inferred from those in the CFT.

It is of some interest to consider how our results generalize to other spin chains
and sigma models, especially those with Lie symmetry groups other than $SO(3)$ or $SU(2)$. 
There are two possible directions in which to generalize: 
\newline
1) While remaining with a spin that is the spin-$S$ representation of $SU(2)$ at each site, we could consider
Hamiltonians with less than the full $SU(2)$ symmetry. A particular case is when the symmetry is reduced to $SO(2)$ 
or $O(2)\subset SU(2)$, as occurs in the $XXZ$ (and in ``easy-axis'' or ``easy-plane'' anisotropy models) spin chains. 
In this kind of model, the sigma model would have the same target space but have reduced symmetry that still includes inversion 
symmetry. We discussed cases of the $O(3)$ sigma model in which the non-topological term has reduced symmetry, 
and found that all basic results are similar, as long as inversion 
symmetry is present. We emphasize again that our basic anomaly argument in Sec.\ \ref{subsec:pathintmodtrans} in the continuum sigma 
model used topology and translation invariance of the sigma model, but the only internal symmetry required was inversion.
Our semiclassical texture arguments in the later sections did use a spontaneously-broken continuous symmetry,
but a continuous symmetry was not crucial in the case of the sigma model. (In this vein, it is interesting that a version 
of the LSM theorem has been proved assuming time-reversal symmetry but without assuming a 
continuous symmetry in the spin chain; see Ref.\ \cite{OgataTasaki}, and references therein for earlier results in the same direction.) 
Other results, including some of those for spin chains, are more delicate.  
\newline
2) There is a class of generalizations of the spin chains leading 
to sigma models similar to the $O(3)$ model, with one or more similar topological term(s) with coefficient(s) $\Theta$ that can be $0$ or
$\pi$ (mod $2\pi$), and with a Lie symmetry group ``larger'' (higher dimensional or higher rank) than $SU(2)$;
examples have long been known \cite{AFFLECK1985397}. For simplicity, we discuss these here as if there is a single parameter $\Theta$.
In many such cases, rather than being a CFT, 
the IR limit for $\Theta=\pi$ is expected to be massive, with spontaneous breaking of translation symmetry (by dimerization) 
in the spin chain, and so of inversion symmetry in the sigma model. Hence it is valuable to know that 
anomalies in gauging inversion symmetry and in modular invariance already arise in these sigma models, without reference to 
the existence of a CFT in the IR. In Appendix \ref{OtherSp}, we show that all our main results on the spin chains and related
sigma models extend to every member of this class of models, in an almost identical form in every case. The anomaly that prevents
gauging inversion symmetry occurs exactly when the time-reversal operation ${\cal T}'$ on each spin in the spin chain squares 
to ${\cal T}'^2=-1$, which implies that the center acts nontrivially on the spin, but the converse may not hold.
It seems likely that our result that the pure $\mathbb{Z}_2$ anomaly
in inversion symmetry in the sigma models is related to ${\cal T}'^2$ on a single spin can be viewed as a direct manifestation of 
the ``Smith isomorphism'', for which see e.g.\ Ref.\ \cite{10.21468/SciPostPhys.8.4.062} (we thank M. Cheng for this remark).
[When inversion symmetry is 
spontaneously broken, the domain wall between regions in the two phases mapped to each other by inversion is a massive particle 
that can be identified with a branch cut in the sigma model, and carries a spin in the ``simplest'' irreducible representation 
consistent with the target space and $\Theta=\pi$ (mod $2\pi$), as discussed in Appendix \ref{OtherSp}, corresponding to spin $1/2$ 
for $SU(2)$; again see Ref.\ \cite{10.21468/SciPostPhys.8.4.062}.] 
In these systems, the symmetry can also be reduced, as in 1).

\bigskip
\noindent {\bf Acknowledgments} We thank N. Benjamin, Z. Komargodski, G. Sierra, N. Seiberg, T. Sulejmanpasic
and especially M. Cheng for many helpful discussions. 
The work of H.S.\ was supported by the French Agence Nationale de la Recherche (ANR) under grant ANR-21-CE40-0003 (project CONFICA). 

\begin{appendix}
\section{\texorpdfstring{$Pin$}{Pin} groups}
\label{PinG}

Here we give a simple description of the $Pin_\pm$ groups, 
and begin with $SO(n)$ and its double cover. (We will not require the use of Clifford algebras, which are frequently used 
for this topic in textbooks, e.g.\ Ref.\ \cite{Dieck}, but for the most part only some simple topology and algebra.) 
For $n\geq 3$, the fundamental group of $SO(n)$ is $\pi_1(SO(n))=\mathbb{Z}_2$,
which follows from the case $n=3$ by an inductive argument using homotopy theory \cite{Dieck}, while
$\pi_1(SO(3))=\mathbb{Z}_2$ was discussed in the main text, using the fact that $SO(3)\cong\mathbb{R}\mathbb{P}^3$ as 
topological spaces. [$SO(2)\cong \mathbb{S}^1$ is a circle, so $\pi_1(SO(2))=\mathbb{Z}$.] Hence, viewed as a topological space, 
for $n\geq 3$ $SO(n)$ has a simply-connected double cover, which is the simply-connected Lie group denoted $Spin(n)$, and is unique 
because it is determined by its Lie algebra. Thus for example, $Spin(3)\cong SU(2)$ and $Spin(4)\cong SU(2)\times SU(2)$. 
There is a similar double cover 
$Spin(2)$ of $SO(2)$, and both of these groups are isomorphic to $U(1)$, which is not simply connected. In $Spin(n)$, we will write
$\pm 1$ for the two elements that map to the identity $1$ in $SO(n)$. We call any element of $SO(n)$ a (proper) rotation,
and say a rotation is ``two dimensional'' if it rotates 
the vectors in a two dimensional plane in $\mathbb{R}^n$ (equipped with its standard inner product), but leaves vectors 
orthogonal to that plane invariant. It is well known that, for all $n\geq 2$, a rotation by $2\pi$ in a two-dimensional plane, 
which is $1$ in $SO(n)$, lifts to $-1$ in $Spin(n)$.

Next, $SO(n)$ is a normal subgroup of $O(n)$, $SO(n)\lhd O(n)$, and $O(n)/SO(n)\cong \mathbb{Z}_2$.
$O(n)$ includes the reflections of $\mathbb{R}^n$, which are orthogonal 
linear maps of $\mathbb{R}^n$ for each of which there is a nonzero vector $\vec{v}\in\mathbb{R}^n$ that is mapped 
by the reflection to $-\vec{v}$, while all vectors orthogonal to $\vec{v}$ are invariant.
Any single reflection, say $r\in O(n)$ such that a defining vector $\vec{v}$ is in the $1$ direction in $\mathbb{R}^n$, 
squares to $1$ and so generates a $\mathbb{Z}_2$ subgroup; $r$ does not lie in $SO(n)$ as it has determinant $-1$,
and the $\mathbb{Z}_2$ subgroup that it generates maps onto the quotient $O(n)/SO(n)\cong \mathbb{Z}_2$ under the quotient map. 
Hence, using the fixed choice of $r$, any element of $O(n)$ not in $SO(n)$ can be expressed as $gr$, $g\in SO(n)$, 
that is, $O(n)$ is (a good example of) a semidirect product (see a text on abstract algebra, for example Ref.\ \cite{dummit2004}), 
here of $SO(n)$ with $\mathbb{Z}_2$, written $O(n)\cong SO(n)\rtimes \mathbb{Z}_2$.
For $n=2$, the elements of $SO(2)$ are rotations $t(\theta)$ by angle $\theta$ [$t(0)=t(2\pi)=1$]. There is a relation 
$rt(\theta)r^{-1}=t(-\theta)=t(\theta)^{-1}$, which together with $r^2=1$ and $t(\theta)t(\theta')=t(\theta+\theta')$ 
for $SO(2)$ determines the structure of the group; for $O(n)$ it is similar but a little more complicated.
For general $n$, conjugation by $r$ maps a two-dimensional rotation in any plane that contains the $1$ axis
to the inverse rotation, and clearly commutes with all rotations that leave all vectors in the $1$ direction invariant;
in particular $r$ commutes with two-dimensional rotations in planes orthogonal to the $1$ axis.
Overall, conjugation by $r$ gives an automorphism of $SO(n)$, and hence also of its Lie algebra
(these hold automatically, as $SO(n)$ is a normal subgroup). [The identification of a group as a semidirect product
can also be reversed: given a group $N$ and a group $H$ of automorphisms of $N$ that act as $n\mapsto \beta_h(n)\in N$
for all $n\in N$, $h\in H$, one can construct the semidirect product $G=N\rtimes H$, with elements expressed as $g=nh$
for unique $n$ and $h$ (so $N$ and $H$ become subgroups of $G$, with $N\cap H=\{1\}$), and
the product in $N\rtimes H$ determined by the formula $hnh^{-1}=\beta_h(n)$ for all $h\in H$, $n\in N$;
then $N\lhd G$, $H\cong G/N$. We use this construction in the next paragraph. Incidentally, a direct product
$G=N\times H$ is a special case of a semidirect product, in which $H$ acts trivially on $N$, in other words elements
of the two subgroups commute, and then $H$ is also a normal subgroup.]
Another more elementary fact is that a reflection in $n$ dimensions can be expressed as a two-dimensional rotation by $\pi$ 
in $n+1$ dimensions, so that $O(n)\subset SO(n+1)$.

In a double cover of $O(n)$ that contains a subgroup $\cong Spin(n)$ as the double cover of the $SO(n)$ subgroup (if one exists), we 
will let $\widehat{r}$ be one of the lifts of the reflection $r$ to the double cover, and use the elements 
$\widehat{g}\in Spin(n)$, or $\widehat{t}(\theta)\in Spin(2)$, where $\widehat{t}(0)=1$. In the example $n=2$, 
we must now have $\widehat{r}\widehat{t}
(\theta)\widehat{r}^{-1}=\widehat{t} (\theta)^{-1}$ for all $\theta$, by continuity. Further, $\widehat{r}^2$ must be either 
$+1$ or $-1$, and must be the same for all lifts of 
all reflections [because for any $n\geq 2$, all reflections can be obtained by conjugation, as $r'=grg^{-1}$ for some $g\in SO(n)$]. 
If $\widehat{r}^2=+1$, then $\widehat{r}$ generates a $\mathbb{Z}_2$ subgroup of the double cover, and the double cover is 
again a semidirect product. A similar argument goes through for all $n$, and so we have a double cover which we call $Pin_+(n)\cong 
Spin(n)\rtimes \mathbb{Z}_2$. For $\widehat{r}^2=-1$, $\widehat{r}$ generates a group $\cong \mathbb{Z}_4$, and 
we almost obtain a semidirect product with $\mathbb{Z}_4$,
but the $\mathbb{Z}_2$ subgroup in $\mathbb{Z}_4$ must be identified with that generated by the element $-1$ in $Spin(n)$,
so we have $Pin_-(n)\cong [Spin(n)\rtimes \mathbb{Z}_4]/\mathbb{Z}_2$. These remarks show that, for each $n\geq 1$, 
there exist just two double covers of $O(n)$ of the required form. Here, for $n=1$, $SO(1)$ is trivial and $Spin(1)\cong\mathbb{Z}_2$; 
the two double covers of $O(1)\cong \mathbb{Z}_2$ are the two Abelian groups of order $4$, namely 
$Pin_+(1)\cong\mathbb{Z}_2\times\mathbb{Z}_2$ and $Pin_-(1)\cong\mathbb{Z}_4$.
For $n=2$, we can already determine that $Pin_+(2)\cong Spin(2)\rtimes \mathbb{Z}_2\cong O(2)$, as well as being a double cover of 
$O(2)$, similar to the case for $Spin(2)$. 
We can also see that, for all $n$, $Pin_-(n)\subset Spin(n+1)$. We give more details for $n=2$ and $n=3$ below.

In this paper we are particularly concerned with the lifts of inversion from $O(n)$ to $Pin_\pm(n)$.
The inversion $I$ in $O(n)$, which sends $\vec{v}\to-\vec{v}$ for all $\vec{v}\in\mathbb{R}^n$, 
is a product of $n$ reflections that each reverse one of a set of $n$ orthogonal vectors, 
and commutes with all rotations [including improper ones, the elements of $O(n)$ with determinant $1$], 
and in fact is the only non-identity element of $O(n)$ that does so when $n>2$; that is, for $n>2$ the center of $O(n)$ is 
$\cong \mathbb{Z}_2$ (note that, for $n>1$, this $\mathbb{Z}_2$ is not the same as the one generated by a single reflection 
in the earlier discussion). Inversion has determinant $-1$ when $n$ is odd, $+1$ when 
$n$ is even. For $n$ odd, such as for $O(3)$, any orthogonal transformation is a product of proper rotations
and an element of the center, so $O(n)\cong SO(n)\times \mathbb{Z}_2$ is a direct product for $n$ odd.
For $n$ even, inversion lies in $SO(n)$, and $O(n)$ is not a direct product.

We can calculate properties of either choice of a lift $\widehat{I}$ of $I$ to $Pin_\pm(n)$ using the elementary methods 
outlined here. For any $n$, $\widehat{I}$ commutes with all elements of $Spin(n)$ (it is in the center), by a continuity argument
similar to that for reflections above, because it commutes with the identity.
For $n=3$, $I$ can be viewed as the product of rotation by $\pi$ in the $23$ two-dimensional plane, and the reflection 
$r$ that reverses the $1$ axis; note that these commute, and so do their lifts to $Pin_\pm(3)$. Then we find that 
$\widehat{I}^2=-1$ in $Pin_+(3)$, $+1$ in $Pin_-(3)$,
essentially because the $2\pi$ rotation in the $23$ plane gives an additional $-1$. 
Then as $O(n)$ is a direct product for $n$ odd, these results imply (similarly to the semidirect products
above) that $Pin_+(3)\cong[Spin(3)\times\mathbb{Z}_4]/\mathbb{Z}_2$, while $Pin_-(3)\cong Spin(3)\times \mathbb{Z}_2$, 
generated in both cases by $Spin(3)\cong SU(2)$ and the lift $\widehat{I}$ of the inversion. 
From this we find that $Pin_+(3)$ can be represented as the set of matrices in $SU(2)$, together with the set of the 
same matrices multiplied by $i$; these can alternatively be described as the unitary matrices with determinant $\pm 1$.
Note also that the center of $Pin_+(3)$ is $\mathbb{Z}_4$, while the center of $Pin_-(3)$ is $\mathbb{Z}_2\times\mathbb{Z}_2$.
All this can be generalized easily to all odd values of $n$, say $n=2k+1$ and $k>0$, by using $k$ two-dimensional rotations 
by $\pi$ in orthogonal planes and one reflection. Then $\widehat{I}^2=(-1)^k$ in $Pin_+(2k+1)$, $\widehat{I}^2=(-1)^{k+1}$ 
in $Pin_-(2k+1)$. Then similar structures $[Spin(n)\times\mathbb{Z}_4]/\mathbb{Z}_2$ and $Spin(n)\times \mathbb{Z}_2$ of the groups, 
and the same two forms for the centers, as for $n=3$ are also found for $Pin_\pm(n)$ for all odd values of $n$, but the $\pm$ 
cases are exchanged each time $n$ is increased by $2$ [cf.\ $Pin_\pm(1)$ above, for example].

In a similar way, for $n$ even we find that $\widehat{I}^2=(-1)^{n/2}$ in $Spin(n)$, and hence in both $Pin_+(n)$ and $Pin_-(n)$. 
From these and preceding results, we can also determine the center of $Spin(n)$ ($n>0$), 
which is $\mathbb{Z}_2=\{\pm 1\}$ for $n$ odd, but for $n$ even and $n>2$ is $\mathbb{Z}_2\times \mathbb{Z}_2$ for $n=0$ (mod $4$), 
and $\mathbb{Z}_4$ for $n=2$ (mod $4$) (due to the lift of inversion squaring to $+1$ or $-1$ in the respective cases). However, 
the lift $\widehat{I}$ is not always in the center of $Pin_\pm(n)$. While both lifts commute with any reflection
when $n$ is odd, for $n$ even they anticommute (conjugation of $\widehat{I}$ by $\widehat{r}$ gives 
$\widehat{r}\widehat{I}\widehat{r}^{-1}=-\widehat{I}$, by another easy calculation). 
By continuity, the same is true in either case for all elements in the second connected component of $Pin_\pm(n)$,
and not only for reflections. 
For $n$ even, $n\geq 2$, the center of $Pin_\pm(n)$ 
is $\mathbb{Z}_2$, for both choices $\pm$, and is a proper subgroup of that of $Spin(n)$. 
We see that, for $n$ even, consideration of inversion does not appear to 
give a description of the structure of $Pin_\pm(n)$ simpler than that above using semidirect products.

The structure of $Pin_+(2)$ was already discussed. $Pin_-(2)\cong [Spin(2)\rtimes \mathbb{Z}_4]/\mathbb{Z}_2$ can be expressed 
as the group generated by the $2\times 2$ unitary matrices $\widehat{t}(\theta)={\rm diag}(e^{i\theta/2},e^{-i\theta/2})$ and
$\widehat{r}=i\sigma_x$, where $\sigma_x$ is one of the standard Pauli matrices. This exhibits it as a subgroup of $SU(2)$, 
and as the double cover of $O(2)$, where $O(2)\subset SO(3)$ is generated by rotations around the $3$ axis and a reflection 
is a rotation by $\pi$ about the $1$ axis. For the lifts of inversion, we have $\widehat{I}=\widehat{t}(\pi)$ (or its inverse) 
in $Spin(2)$ and hence in both $Pin_+(2)$ and $Pin_-(2)$. 

Turning to representations of the $Pin_\pm(3)$ groups, we can start from the spin-$j$ irreducible representations of the $Spin(3)$ 
subgroup. When the spin $j$ is integer, $Pin_\pm(3)$ reduces to $O(3)$, and because $O(3)$ is a direct product, there is a 
representation in which the inversion $I$ acts as multiplication by $+1$ and another representation in which it acts 
by $-1$, for any $j$. For $Pin_+(3)$ and $j$ half-integer, one lift of inversion acts as multiplication by $i$, the other by $-i$. 
[For spin $1/2$, one of these is the defining representation of $Pin_+(3)$ mentioned above.] These representations are irreducible 
and also faithful. On the other hand, for $Pin_-(3)$ and $j$ half integer, a lift of inversion squares to the identity, so 
the lift acts as multiplication either by $1$ or by $-1$. These are the same as the action of the center of $Spin(3)$, 
which implies that, while irreducible, these representations are not faithful. A faithful half-integer spin-$j$ 
representation of $Pin_-(3)$ would have dimension $2(2j+1)$.

%%%%%%%%%%%%%%%%%%%%%%%%%%%%%%%%%%%%%%%%%
\section{Rigid body motion and the Peter-Weyl theorem}
\label{PetWeyl}

In this Appendix, we give detailed definitions and prove some results that relate to the quantum mechanics
of a rigid body, including a simple proof of the Peter-Weyl theorem, which appear in Sec.\ \ref{Subsec:semicl}. 
A rigid body in three dimensions
with one point fixed can be described by a rotation matrix $O$ in $SO(3)$ that specifies a rotation from one reference
orientation of the body to any orientation. For a quantum mechanical treatment, we need to understand the Hilbert space of
square-integrable functions of $O$, and how rotation symmetries act on the Hilbert space. The most important points that require
explanation are why the left and right quantum numbers $j_L$, $j_R$, must be equal, and the fact that the matrix elements
of $SO(3)$ in finite-dimensional irreducible representations can be used as a basis for the Hilbert space (the Peter-Weyl theorem).

The first part of our treatment can handle groups more general than $SO(3)$, so we consider any compact group $G$. 
First, it may be helpful to recall that, given a space $X$ and a group $G$, a left action of $G$ on $X$
means that for every $g\in G$ there is an invertible map of $X$ into itself, taking $x\in X$ to, say $^g\!x$,
with the properties that, for all $x\in X$, $^1\!x=x$ and $^{g_1}\!(^{g_1'}\!x)=\,^{g_1g_1'}\!x$ for all $g_1$, $g_1'\in G$. 
In particular, for $X=G$, from associativity of multiplication in any group we obtain a left action of a group $G$ on itself
by left multiplication, given by $^{g_1}\!g=g_1 g$ for all $g$, $g_1\in G$. We can also obtain another left action of $G$ on itself,
which we write as $g\mapsto g^{g_2}$ [again with $g^1=g$ and $(g^{g_2'})^{g_2}=g^{g_2g_2'}$], given by right multiplication by 
$g_2^{-1}$, that is $g^{g_2}=gg_2^{-1}$; notice the need for the inverse in the definition. (Hopefully no confusion will arise 
from a left action that acts by right multiplication.) The two left actions commute, that is 
$^{g_1}\!(g^{g_2})=(^{g_1}\!g)^{g_2}$, so we may write these as $^{g_1}\!g^{g_2}$. Our exclusive use of left actions of groups
on spaces agrees with the usual conventions in physics, both for groups and for the action of operators on vector spaces, which
correspond to multiplication of a {\em column} vector by a matrix on the {\em left}.

Next, we show that $G$ can be regarded as a homogeneous space, with left and right multiplication by elements of $G$
as two commuting sets of symmetry operations. For this, we begin with two copies of $G$ and form the direct product $G\times G$,
with elements $(g_1,g_2)$ where both $g_1$, $g_2\in G$; the binary operation in the product group is
\beq
(g_1,g_2)(g_1',g_2')=(g_1 g_1',g_2g_2').
\label{gpmult}
\eeq
Like any direct product, $G\times G$ contains two subgroups isomorphic to $G$, the first with elements $(g_1,1)$, the second with 
elements $(1,g_2)$; we denote these subgroups by $G_L$, $G_R$, and typical elements in each (with the forms just given)
by $g_L$, $g_R\in G\times G$, respectively. [By a standard abuse of notation, we can also write $G\times G$ as $G_L\times G_R$.]
We have the usual left action of $G\times G$ on itself by left multiplication. 
We will also refer to the left action of a subgroup $G_{\rm diag}$ of $G\times G$ by right multiplication; $G_{\rm diag}$ 
consists of elements of the form $(h,h)\in G\times G$. Then the actions of $G\times G$ and of $G_{\rm diag}$ on $G\times G$ commute.

The quotient or coset space $(G\times G)/G_{\rm diag}$ can be shown to be homeomorphic to $G$ as a topological space
(we usually suppress mention of the topologies), as follows. From $(g_1,g_2)\in G\times G$, we form $g=g_1 g_2^{-1}$,
and we see immediately that we obtain the same element $g$ if (and also only if) we use $(g_1h^{-1},g_2h^{-1})=(g_1,g_2)^h $
in place of $(g_1,g_2)$, which essentially proves the statement. Further, as the left action of $G\times G$
on itself commutes with that of $G_{\rm diag}$ on $G\times G$, we have a left action of $G\times G$
on the coset space, given by $^{(g_1,g_2)}\!g=g_1g g_2^{-1}=\,^{g_1}\!g^{g_2}$ (i.e.\ left multiplication by $G_L$, and simultaneous
right multiplication by $G_R$). The diagonal action in $G_L\times G_R$ of the center $Z(G)$ of $G$, that is the set of $(g',g')$ 
where $g'\in Z(G)$, leaves any point in the quotient space invariant, so the group that acts faithfully on $G$ 
is in fact $[G\times G]/Z(G)$.

The subgroup $G_{\rm diag}$ appears to have disappeared from the coset space. However, there is an isomorphic
subgroup of $G\times G$ (and isomorphic to $G$) associated to any $g\in G\cong (G\times G)/G$. Namely, 
it is the set of elements of $G\times G$ that leave $g$ invariant, the so-called isotropy subgroup of $g$.
The elements are the pairs $(g_1,g_2)$ such that $g_1 g g_2^{-1}=g$, so $g_1=gg_2g^{-1}$, and these form
a subgroup $G_g$, $G_g\subset G\times G$, isomorphic to $G$. For $g=1$, $G_1$ has the same form as $G_{\rm diag}$ 
that we had before.
(The isotropy subgroup will effectively take the place of the traditional
statement that rotations in the laboratory and body frames correspond inversely, see e.g.\ Ref.\ \cite{landau1981quantum}, 
which appears to mean that the two left actions on $g$ by left multiplication by $g_1$ and by right multiplication by $g_2$ correspond 
if $g_1g=gg_2^{-1}$; differentiating with respect to the parameters in $g_2$ at $g_1=g_2=1$, with $g$ held fixed,
then gives a relation between generators, but it holds only when they are evaluated at $g$; we will not use this.) 

If we wish to consider the representation of $G\times G$ on the Hilbert space
of functions on $G$, then to obtain a left action $\lambda$ of $G\times G$ we must define, for any function 
$\phi$ on $G$, 
\beq
[\lambda(g_1,g_2)\phi](g)=\phi(g_1^{-1} g g_2),
\eeq
in order that $\lambda(g_1,g_2)\lambda(g_1',g_2')=\lambda(g_1g_1',g_2g_2')$. 
Forms analogous to this appear in basic quantum mechanics, for example for translations and rotations on functions
on $\mathbb{R}^3$, as we will see.

Although we could continue in this generality for a general compact group, for now we assume that $G=SO(3)$,
so that we can use notation as in the main text [with typical element $O\in SO(3)$], and make occasional comments 
on more general cases. For the left action of $SO(3)_L\times SO(3)_R$ on the Hilbert space of functions on $SO(3)$,
we can make use of general representation-theoretic principles such as full reducibility of representations,
which holds for any compact group (it will be convenient to use standard notation for finite-dimensional irreducible representations
of $SO(3)$, though we do not need to assume at this stage that all irreducible representations are finite dimensional;
the fact that they are is a consequence of the Peter-Weyl theorem which we discuss later). An irreducible representation 
of a direct product of two groups is a (tensor) product of irreducible representations of the two groups, 
so an irreducible $SO(3)_L\times SO(3)_R$ multiplet has a basis that is a collection of complex functions, 
say $Y^{j_Lj_R}_{m_Lm_R}(O)$, of $O\in SO(3)$ [or more generally, $O\in Spin(3)\cong SU(2)$, similarly], where the pair 
$(j_L,j_R)$ labels the choices of irreducible representation of $SO(3)_L\times SO(3)_R$ (we refine our definition below). 
We will study its precise transformation behavior. We should also note our convention that 
$SO(3)_L$ and $SO(3)_R$ are the same Lie group,
and that their generators, $\vec{J}_L$, $\vec{J}_R$ respectively, obey the same commutation relations.

We will refer to a fixed set of orthonormal basis vectors $\{|jm\rangle\}$ in the spin-$j$ irreducible representation of $SU(2)$.
If every rotation [or $SU(2)$ element] $O$ is represented by a corresponding unitary operator $\widehat{O}$, then in this basis 
the matrix elements of $\widehat{O}$ are $D^j_{mm'}(O)=\langle jm|\widehat{O}|jm'\rangle$. Because (with our conventions) the 
representation must obey
$\widehat{(O_1O_2)}=\widehat{O_1}\widehat{O_2}$, the matrices obey 
\beq
D^j_{mm'}(OO')=\sum_{m''}D^j_{mm''}(O)D^j_{m''m'}(O')
\label{gphom}
\eeq
for each $j$. These matrices $D^j$ are the representation or Wigner matrices, and this equation says that they give a group 
homomorphism of $Spin(3)\cong SU(2)$ into the unitary matrices [for fixed $j$, these are $(2j+1)\times (2j+1)$ matrices]. 

The case of the semiclassical states 
in the spatial P b.c.\ case, or simply the space of complex functions of a unit vector $\vec{n}\in\mathbb{S}^2$, 
can serve us as a guide. A standard choice of basis is the set of spherical harmonics $Y_{jm}(\vec{n})=\langle\vec{n}|jm\rangle$,
where in the present case $j$ must be integer. The orthonormality and completeness of the spherical harmonics can be 
expressed by saying that they determine a unitary transformation between the Hilbert space of functions of $\vec{n}$
and the space spanned by the set of $|jm\rangle$. Under a rotation $O_1$ they transform to linear combinations, as 
\beq
(\widehat{O_1}Y_{jm})(\vec{n})\equiv\langle\vec{n}|\widehat{O_1}|jm\rangle= Y_{jm}(O_1^{-1}\vec{n}) 
= \sum_{m'} D^j_{m'm}(O_1) Y_{jm'}(\vec{n}).
\eeq
Then if $\psi^j_m=\langle jm|\psi\rangle$ are the components of a state vector $|\psi\rangle$ in the basis of 
$|jm\rangle$, the components of $|\psi'\rangle = \widehat{O}|\psi\rangle$ are $\psi'^j_m=\sum_{m'}D^j_{mm'}(O)\psi^j_{m'}$ 
as they should be.

Similarly, for the rigid body, or the semiclassical A b.c.\ case, so far we have the basis functions
$Y^{j_Lj_R}_{m_Lm_R}(O)=\langle O|j_Lm_L,j_Rm_R\rangle$ for an orthonormal basis set $\{|j_Lm_L,j_Rm_R\rangle\}$
and they determine a unitary transformation between the space spanned by the latter basis set and the space of functions
on $SO(3)$. If $|\Psi\rangle$ is a general element of the former space, then we have that the components of $|\Psi\rangle$ 
are $\psi^{j_Lj_R}_{m_Lm_R}=\langle j_Lm_L,j_Rm_R|\Psi\rangle$,
or as a function $\psi(O)=\langle O|\Psi\rangle=\sum_{j_Lm_L,j_Rm_R}\psi^{j_Lj_R}_{m_Lm_R}\langle O|j_Lm_L,j_Rm_R\rangle$. Under a 
$Spin(3)_L\times Spin(3)_R$ transformation by $(O_1,O_2)$, we will require our basis functions to transform as 
\begin{eqnarray}
\left(\widehat{(O_1,O_2)} Y^{j_Lj_R}_{m_Lm_R}\right)(O) &\equiv& Y^{j_Lj_R}_{m_Lm_R}(O_1^{-1}OO_2) \\
&=& \sum_{m_1,m_2}D^{j_L}_{m_1m_L}(O_1)
Y^{j_Lj_R}_{m_1m_2}(O)D^{j_R}_{m_Rm_2}(O_2^{-1}).
\label{transyjjmm}
\end{eqnarray}
One can see that, if $j_L=j_R=j$, then $Y^{jj}_{m_Lm_R}(O)\propto D^j_{m_Rm_L}(O^{-1})$ is a solution
to this equation, using the group homomorphism property, eq.\ (\ref{gphom}). We will now show that 
these are the only possible nonzero solutions (up to redefinitions). 

First, we examine the transformation behavior.
Eq.\ (\ref{transyjjmm}) implies that the components of $|\Psi'\rangle = \widehat{(O_1,O_2)}|\Psi\rangle$
are 
\beq
\psi'^{j_Lj_R}_{m_Lm_R}=\sum_{m_1m_2}D^{j_R}_{m_Lm_1}(O_1)\psi^{j_Lj_R}_{m_1m_2}D^{j_R}_{m_2m_R}(O_2^{-1}),
\eeq
which makes the left action by left and right multiplication on the matrix $\psi^{j_Lj_R}_{m_Lm_R}$ evident.
The transformation on the right index under the transpose of the inverse for $O_2$ (as compared with that for $O_1$
on the left index) requires a little discussion.
Rather than viewing the matrix $Y^{j_Lj_R}_{m_Lm_R}(O)$ [for fixed $(j_L,j_R)$] as an element of a tensor product
of the two irreducibles $j_L$ and $j_R$, we choose (in order to simplify statements that follow) to view it as 
an element of the space of linear maps from the space
spanned by $|j_Rm_R\rangle$ to that spanned by $|j_Lm_L\rangle$. (This is natural, as $O$ is viewed as a map taking 
one frame to another.) That is, we can think of a linear map or operator $\Psi$ between two Hilbert spaces,
namely two copies of the space spanned by $\{|jm\rangle\}$, and turn the space of such maps $\Psi$ into a Hilbert space
by using the Hilbert-Schmidt norm $||\Psi||_2^2= {\rm Tr}\,\Psi^\dagger \Psi$; the states $|j_Lm_L,j_Rm_R\rangle$ 
can be viewed as a basis for this space. In these terms, what we have said is that 
\beq
\Psi'=\widehat{O_1}\Psi\widehat{O_2^{-1}}
\eeq
[where the operator $\widehat{O_1}$ (resp., $\widehat{O_2^{-1}}$) acts by left multiplication, as usual, on the space spanned 
by $|j_Lm_L\rangle$ (resp., $|j_Rm_R\rangle)$]. The same form applies for the case of general $G_L\times G_R$
action on functions on $G$ (where $j$ is considered as a label set that characterizes an irreducible representation).
Then a linear map from the space spanned by $|j_Rm_R\rangle$ to that spanned by $|j_Lm_L\rangle$ (that is, any $\Psi$) 
transforms under $G_L$ as the representation $j_L$, and under $G_R$ as the {\em dual} of the representation $j_R$, as follows.

To be specific, the dual space of a vector space of column vectors is the vector space of row vectors,
and matrices acting on these from the right can be represented as the transpose matrices acting on the transpose
(column) vectors from the left (note this is {\em not} the adjoint, and neither complex conjugation nor an inner product are involved).
In terms of matrices, if a representation of a group $G$ is given in some orthonormal basis by matrices $D(g)$ ($g\in G$), 
then the dual representation is given in the dual basis by $D(g^{-1})^T=D(g)^{T-1}$, the transpose matrices of the inverse element, 
in order that simultaneous transformation of both the column and row vector leaves the natural multiplication of row 
with column, which gives a scalar, invariant \cite{fulton1991}. On using the inner product on the Hilbert space,
the representations are always unitary, and assuming the basis is orthonormal, the transpose 
of the inverse is then the (element-wise) complex conjugate $\overline{D(g)}=D(g^{-1})^T$ of the matrix in the original representation, 
and so the dual representation is sometimes called the ``conjugate'' representation. [In the case of $SU(2)$ and some other groups, 
any representation is self-dual, that is, isomorphic to its dual, but we still need the correct transformation law, 
and we keep the discussion general by not making use of such an isomorphism.] For a Lie group, the matrices
that represent the generators, say $\vec{J}$, in the dual of a representation are minus the transpose of those in the original
representation, and hence obey the same commutation relations, as required. In another notation that involves tensors, 
if vectors in a space are represented in components by tensors with a single index, which is an upper index, then 
vectors in the dual space (covectors) are represented by tensors with one lower index, and a linear map from
vectors to vectors is a tensor with one upper and one lower index, and transforms accordingly, exactly as discussed here
(but note that our transformations are active, not merely a passive change in basis). Finally, we emphasize that,
even if different definitions of a transformation are used initially, in the end the results will be the same, including
the appearance (explicitly or not) of a dual representation.

We now obtain consequences of the transformation law. 
First, recall the definition of the isotropy subgroup $SO(3)_O$ of $O$, 
which can be rewritten as the set of pairs $(O_1,O_2)$ such that $O_1^{-1}O=OO_2^{-1}$. For given $O$ and any such $O_1$, $O_2$, 
we then have $\psi(O_1^{-1}OO_2)=\psi(O)$. For the basis states $Y^{j_Lj_R}_{m_Lm_R}$, 
this implies 
\beq
\sum_{m_1}D^{j_L}_{m_1m_L}(O_1)Y^{j_Lj_R}_{m_1m_R}(O)
=\sum_{m_2}Y^{j_Lj_R}_{m_Lm_2}(O)D^{j_R}_{m_Rm_2}(O_2).
\eeq
Suppose for example that $O=1$. Then $O_1=O_2$, and $Y^{j_Lj_R}_{m_Lm_R}(1)$ would be an intertwiner between the
irreducible representations labeled by $j_L$, $j_R$. (To see this, it may help to write the $D$ matrices using the transpose,
and replace $O_1$, $O_2$ by their inverses, so that we have transformations in the duals of $j_L$, $j_R$, and the matrices
are multiplied in the usual way.) That is, these numbers are the elements of the matrix of a linear map that commutes with the action
of the group $SO(3)_1\cong SO(3)$ on the two vector spaces. Schur's lemma \cite{Dieck,fulton1991,dummit2004} says that such a map is 
zero unless the irreducible representations
labeled $j_L$, $j_R$ are isomorphic, that is, $j_L=j_R$. (The expressions using the dual were crucial in arriving 
at such a simple statement.) Further, for $j_L=j_R$, $Y^{j_Lj_R}(1)$ must be a multiple of the 
identity: $Y^{j_Lj_R}_{m_Lm_R}(1)\propto \delta_{m_Lm_R}$ (because we use a fixed basis in any irreducible representation),
and if nonzero it gives the isomorphism. For general $O$, for the isotropy subgroup we have $O_1=OO_2O^{-1}$, and $O$ determines a
group isomorphism between two groups, both isomorphic to $SO(3)$, and we have a slightly more general case of the same phenomenon
(but no additional restrictions emerge). This argument clearly generalizes to any compact (including any finite) group. 
From here on, we can simplify notation by writing simply 
$Y^j_{m_Lm_R}(O)$ in place of $Y^{jj}_{m_Lm_R}(O)$ when $j_L=j_R=j$, as other cases do not occur.

Starting from $O=1$, we can transform by $(O_1,O_2)$ to reach any $O$, and then using the homomorphism property, eq.\ (\ref{gphom}), 
of the $D$ matrices, we find $Y^j_{m_Lm_R}(O)\propto D^j_{m_Rm_L}(O^{-1})$ (again, the dual). Using eq.\ (\ref{gphom}), we see 
that this satisfies eq.\ (\ref{transyjjmm}) for $j_R=j_L$, and we have now shown that it is the only nonzero form that 
does so. As the $D$ matrices are unitary, it is the same as 
\beq
Y^j_{m_Lm_R}(O)\propto\overline{D^j_{m_Lm_R}(O)},
\eeq
a result that seems to be well known (again, as at most places in the paper, the overline denotes complex conjugate). 
These results all take the same form in any orthonormal basis,
and for any compact group.

Now we turn to the Peter-Weyl theorem, already stated in Sec.\ \ref{Subsec:semicl}. 
To illustrate some aspects of a proof, first consider the case of 
complex functions on $\mathbb{S}^2$, as for the P b.c.\ case. The components of $\vec{n}$ can be viewed as coordinate functions 
on $\mathbb{S}^2$, which satisfy the algebraic relation $\vec{n}^2=1$; under multiplication and linear combination
(with complex coefficients), and including the constant function, they generate an algebra of polynomial functions 
on $\mathbb{S}$, with unit element $1$. The monomials of each degree $j$ (after simplifying by using $\vec{n}^2=1$) 
are the so-called cubic harmonics of angular momentum $j$, related to the spherical harmonics by a change of basis. 
[Similar cubic harmonics exist, and span the space of functions, on the sphere $\mathbb{S}^{n-1}$ in $n$ dimensions for any $n>0$.]
For functions on $SO(3)$, we can similarly use the components of $O$ as coordinate functions, subject to the algebraic 
relations $O^TO={\bf 1}$ (where $\bf 1$ denotes the identity matrix), and obtain an algebra of polynomial functions.
In both cases, the fact that these functions span (i.e.\ more formally, are ``dense in'') the space of all continuous functions 
(and hence the Hilbert space of all square-integrable functions) on $\mathbb{S}^2$ or $SO(3)$, respectively, follows from 
the Stone-Weierstrass theorem (included in Ref.\ \cite{Dieck}).
The degree-$j$ monomials in the elements of $O$ are the matrix elements of $O$ acting in the $j$-fold tensor product 
of the $3$-dimensional (i.e.\ $j=1$) defining representation of $SO(3)$, which contains one copy of the angular momentum 
$j$ irreducible as a direct summand in its Clesch-Gordan decomposition, along with copies of irreducibles with lower $j$ 
(e.g.\ see any of Refs.\ \cite{Schiff,landau1981quantum,Dieck}; the Peter-Weyl theorem is not used in this). 
As all irreducible representations can be obtained by decomposing tensor products of the defining representation in this way, an 
inductive argument proves the Theorem. Ref.\ \cite{Dieck} contains a more direct proof. Then finally, the matrix elements
of $O$ in the spin-$j$ irreducible representation, as $j=0$, $1$, \ldots, runs over all possible values, gives a basis for
the Hilbert space of complex functions on $SO(3)$. 

The argument we gave for $SO(3)$ also goes through at least for the ``classical'' groups, which can be defined 
in terms of matrices, as in our example of $SO(3)$; this too is discussed in Ref.\ \cite{Dieck}. 
The Peter-Weyl theorem for a general compact group $G$ again implies that
the states (in the Hilbert space of functions on the group) that form a given irreducible representation 
under $G_L$ transform in the dual of that representation under $G_R$ (as we already discussed). For example, 
for the representation matrices of the defining representation of a classical group, the left action on the representation 
matrices by right multiplication is described by the dual of the defining representation, a fact which is probably well known. 
The set of all equivalence classes of irreducible representations under isomorphism includes the dual of any irreducible,
so in the statement of the Peter-Weyl theorem we could equally say that the matrix elements of the representation matrices
of the duals of all of the irreducible representations span the Hilbert space of functions on $G$, with the set of 
$\overline{D^j_{m_L,m_R}(g)}$ (where $g\in G$) as a basis. This behavior is also found in the 
representation theory of finite groups. By the Wedderburn structure theorem, the group algebra 
(i.e.\ the algebra of complex functions on the group, with the multiplication induced from that in the group) can be decomposed 
as a direct sum of matrix algebras, with the sum indexed by the isomorphism classes of irreducible representations, and
each summand isomorphic to the tensor product of that irreducible with its dual; see for example Ref.\ \cite{dummit2004}.
(If we begin by ignoring the multiplication in the algebra, then a proof of this follows easily from the arguments 
earlier in this Appendix. Thus the dimension of the group algebra, which is equal to the order of the group, 
is also equal to the sum of the squares of the dimensions of the irreducible representations. The algebra structure 
can be recovered at the end.) This is a special case of the Peter-Weyl theorem (finite groups are compact).

%%%%%%%%%%%%%%%%%%%%%%%%%%%%%%%%%%%%%%%%%%%%
\section{Vertex models}
\label{sixver}

While our discussion in the main text centered on the relationship between continuum field theory and spin chains --- 
hence, systems discrete in space but continuous in time --- one should expect similar features to be present also in  
two-dimensional statistical mechanics models 
such as vertex models, which are discrete in both directions, and have the same continuum limit. 
Let us discuss this by considering to start the case of spin $1/2$ and the 6-vertex model, for which many exact results are 
available due to integrability. 

The 6-vertex model is defined on the square lattice, with every edge carrying a spin $1/2$ degree of freedom conveniently represented 
by an arrow. By convention we will associate the value $S_3=+\frac{1}{2}$ to arrows going up or right, and the opposite for those going 
down or left. The Boltzmann weights can be encoded into a so-called $R$-matrix propagating in the south-west to north-east direction:
\begin{equation}
\check{R}=\left(\begin{array}{cccc}
a&0&0&0\\
0&c&b&0\\
0&b&c&0\\
0&0&0&a
\end{array}\right)\label{weights}
\end{equation}
where the possible vertex configurations and their weights are represented on Figure  \ref{6vertex} below.  
\begin{figure}[h]
    \centering
    \resizebox{1\textwidth}{!}{%
    \begin{minipage}{0.1\textwidth}
        \centering
        \begin{tikzpicture}
            \draw[blue, arrows={-Stealth[inset=0pt, angle=90:2pt]}] (0,0) -- (.25, 0);
            \draw[blue] (.25,0)--(.5,0);
            \draw[blue, arrows={-Stealth[inset=0pt, angle=90:2pt]}] (.5,0) -- (.75, 0);
            \draw[blue] (.75,0)--(1,0);
            \draw[blue, arrows={-Stealth[inset=0pt, angle=90:2pt]}] (.5,-.5) -- (.5, -.25);
            \draw[blue] (.5,-.25)--(.5,0);
            \draw[blue, arrows={-Stealth[inset=0pt, angle=90:2pt]}] (.5,0) -- (.5, .25);
            \draw[blue] (.5,.25)--(.5,.5);
        \end{tikzpicture}
      %  \caption{a}
    \end{minipage}%
    \hspace{0.01\textwidth} % Small horizontal space between the figures
    
    \begin{minipage}{0.1\textwidth}
        \centering
        \begin{tikzpicture}
            \draw[blue, arrows={-Stealth[inset=0pt, angle=90:2pt]}] (0.5,0) -- (.25, 0);
            \draw[blue] (.25,0)--(0,0);
            \draw[blue, arrows={-Stealth[inset=0pt, angle=90:2pt]}] (1,0) -- (.75, 0);
            \draw[blue] (.75,0)--(0.5,0);
            \draw[blue, arrows={-Stealth[inset=0pt, angle=90:2pt]}] (.5,0) -- (.5, -.25);
            \draw[blue] (.5,-.25)--(.5,-.5);
            \draw[blue, arrows={-Stealth[inset=0pt, angle=90:2pt]}] (.5,.5) -- (.5, .25);
            \draw[blue] (.5,.25)--(.5,0);
        \end{tikzpicture}
      %  \caption{a}
    \end{minipage}%
    \hspace{0.01\textwidth} % Small horizontal space between the figures
    
    \begin{minipage}{0.1\textwidth}
        \centering
        \begin{tikzpicture}
            \draw[blue, arrows={-Stealth[inset=0pt, angle=90:2pt]}] (0,0) -- (.25, 0);
            \draw[blue] (.25,0)--(.5,0);
            \draw[blue, arrows={-Stealth[inset=0pt, angle=90:2pt]}] (.5,0) -- (.75, 0);
            \draw[blue] (.75,0)--(1,0);
            \draw[blue, arrows={-Stealth[inset=0pt, angle=90:2pt]}] (.5,0) -- (.5, -.25);
            \draw[blue] (.5,.25)--(.5,-.5);
            \draw[blue, arrows={-Stealth[inset=0pt, angle=90:2pt]}] (.5,.5) -- (.5, .25);
            \draw[blue] (.5,.25)--(.5,0);
        \end{tikzpicture}
      %  \caption{b}
    \end{minipage}%
    \hspace{0.01\textwidth} 
    
    \begin{minipage}{0.1\textwidth}
        \centering
        \begin{tikzpicture}
    
          \draw[blue, arrows={-Stealth[inset=0pt, angle=90:2pt]}] (0.5,0) -- (.25, 0);
            \draw[blue] (.25,0)--(0,0);
            \draw[blue, arrows={-Stealth[inset=0pt, angle=90:2pt]}] (1,0) -- (.75, 0);
            \draw[blue] (.75,0)--(0.5,0);
        
            \draw[blue, arrows={-Stealth[inset=0pt, angle=90:2pt]}] (.5,-.5) -- (.5, -.25);
            \draw[blue] (.5,-.25)--(.5,0);
            \draw[blue, arrows={-Stealth[inset=0pt, angle=90:2pt]}] (.5,0) -- (.5, .25);
            \draw[blue] (.5,.25)--(.5,.5);
        \end{tikzpicture}
      %  \caption{b}
    \end{minipage}
       \hspace{0.01\textwidth} % Small horizontal space between the figures

    \begin{minipage}{0.1\textwidth}
        \centering
        \begin{tikzpicture}
            \draw[blue, arrows={-Stealth[inset=0pt, angle=90:2pt]}] (0,0) -- (.25, 0);
            \draw[blue] (.25,0)--(.5,0);
            \draw[blue, arrows={-Stealth[inset=0pt, angle=90:2pt]}] (1,0) -- (.75, 0);
            \draw[blue] (.75,0)--(.5,0);
            \draw[blue, arrows={-Stealth[inset=0pt, angle=90:2pt]}] (.5,0) -- (.5, -.25);
            \draw[blue] (.5,.25)--(.5,-.5);
            \draw[blue, arrows={-Stealth[inset=0pt, angle=90:2pt]}] (.5,0) -- (.5, .25);
            \draw[blue] (.5,.25)--(.5,.5);
        \end{tikzpicture}
    %  \caption{c}
    \end{minipage}
    
     \hspace{0.01\textwidth} 

      \begin{minipage}{0.1\textwidth}
        \centering
        \begin{tikzpicture}
            \draw[blue, arrows={-Stealth[inset=0pt, angle=90:2pt]}] (.5,0) -- (.25, 0);
            \draw[blue] (.25,0)--(0,0);
            \draw[blue, arrows={-Stealth[inset=0pt, angle=90:2pt]}] (.5,0) -- (.75, 0);
            \draw[blue] (.75,0)--(1,0);
            \draw[blue, arrows={-Stealth[inset=0pt, angle=90:2pt]}] (.5,-.5) -- (.5, -.25);
            \draw[blue] (.5,-.25)--(.5,0);
            \draw[blue, arrows={-Stealth[inset=0pt, angle=90:2pt]}] (.5,.5) -- (.5, .25);
            \draw[blue] (.5,.25)--(.5,0);
        \end{tikzpicture}
      %  \caption{c}
    \end{minipage}%
   
    }
    \caption{Vertices of the 6 vertex model. The corresponding weights are $a,a,b,b,c,c$}\label{6vertex}
\end{figure}
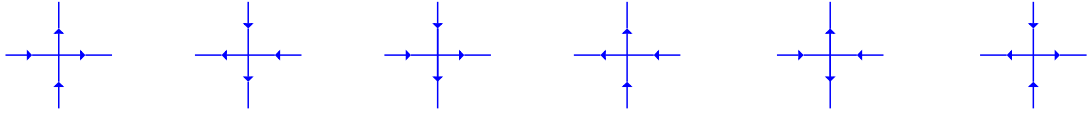
These  Boltzmann weights can be parametrized in terms of a spectral parameter $u$, and  logarithmic derivatives of the transfer matrix 
at $u=0$ give rise to  $XXZ$ spin chains Hamiltonians. The parameter $\Delta=(a^2+b^2-c^2)/2ab$ is independent of $u$ and parametrizes 
the (spin space) anisotropy of the $XXZ$ interaction. It is standard to set $\Delta=-\cos\gamma$. 

Now an important point is that, if we focus on the regime of the $6$-vertex model for which weights are real and positive, the 
Hamiltonians so obtained are all ferromagnetic (i.e., the coefficient of their interaction $S_{1}S_{1}'+S_{2}S_{2}'$ is negative). 
Hence, in particular, this procedure does not give rise directly to the AF $XXX$ Hamiltonian.  Indeed, the parametrization of the 
weights in the regime of the 6-vertex model known to be in the same universality class as the AF $XXX$ chain is $a=1-u,b=u,c=1$, 
giving
\begin{equation}
\check{R}=\left(1-\frac{u}{2}\right) \mathbb{I}- 2~u~ h\label{Rlimit}
\end{equation}
with 
\begin{equation}
h=-\left(S_1S'_1+S_2S'_2-S_3S'_3\right)\label{smallh}
\end{equation}
where $\vec{S},\vec{S}'$ are the spin $1/2$ operators for the west/north and south/east edges respectively). 
This is in fact the Hamiltonian of the ferromagnetic $XXZ$ spin chain at $\Delta=-1$. While eq.\ (\ref{Rlimit}) corresponds 
to the limit $\gamma\to 0$, similar results hold for $\gamma\in [0,\pi]$. In particular, the limit $\gamma\to \pi$ gives rise  
to the ferromagnetic $XXX$ spin chain. 

Starting from $N$  $\check{R}$ matrices, the transfer matrix propagating say vertically from bottom to top is obtained as follows. 
We give the special label $\alpha$ to spins on the horizontal edges, and introduce new matrices $R=P\check{R}$ where $P$ is the 
permutation operator. These matrices now send vertical (resp. horizontal) edges to vertical (resp. horizontal) edges and thus can 
be given labels $R_{\alpha,i}$ with $i=0,\ldots N-1$. The product 
\begin{equation}
T(u)=R_{\alpha,N-1}(u)\ldots R_{\alpha,0}(u)\label{monomat}
\end{equation}
is the monodromy matrix. It reduces to the right translation operator in the limit $u\to 0$: $T(0)=\tau$. The transfer matrix itself 
is the trace $t(u)=\hbox{Tr}_\alpha T(u)$. 

For a periodic 6-vertex model, the Hamiltonian is then obtained as the logarithmic derivative of the transfer matrix at $u=0$ (it 
commutes with it for arbitrary $u$; the sign in (\ref{ferro$XXZ$}) is of course such  that the ground state of the Hamiltonian and 
transfer-matrix are the same) 
\begin{equation}
H=-\sum_{\ell=0}^{N-1} S_{\ell 1}S_{\ell+1, 1}+S_{\ell 2}S_{\ell+1, 2}-S_{\ell 3}S_{\ell+1, 3}\label{ferro$XXZ$}
\end{equation}
with $\vec{S}_{N}=\vec{S}_0$.  When $N$ is even, this Hamiltonian is related to the periodic AF $XXX$ spin chain by a simple unitary 
transformation (a product of rotations by $\pi$ around the $3$ axis for every other spin \cite{gaudin2014bethe}; note that this is 
equivalent to the transformation that produces the Marshall sign rule \cite{Marshall1955}). When $N$ is odd 
however, the two Hamiltonians have different spectra. To recover the AF $XXX$ chain in the limit $u\to 0$ --- and thus to guarantee 
that the vertex model provides the Euclidian equivalent of the spin chain (meaning, the associated  low-energy, long distance limits are the same) 
---  we need to act with the same unitary transformation on  a {\sl twisted} ferromagnetic $XXZ$ chain, 
i.e.\ (\ref{ferro$XXZ$}), 
but with boundary conditions $S_{N1}=-S_{01}$, $S_{N2}=-S_{02}$, while $S_{N3}=S_{03}$.  Note that this  
preserves the commutation relations of the operators.

This, in turn, is obtained in the $u\to 0$ limit of the 6 vertex model with weights as above but with twisted boundary conditions. 
The Hamiltonian is now the logarithmic derivative of a modified transfer matrix \cite{Klumper_1991} 
$t(u)=\hbox{Tr}_\alpha \left[ e^{\epsilon i\pi S_{\alpha 3}} t(u)\right]$, $\epsilon=\pm1$, where $\epsilon=\pm 1$ can be chosen 
arbitrarily, but has to be constant throughout the calculations. This twist corresponds to giving to arrow pointing right (resp., left) 
at the edge an extra weight $\pm\epsilon~ i$. Note that now $t(0)$ is a modified translation operator  
$t(0)=e^{\epsilon i\pi S_{03}} \tau$, and commutes with the twisted Hamiltonian.

Bethe-ansatz  calculations \cite{DEVEGA1987619,ALCARAZ} show that the critical properties of the 6-vertex model with weights 
as given above are the same as those obtained in  the limit $u\to 0$ for the whole region $0<u<1$, including the (space) isotropic 
point $u=\frac{1}{2}$. It follows from this that the two-dimensional statistical mechanics equivalent of the partition function 
(\ref{partfn}) --- in particular, the (logarithms of the) eigenvalues of the transfer matrix have the same scaling as those of the 
eigen-energies of the AF $XXX$ Hamiltonian --- is, for $N$ even, the partition function of the 6-vertex model  with $u=\frac{1}{2}$ and 
an even length $M$ in the time direction. 

The constraint on the length $M$ can be understood by thinking of the model in the crossed channel with space and time exchanged. 
In this case, we would like a model equivalent in the Hamiltonian  limit (which is now at $u\to 1$) to the AF $XXX$ spin chain; 
but as we have seen this only works when the length $M$ is even. 

The continuum limit of the 6-vertex model partition function with doubly periodic boundary conditions and $N,M$ even is indeed well known 
to reproduce $Z_{PP}(\uptau)$ in the continuum limit, with $\uptau=i \frac{M }{N}$ \cite{fsz87}. Partition functions with a more general 
$\uptau$ are obtained by moving away from the value $u=\frac{1}{2}$: in general we have $\frac{\uptau_2}{\uptau_1}=\tan (\pi u)$ 
\cite{KOO1994459}. Note that taking $u\neq \frac{1}{ 2}$ makes the weights $a,b$ different but leaves $\Delta$ invariant.

The partition function  $Z_{AP}$ from the main text is then obtained by taking the 6-vertex model with $N$ odd and $M$ even, 
together with  --- as discussed above ---
 twisted boundary conditions in the space direction. The twist means that every configuration of arrows is weighed, on top of the 
product of local vertex weights $a,b,c$, by a factor $(\epsilon_x i)^{\N_r-\N_l}$ where $\N_r$ and $\N_l$ are the number of right and left 
arrows crossing the column $x=0$. (In general we dub the  column or row  where $\N_r$ and $\N_l$ are evaluated a frustration line. 
This line can be oriented so that arrows crossing it from left to right get a weight $i$. The choice of $\epsilon_x=\pm 1$ then amounts 
to a choice of orientation). Since  $\N_r+\N_l=M$ is an even number for $M$ even, $\N_r-\N_l$ is also even, so the weight is not 
affected by the choice of $\epsilon_x=\pm 1$ - and is real.  Moreover, by charge (magnetization) conservation in the crossed channel,  
$\N_r-\N_l$ does not depend on the column where it is evaluated \cite{bax85}.

The PA partition function would be reproduced from a vertex model with $N$ even and $M$ odd, and twisted boundary conditions in the 
time direction. In this case, the phase factors would be $(\epsilon_y i)^{\N_u-\N_d}$ where $\N_u$ and $\N_d$  are the number of up 
and down arrows crossing a given row (with $\N_u-\N_d$ independent of the row) and the same arguments would apply. 

Now for the AA partition function two choices are possible: we can take $\epsilon_x=\epsilon_y\equiv \epsilon$ (case (i)) or 
$\epsilon_x=-\epsilon_y\equiv\epsilon$ (case (ii)). In  case (i), up moving arrows crossing any given row and right moving arrows 
crossing any given column get a factor $\epsilon i$, and the opposite for down and left. So the total factor is 
$(\epsilon i)^{\N_r-\N_l+\N_u-\N_d}$. In  case (ii) we get a total factor  $ (\epsilon i)^{\N_r-\N_l+\N_u-\N_d}(-1)^{\N_u-\N_d}$. 
Since now both $\N_r=\N_l$ and $\N_u-\N_d$ are odd, their sum is even, and once again the phase factor doesn't depend on $\epsilon$ 
(and is real). But because of the $(-1)^{\N_u-\N_d}$ this factor takes opposite values for choices (i) and (ii).

The two choices can be interpreted as two different splits of the crossing of the two  (row and column) frustration lines. 
In case (i), up and right  (resp. down and left) moving arrows get the same phase factor, so this can be interpreted as a split 
where the frustration line comes vertically, veers to the left, goes around the (space) periodic boundary condition, veers to the 
right and continues vertically until it wraps around the (time) periodic boundary condition.

Note that in the case $u=\frac{1}{2}$ and thus $\uptau_1=0$, $\uptau$ purely imaginary, the local Boltzmann weights are invariant under 
reflection symmetry through the column $x=0$ or the row $y=0$. This symmetry switches the sign of $\N_r-\N_l$ or $\N_u-\N_d$. Since 
either of these numbers is odd, the reflection changes on the other hand the sign of the contribution of every arrow configuration to 
the partition function. The partition function is thus equal to its opposite, so it vanishes identically in this case. 

The different partition functions of the $SU(2)$ level 1 WZW CFT (\ref{ZPPZAP},\ref{ZPA},\ref{ZAA}) can then be obtained using 
standard Bethe-ansatz results to analyze the low-energy spectrum. 

Extension of the previous discussion to the $XXZ$ case is straightforward. The vertex model weights are now given by 
$a=\sin(\gamma-u)$, $b=\sin u$, $c=\sin\gamma$ (an overall rescaling by $\gamma$ is necessary to compare with results at 
$\gamma=0$), so the $\check{R}$ matrix near $u\to 0$ reads now
\begin{equation}
\check{R}\approx \left(\sin \gamma- \frac{u}{2}\cos\gamma\right)\mathbb{I} -2~u~h_\gamma
\end{equation}
where now 
\begin{equation}
h_\gamma=-\left(S_1S'_1+S_2S'_2-\cos\gamma S_3S'_3\right)\label{smallhgamma}
\end{equation}
To map this onto the antiferromagnetic $XXZ$ spin chain, the same unitary transformation as in the $XXX$  (product of rotations by 
$\pi$ around the 3 axis for every other spin) is necessary, and it can in turn be implemented in the 6-vertex model using the 
same twisted boundary conditions. The discussion then carries over straightforwardly, as the twist and the parity constraints on 
the numbers of arrows ${\cal N}$ are independent of $\gamma$.  

Moving now to higher spin, little is known about (non-integrable) vertex models giving rise to higher spin Heisenberg chains in the Hamiltonian  limit 
--- in particular concerning their phase diagrams, the existence of a (space) isotropic point etc. Nonetheless, we expect that the same 
phenomenon we observed in eqs.\ (\ref{Rlimit}), (\ref{smallh}) should generalize. The point is that, for $u$ positive and small (so 
$\check{R}\propto e^{-uh}$ as $u\to 0$ gives rise to a  transfer matrix with same ground-state as the Hamiltonian), the (in particular, 
off-diagonal) elements 
of $\check{R}$ should all be positive in order to obtain physical (positive, real) Boltzmann weights for the vertex model [like 
$a,b,c$ in eq.\ (\ref{weights})]. This is the case for the same choice (\ref{smallh}) but now for higher spin (Refs.\ \cite{Marshall1955},
\cite{BARWINKEL2000227}). It follows that, just like for spin $1/2$, the physical Euclidian version of the spin $S$ $XXX$ 
Heisenberg chain involves a vertex model with arrows now taking $2S+1$ values, and with the limit $u\to0$ (eq.\ \ref{ferro$XXZ$}), which is 
obtained like in the spin $1/2$ case by a product of rotations by $\pi$ around the $z$ axis for every other spin. Like for spin $1/2$, 
in the case $N$ odd, the boundary conditions are twisted: $S_{N1}=-S_{01}$, $S_{N2}=-S_{02}$, $S_{N3}=S_{03}$. 

While the vertex model will not be integrable, the $\check{R}$ matrix (\ref{Rlimit}) generalized to higher spin will still give 
rise to the required Hamiltonian by the same construction as in the spin $1/2$ case. In particular, we can still introduce the 
monodromy matrix (\ref{monomat}), and the transfer matrix can still be written as its trace over $\alpha$ --- even though now two 
transfer matrices with different spectral parameters will not commute in general. 
For $N$ odd, the twisted boundary conditions for the Hamiltonian  are still  obtained by inserting in the monodromy matrix 
$e^{\epsilon i\pi S_{\alpha 3}}$. The trace at $u=0$  gives the modified translation operator, which still commutes with the twisted 
Hamiltonian for higher spin as well. Horizontal arrows crossing the frustration line now should get a weight $e^{\epsilon i\pi m}$, 
where $m=-S,-S+1,\ldots,S-1,S$.  In turn, the same construction as for $S=\frac{1}{2}$ carries over to the case of $N$ even $M$ odd, 
and then $N,M$ both odd.

Now we see that for spin $S$ an integer, the extra weights for arrows crossing the frustration line do not depend on $\epsilon$, since 
$e^{i\pi m}=e^{-i\pi m}$ for $m$ an integer. The partition functions of the vertex model are thus uniquely defined independently of 
the choice of $\epsilon_x,\epsilon_y$. In contrast, when $S$ is half an odd-integer, we have $e^{i\pi m}=-e^{-i\pi m}$: two possible 
signs for the AA partition function are thus obtained depending on how we split the crossing of the frustration lines, 
exactly like for  $S=1/2$. 

%%%%%%%%%%%%%%%%%%%%%%%%%%%%%%%%%%%%%%%%%%%%%%%%%%%%
\section{Other spin chains and sigma models}
\label{OtherSp}

Here we will sketch (as briefly as we can) how our results generalize to symmetry groups beyond $SU(2)$ and $SO(3)$; we will assume somewhat
more knowledge of Lie groups and their representations \cite{Dieck,fulton1991}, and also of algebraic topology \cite{Hatcher2002}, 
than in the remainder of the paper (those references 
should be consulted whenever an immediate citation is not given). We also assume that the reader is familiar with the arguments in the main text.
We mainly focus on the cases that lead to sigma models with inversion symmetry, parallel to the $O(3)$ model
in the main text, but we also include some analysis of the topological term for more general cases.

%%%%%%%%%%%%%%%%%%%%%%%%%%%%%%%%%%%
\subsection{Introductory remarks on symmetry groups, spins, and target spaces}

We will begin from a spin chain, in which each site is a nontrivial irreducible, and necessarily finite-dimensional, 
representation (not necessarily the same at each site)
of a compact, connected, simply-connected semisimple Lie group $\cal G$ with simple Lie algebra that acts on the irreducible 
representation (examples of the possible $\cal G$s are given below). [The assumption of a simple Lie algebra is not crucial
for most results. If $\cal G$ is semisimple but the Lie algebra is not simple, then $\cal G$ is a direct product of groups, 
each with a simple Lie algebra, 
and the irreducible representations of $\cal G$ are tensor products of those of the simple factors of the Lie algebra of 
$\cal G$ or of the corresponding direct factors in $\cal G$. Then the sigma model target space is a direct product of those 
of the direct factors of $\cal G$ also,
and the problem essentially reduces to that with a simple Lie algebra, except at certain points that we will mention.
For now, we will continue to assume a simple Lie algebra unless stated otherwise, but the modifications if that 
assumption were dropped should usually be evident.] $\cal G$ may not act faithfully on the irreducible representation,
but $\cal G$ has a finite center [for a group $K$, its center, denoted by $C$ or $C(K)$, 
is the subgroup of elements that commute with every element of $K$],
and for any nontrivial representation of $\cal G$ there is a quotient group of $\cal G$ by a subgroup
(possibly, the one-element trivial subgroup, or $C$ itself) of the center  
that acts faithfully. We let $G$ be the adjoint form of the group, that is, $\cal G$ modulo its center, $G={\cal G}/C({\cal G})$; 
the center of $G$ is therefore trivial, $C(G)=0$ (if the Abelian group is written additively). [In the main text, $G$ is $SO(3)$, 
and $\cal G$ is $Spin(3)\cong SU(2)$.] We usually (but not always) wish to ensure the existence of a translation-invariant 
Hamiltonian on a chain of $N$ sites, and for that we require that each site be the same (i.e.\ isomorphic) complex (and unitary) 
representation $V$ of $\cal G$. To make it a Heisenberg Hamiltonian, we use nearest-neighbor terms bilinear in the generators 
of $\cal G$ on the two sites and invariant under the group, as in eq.\ (\ref{spinchHam}); 
such a term is unique up to a constant factor (though additional $\cal G$-invariant terms do not necessarily 
affect the physics that we discuss). To make it truly antiferromagnetic, 
we require that the spin at each odd-numbered site be the dual of the one at each even-numbered site,
because then, for one choice of sign of the coefficient, the ground state
of a nearest-neighbor pair is a singlet state. For this case, the generalization of N\'eel order also works out; 
for a self-dual irreducible representation, the ``opposite'' weight of the highest 
(or ``dominant'') weight is a ``lowest'' weight that also occurs in the representation 
(see e.g.\ Ref.\ \cite{Dieck}, page 261), and for the non-self-dual version, the ``opposite'' of the highest weight 
is a lowest weight in the dual. Then these highest- and lowest-weight states can be alternated to obtain a N\'eel state.
Hence, to obtain both antiferromagnetism and translation 
invariance in the same model, we require that $V$ be self-dual. [We will sometimes consider the more general situation of
a non-self-dual irreducible representation $V$ on the even sites that alternates with 
its dual on the odd sites; in this case, translation by one site is not a symmetry. If the representations alternate, but 
are not dual to each other, the likely result is ferromagnetism.] 
For a self-dual irreducible representation $V$, we let
$\widehat{G}$ be the quotient of $\cal G$ that acts faithfully on $V$, which then is the same group for all sites. 
Finally we note that, for a self-dual representation $V$, there is no difficulty considering a translation-invariant chain 
with $N$ odd, as well as $N$ even. 

The coherent states of a single non-trivial irreducible representation $V$ (here not necessarily self-dual) of $\cal G$, 
generated by beginning with a highest weight state and applying an element of $\cal G$, 
form a compact manifold (a homogeneous space) homeomorphic 
to ${\cal G}_\mathbb{C}/{\cal P}$, where ${\cal G}_\mathbb{C}$ is the complexification of $\cal G$, 
and ${\cal P}$ is a ``parabolic'' (and proper) subgroup of ${\cal G}_\mathbb{C}$ (for background to this paragraph, see 
Ref.\ \cite{fulton1991}, section 23.3;
we will also explain more later). (We note that for the trivial one-dimensional representation, this space is a single point 
$\cong {\cal G}_\mathbb{C}/{\cal G}_\mathbb{C}$.)
In the most general cases, these are so-called flag manifolds.
[For a discussion of sigma models with target space a flag manifold, see Ref.\ \cite{AFFLECK20221}; note that this paper is mainly 
concerned with spin chains that have the same, not necessarily self-dual, representation on each site (in which case 
interactions beyond nearest neighbor are sometimes required), and with the corresponding sigma 
models, thus differing from our work, though there is some overlap.] On the other hand, the ``nicest'' spaces occur for 
${\cal P}$ a {\em maximal} parabolic subgroup,
that is, one that is not contained in any larger parabolic subgroup (again, we explain the meaning of ``maximal parabolic'' 
further below). [For some purposes, we need to know that these spaces are simply connected. This follows: 
$\pi_1({\cal G}_\mathbb{C}/{\cal P}) \cong \pi_0({\cal P})=0$, because any parabolic subgroup of ${\cal G}_\mathbb{C}$ 
is necessarily connected.]
Alternatively, we can view the same space as the compact group $\cal G$, modulo the (compact) isotropy subgroup ${\cal P}\cap {\cal G}$
of the highest weight vector
in the projective space $\mathbb{P}(V)$ of $V$ [for example, $\mathbb{P}(\mathbb{C}^n)=\mathbb{C}\mathbb{P}^{n-1}$], 
that is, the subgroup that leaves the highest weight vector in $V$ invariant up to a phase factor; 
thus for example, $\mathbb{C}\mathbb{P}^{n-1}\cong SU(n)/S[U(1)\times U(n-1)]$, and we give further examples below.
(Here we could replace the group $\cal G$ by $G$, and a similar quotient for the isotropy subgroup, as the center belongs to the 
isotropy subgroup of every point in $\mathbb{P}V$.) 
The discussion in Section \ref{SpinCh} generalizes to this situation; as we explain more fully later, the integer $2S>0$ 
must be replaced by a finite set of non-negative integers in general, which determine the highest weight in $V$,
and represent the Chern class of a line bundle on ${\cal G}_\mathbb{C}/{\cal P}={\cal G}/({\cal P}\cap {\cal G})$, 
uniquely determined (up to isomorphism) by $V$. 
(For a case of $\cal P$ maximal parabolic, only one of the integers is nonzero. In the more general case of 
$\cal G$ a direct product of more than one factor, this implies in fact that only one factor is represented nontrivially, 
and then $\cal G$ can be replaced by a group with simple Lie algebra.) Moreover, the
representation $V$ can be recovered uniquely (up to isomorphism) from the space ${\cal G}_\mathbb{C}/{\cal P}$ and the line bundle 
(the Borel-Weil construction). The passage from the spin chain to the sigma model in the semiclassical continuum limit works 
as before (where in general the sites alternate between $V$ and its dual), provided we consider
a sequence of spins $V$ (on the even sites, and its dual on the odd sites) in the same class, such that the homogeneous space 
is the same for all of them; then the target space 
of the sigma model is the same homogeneous space ${\cal G}_\mathbb{C}/{\cal P}$ 
[we will discuss the topological term(s) in detail below,
but just state here that the parameters are an ordered set $\Theta_l$ for a finite set of $l$ (the index $l$ was already used for branch cuts
in Sec.\ \ref{subsec:pathintmodtrans}, but as we will not write explicit formulas that involve both types of objects this should 
not cause any confusion), with each $\Theta_l$ 
equal to $\pi$ times one of the non-negative integers mentioned; thus for a maximal parabolic case, only one $\Theta_l$ is nonzero].
It may be instructive in practice to consider the ``simplest'', or lowest-dimensional, representation for which the 
homogeneous space is the one given. For $G=SO(3)$, that is spin $1/2$; we mention some other examples below, and give 
a general discussion later in this Appendix. Once again, from the target space ${\cal G}_\mathbb{C}/{\cal P}$ 
and the knowledge of the full set of topological-term parameters $\Theta_l$ (which correspond to the line bundle, 
as we describe below), the representation $V$ can be recovered.
Further, self-duality of $V$ implies that there is an inversion symmetry $I$ (with $I^2=1$) of the target space, which commutes 
with the Lie group symmetry, but just as in the case of $SU(2)$ or $SO(3)$, $I$ does not belong to either $G$ 
or ${\cal G}$. Then the group of symmetries that act on the sigma model target space is a direct product, $G\times \mathbb{Z}_2$,
which has two connected components, like $O(3)\cong SO(3)\times\mathbb{Z}_2$. 

%%%%%%%%%%%%%%%%%%%%%%%%%%%%%%%%%%%%%%%
\subsection{Examples of spin chains and target spaces}
\label{sec:extarget}

We now give some examples of these considerations, using cases in which the parabolic subgroup is maximal; 
in these cases there is only a single nonzero $\Theta_l$ parameter, as in the $O(3)$ sigma model. Also, we describe only
cases in which the (irreducible) representation $V$ is self dual, the case of main interest in this paper. For many cases, we can describe the
irreducible representations in the language of (Ferrers-) Young diagrams and tableaux, with which the reader may be familiar
(this is Weyl's construction). 
We will also describe the action of the center on some of the representations considered. The values of the $\Theta_l$ parameters 
will be determined later in general.
(For the mathematical background to this paragraph, again see Ref.\ \cite{fulton1991}, 
section 23.3 and also page 447, and we explain further below). 

First we have ${\cal G}=SU(n)$ spin chains with $n$ even, 
$n\geq 2$. (If $n$ is odd, there is no self-dual irreducible representation with maximal parabolic subgroup.)
Note that the center of $SU(n)$ is $\mathbb{Z}_n$ (if $n$ is odd there is no $\mathbb{Z}_2$ quotient of the 
center $\mathbb{Z}_n$). Then $G\cong SU(n)/\mathbb{Z}_n$, which is known as $PSU(n)$, the projective special unitary group.
The irreducible representations $V$ described by a rectangular Young diagram with $n/2$ rows of boxes are self-dual. For these, 
the center acts trivially if the number of columns is even (because then the total number of boxes is a multiple of $n$)
and, if the number of columns is odd, there are $n/2$ elements of the center that act as $-1$. Then when the number of columns is 
even we have $\widehat{G}\cong G$, but when the number of columns is odd the group $\widehat{G}$ is 
$\widehat{G}\cong SU(n)/\mathbb{Z}_{n/2}$, which is a double cover of $G$ (the center of this $\widehat{G}$ is 
$\mathbb{Z}_2\cong \mathbb{Z}_n/\mathbb{Z}_{n/2}$). It has long been understood \cite{AFFLECK1985397} 
that the continuum limit of these spin chains is a sigma model with target space the complex Grassmannian manifold 
${\cal G}_\mathbb{C}/{\cal P}=\mathbb{G}_{n,n/2}$
[for which there are equivalent definitions
\begin{equation}
\mathbb{G}_{n,k}\cong U(n)/[U(k)\times U(n-k)]\cong SU(n)/S[U(k)\times U(n-k)],
\label{GrassmanDef}
\end{equation}
and as usual the prefix $S$ means the subgroup of matrices with determinant $1$],
and with the single nonzero $\Theta_l$ parameter equal to $\pi$ 
times the number of
columns in the diagram. If the number of columns is odd, these representations are real if $n=0$ (mod $4$), but quaternionic 
if $n=2$ (mod $4$); later we explain the meaning of these terms, which also relate to time reversal and will be crucial
for the results. The simplest $V$ for the $n=4$ case has a single column in the diagram,
and can also be viewed as the six-dimensional vector representation of $SO(6)\cong SU(4)/\mathbb{Z}_2$. 

There are numerous other examples. For ${\cal G}=Spin(n)$ with $n>1$ odd, we saw in Appendix \ref{PinG} that its center is 
$\mathbb{Z}_2$, and $G=SO(n)$. On the other hand, for $Spin(n)$, $n>2$ even, the adjoint form $G$ is denoted $PSO(n)$.
Here the center of $Spin(n)$ is $C=\mathbb{Z}_4$ if $n=2$ (mod $4$), while $C=\mathbb{Z}_2\times\mathbb{Z}_2$ if $n=0$ (mod $4$)
(see Appendix \ref{PinG}). For $n$ even, the center $C=\mathbb{Z}_2$ of $SO(n)$ is generated by the inversion denoted $I$ 
in Appendix \ref{PinG}; then $PSO(n)\cong SO(n)/\mathbb{Z}_2$. For any even $n$, $SO(n)\cong Spin(n)/\mathbb{Z}_2$,
where $\mathbb{Z}_2$ is a particular $\mathbb{Z}_2$ subgroup of the center $C(Spin(n))$ [for $n=2$ (mod $2$), 
there is only one possibility for the $\mathbb{Z}_2$ subgroup, but for $Spin(n)$ for $n=0$ (mod $4$) there are two 
$\mathbb{Z}_2$ subgroups other than that which yields $SO(n)$ as the quotient].

Still for $n$ even, inversion $I$ obviously acts non-trivially on the $n$-dimensional vector representation $V$ (which exists for any 
$n$), and the target space for these cases is $Spin(n)/[U(1)\times Spin(n-2)]$ 
[this is almost the same as $SO(n)/[U(1)\times SO(n-2)]$, and $n=6$ here agrees with $n=4$ in the preceding example of $SU(4)$]; 
the inversion in the center of $SO(n)$ acts trivially on this 
target space, so is {\em not} the inversion of the target space referred to above. The general irreducible representations
that give to rise to this target space can be described as a Young diagram with only one row of boxes, and the only nonzero $\Theta_l$ 
parameter is the number of boxes times $\pi$.

For ${\cal G}\cong Spin(n)$ with any odd $n\geq 3$, and for the symplectic groups ${\cal G}\cong Sp(2n)$ with any $n\geq 1$, 
in each case all irreducible representations are self-dual, the center of $\cal G$ is $\mathbb{Z}_2$, 
and the situation resembles the special cases $Spin(3)\cong Sp(2)\cong SU(2)$: there are many irreducible representations 
on which the center acts non-trivially. For $Spin(n)$, $n$ odd, which we write as $n=2r+1$, the simplest $V$ with nontrivial 
action of the center is of course the spinor representation, which has dimension $2^r$. 
The sigma model target space is $Spin(2r+1)/U(r)$, the so-called spinor variety, and the only nonzero $\Theta_l$ is equal to 
$\pi$ for the spinor. 
For $Sp(2n)$, the simplest $V$ that can be described uniformly for all $n>0$ is the $2n$-dimensional fundamental representation.
The center $C(Sp(2n))$ is a $\mathbb{Z}_2$ subgroup of the center of $SU(2n)$ [note that $Sp(2n)\subset SU(2n)$], 
and acts nontrivially on the $2n$ dimensional $V$ 
(also $V$ is quaternionic for all $n$). The corresponding sigma model target 
space is $Sp(2n)/[U(1)\times Sp(2n-2)]$ \cite{PhysRevLett.66.1773,SachdevRead}. The same target space is obtained 
for all the irreducible representations that correspond to a Young diagram with only one row, and the 
only nonzero $\Theta_l$ parameter is $\pi$ times the number of boxes in the diagram. 
For the cases $Spin(5)\cong Sp(4)$, these spaces coincide; the spinor of $Spin(5)$ is the fundamental of $Sp(4)$, 
and the target space is $\cong \mathbb{C}\mathbb{P}^3$. 
[We note that there is ambiguity here in the familiar coset-space notation we use. There are two spaces that might be 
called $Sp(4)/[U(1)\times SU(2)]$, on which $Sp(4)$ acts differently. 
One space is $\mathbb{C}\mathbb{P}^3$ \cite{fulton1991}, while the other arises as the space of coherent states of the $5$-dimensional 
vector representation of $Sp(4)/\mathbb{Z}_2\cong SO(5)$, suggesting the alternative notation $SO(5)/[SO(2)\times SO(3)]$.
The difference between the spaces arises because the Lie algebras of $Spin(n)$ (for odd $n>3$) and $Sp(2n)$ (for all $n>1$) 
have both ``long'' and ``short'' roots \cite{Dieck,fulton1991} (they are said to be ``not simply laced''), so each has 
two types of $SU(2)$ subgroups generated by these roots. There is no ambiguity if
the form ${\cal G}_\mathbb{C}/{\cal P}$ is used, because the ${\cal P}$s are different (not isomorphic) in the two cases.]

Hence we have given some examples associated with the classical groups $SU(n)$ ($n$ even), $SO(n)$, and $Sp(2n)$
[it will be convenient, though not standard terminology, to refer to the double cover $Spin(n)$ of $SO(n)$ 
as classical also]; for most of these values of $n$, there are also other examples of target spaces and sigma models 
with a single nonzero $\Theta_l$ apart from those mentioned here (as we explain later).
In all of the examples given here, the center of $\cal G$ acts nontrivially on the irreducible representation $V$ 
if and only if the single nonzero $\Theta_l$ parameter equals $\pi$ (mod $2\pi$); for the cases of $SU(n)$, $n$ even, and $Spin(n)$, 
$n$ odd, we have given the unique examples of such target spaces. In general, there will be more than 
one nonzero $\Theta_l$ parameter, and the relation with the center will be more complicated.
We might point out that, for each group 
$G\times\mathbb{Z}_2$, the double cover $[\widehat{G}\times\mathbb{Z}_4]/\mathbb{Z}_2$, which may arise when $\Theta_l=\pi$ 
(mod $2\pi$) and the b.c.\ is A, does not in general have a
well-known name [in particular, for $G=PSO(n)$, $n$ even, $PSO(n)\times\mathbb{Z}_2\not\cong SO(n)$, 
and $[SO(n)\times \mathbb{Z}_4]/\mathbb{Z}_2\not\cong O(n)$ (cf.\ Appendix \ref{PinG})]. However, for $n$ odd,
$O(n)\cong SO(n)\times \mathbb{Z}_2$, and the spinor is an irreducible (but possibly not faithful) representation 
of either of the double covers $Pin_\pm(n)$, which were discussed in Appendix \ref{PinG}. 

%%%%%%%%%%%%%%%%%%%%%%%%%%%%%%%%%%%%%
\subsection{Invariant \texorpdfstring{$\cal C$}{\cal C} and time reversal: Theorem 1}
\label{subsec:thm1}

Now we consider the main results from the text in the present general setting of a covering group $\cal G$, 
a $\cal G$-invariant, translation-invariant spin chain with a self-dual irreducible representation $V$ of $\cal G$ at each site, 
or the corresponding sigma model. Later we will describe the representations and the topological term(s) in detail, 
in which the highest weight $\lambda$ of the representation $V$ is represented by an indexed set of non-negative integers $(a_l)_l$
(note our notation $(\cdot)_i=(\cdot)_{i\in A}$ for an indexed set, where the index $i$ is an element of a set $A$),
and the coefficients $\Theta_l=\pi a_l\geq 0$, which means that from the $\Theta_l$ coefficients as real numbers (not modulo $2\pi$)
the representation or ``spin'' $V$ with highest weight $\lambda$ can be recovered within the sigma model. Those details are 
not essential at present, but they mean that we can determine all necessary data from a given sigma model of the present type. 
In general, self-duality may impose additional restrictions on the values of $(\Theta_l)_l$, which we explain later,
and an understanding of those is not required at the moment, though the restrictions are necessary for the validity of what follows..

In this setting, we can extend the construction of the topological invariant $\cal I$ to the group $\cal G$ and the sigma model
with target space ${\cal G}_\mathbb{C}/{\cal P}$ with given values of the topological coefficients $\Theta_l\geq 0$ for all $l$. 
At the most basic level, 
in the $O(3)$ case, we observed that the integrand of the topological term is odd (or ``antisymmetric'') under inversion. 
If we view the (sum of the) topological term(s) in general as the integral of a $2$-form [the curvature or field strength of 
a line bundle or connection
(vector potential) on the target space], then defining the integral of any $2$-form involves a choice of orientation on spacetime. 
The statement of inversion (anti-) symmetry of the $2$-form must more formally be stated as saying that the pullback of the 
$2$-form by the inversion map gives minus the same $2$-form. These properties were the starting point, and they do hold for 
the general cases we consider, provided $V$ is self-dual. (We will check this below.) As we observed at the end of 
Sec.\ \ref{subsec:pathintmodtrans}, the anomaly is purely topological, and in fact the result holds whenever the target space and the topological data 
(the $\Theta_l$s) are as given here (and in more detail below).

It was the antisymmetry of the $2$-form that necessitated the attention paid to 
branch cuts, across which the sigma model field jumps from a point in the target space to its antipode (image under inversion). 
For each cut, in the first formulation of $\cal I$ we must choose a side of the cut (or assign an orientation), 
and evaluate the Berry phase of the spin $V$ with highest weight $\lambda$ corresponding to the topological terms, 
to obtain the ``boundary'' terms for the cuts. [Alternatively, we can use the second form of the invariant, for which it is helpful to 
notice that because the target space ${\cal G}_\mathbb{C}/{\cal P}$ is always simply connected, it is always orientable.] Together with 
the ``bulk'' contribution of the topological terms, this produces $\pi {\cal N}'$, 
which a multiple of $\pi$ that is well-defined modulo $2\pi$. Hence ${\cal I}=e^{-i\pi{\cal N}'}=\pm 1$ 
is a topological and isotopy invariant just as in Sec.\ \ref{subsec:pathintmodtrans}. We note here that, as for the $SU(2)$ case,
while the use of the same set of $\Theta_l$ (or $a_l$) for both bulk and boundary terms was crucial, the final result
depends on each $\Theta_l$ only modulo $2\pi$ (or on each $a_l$ modulo $2$).

We can now repeat all the main arguments concerning gauge invariance and modular transformations 
in terms of branch cuts as in Sec.\ \ref{subsec:pathintmodtrans}. As we saw there, $\cal I$ will be gauge invariant
(and modular invariance of the sum over boundary conditions will hold) if and only if the Berry phase factor, which we will denote 
by $\cal C$, for the spin in the irreducible representation $V$, evaluated on a loop obeying a certain ``equivariance'' property, 
is $+1$. In the notation for the $O(3)$ sigma model, the target space was $\mathbb{S}^2$, and the field was represented by a 
unit vector $\vec{n}$. The property in question was that, if the loop is represented by $\vec{n}(x)$ with $x$ viewed as periodic 
with period $2L$, then $\vec{n}(x+L)=-\vec{n}(x)$ for all $x$. In the present case, the condition is that the sigma model field, 
or point in the target space, for $x+L$ is the antipode of that for $x$. In particular, the loop passes through the antipode
of its starting point. 

We will evaluate this Berry phase factor directly, which will show incidentally that it must be $\pm 1$.
Let us continue to use the notation $\vec{n}$ for points in the target space, and $-\vec{n}$ for the antipode (or inversion)
of $\vec{n}$, even in the general case. It is convenient to consider a discrete loop; the result holds for its phase even for this case,
as long as the phase factor is well-defined. 
Thus we will consider a sequence of points $\vec{n}_\ell$, $\ell=0$, $1$, \ldots, $2M$, with $\vec{n}_{\ell+M}=-\vec{n}_\ell$. 
For each such point, there is a corresponding coherent state $|\vec{n}_{\ell}\rangle$ of the spin, uniquely determined by $\vec{n}$ 
up to a phase factor. The Berry product (also known as a ``Bargmann invariant'') associated to the discrete loop is then 
\beq
\prod_{\ell=0}^{2M-1}\langle \vec{n}_{\ell+1} |\vec{n}_{\ell}\rangle,
\eeq
in which we recall that the arbitrary choice of phase in each $|\vec{n}_\ell\rangle$ cancels.

We recall that, for self-dual $V$, the lowest weight ``opposite'' to the highest weight is also in the representation.
Because the lowest weight vector can be obtained by applying a $\cal G$ element to the highest weight vector,
it too is a coherent state and, up to a phase factor, is the time reverse of the highest weight state. The same is true for 
any coherent state generated from the highest weight: the coherent state $|-\vec{n}\rangle$ can be viewed as ${\cal T}'|\vec{n}\rangle$,
up to a phase factor. (Here ${\cal T}'$ is time reversal acting on the spin $V$, as in the spin chains in the main text.) 
We write the latter as ${\cal T}'|\vec{n}\rangle=|{\cal T}'\vec{n}\rangle$. Then for $\ell=0$, $1$, \ldots, $M-1$, we define
$|\vec{n}_{\ell+M}\rangle=|{\cal T}'\vec{n}_\ell\rangle$. Then in the Berry phase factor, we consider 
$\langle \vec{n}_{\ell+M+1} |\vec{n}_{\ell+M}\rangle=\langle {\cal T}'\vec{n}_{\ell+1} |{\cal T}'\vec{n}_{\ell}\rangle$ 
for $\ell=0$, \ldots, $M-1$. Now we need to make use of the adjoint operation on the antiunitary operator ${\cal T}'$,
which is not always defined in quantum mechanics textbooks. In contrast to a linear operator $A$, for which the adjoint $A^\dagger$ 
is defined by requiring that $\langle A\psi|\psi'\rangle=\langle \psi|A^\dagger\psi'\rangle$ for any two state vectors $|\psi\rangle$, 
$|\psi'\rangle$, for an antilinear operator $S$ the definition of $S^\dagger$ is 
$\langle S\psi|\psi'\rangle=\overline{\langle \psi|S^\dagger\psi'\rangle}$ for any two state vectors $|\psi\rangle$, 
$|\psi'\rangle$. (As for linear operators, $S^{\dagger\dagger}=S$
for any antilinear $S$. We note in passing that Dirac's notation $\langle \psi|A^\dagger|\psi'\rangle\equiv 
\langle A\psi|\psi'\rangle=\langle \psi|A^\dagger\psi'\rangle$ fails to make sense at all for an antilinear operator $A$.) 
For us, ${\cal T}'$ is antiunitary, which implies ${\cal T}'^\dagger={\cal T}'^{-1}$.
Then we obtain $\langle \vec{n}_{\ell+M+1} |\vec{n}_{\ell+M}\rangle=\langle {\cal T}'\vec{n}_{\ell+1} |{\cal T}'\vec{n}_{\ell}\rangle
=\overline{\langle\vec{n}_{\ell+1} |\vec{n}_{\ell}\rangle}$ (the final equality is actually the definition of anti-unitarity 
of ${\cal T}'$). If we divide by the magnitude of the product (assuming it is nonzero), 
then all but two of the phase factors in the product cancel in pairs, leaving 
\begin{eqnarray}
\frac{\langle\vec{n}_{M} |\vec{n}_{M-1}\rangle\langle\vec{n}_{0} |\vec{n}_{2M-1}\rangle}
{|\langle\vec{n}_{M} |\vec{n}_{M-1}\rangle\langle\vec{n}_{0} |\vec{n}_{2M-1}\rangle|}
&=&\frac{\langle{\cal T}'\vec{n}_{0} |\vec{n}_{M-1}\rangle\langle\vec{n}_{0} |{\cal T}'\vec{n}_{M-1}\rangle}
{|\langle{\cal T}'\vec{n}_{0} |\vec{n}_{M-1}\rangle\langle\vec{n}_{0} |{\cal T}'\vec{n}_{M-1}\rangle|}\nonumber\\
&=&{\cal T}'^2\frac{|\langle\vec{n}_{0} |{\cal T}'\vec{n}_{M-1}\rangle|^2}
{|\langle{\cal T}'\vec{n}_{0} |\vec{n}_{M-1}\rangle\langle\vec{n}_{0} |{\cal T}'\vec{n}_{M-1}\rangle|}\nonumber\\
&=&{\cal T}'^2
\end{eqnarray}
where, at the end, ${\cal T}'^2$ means the eigenvalue $\pm 1$ of ${\cal T}'^2$ acting on the self-dual irreducible representation $V$. 
We can take a limit in which a continuous loop in the target space 
is obtained, and in that limit the magnitude of the Berry product tends to unity. That gives the Berry phase factor denoted 
$\cal C$, and we conclude that 
\beq
{\cal C}={\cal T}'^2.
\eeq
This shows that ${\cal C}=\pm 1$, and of course, if ${\cal G}=SU(2)$, then ${\cal C}=(-1)^{2S}$, and so it agrees 
with the main text for this case. It also shows that it is the same for any loop with the stated property. 
We note that the target space is simply connected, and so its quotient by the inversion operation has fundamental group 
$\pi_1=\mathbb{Z}_2$. Thus there is only a single homotopy class of nontrivial loops in the quotient. Any loop in the target 
space with the given property can be viewed as the lift of a nontrivial loop in the quotient space, and hence in the target space 
there is only a single nontrivial homotopy class of loops with the desired equivariance property.

We have established that $\cal C$ is $\pm 1$, and by continuity of the Berry phase for a closed loop as the loop is varied,
it follows that it is a topological invariant. Then a small change in one part of the loop must be mirrored by a corresponding change
in the opposite part of the loop, and the flux threading the small loop representing each of the changes must be opposite so
that the total is unchanged. As the change in one small part of the loop is arbitrary, this confirms the antisymmetry
of the $2$-form under pullback by inversion, mentioned above. Note that this is similar to the argument in Section 
\ref{subsec:pathintmodtrans}, but run in reverse.

Hence we have shown that, in terms of the time-reversal behavior 
of the spin (i.e.\ the self-dual irreducible representation of $\cal G$) $V$ that corresponds to the topological terms, 
there is an 't Hooft anomaly in the inversion symmetry, and an anomaly in modular invariance,
in the sigma model in the path-integral formulation if and only if ${\cal T}'^2=-1$, where 
${\cal T}'^2$ can only be $\pm 1$. 
We note here that the gauge (non-)invariance of $\cal I$, depending on whether ${\cal C}=+1$ or $-1$, is a 
topological result that does not involve any assumption of translation invariance of the non-topological action 
or the Hamiltonian, however, the modular (non-)invariance arguments did use translation invariance in both space and time,
and hence the existence of a time-independent Hamiltonian $H$ and a conserved total momentum $P$ in the sigma model.
In general, we may as well assume the latter translation symmetries are valid. We also note that, 
for a self-dual irreducible representation $V$, one says that $V$ is real if ${\cal T}'^2=+1$, quaternionic if ${\cal T}'^2=-1$ 
(we discuss this further later).

We summarize these results in a Theorem, the main result of this Appendix (and of the paper for the general case).
Notation: we use the groups $\cal G$ and $G$, the sigma model with field $\vec{n}$ in the target space ${\cal G}_\mathbb{C}/P$ and 
parameters $\Theta_l\geq 0$ (always multiples of $\pi$), the irreducible self-dual representation $V$ of $\cal G$ that appears 
in the spin chain and which is associated to the indexed set of parameters $(\Theta_l)_l$ in the corresponding sigma model, 
and the invariants ${\cal I}[\vec{n}]$ and $\cal C$:

{\bf Theorem 1}: Consider a sigma model of the form discussed here. There is an invariant $\cal C$, defined as the Berry phase factor
for an equivariant loop in the target space of the sigma model and taking the values $\pm 1$. Then the topological invariant 
$\cal I$ and all of the partition functions with the four combinations of b.c.s untwisted and twisted by the inversion symmetry are
gauge invariant (so there is no 't Hooft anomaly, i.e.\ the inversion symmetry can be gauged, and hence modular invariance 
of the gauged model can be obtained) if and only if ${\cal C}=+1$. Further, ${\cal C}={\cal T}'^2$, where ${\cal T}'$ is 
the time reversal operator in the representation $V$; in other words, there is an 't Hooft anomaly if and only if $V$ is 
quaternionic, rather than real.

We comment here that this continues to hold when the Lie algebra of $\cal G$ is not simple, so that the sigma model target space
is a direct product, and the inversion map is the product of the inversion maps in each direct factor. The whole argument 
extends immediately to this case, and the value of $\cal C$ is the product of its values for each of the irreducible representations 
of the direct factors of $\cal G$ that appear as tensor factors in $V$; the same clearly holds for ${\cal T}'^2$, and again 
${\cal C}={\cal T}'^2$. We also comment that the first statement, that there is an 't Hooft anomaly exactly when ${\cal C}=-1$,
also holds for more general sigma models with inversion symmetry of the target space, provided that the $2$-form involved
in the topological terms is odd under pullback by inversion, as in the examples that are our main interest. There may be some deeper
topological argument that would determine the value of $\cal C$ in these more general cases, but we will not consider that further.

%%%%%%%%%%%%%%%%%%%%%%%%%%%%%%%%%%%%%%
\subsection{Consequences of Theorem 1}

While Theorem 1 is the main result, we will now flesh out in some detail how $\cal C$ determines results for translation and discrete 
symmetries in the sigma models and the spin chains. The discussion of the spin quantum numbers, especially for the spatial A b.c.\ sector, 
are more involved and best left for a later subsection.

First, for the sigma models, $\cal C$ determines the modular transformation 
behavior of the four partition functions.
As explained in the Introduction, for the spatial P b.c., the modular $T$ operation leaves the partition functions invariant,
and as $T$ is implemented by insertion of $e^{-iPL}$ inside the trace, it follows that all states in the Hilbert space 
have momenta of the form $P=2\pi m/L$ (where $m$ will denote a generic integer throughout) for either value of $\cal C$. For the 
spatial A b.c., we instead have  $T^2={\cal C}={\cal T}'^2$. $T^2$ corresponds to applying translation by $2L$, and then
for ${\cal C}=+1$, all states have momenta $P=\pi m/L$ (not only $2\pi m/L$), while for ${\cal C}=-1$, all states have momenta
$2\pi (m\pm 1/4)/L$. [As a side remark, we should explain that for the A b.c.\ and either value of $\cal C$, some pairs of states 
have momenta differing by $2\pi(m+1/2)/L$, which is connected with the fact that some fields, for example that representing 
$\vec{n}$ in the target space, have Fourier components with these momenta as a direct result of the A b.c., but to see this 
we would need to represent $\vec{n}$ with some coordinates, most likely matrices, with the property that the inversion operation 
reverses the sign of all the components, exactly as for the vector $\vec{n}$ in the $O(3)$ case.] We also find from modular 
transformations that inversion $I$ acts as $I^2=1$ in the spatial P b.c.\ sector, but as $\widehat{I}^2={\cal C}$ in the spatial
A b.c.\ sector, where (as indicated) $I$ must be lifted to one of $\pm \widehat{I}$ in the ${\cal C}=-1$ case.

For the discrete symmetries of spatial reflection (parity) and temporal reflection (time reversal) in the sigma model, 
we first considered the semiclassical quantization of the motion (on ``collective coordinate space'') of the classical texture 
for either the P or the A spatial b.c., and it will be useful to do the same here. Although we are concerned with properties
of how discrete symmetries act on the Hilbert space, which do not involve reference to the Hamiltonian,
it is convenient to assume the full internal symmetry under $\cal G$ of the Hamiltonian (in addition to translation, reflection, 
and time-reversal invariance), and then argue that the results will not change if the global internal symmetry is broken by 
continuously varying some parameters, as we did in the main text near eq.\ (\ref{eq:HA}). First, for the spatial P b.c., 
in the classical ground state $\vec{n}(x)$ is constant in space
(again letting $\vec{n}$ represent the field in the general case, with $-\vec{n}$ for the antipode of $\vec{n}$),
the collective coordinate space is the target space ${\cal G}_\mathbb{C}/{\cal P}$, and the wavefunctions
in semiclassical quantization can be chosen real. Then it is straightforward to see that reflection $\cal R$ acts trivially
and that time reversal $\cal T$ squares to ${\cal T}^2=1$ as in the $O(3)$ case in the main text. For the full Hilbert space, 
we can then use the arguments involving symmetry properties of the field $\vec{n}(x)$ and the conjugate angular-momentum 
density $\vec{j}(x)$ to obtain the results, which have the same form as for the $O(3)$ case.

In the case of a spatial A b.c., classically the lowest energy configuration is a texture in the sigma model target space field, 
such that $\vec{n}(0^+)=-\vec{n}(L^-)$, where $0$ and $L$ in $\mathbb{R}$ are identified to form a circle. Using a particular such 
texture as a reference (the details of which will not be needed at the moment), the orbit of this texture under the global symmetry $G$
is the collective-coordinate
space that enters semiclassical quantization. In general, this space is a homogeneous space, a quotient of $G$ by a subgroup
of $G_R$, and may not be simply connected; we postpone detailed discussion of these spaces until later.
[The spatial translations and reflections form $O(2)_{\rm sp}$. We can also view $O(2)_{\rm sp}$ as a subgroup of $G_R$ in the group
$G_L\times G_R$ of symmetries of the homogeneous space, as in the main text near eq.\ \ref{eq:HA}, and Appendix \ref{PetWeyl}.
$O(2)_{\rm diag}$ involves a subgroup $\cong O(2)$ in $G$, acting 
simultaneously with the corresponding element of the group $O(2)_{\rm sp}$. Because the reflection can be reached by a path of 
rotations, beginning from the identity, in $G$, this $O(2)$ subgroup of $G$ is embedded in an $SO(3)$ 
subgroup, as in the case of $G=SO(3)$.] 
Then as for $SO(3)$, the semiclassical motion 
on $G$ may involve a nontrivial connection or vector potential that represents a flux threading a nontrivial loop
in the homogeneous space, with values in a one-dimensional representation of the finite fundamental group of the homogeneous space. 
We discuss the general form of this connection in detail later, 
but here we will only need the fact, as discussed in the main text, that if the texture changes by continuous translation until 
translation by $2L$ is reached, then the invariant $\cal I$ contributes a phase factor or holonomy $\cal C$. This is the 
exponentiated flux or Berry phase factor for this loop in the homogeneous space, and if it is nontrivial for some nontrivial loop
in $\pi_1$ then $G$ is replaced by a covering group $\widehat{G}$ that acts linearly on the semiclassical quantum states. 
(We show later that, for any given $\cal G$, there is an element 
of $C({\cal G})$ such that time reversal squared ${\cal T}'^2$ on any self-dual irreducible $V$ is equal to the action of that 
element on $V$ as multiplication by a scalar, which relates these points.) Then if ${\cal C}=+1$, the $SO(3)$ subgroup of $G$
in which the copy of $O(2)_{\rm sp}$ lives remains $SO(3)$ in whatever covering group $\widehat{G}$ is obtained 
(possibly $\widehat{G}=G$), but if ${\cal C}=-1$ then $\widehat{G}\neq G$ definitely, and this $SO(3)$ [$O(2)_{\rm sp}$] 
is lifted to $Pin_+(3)$ [resp., $Pin_-(2)$], just as in the main text. We emphasize here that, even when $G$ is lifted 
to a nontrivial cover $\widehat{G}$, an $SO(3)$ subgroup in $G$ may lift to an $SO(3)$ subgroup in $\widehat{G}$. 
From this it now follows that in the spatial A sector, if $\widehat{\cal R}$ denotes either $\cal R$ (if 
${\cal C}=+1$) or one of its lifts $\widehat{\cal R}$ (if ${\cal C}=-1$), then $\widehat{\cal R}^2={\cal C}$.
Further, the group-theoretic commutator with $\widehat{I}$ (defined likewise from $I$) is $[\widehat{I},\widehat{\cal R}]={\cal C}$.

Finally, for time-reversal symmetry in the case of spatial A b.c., the arguments are again very similar to those in the main text,
in particular ${\cal T}^2=1$. 
The results can again be extended to the full Hilbert space by the same means as before. The 
conclusion is that all relations in the display (\ref{arr:discrelssigma}) also hold for the more general sigma models discussed
here, where again $(-1)^{\Theta/\pi}$ must be replaced by $\cal C$, and for the spatial P b.c., or when ${\cal C}=+1$, hats can be 
dropped.

Turning to the spin chains, we consider the crystal momentum of the ground state(s) by semiclassical methods.
For length $N$ even, we can view the N\'eel state as a product state, with each tensor factor the same 
coherent state $|\vec{n}\rangle$ on the even sites, alternating with its time reverse (rather than a rotation) on the odd sites. 
Then a simple calculation shows that under translation $\tau$ by one site, the state becomes the global time reverse of the product 
state, times a phase factor $({\cal T}'^2)^{N/2}$. In the semiclassical quantization picture, the ground state for a particle
moving on the target space (without any vector potential) will be nondegenerate with constant wavefunction that
can be taken real. It follows that the crystal momentum $P'$ of the ground state will be $0$ if ${\cal C}=+1$ or $N/2$ 
is even, but $\pi/a$ if ${\cal C}=-1$ and $N/2$ is odd, exactly as for the ${\cal G}=SU(2)$ spin $S$ chains. For $N$ odd, 
we should consider a lowest-energy texture of the N\'eel state. To calculate the effect 
of translation by two sites, $\tau^2$, this reduces as in the main text to a Berry product that as $N\to\infty$ tends to 
$\tau^2={\cal C}$. 
So the crystal momentum of the ground states will be close to $\pm \pi/(2a)$ if ${\cal C}=-1$, but to $0$ or $\pi/a$ if ${\cal C}=+1$. 
The calculation of the precise values involves more detailed work, and will not be done here. We expect that the result (for $N$ odd)
is that for ${\cal C}=+1$, the crystal momentum is $P'=0$, while for ${\cal C}=-1$ it is a multiple of $2\pi/L$ 
as close as possible to the stated values, as for the $O(3)$ case.

Making use of the semiclassical results in the sigma model, we can also obtain the behavior of the discrete symmetries 
of spatial reflection (fixing one site) ${\cal R}'$ and time reversal ${\cal T}'$ in the spin chain. The results 
are the same as in the display (\ref{discrels}), where $(-1)^{2NS}$ must be replaced
by ${\cal C}^N$ in general. 
Finally, the discrete symmetries of the spin chain can be related in the continuum limit to those of the sigma model 
exactly as in Section \ref{sec:spchsig} in the main text, where at the end the groups $O(3)$ and $Spin(3)$ must be replaced 
by $G\times \mathbb{Z}_2$ and $\widehat{G}$, respectively, in general. 
Thus the same quantity ${\cal C}$ controls all of the related anomaly effects, including the ground-state crystal momenta
in the spin chain.

%%%%%%%%%%%%%%%%%%%%%%%%%%%%%%%%%%%%%%%%%%%%%%%
\subsection{Relations between \texorpdfstring{$\Theta_l$} {Theta} parameters, the center, and \texorpdfstring{${\cal T}'^2$} {T} }
\label{coroots}

Next we go into detail about the coefficients $\Theta_l$ in the topological term in the sigma models,
and their relation to the center and to ${\cal T}'^2$ for self-dual irreducible representations.
For a general reference, again see Ref.\ \cite{fulton1991}. At the beginning, some additional generality will be useful technically; 
we return to self-dual representations afterwards.
So we will have in mind an irreducible representation $V$ that may not be self-dual, 
in which case the spin chain is the one with $V$ at the even sites, and the dual of $V$ at the odd sites; for such a chain 
with periodic b.c.\ (and Hamiltonian invariant under $\cal G$), the number of sites $N$ must be even. These chains still allow
construction of antiferromagnetic Hamiltonians, a N\'eel state, and coherent states as before, but translation symmetry by one site, and
inversion symmetry acting on the target space, are lost. [An example is a chain with the defining representation
of $SU(n)$, for some $n>2$, at each site; the target space is then $\mathbb{C}\mathbb{P}^{n-1}$ and $\Theta=\pi$, as is well known.] 
Thus we consider any complex simply-connected semisimple Lie group $\cal G$ with simple Lie algebra. 

First we will be interested
in the second homotopy group $\pi_2$ of the target space of the sigma model, because $\pi_2$ classifies maps of $\mathbb{S}^2$ into
the target space, and when $\pi_2$ is, say, $\mathbb{Z}^r$ for $r>0$, there are topological 
or ``$\Theta$'' terms that can be included in the action of the (two-dimensional) sigma model (at least when $\pi_1$ is trivial), 
as in the $O(3)$ sigma model. As we have seen, the target space of the sigma model is the space of coherent states of the 
representation $V$, 
which is the homogeneous space ${\cal G}_\mathbb{C}/{\cal P}$, where ${\cal P}$ is again the parabolic subgroup subgroup 
determined by the highest weight of $V$. A minimal parabolic subgroup is called a Borel subgroup, generically denoted $B$; 
$B$ is not unique, but different $B$ for given ${\cal G}_\mathbb{C}$, $B\subset {\cal G}_\mathbb{C}$, are conjugate in 
${\cal G}_\mathbb{C}$ (so isomorphic). ${\cal G}_\mathbb{C}/B$ can be considered the flag manifold of ${\cal G}_\mathbb{C}$. 
For general ${\cal P}$, there is always a Borel subgroup contained in it, 
so we can assume $B\subset{\cal P}$, and then there is a quotient map ${\cal G}_\mathbb{C}/B\to{\cal G}_\mathbb{C}/{\cal P}$. 
This will mean that we can derive what we need from considering ${\cal G}_\mathbb{C}/B$. Like ${\cal G}_\mathbb{C}/{\cal P}$, 
${\cal G}_\mathbb{C}/B$ is simply connected; its $\pi_1=0$. 
Then by the Hurewicz theorem, $\pi_2({\cal G}_\mathbb{C}/{\cal P})\cong H_2({\cal G}_\mathbb{C}/{\cal P})$, the second homology 
group (in particular, for ${\cal P}=B$; note that all homology and cohomology groups will be taken with integer coefficients), 
and further $\pi_2({\cal G}_\mathbb{C}/{\cal P})\cong \pi_1({\cal P})$. As mentioned already, the space of coherent states 
of a highest weight $\lambda$ in an irreducible representation $V$ of $\cal G$ give rise to a complex line bundle over 
${\cal G}_\mathbb{C}/{\cal P}$, where ${\cal P}$ is the isotropy subgroup (in ${\cal G}_\mathbb{C}$) of the image of 
the highest weight vector in the (complex) projective space $\mathbb{P}(V)$. Any complex line bundle over 
${\cal G}_\mathbb{C}/{\cal P}$ can be 
pulled back to one over ${\cal G}_\mathbb{C}/B$, and the group of isomorphism classes of line bundles over ${\cal G}_\mathbb{C}/B$ is 
generated by the equivalence classes of these line bundles; note that the group of isomorphism 
classes of complex line bundles over any space $X$ is isomorphic to the second cohomology group $H^2(X)$. Hence the group of 
line bundles over ${\cal G}_\mathbb{C}/B$, $H^2({\cal G}_\mathbb{C}/B)$, is isomorphic to the weight lattice of $\cal G$, which is the 
free Abelian group $\mathbb{Z}^r$, where $r>0$ is the rank of $\cal G$ (or of ${\cal G}_\mathbb{C}$). The integer dual of this group, 
that is, the group of $\mathbb{Z}$-linear maps from the group into the integers, is then also $\mathbb{Z}^r$, and because 
${\cal G}_\mathbb{C}/B$ is simply connected, this integer dual is isomorphic
to the second homology group, $H_2({\cal G}_\mathbb{C}/B)\cong \mathbb{Z}^r$; then $H_2$ and $H^2$ are integer duals of each other,
and we have determined that $\pi_2({\cal G}_\mathbb{C}/B)\cong \mathbb{Z}^r$ (which can also be shown more directly). 
Homotopy and homology groups are covariant under maps of spaces, so there are surjections $\pi_2({\cal G}_\mathbb{C}/B)\to 
\pi_2({\cal G}_\mathbb{C}/{\cal P})$ and $H_2({\cal G}_\mathbb{C}/B)\to H_2({\cal G}_\mathbb{C}/{\cal P})$. 
These quotient groups $\pi_2({\cal G}_\mathbb{C}/P)\cong H_2({\cal G}_\mathbb{C}/P)\cong \pi_1(P)$ are again free Abelian groups, 
with rank less than $r$ except when ${\cal P}\cong B$. 

Next we will express the topological term(s) arising in the sigma model with target space ${\cal G}_\mathbb{C}/{\cal P}$ that arises 
from the spin chain based on the representation $V$ in terms of the data of the irreducible representation $V$. Here it will 
be useful to introduce some bases of the lattices involved, based on Lie algebra representation theory. First, we can choose 
a Cartan subalgebra of ${\cal G}_\mathbb{C}$, which we denote $\mathfrak{H}$ (again, without loss of generality, we can assume 
the Lie algebra of ${\cal G}_\mathbb{C}$ is simple); 
it can be viewed as $\cong \mathbb{R}^r$
(strictly speaking, the Cartan subalgebra of ${\cal G}_\mathbb{C}$ should be $\mathbb{C}^r$, but we will refer only to
this more familiar real version, relevant to the finite-dimensional representations and to the compact form $\cal G$).
The dual space (over $\mathbb{R}$), $\mathfrak{H}^*$, of $\mathfrak{H}$ is the space of $\mathbb{R}$-linear maps from $\mathfrak{H}$ 
into $\mathbb{R}$, and then $\mathfrak{H}$ can also be viewed as the dual of $\mathfrak{H}^*$. 
This gives the natural pairing of $h\in\mathfrak{H}$ with $h'\in\mathfrak{H}^*$ which we denote by $\langle h,h'\rangle$ 
(a real number). The weights of the finite-dimensional representations form a discrete set of vectors (the weight lattice) 
in the dual space $\mathfrak{H}^*\cong\mathbb{R}^r$ also. The nonzero weights $\alpha$ of the adjoint representation are called roots;
they generate (over $\mathbb{Z}$) a lattice, the root lattice, which is a subset of the weight lattice.
The Lie algebra structure motivates the definition of distinguished elements $h_\alpha\in\mathfrak{H}$, one for each root $\alpha$,
called coroots or inverse roots (see e.g.\ Refs.\ \cite{Dieck,fulton1991} for the definition). The set of coroots generates
a lattice in $\mathfrak{H}$, which we call the coroot lattice (it must be remembered that this lattice 
is $\subset\mathfrak{H}$, not $\mathfrak{H}^*$). The weights $\lambda$ in the weight lattice have integer-valued pairing with 
the coroot $h_\alpha$ for any root $\alpha$, $\langle h_\alpha,\lambda\rangle \in\mathbb{Z}$, and hence with any vector in the 
coroot lattice.
Following Lie algebra theory, we choose a set of ``simple'' roots $\alpha_l$, $l=1$, \ldots, $r$, which form a basis 
for the root lattice (the simple roots are ``positive''; we discuss this below). Then the set of $h_l=h_{\alpha_l}$ forms 
a basis for the coroot lattice in $\mathfrak{H}$. In the weight lattice,
there is a unique basis set of vectors $\omega_l$, $l=1$, \ldots, $r$, dual to the former basis, that is 
$\langle h_l,\omega_{l'}\rangle=\delta_{ll'}$. The weights $\omega_l$ are called the fundamental weights. In $\mathfrak{H}^*$, 
there is a wedge-shaped region with $r$ edges, called the Weyl chamber. Each fundamental weight is the first nonzero 
weight on an edge of the Weyl chamber. The highest weight $\lambda$ of any finite-dimensional irreducible representation $V$ 
lies in the Weyl chamber, and can be expressed as a linear combination of the fundamental weights, with non-negative 
integer coefficients: $\lambda=\sum_{l=1}^r a_l \omega_l$, $a_l\geq 0$. For any such weight, there is an irreducible 
representation with that as its highest weight. The irreducible representations with highest weight a fundamental weight
are called fundamental representations, and can be denoted $V_l$. We will assume the standard choices for, and indexing, of the fundamental weights,
for which see the references.

As an aside, we are now in a position to define parabolic subgroups. In the Lie algebra of ${\cal G}_\mathbb{C}$, 
in addition to the Cartan subalgebra, there are elements that, acting in a representation, map the eigenspace of vectors of a given
weight into that of another weight (the ``raising'' and ``lowering'' operators). The vector in $\mathfrak{H}^*$ by which one of 
these elements changes any weight is simply a corresponding root, and hence these elements correspond to the roots. 
The roots come in pairs, differing by a sign. A subset of the roots, with certain properties (namely, containing exactly one from 
each pair of opposite roots, and closed under addition; any positive root can be expressed as a sum of simple roots), 
is called the set of positive roots (the ``raising'' operators; 
the set of positive roots is not uniquely defined, but different choices are related by conjugation in 
${\cal G}_\mathbb{C}$, and give the same results), and those roots differing by a minus sign from a positive root 
are called negative. Then for a choice of the set of positive roots, the Borel subalgebra 
is generated by the Cartan subalgebra and the Lie algebra elements corresponding to the positive roots; the group generated 
by this Lie algebra is called a Borel subgroup. More generally, for our purposes we can define a parabolic subgroup to be one
generated by the Cartan subalgebra together with the elements corresponding to a proper subset of the roots (and containing 
all the positive roots) such that the sum of any two roots in the subset is also in the subset.  
Hence, any parabolic subgroup contains the Borel subgroup. In fact, it is sufficient 
to use, as generators for a parabolic subgroup, a proper subset of the opposites of the simple roots, 
together with all the simple roots. It follows that the parabolic subgroups of ${\cal G}_\mathbb{C}$ 
can be easily read off from the Dynkin diagram of the Lie algebra of ${\cal G}_\mathbb{C}$.

Returning to our discussion of the topology of ${\cal G}_\mathbb{C}/B$, we now see that $H_2\cong \pi_2\cong \mathbb{Z}^r$ 
can be identified with the coroot lattice, with a basis $h_l$, so any second homology or homotopy class
can be viewed as an integer combination $\sum_l b_l h_l$ of the coroots $h_l$. For a given irreducible representation $V$ 
with highest weight $\lambda=\sum_l a_l\omega_l$, the pairing of a homology class $\sum_l b_l h_l$ with a cohomology class $\lambda$
is the integer ${\cal N}=\sum_l a_l b_l$. For a line bundle on ${\cal G}_\mathbb{C}/B$, or on ${\cal G}_\mathbb{C}/{\cal P}$ 
for some parabolic ${\cal P}$, the flux of the curvature of the connection through the homology or homotopy class is 
then $2\pi\sum_la_l b_l$. We saw in Section \ref{SpinCh} that the total topological term in the corresponding sigma model 
is half of this (the derivation generalizes easily to our more general cases of $\cal G$ and self-dual $V$, 
given the existence of N\'eel states). Hence we can view the total topological term as 
\beq
\sum_l\Theta_l b_l,
\eeq
with parameters 
\beq
\Theta_l=\pi a_l
\eeq
that depend on the highest weight $\lambda$ of the irreducible representation $V$ chosen. In particular, 
if $V$ is a fundamental representation, then only one $\Theta_l$ is nonzero, and that one is equal to $\pi$.
As the fundamental weights correspond to a basis of the group of line bundles, for each fundamental we have a curvature $2$-form 
or field strength, which we can also denote $\omega_l$. Then the topological term can also be expressed as the integral 
\beq
\pi{\cal N}=\pi \int_{\sum_l b_l h_l}\sum_l a_l\omega_l,
\eeq
where $\int_{h_l} \omega_{l'}=\delta_{ll'}$ and $\sum_l b_l h_l$ stands is the homotopy
(or homology) class in $\pi_2$ (or $H_2$) in which the image of the spacetime torus $\mathbb{T}^2$ lies
(i.e.\ the integral is taken over that two-dimensional surface, or pulled back to spacetime), 
and this generalizes the integral for the $O(3)$ sigma model. The $2$-forms can be chosen 
to be invariant under the compact form $\cal G$. 

All of this extends naturally to the case in which the parabolic subgroup 
$\cal P$ is larger than $B$, $B\subset {\cal P}$; 
these are cases in which some $a_l$ are zero, so the highest weight lies in a face (of dimension $<r$) of the Weyl chamber. 
(This corresponds to the classification of isomorphism-classes of parabolic subgroups $P$ as corresponding to subsets 
of the simple roots.)
Then there are correspondingly fewer nonzero $\Theta_l$ parameters. For the maximal parabolic case, there is only a 
single $\Theta_l$ remaining. In these cases, the ``simplest'' irreducible representation is just the corresponding fundamental weight.
Notice that this also gives us, by using again the duality between $H_2$ and $H^2$, a description of the second homology $H_2$
or homotopy $\pi_2$ group of a space ${\cal G}/{\cal P}$. $H_2({\cal G}_\mathbb{C}/{\cal P})$ is $H_2({\cal G}_\mathbb{C}/B)$, 
modulo the sublattice of the coroot lattice generated by the coroots orthogonal to 
(i.e.\ giving zero when paired with) the highest weight vector.

For a different simple example, we consider $Sp(4)\cong Spin(5)$, which as we mentioned earlier is not simply laced. 
We work in $\mathfrak{H}^*=\mathbb{R}^2$. We can represent the weight lattice of vectors as the set of all vectors 
of integers, and the roots are the vectors of the form $(\pm 1,\pm 1)^T$, $(\pm 2, 0)^T$, or $(0,\pm 2)^T$, 
where all combinations of signs are allowed. The simple roots are $\alpha_1=(1,-1)^T$, $\alpha_2=(0,2)^T$,
and the fundamental weights are $\omega_1=(1,0)^T$, $\omega_2=(1,1)^T$. 
The fundamental weights lie on the two edges of the Weyl chamber (the wedge with interior angle $\pi/4$) for this case.
The irreducible representation with highest weight $\omega_1$ is the four-dimensional defining representation of $Sp(4)$
[or the spinor of $Spin(5)$], while that with highest weight $\omega_2$ is the five-dimensional representation
[or the defining ``vector'' representation of $SO(5)\cong Spin(5)/\mathbb{Z}_2$]. The sigma model target spaces of these 
were discussed already, and in both cases the topological term has a single nonzero $\Theta$, namely $\Theta_1=\pi$ 
in the first case, $\Theta_2=\pi$ in the second; for the case of $\omega_1$, this confirms 
our earlier conclusion. Note that the highest weight in the adjoint representation is $2\omega_1$.
Note also that we used the standard labeling for $Sp(4)$, which differs from the standard one for $Spin(5)$ (in which the 
two values $1$, $2$ of the labels on the simple roots and fundamental weights are interchanged).

Next we turn to the center $C=C({\cal G})$ of $\cal G$. Elements of the center act on an irreducible representation $V$ 
by multiplication by a scalar and, under tensor products of representations, the scalars multiply. As the center is a subgroup 
of any Cartan subgroup of $\cal G$, it can be viewed as acting on the weight spaces of any representation, and so 
each element of the center gives a homomorphism (of Abelian groups) of the weight lattice into $U(1)$. Any element of $C$ acts 
trivially on the adjoint representation (because $\cal G$ acts by conjugation on the Lie algebra, and all elements of $C$ commute 
with all generators of $\cal G$), and hence on the root lattice. 
Then the weight lattice modulo the root lattice 
is a finite Abelian group $C^*$; $C^*$ is isomorphic to the center (but not in a canonical or intrinsic way, so this isomorphism is not 
useful), and $C^*$ is a sort of dual of the center $C$ (i.e.\ it is the space of representations of $C$). Then the set of irreducible representations
can be partitioned into conjugacy classes, each of which corresponds to an element of the dual group $C^*$.
The centers, or their duals, of the classical simply-connected semisimple Lie groups with simple Lie algebra are listed 
in Table \ref{Liecenter}. Note that $SU(n)$ has rank $r=n-1$, and $Sp(2n)$ has rank $r=n$.

\begin{table}[h!]
\centering
\begin{tabular}{ l|ll }
$\cal G$&$C$ or $C^*$& \\ \hline
$SU(n)$ & $\mathbb{Z}_n$& \\ \hline
$Sp(2n)$ & $\mathbb{Z}_2$& \\ \hline
$Spin(2r+1)$ & $\mathbb{Z}_2$& \\ \hline
$Spin(2r)$ & $\mathbb{Z}_2\times\mathbb{Z}_2$ &($r$ even) \\
           & $\mathbb{Z}_4$ &($r$ odd)\\\hline
\end{tabular}
\caption{Centers $C$ and dual centers $C^*$ of the classical Lie groups}
\label{Liecenter}
\end{table}

The action of the center on any irreducible representation can be determined from the decomposition of its
highest weight as a combination of fundamental weights, together with the action of the center on the fundamental representations.
Further, there is a subset of the fundamental weights that generates the group $C^*$. 
Thus for $SU(n)$, the conjugacy class of the first (or defining) fundamental representation, which is $n$-dimensional with 
highest weight $\omega_1$, can be taken as the generator of $C^*$, in a well-known way. Likewise, for $Sp(2n)$, the first 
fundamental weight $\omega_1$, that of the defining representation, is again a generator of $C^*$. For $Spin(2r+1)$, 
the only fundamental weight that is nontrivial under the center is $\omega_r$, that of the spinor representation. For $Spin(2r)$, 
the first fundamental weight $\omega_1$ is that of the $2r$-dimensional vector representation, and for $r$ even that and the highest 
weight of either of the spinors, $\omega_{r-1}$ or $\omega_r$, generate $C^*$; for $r$ odd, either of the two spinor fundamental 
weights suffices (cf.\ Appendix \ref{PinG}).

It will be useful to understand better the duality between $C$ and $C^*$. $C$ is a finite Abelian group, so it acts on
any irreducible representation of $\cal G$ by multiplication by a scalar that is a (generally, complex) root of unity, and under tensor products
these scalars multiply. So if the groups $C$ and $C^*$ are written additively, the action of an element $c$ of $C$
on a conjugacy class of irreducible representations, or element of the dual group, $c'\in C^*$ can be written as 
$e^{2\pi i \langle c,c'\rangle}$, which defines the pairing $\langle c,c'\rangle$ as a rational number modulo an integer.
The latter quantities lie in the less-familiar group $\mathbb{Q}/\mathbb{Z}$ (where $\mathbb{Q}$ is the additive group 
of rational numbers). This pairing is ``perfect'', meaning that it is non-degenerate [no nonzero $c$ has $\langle c,c'\rangle=0$ 
(mod $\mathbb{Z}$) for all $c'$, and the same with $c$ and $c'$ interchanged; see Ref.\ \cite{fulton1991}]. 
(Note that when the Lie algebra of $\cal G$ is not simple, the pairing $\langle\cdot,\cdot\rangle$ is the 
sum of those for each simple factor.) It is then in this sense that $C$ and $C^*$ are dual.

Now we turn again to self-dual irreducible representations of $\cal G$. In the dual of a representation of
a group, any group element acts as the transpose of the representation of the inverse of that element 
on the original representation. As all our representations are unitary, this is also the complex conjugate representation. 
Vectors in the dual have, by definition, a natural $\cal G$-invariant pairing with vectors in the original representation. 
If the dual is isomorphic to the original, then we obtain a pairing of the vectors in the original with itself. That is, 
there is a $\cal G$-invariant bilinear form on the original representation ($V$, say). This bilinear form may be either 
symmetric or antisymmetric,
and the two cases are referred as real and quaternionic, while a non-self-dual irreducible representation is said to be complex. 
These terms reflect the fact that in the real case, a basis of $V$ can be found in which the representation of any group element 
is real [as for integer-spin representations of $SU(2)$, for example], so that the {\it a priori} complex vector space
can be replaced by a real one (i.e.\ with real scalars). In the quaternionic case, a basis in which matrix elements are all real
cannot be found, but the vector space can be given the structure of a quaternionic vector space, in which the scalars
are the quaternions $\mathbb{H}$ [this occurs for the half-integer spin representations of $SU(2)$]. 
When an irreducible representation is complex, neither a real nor a quaternionic structure can be introduced. 
[For a reducible representation, there may be various ways of introducing a ${\cal G}$-invariant real or quaternionic 
structure, or combinations of both in orthogonal subspaces.]
In quantum mechanics, complex conjugation is related to time reversal. Time reversal invariance of an irreducible representation 
means that the conjugate representation is isomorphic to the original one, by using some unitary map
after complex conjugation (of the components relative to some choice of basis), and so corresponds to a self-dual representation.
The unitary map corresponds closely with the invariant pairing, and then time reversal squares to $+1$ for a real
representation, and to $-1$ for a quaternionic one. (Ref.\ \cite{Dieck}, p.\ 261 ff., is very helpful on this whole topic, 
and there ${\cal T}'$ is called the structure map, and the eigenvalue of ${\cal T}'^2$ is called the ``index''.) From the construction of the 
time reversal operator ${\cal T}'$, 
it is clear that on a tensor product of self-dual irreducible representations, the values of ${\cal T}'^2$ can be multiplied. 

Further, duality or conjugation maps each fundamental representation either to itself (if it is self-dual) 
or else exchanges it with another fundamental representation of the same dimension (its dual), 
and from this its effect on any irreducible representation
can be obtained in terms of the highest weight $\lambda=\sum_i a_l\omega_l$. That is, if we define $\overline{l}$ so that 
the dual of the $l$th fundamental is the $\overline{l}$th (and $\overline{\overline{l}}=l$), 
then duality maps $\omega_l$ to $\omega_{\overline{l}}$; $\omega_{\overline{l}}=\omega_l$
if and only if the corresponding fundamental representation $V_l$ is self-dual. Then an irreducible representation is self-dual
if and only if $a_{\overline{l}}=a_l$ for all $l$; clearly, the condition is trivial for those $l$ such that $V_l$ is self-dual. 

To fully characterize ${\cal T}'^2$ for self-dual irreducible representations, we first point out that the conjugacy class 
of a self-dual irreducible representation is an element in $C^*$ of order at most $2$, because the trivial representation 
occurs in the direct-sum decomposition of its tensor product with itself, by the discussion of self-duality.  
The elements of order $2$ in $C^*$ generate a subgroup, which we denote by $C^*_{{\rm ord}\; 2}$, and we will see that 
the conjugacy classes of self-dual irreducible representations generate this group, rather than a proper subgroup of it. 
(Note that a conjugacy class in $C^*$ may contain irreducible representations that are complex.)
The elements of the center act as $\pm 1$ on the subgroup $C^*_{{\rm ord}\;2}$, 
and this action factors through a quotient group, the quotient of $C$ by the group of elements $c\in C$ that (for $C$ written
additively) are twice another element, $\{c\in C: \exists c'\in C\; c=c'+c'\}$. We write the quotient group as $C/2C$, and call it 
the modulo $2$ reduction of $C$. These two groups $C^*_{{\rm ord}\; 2}$, $C/2C$ are (noncanonically) isomorphic 
and the previous pairing reduces to a perfect pairing of these two also,
with values of $e^{2\pi i\langle k,k'\rangle}$ in $\pm 1$ (it is sufficient to see this for cyclic groups, 
for which it is easy to obtain). 

Further results allow us to give a characterization of the values of ${\cal T}'^2$; see Ref.\ \cite{Bourbaki}, Ch. VIII, page 135.
 It follows from the proof of Prop.\ 12 there that the self-dual
irreducible representations with highest weight in the root lattice (i.e.\ those in the conjugacy class of the identity in $C^*$)
are real (because $0$ is a weight in the representation; for example, the adjoint, for which the symmetric bilinear form is 
the Killing form), and further that the index ${\cal T}'^2$ is constant on the conjugacy classes of self-dual irreducible 
representations (which are in $C^*_{{\rm ord}\;2}$). Hence ${\cal T}'^2$ determines a group homomorphism from the Abelian group 
of conjugacy classes of self-dual irreducible representations into the group $\{\pm 1\}$, and there is a formula for its action 
at the reference given. Then we can say that, on a self-dual representation, because the pairing of 
$C^*_{{\rm ord}\;2}$ with $C/2C$ is perfect, ${\cal T}'^2$ can be viewed as an element of $C/2C$, the reduction modulo $2$ 
of the center (it could be the identity in $C/2C$). In addition to these abstract statements, we will see,
mostly by direct calculation, that this holds for all cases of the (covers of) the classical Lie groups.
[Note that the statement ${\cal T}'^2\in C/2C$ continues to hold even when the representation is not irreducible,
when the center may not act by scalar multiples of the identity.] Then with Theorem 1, we have established that,
for given $\cal G$, there is an element of $C/2C$ such that its action on $V$ is $\cal C$. 

%%%%%%%%%%%%%%%%%%%%%%%%%%%%%%%%%%%%%%%%%%
\subsection{Explicit results and examples} 
\label{sec:centerexpl}

In this subsection, we combine results of the preceding parts, to describe explicitly for which cases
of $\cal G$ and $V$ there is an 't Hooft anomaly in inversion symmetry, and for which there is not.
We also give details about the action of the center, to elucidate the relations between $\cal C$,
the center, and the values of the $\Theta_l$ parameters.
Again, Ref.\ \cite{fulton1991} is the background reference for most of this.

In all cases, we describe the self-dual irreducible representations in terms of the non-negative
integers $a_1$, $a_2$, \ldots, $a_r$, or $a_l$, where the range $l=1$ to $r$ will always be assumed.
Alternatively, we could express them in terms of the $\Theta_l$ parameters, 
using $a_l=\Theta_l/\pi$ for all $l$. We note that, because only information modulo $2$ will be needed
to calculate $\cal C$, for that purpose it will be sufficient to specify each $\Theta_l$ only modulo $\pi$. 

For $SU(n)$, an irreducible representation is self-dual if and only if $a_l=a_{n-l}$ for all $l$. 
If $n$ is even, $V_{n/2}$ is self-dual, and $a_{n/2}$ is unconstrained (except for being a non-negative integer). 
The action of the most natural generator 
of the center $\mathbb{Z}_n$ on an irreducible representation is given by the exponential of $2\pi i/n$ times $\sum_l l a_l$,
which is the number of boxes in the corresponding Young diagram ($a_l$ is the number of columns of length $l$),
or to the conjugate of this for the inverse of this generator.
For a self-dual irreducible representation $V$, if $n$ is even the action of any generator of the center reduces to 
$(-1)^{a_{n/2}}$, which is $-1$ if $a_{n/2}$ is odd, and $+1$ if $a_{n/2}$ is even,
and it is also $+1$ if $n$ is odd. Note that $C/2C$ and $C^*_{{\rm ord}\;2}$ are isomorphic to $\mathbb{Z}_2$ if $n$ is even, 
but are trivial if $n$ is odd. For $n$ even, we only need the action of ${\cal T}'^2$ on the 
$n/2$th fundamental representation. This representation is real, ${\cal T}'^2=+1$, if $n/2$ is even,
and quaternionic, ${\cal T}'^2=-1$, if $n/2$ is odd, as is well explained in Ref.\ \cite{fulton1991}. 
Hence when $n=0$ (mod $4$), ${\cal C}={\cal T}'^2$ corresponds to the identity element of $C/2C$, but to the nonidentity
element when $n=2$ (mod $4$). We can summarize our conclusions by saying that, for ${\cal G}=SU(n)$, $n$ even or odd, 
${\cal C}=+1$ if $n=0$ or $\pm 1$ (mod $4$), while ${\cal C}=-1$ if $n=2$ (mod $4$) and $a_{n/2}$ is odd. 

For ${\cal G}=Sp(2n)$, all fundamental, and hence all irreducible, representations are self dual for all $n$. 
The center is $C=\mathbb{Z}_2$,
and acts nontrivially on the first fundamental representation. (When the center is a product of $\mathbb{Z}_2$ factors, as here,
reduction modulo $2$ has no effect.) In general, the nonidentity element in $C$
acts as $(-1)^{a_1+a_3+\ldots}$, where in the exponent the sum $a_1+a_3+\ldots$ ends with $a_r$ if $r$ is odd,
$a_{r-1}$ if $r$ is even. The first fundamental, or defining, representation is quaternionic, which is connected
with the fact that the symplectic groups are defined as the groups of isometries of a quaternionic inner product 
space of dimension $n$ (over the quaternions). As all irreducible representations can be obtained for this case by 
decomposing tensor products of the first fundamental into direct sums, we have ${\cal C}={\cal T}'^2=(-1)^{a_1+a_3+\ldots}$ also 
(there is a mistake in Ref.\ \cite{fulton1991} when on page 447 they discuss self-dual representations for this case). 

When $\cal G$ is one of the spin groups, the labeling of irreducible representations and the self-duality properties
are somewhat more complicated. The irreducible tensor representations of $SO(n)$ that arise from decomposing the 
tensor products of the $n$-dimensional vector representation (which is the first fundamental representation of $Spin(n)$
for all $n\geq 5$) are in most cases real (we will describe the exceptions in a moment). The only
fundamental representations of $Spin(n)$ that might be quaternionic are the spinor representations, for which there is always
some element of the center that acts nontrivially [namely the nontrivial element in the kernel of the map to 
$SO(n)\cong Spin(n)/\mathbb{Z}_2$]. In more detail,
for $Spin(2r+1)$, the first $r-1$ fundamental representations are antisymmetrized tensor products of $i$
copies  of the vector, but the $r$th is the spinor, with highest weight $\omega_r$. For $Spin(2r)$, the first $r-2$ fundamental 
representations are antisymmetrized tensor products of $i$ copies of the vector, but the $r-1$th and $r$th are the two
nonisomorphic spinors. The first $r-1$ or $r-2$ (respectively) tensor fundamentals are real in all cases. The type of the spinor 
representation(s) (the same for both when $n$ is even) depends on $n$ modulo $8$, which is a manifestation of Bott periodicity; 
see Table \ref{typespinors}.  

\begin{table}[h!]
\centering
\begin{tabular}{ l|c|c|c|c|c|c|c|c| }
$n$ (mod 8)&0&1&2&3&4&5&6&7\\ \hline
type & $\mathbb{R}$ & $\mathbb{R}$ & $\mathbb{C}$ & $\mathbb{H}$ & $\mathbb{H}$ & $\mathbb{H}$ & $\mathbb{C}$ & $\mathbb{R}$ \\
%\hline
\end{tabular}
\caption{Representation type for spinor(s) of $Spin(n)$ for $n$ modulo $8$; real is denoted by $\mathbb{R}$, complex by $\mathbb{C}$,
and quaternionic by $\mathbb{H}$ (after Ref.\ \cite{Dieck}).}
\label{typespinors}
\end{table}

Then we have the following. For $Spin(n)$, $n=2r+1$, the center $C$ is $\mathbb{Z}_2$, and for any $r$
the nonidentity element of $C$ acts as $(-1)^{a_r}$. All irreducible representations are self dual. 
The antisymmetrized tensor product of $r$ copies of the vector representation
is irreducible with highest weight $2\omega_r$ (i.e.\ $a_r=2$), and is real. We then have 
${\cal C}=+1$ for all irreducible representations of $Spin(n)$ if $n=\pm 1$ (mod $8$) ($\cal C$ is equal to 
the identity in $C$), but ${\cal C}=(-1)^{a_r}$
if $n=\pm 3$ (mod $8$). 

For the groups $Spin(n)$, $n$ even, we begin with a general discussion. 
The weights $\omega_1$, $\omega_2$, \ldots, $\omega_{r-2}$, $2\omega_{r-1}$,
$2\omega_r$, $\omega_{r-1}+\omega_r$ are the highest weights of a set of irreducible representations that are the most basic
for these cases; here $\omega_1$, \ldots, $\omega_{r-2}$ are the highest weights of antisymmetrized tensor products of 
$1$, \ldots, $r-2$ copies of the vector representation (respectively), $\omega_{r-1}+\omega_r$ is the highest weight 
of the antisymmetrized tensor product of $r-1$ copies of the vector, and the antisymmetrized tensor product of $r$ copies decomposes 
as the direct sum of the irreducible representations with highest weights $2\omega_{r-1}$, $2\omega_r$ (all this holds for $n=2r$, 
$r$ odd as well as even). In general, the irreducible representations of $SO(n)$, $n$ even, are all those with $a_{r-1}+a_r$ even. 
The center of $SO(n)$, $n$ even, is $\mathbb{Z}_2$, but the details of its action depend on whether $r$ is odd or even.

For $Spin(n)$ ($n=2r$) with $n=2$ (mod $4$) (i.e.\ $r$ odd), the center $C$ is $\mathbb{Z}_4$, so $C/2C$ and $C^*_{{\rm ord}\;2}$ 
are both $\mathbb{Z}_2$. These cases resemble $SU(n)$ with $n$ a multiple of $4$ and, in particular, $Spin(6)$ and $SU(4)$ 
are isomorphic.  One generator of the center acts on a general 
irreducible representation as $(-1)^{a_1+a_3+\ldots a_{r-2}}\, i^{a_{r-1}-a_r}$, 
and the other as the complex conjugate of this. The two spinor representations are complex, and each is the dual 
(or time reverse) of the other; more generally, duality or time reversal exchanges the values of $a_{r-1}$ and $a_r$. 
The non-spinor irreducible representations with highest weight $2\omega_{r-1}$ or $2\omega_r$ are again complex; 
as mentioned above, the antisymmetrized tensor 
product of $r$ copies of the vector representation decomposes as the direct sum of these two (hence it is self-dual,
but its irreducible direct summands are not). To obtain a self-dual irreducible representation, we must have $a_{r-1}=a_r$;
such an irreducible representation is real, and is a representation of $SO(n)$. So for these cases, ${\cal C}={\cal T}'^2=+1$ 
for any self-dual irreducible representation. Thus $\cal C$ corresponds to the identity element of $C/2C$. [$C/2C$ is still nontrivial,
and acts on a general self-dual irreducible representation of $SO(n)$, $n=2$ (mod $4$), as $(-1)^{a_1+a_3+\ldots a_{r-2}}$.]

For $Spin(n)$ with $n=0$ (mod $4$), or $n=2r$ ($r$ even), the situation is more complicated because the center is 
$\mathbb{Z}_2\times \mathbb{Z}_2$,
so it is necessary to specify the action of two independent generators to describe its action on an irreducible representation,
and there is no canonical choice for which two nonidentity elements should be chosen as generators.
This resembles the case of $Spin(4)\cong SU(2)\times SU(2)$ case, which does not have a simple Lie algebra,
but otherwise is a good example; there a natural choice might be to use the generators of the center
of each $SU(2)$ factor. We will do likewise for the general case. For general irreducible representations 
of $Spin(n)$, $n=0$ (mod $4$), let us define two generators of the center, the first acting as $-1$ on the spinor with 
highest weight $\omega_{r-1}$ and trivially (as $+1$) on that with highest weight $\omega_r$, and the second similarly 
but with the roles of $\omega_{r-1}$, $\omega_r$ reversed. Using iterated tensor products of either or both spinors,
and decomposing into direct sums of irreducible representations, all irreducible representations of $Spin(n)$ can be obtained
(see Refs.\ \cite{fulton1991,Dieck}), and in particular the tensor product of the two different spinors decomposes as a 
direct sum of the irreducible representations with highest weights 
$\omega_1$, $\omega_3$, \ldots, $\omega_{r-3}$, $\omega_{r-1}+\omega_r$. Hence these irreducible representations transform
by $-1$ under either of the two generators defined. Similarly, the tensor product of two copies of the same spinor
decomposes as the direct sum of irreducible representations with highest weights $\omega_2$, $\omega_4$, \ldots, $\omega_{r-2}$, 
and either $2\omega_{r-1}$ or $2\omega_r$, corresponding to the spinor used, so these are invariant under either generator. 
This agrees with the action of the inversion in $SO(n)$ on these representations, and hence both generators map to the inversion, 
which is the generator of the quotient $\mathbb{Z}_2$, the center of $SO(n)$. For these cases, all representations of $SO(n)$ 
are real (this is because both the spinors are self-dual 
and of the same type; see Table \ref{typespinors}). Hence the inversion operation in $SO(n)$ acts 
on any irreducible representation with $a_{r-1}+a_r$ even [i.e.\ any irreducible representation of $SO(n)$ 
for $n=0$ modulo $4$] as $(-1)^{a_1+a_3+\ldots +a_{r-3}+a_{r-1}}$ [note that, under the stated conditions, 
$(-1)^{a_{r-1}}=(-1)^{a_r}$]. 
This discussion shows that the first of the generators of the center defined above acts on
the general irreducible representation of $Spin(n)$, $n=0$ (mod $4$), as $(-1)^{a_1+a_3+\ldots +a_{r-3}+a_{r-1}}$,
while the second acts as $(-1)^{a_1+a_3+\ldots +a_{r-3}+a_r}$. Finally, for time reversal squared, from Table \ref{typespinors}
we have ${\cal C}={\cal T}'^2 = (-1)^{a_{r-1}+a_r}$ if $n=4$ (mod $8$), but ${\cal C}=+1$ for all irreducible representations
if $n=0$ (mod $8$). We see that, in the first case, $\cal C$ corresponds to the product of our two generators of the center, while
in the second case it corresponds to the identity element in the center. 

For completeness, we briefly mention the cases of exceptional Lie groups (with simple Lie algebras), 
denoted $E_6$, $E_7$, $E_8$, $F_4$, and $G_2$ (the subscript is the rank $r$). For these, the centers are as follows:
trivial for $E_8$, $F_4$, $G_2$; $\mathbb{Z}_3$ for $E_6$; and $\mathbb{Z}_2$ for $E_7$. 
Hence the modulo $2$ reduction of the center is trivial in all cases with the sole exception of $E_7$. 
Representations of $E_8$, $F_4$, and $G_2$ are all real, while two of the three conjugacy classes of irreducible representations 
of $E_6$ consist only of complex representations (and the dual of an irreducible representation in one of those 
conjugacy classes belongs to the other). For $E_7$, all representations are self-dual, and those in the non-identity
conjugacy class are quaternionic (Ref.\ \cite{Bourbaki}, pp. 217, 218). One of the latter is a $56$-dimensional fundamental 
representation (it is the lowest-dimensional fundamental representation), with highest weight $\omega_7$; it and its antisymmetric 
invariant bilinear form are constructed in Ref.\ \cite{adams1996}. So for $E_7$ and any irreducible representation in the 
non-identity conjugacy class, ${\cal C}={\cal T}'^2$ is the nontrivial element of the center $C(E_7)$, while for the other four exceptional
groups, ${\cal C}=+1$, the identity in $C/2C$. 

To conclude this part, we summarize by giving the logical relations among the properties, with a few particular examples, 
some of them discussed earlier. First, we 
have given examples of a group $\cal G$ and a self-dual irreducible representation $V$ used to construct a spin chain, 
or of a sigma model with a corresponding target space and some values of the $\Theta_l$ parameters, 
in which ${\cal C}={\cal T}'^2=-1$, so that by Theorem 1 there is an anomaly that prevents gauging of the inversion
symmetry in the sigma model. This occurs, 
for example, in the case of $SU(n)$, $n$ even, and the representations
corresponding to a rectangular Young diagram with $n/2$ rows, for which the target space is a Grassmannian manifold, 
as described earlier. For these, the nonzero parameter is $\Theta_{n/2}$, equal to $\pi$ times the number of columns
($=a_{n/2}$) in the Young diagram, for which there is an anomaly, ${\cal C}=-1$, if and only if both $n/2$ and $a_{n/2}$
are odd. 
This corresponds to $\Theta=\pi$ (mod $2\pi$) in the main text, which is a special case.
Other examples include the case of the conjugacy class of a spinor of $Spin(n)$, $n=2r+1$, for which $\Theta_r=\pi$, 
where ${\cal C}=-1$ if and only if $2r+1=\pm 3$ (mod $8$), and the conjugacy class of a spinor of $Spin(n)$, $n=4$ (mod $8$).

As we have seen, $\cal C$ is related to time reversal squared, which for given $\cal G$ corresponds to the action
on $V$ of an element of the reduction modulo $2$ of the center of $\cal G$. So when there is an anomaly,
it implies that there is some element of the center that acts nontrivially on $V$. For spin chains with the
latter property, this is the case in which the LSM-Affleck theorem applies, so that (in our cases) for $N$ even 
there is a state nearly degenerate with the ground state. So ${\cal C}=-1$ implies that the center acts non-trivially (so that
the LSM-Affleck theorem
applies), but the converse does not hold: if the center acts non-trivially, there may be no anomaly that prevents gauging
of the inversion symmetry. This occurs in examples such as the $SU(n)$ rectangular Young diagram case with $a_{n/2}$ odd but $n/2$ even,
and for $Spin(n)$ spinors and $n=0$, $\pm 1$ (mod $8$). This is one reason why we have given some detail
about the action of the center on various irreducible representations. Whenever ${\cal C}=+1$, the invariant $\cal I$
has a gauge-invariant extension to allow A b.c.s, and when some $\Theta_l$ are nonzero (mod $2\pi$), it is nontrivial: 
it does not only take the constant value $1$, but is $-1$ for some configurations.

Nontrivial action of the center on the self-dual representation $V$ implies that some of the $\Theta_l$ parameters are equal
to $\pi$ (mod $2\pi$), but not conversely.
A third class of examples is those in which the center acts trivially on $V$ (and in particular, ${\cal C}=+1$),
but there are nonetheless some $\Theta_l$ parameters equal to $\pi$ (mod $2\pi$). Then again $\cal I$ is gauge invariant,
and may not be constant. This occurs, for example, for cases such as some of the tensor representations of $SO(n)$,
$n$ odd, for which any irreducible (including the vector) representation $V$ is real and the center is trivial. 
We note that the center always acts trivially on the adjoint representation, but that for example for $Spin(n)$, $n\geq 7$, 
the adjoint is one of the fundamental representations, with highest weight $\omega_2$, so $\Theta_2=\pi$ (and all other 
$\Theta$ parameters vanish) in this case.

Finally of course, there are examples in which all data are trivial, that is, all $\Theta_l=0$ (mod $2\pi$) for all $l$,
which implies the trivial action of the center on $V$, and that ${\cal C}=+1$.
Clearly, this occurs for irreducible representations whose highest weight lies in a sublattice of the lattice
of weights of self-dual representations, defined in terms of fundamental weights by $a_l=0$ (mod $2$) for all $l$. 
This corresponds to $\Theta=0$ (mod $2\pi$) in the main text. An interesting example (one of those mentioned earlier) is 
the case of ${\cal G}=Sp(2n)$, $n\geq 1$, and the adjoint representation, for which $a_1=2$, $a_l=0$ for $l\neq 2$.

Comparing these cases with the main text, we see that the general situation discussed here is much richer. 
For ${\cal G}=SU(2)$, there were only the cases of $S$ half integer or $\Theta=\pi$ (mod $2\pi$), 
for which ${\cal C}=-1$, and $S$ integer or $\Theta=0$ (mod $2\pi$), for which ${\cal C}=+1$. Here we have seen
that in general there are also cases intermediate between these two extremes, and that some of these may occur for the same
group $\cal G$. For example, for $Spin(5)\cong Sp(4)$,
the fundamental representations are [again, in the ordering standard for $Sp(4)$] the four-dimensional spinor
of $Spin(5)$, and the five-dimensional vector of $Spin(5)$ or $SO(5)$. These have $\Theta_1=\pi$, $\Theta_2=0$
for the first, and the values are exchanged for the second. ${\cal C}=-1$ (so there is an 't Hooft anomaly), 
and the center acts nontrivially, on the first, while ${\cal C}=+1$ (there is no anomaly), and the center acts trivially, 
on the second, yet one of the two $\Theta$ parameters equals $\pi$ 
in each case.

%%%%%%%%%%%%%%%%%%%%%%%%%%%%%%%%%%%%%%%%%%%%%%%%%%%%%
\subsection{Representations carried by states in the P and A b.c.\ sectors}

In Section \ref{Subsec:semicl}, we also determined the spins of the states in Hilbert space in the spatial A b.c.\
sector of the $O(3)$ sigma model using semiclassical quantization, with agreement with the results expected in the 
odd-length $SU(2)$ spin chain. So far in this Appendix we have not extended that result to the general models
considered; in this final subsection we discuss this problem. We discuss
the topology (the fundamental group) of the collective-coordinate spaces, which is part of the semiclassical
quantization analysis, but also confirms that $\cal I$ is the only topological invariant that can be constructed
when one or more b.c.\ is antiperiodic. Then we consider the decomposition into irreducible components of the Hilbert
spaces of function on the collective coordinate spaces, focused mainly on the case of ${\cal G}=SU(n)$, without
giving a fully detailed analysis of all cases.

First, for the spatial P b.c.\ sector of the general sigma models discussed here, the collective-coordinate space
is the sigma-model target space ${\cal G}_\mathbb{C}/{\cal P}$, and the quantum mechanics does not involve a connection (vector potential).
The center of $G$ acts trivially on the target space, and so also on the Hilbert space of complex functions
on the target space. It follows that the collective-coordinate Hilbert space can be decomposed as a direct sum of subspaces, 
each of which transforms as an irreducible representation of $G$ that lie in the conjugacy class of the identity element of $C$ (or of $C/2C$);
this agrees with the results in the spin chain for even length (note that these statements also hold when $V$ is not self-dual,
but alternates with its dual on alternate sites of the chain). A full analysis would determine which irreducible representations
actually occur in that decomposition; we will see that it is not always the case that all irreducible representations
in the conjugacy class of the identity in $C^*$ appear, though we can show that they do for any flag manifold, and we determine their multiplicities
also (this analysis will be given later).
For the spatial A b.c., we again consider a minimum-energy texture
that obeys the b.c.. For $G=SO(3)$, the orbit of this texture under $G$ filled out $G$ itself, with fundamental group
$\pi_1(SO(3))=C(SU(2))=\mathbb{Z}_2$. In this case we could show that the collective coordinate Hilbert space decomposed
as a direct sum into representations, where for $\Theta=0$ (mod $2\pi$) all and only integer spins occurred, while 
for $\Theta=\pi$ (mod $2\pi$) all and only half-integer spins occurred. For more general groups, the collective-coordinate 
space is not $G$, but a homogeneous space for $G$ (not the same as the target space), as we will explain. 
Then the fundamental group might differ from the
center of $G$. We would like to characterize which irreducible representations of functions on the collective-coordinate Hilbert space
occur for each equivalence class of
connections, or of $\Theta_l$ values (mod $2\pi$) for all $l$. We expect that the fundamental group of the collective-coordinate space
will be $C/2C$, and that only irreducible representations in the same conjugacy class in $C^*_{{\rm ord}\;2}$
as $V$ can occur. For this part also, we will later give some results for special cases.

In connection with this problem for both spatial A or P  b.c.s, we make some remarks on the general form
of the representation content of the full (not only the collective-coordinate) Hilbert space.
We point out again that, although any self-dual irreducible representation belongs to a class in $C^*_{{\rm ord}\;2}$,
it is not always the case that all irreducible representations in such a class are self dual. Even for a tensor product of 
real irreducible representations, which would have a natural real structure, the irreducible components are not necessarily
real. For example, if $V$ is the six-dimensional irreducible representation of $Spin(6)\cong SU(4)$ 
[or of $SO(6)\cong SU(4)/\mathbb{Z}_2$], the Hilbert space of a spin chain of $N=3$ sites is the tensor product 
$V\otimes V \otimes V$, which contains a totally-antisymmetrized subspace of dimension $20$, and the latter decomposes 
into a direct sum of the ten-dimensional irreducible representation of highest weight $2\omega_2$ 
[in the $Spin(6)$ labeling; in the $SU(4)$ labeling, it would be $2\omega_1$] with its dual (conjugate) \cite{fulton1991},
which thus are complex, but lie in the same conjugacy class as $V$. This counterexample 
shows that we cannot necessarily expect to exclude such irreducible representations from appearing in the 
Hilbert spaces of the spin chains or sigma models. As in the example, such an irreducible representation will
appear along with its dual, and for a Hamiltonian that is time-reversal invariant as well as $\cal G$ invariant, 
their energies will be the same. A similar phenomenon occurs even in the conjugacy class of the identity, as for 
the P b.c.\ or the even-length spin chain. Thus, in the same example, for the tensor product of six copies of $V$,
the irreducible representation with highest weight $4\omega_2$ occurs in the tensor product of two copies of the 
highest weight $2\omega_2$ irreducible representation, which arise in each product of three copies; that representation 
is again complex, but lies in the trivial conjugacy class. 

Although the collective-coordinate space of minimum-energy textures with values in the target space and A b.c.\
is not entirely trivial to obtain in general (we give some results later), it is not difficult to find directly the fundamental group
of the space of continuous configurations with the spatial A b.c.\ when the minimum-energy condition is dropped. 
We begin with preliminary remarks
about the action of the inversion on the homotopy classes of maps of $\mathbb{S}^2$ into the target space. 
These remarks concern some details that arise from fundamental representations that are complex,
so we consider the space ${\cal G}_\mathbb{C}/{\cal P}$ of coherent states for a representation $V$ (not necessarily self-dual)
for which the highest weight $\lambda=\sum_la_l\omega_l$ includes terms with $a_{\overline{l}}>0$ whenever $a_{l}>0$
(for self-dual fundamental representations, the latter condition is trivial). For this space (which is also the sigma model target space), 
we also denoted by $\omega_l$ the curvature $2$-form of the connection corresponding to each $\omega_l$, 
which exist on the space when $a_l>0$. As the duality map on representations sends $l$ to $\overline{l}$, for such a target space
inversion maps $\omega_l$ to $-\omega_{\overline{l}}$. If $V_l$ is self dual, this is the antisymmetry already discussed,
but we should notice what occurs if $V_l$ is not self-dual. Then for $V$ self dual, $a_{\overline{l}}=a_l$ for all $l$, and then
the total two-form has the antisymmetry property. Similar remarks apply to the homotopy classes of maps of $\mathbb{S}^2$ into
the target space (or similarly homology $2$-cycles). As we can see already for the target space $\mathbb{S}^2$, there is always
a reversal of sign, and in general for not necessarily self-dual fundamental representations $V_l$, 
the corresponding homotopy classes, corresponding to coroots $h_l$, are mapped to $-h_{\overline{l}}$. (Here only $h_l$
for $l$ such that $a_l>0$ occur as nontrivial homotopy classes for our given target space, consistent with
the determination of the homotopy group in Sec.\ \ref{coroots}; the other $h_l$, which occur for the flag manifold ${\cal G}_\mathbb{C}/B$, 
have been mapped to zero in passing to the quotient ${\cal G}_\mathbb{C}/{\cal P}$.) This preserves
the pairing of weights and coroots $\langle h_l,\omega_{l'}\rangle =\delta_{l,l'}$, as it must.

Next we turn to the computation of homotopy classes of configurations of the $\vec{n}$ field 
(representing a point in the target) with the AP b.c., 
which is relevant for understanding the invariant $\cal I$, and less so to the semiclassical analysis; all b.c.\ choices with at least
one A can be mapped to this by a choice of cycles on the torus, and the nonuniqueness of that map, associated
with the possible gauge non-invariance of $\cal I$, is not an issue at the moment. If we cut the configuration
along a branch cut, say at $x=0^+$, then we have a map of a cylindrical strip into the target space,
with the equivariance condition that corresponding (i.e.\ at the same $t$) points on the boundary at either end
are the inversion of each other. (This was discussed and illustrated for the $O(3)$ model in Sec.\ \ref{subsec:pathintmodtrans},
and the following argument extends an argument given there also, though in a slightly different form.)
Then we wish to find the fundamental group of the space of these maps,
for the branch cut fixed in spacetime. [We can also represent these maps by doubling the length in $x$ to $2L$, 
so that $\vec{n}$ can be taken periodic, with $\vec{n}(x+L)=-\vec{n}(x)$. This contains exactly the same
information, but the use of the strip will be convenient.] Because the target space is simply connected, the loop
swept out by each edge of the strip can be contracted to a point, where each point is the antipode of the other.
Then we have a map of a sphere into the target space, with the equivariance property that there
are two particular points, say the north and south poles, which are always mapped to
opposite points in the target space. Again because the target space is simply connected, it is clear that
any map of a sphere into the target can be deformed to this form. Hence the homotopy classes
of such equivariant maps of a sphere is the same as the second homotopy group $\pi_2$ of the target space. 
Because we can always deform (by a homotopy) the two antipodal points to two fixed such points,
these homotopy classes form the same group as $\pi_1$ of the space of paths from one of the fixed
points to the other, and it is a standard result in homotopy theory (mentioned e.g.\ in 
Ref.\cite{milnor1963morse}, Sec.\ 22) that the fundamental group of the space of 
loops in the space of paths in the target space is the same as $\pi_2$ of the target space. 

For the homotopy classes of equivariant maps of strips, we consider a homotopy between two such maps.
It is clear that for the initial, and for the final, such map, we can contract the loop
swept out by either edge to a point, as we just discussed. It is not clear that we can do
so continuously for all intermediate points during the homotopy as well. In fact,
the loop that is the image of one edge of the strip sweeps out a surface during the homotopy,
and if we contract the initial and final loops to points then we have another map of a sphere
into the target space. Again, such maps are classified up to homotopy by $\pi_2$ of the target
space, which is nontrivial in all cases we consider. Further, the other end of the strip
sweeps out the inversion of the same image of a sphere, in a possibly different homotopy class. 
In fact, if the homotopy class associated with one end is $\sum_lb_l h_l$ (where only $h_l$ that correspond
to $a_l>0$ can occur), then after consideration of orientation, that for the other end is $\sum_l b_l h_{\overline{l}}
=\sum_l b_{\overline{l}}h_l$. Hence the initial and final equivariant maps of strips into the target may differ 
by an element $\sum_lb_l (h_l+h_{\overline{l}})$ of $\pi_2$ of the target space. That is, the homotopy classes
of equivariant maps of strips into the target, which is $\pi_1$ of the space of equivariant paths 
in the target (where an equivariant path is any constant $t$ slice of an equivariant map of a strip), are given by the quotient
group $\pi_2({\cal G}_\mathbb{C}/{\cal P})/\overline{2}\pi_2({\cal G}_\mathbb{C}/{\cal P})$, 
where $\overline{2}\pi_2({\cal G}_\mathbb{C}/{\cal P})$ is the subgroup generated by $h_l+h_{\overline{l}}$
(the notation resembles that for the modulo $2$ reduction of an Abelian group, to which the quotient
reduces when $h_l$ for complex $V_l$ are absent; note that in the general case, the Abelian quotient group may 
contain factors $\mathbb{Z}$ as well as factors $\mathbb{Z}_2$). 
This extends the result given for $O(3)$ in Sec.\ \ref{subsec:pathintmodtrans}.
We note that this {\it a priori} determination of the homotopy classes could have been
given before the construction of $\cal I$. The values of $a_l$ were not involved in the calculation of
the homotopy classes (though they played a role in determining the target space).
For our purposes in the sigma model, we impose $a_l=a_{\overline{l}}$ for all $l$, to obtain
inversion symmetry (i.e.\ antisymmetry of the topological term). Then $\cal I$ is in fact precisely
the most general topological (homotopy) invariant, with inversion symmetry, of AP configurations 
that one can construct, and the restriction of each $\Theta_l$ to $0$ or $\pi$ (mod $2\pi$) is necessary because
of topology, as well as inversion symmetry. 

\subsubsection{Example: complex Grassmannian}
\label{App:grass}

Here we determine the fundamental group of the space of minimum-energy textures with the spatial A b.c.\ for the case
when the sigma model target space is the complex Grassmannian $\mathbb{G}_{n,n/2}$ ($n$ even; see Sec.\ \ref{sec:extarget}).
This arises for example as the space of coherent states for the irreducible representation of $SU(n)$ that corresponds
to a Young diagram with one column of $n/2$ boxes, or equivalently for the $n/2$th fundamental representation, with highest
weight $\omega_{n/2}$. The complex Grassmannian $\mathbb{G}_{n,n/2}$ is the space of subspaces in $\mathbb{C}^n$ of dimension $n/2$,
and can be viewed as the orbit under $SU(n)$ of one such subspace
(the general $\mathbb{G}_{n,k}$ arises likewise as the orbit space of a subspace of dimension $k$). It can be represented
non-redundantly by using an $n\times n$ matrix-valued field $Q$, related to the projection operator onto an $n/2$-dimensional
subspace, with the properties that $Q=Q^\dagger$, and that $Q$ has $n/2$ eigenvalues equal to $1$, and $n/2$ equal to $-1$, 
so $Q$ squares to the identity. $U(n)$ acts on $Q$ by conjugation. We can choose a standard $Q$ as a reference point (corresponding 
to the north pole on $\mathbb{S}^2$ in the case $n=2$), say
\begin{equation}
Q_0=\left(\begin{array}{cc} {\rm id}_{n/2}&0 \\ 0&-{\rm id}_{n/2} \end{array} \right),
\end{equation}
where ${\rm id}_{n/2}$ is the $n/2 \times n/2$ identity matrix. A $Q$ close to $Q_0$ differs from $Q_0$ at first order
by a matrix of the form
\begin{equation}
\delta Q=\left(\begin{array}{cc} 0& a \\ a^\dagger& 0 \end{array} \right),
\end{equation}
for some $n/2$ by $n/2$ complex matrix $a$. $\delta Q$ can be viewed as an element of the tangent space of the manifold at $Q_0$.
It can also be viewed as a representative of an equivalence class in the Lie algebra of $U(n)$ modulo that of $U(n/2)\times U(n/2)$
[or of the same with $S$ prefixed to both; see eq. (\ref{GrassmanDef})]; we can fix the diagonal elements as $0$ 
by requiring $\delta Q$ to be orthogonal
to the Lie algebra of $U(n/2)\times U(n/2)$ with respect to the Killing form on the Lie algebra of $U(n)$. Note that
$\delta Q$ contains the correct number of real parameters, $n^2/2$, for these descriptions to hold. Note also that,
for $n=2$, $Q$ is related to $\vec{n}$ by $Q=\vec{n}\cdot\vec{\sigma}$, where $\vec{\sigma}$ is again the vector
of Pauli matrices. Thus we recover the $2$-sphere $\mathbb{S}^2$, with a complex coordinate $a$ used in an infinitesimal neighborhood 
of the north pole $\vec{n}=(0,0,1)^T$ corresponding to $Q_0$.

Now if we wish to characterize a geodesic in the target space that passes through $Q_0$, we can identify it with
its tangent vector at $Q_0$, because a geodesic parallel transports its tangent vector along the path (here we make use
of Riemannian geometry on the target space, with a metric and Levi-Civita connection). For our spaces, which are homogeneous spaces, 
we identify geodesics with the paths produced by acting with a suitable one-parameter group defined by exponentiating
an element of the (real) Lie algebra times a real parameter, say $x$, where the Lie algebra element must be of 
the form of $\delta Q$ above (at $Q_0$), to ensure that it is a geodesic (otherwise, the path could resemble a circle 
of latitude rather than a great circle such as the equator, in the case of $\mathbb{S}^2$ and the group $SU(2)$). 
For a minimum-energy texture, we require such a path, with the additional equivariance condition, or A b.c., 
that at $x=L^-$ it passes through the antipode of the starting
point at $x=0^+$. For the complex Grassmannnian, the antipode of $Q_0$ is $-Q_0$. This should be clear by considering
the latter as representing the subspace that is the orthogonal complement of that represented by the former. 

The isotropy subgroup of $Q_0$, that is the subgroup of $U(n)$ that leaves it fixed, is $U(n/2)\times U(n/2)$ (or the same with the prefix $S$ on 
the product),
so the space of orbits is indeed $\mathbb{G}_{n,n/2}$. Hence an element $(u_1,u_2)$ (where $u_1$, $u_2$ are unitary) of either 
form of isotropy subgroup (with $\det u_1 \det u_2=1$ for the case with $S$) acts on $a$
by $a\mapsto u_1 a u_2^{-1}$. The tangent space at $Q_0$, which we identify with the space of all $a$, forms an irreducible representation 
of the isotropy subgroup (this will not always be true for other target manifolds). On the other hand, the tangent space does not form a single
orbit under the isotropy subgroup, even if we remove $a=0$ and fix a normalization of the nonzero elements by ${\rm tr}\, aa^\dagger=1$. 
The eigenvalues of $aa^\dagger$, which are also the eigenvalues of $a^\dagger a$, and which sum to unity, are non-negative, 
and their positive square roots are the so-called singular values of $a$. The set of singular values, say in non-increasing order, suffices to label
the distinct orbits in the tangent space under the isotropy subgroup; that is, any set of non-increasing non-negative eigenvalues that sum to unity,
together with elements $(u_1,u_2)$ of $S[U(n/2)\times U(n/2)]$, modulo a subgroup
of the latter that leaves the tangent vector invariant, can be used to parametrize the space of normalized tangent vectors. 
We call the latter subgroup the isotropy subgroup of the tangent vector
(it is a subgroup of the isotropy subgroup of the point $Q_0$, and in general depends on the multiplicities of the singular values). 

In general, not every geodesic leaving $Q_0$ along some tangent vector will pass through the antipode $-Q_0$.
[For $n=2$, any geodesic leaving the north pole in $\mathbb{S}^2$ passes through the south pole, so this was not an issue in the $O(3)$ model.]
Hence to identify geodesics that pass from $Q_0$ to its antipode, we need to find the ones that do reach $-Q_0$, 
and we also require them to go to the antipode as directly as possible, or with minimal length,
giving a so-called minimal geodesic which serves as our lowest-energy texture. If we choose $a={\rm id}_{n/2}/\sqrt{n/2}$ 
then these conditions are satisfied. That is because (dropping henceforth the factor $(n/2)^{-1/2}$, for convenience), 
for each pair $(i,n/2+i)$ of indices ($i=1$, \ldots, $n/2$), we have a $2\times 2$ block 
with the form of the Pauli matrix $\sigma_1$, and the path generated by exponentiating this part of $a$ goes from the
initial point, which restricts to $\sigma_3$, to $-\sigma_3$. Our choice ensures that the paths in each $2\times 2$ block 
all reach $-\sigma_3$ at the same ``time'', as required. More general such paths are obtained by acting with $S[U(n/2)\times U(n/2)]$
on this $a$, giving $a=u$ (or $u/\sqrt{n/2}$ on restoring the normalization), where $u$ is any unitary matrix. Further, 
because all eigenvalues of any such $a a^\dagger$ are $1$, any such choice of tangent vector has the largest
possible isotropy subgroup, as befits the restriction to such a special space of geodesics;
namely, when $a={\rm id}_{n/2}/\sqrt{n/2}$, the isotropy subgroup consists of the block-diagonal matrices 
$(u_1,u_2)$ with $u_1=u_2$. So the isotropy subgroup of such a lowest-energy
or minimal geodesic is isomorphic to the diagonal $U(n/2)$ in $U(n/2)\times U(n/2)$. If the form with determinant equal to $1$ is used
then the isotropy subgroup (we call it ${\cal G}_0$) consists of block-diagonal matrices with both $u_1$, $u_2$ on the diagonal equal 
to a unitary $u$, where $(\det u)^2=\det u^2=1$, so $\det u=\pm 1$, or equivalently $u^2\in SU(n)$. 

Now we are ready to carry out some calculations of homotopy groups. First, we consider $\pi_1$ of the space of minimal geodesics
with the endpoints fixed. If we use the form of groups without the prefix $S$ [as in the middle member of eq.\ (\ref{GrassmanDef})], this space is the isotropy subgroup
of a point modulo the isotropy subgroup of a tangent vector to a minimal geodesic, that is, 
\beq
\frac{U(n/2)\times U(n/2)}{U(n/2)},
\eeq
where the denominator is the diagonal $U(n/2)$ as discussed. But this corresponds to exactly what we had in Appendix \ref{PetWeyl};
the quotient space is simply $U(n/2)$. So the fundamental group is $\cong \pi_1(U(n/2))\cong \mathbb{Z}$. By comparison, as we mentioned
already, the fundamental group of the space of {\em all} continuous paths from (say) $Q_0$ to $-Q_0$ is isomorphic to 
$\pi_2(\mathbb{G}_{n,n/2})\cong \mathbb{Z}$. 
So in this particular case, these are the same, however, for most of the target spaces of interest to us, the analogous statement will not hold.
We note that our discussion here is close to the results, and methods used, in Bott's proof of his periodicity theorem for unitary groups, 
as discussed in Ref.\ \cite{milnor1963morse}, Secs.\ 23 and 24, though our calculation does not appear in that reference (however, an analogous
result is obtained in the orthogonal case, on pages 141--142). So here it is safe to say that for each homotopy class of loops
in the space of paths with fixed antipodal endpoints, there is a representative as a loop in the space of minimal geodesics.

For the analysis of semiclassical quantization of motion on the collective-coordinate space, we want to find
the fundamental group of the space of all minimal geodesics, where the endpoints are not fixed but variable.
This space can be best expressed in the form with prefix $S$, as $SU(n)/{\cal G}_0$. As $SU(n)$ is simply connected,
the long exact homotopy sequence tells us that $\pi_1(SU(n)/{\cal G}_0)\cong \pi_0({\cal G}_0)\cong \mathbb{Z}_2$, where
it should be clear that ${\cal G}_0$, the space of unitary $u$ such that $\det u=\pm 1$, has two connected components,
and we note that $\pi_0$ of a group naturally has a group structure, which is a quotient of that of the group. 
We also note that the center $C=\mathbb{Z}_n$ of ${\cal G}=SU(n)$ is a subgroup of ${\cal G}_0$, and its elements, 
which are of the form $e^{2\pi i k/n}\,{\rm id}_n$
($k$ an integer), are divided equally between the two connected components.
Hence this $\pi_0({\cal G}_0)\cong \mathbb{Z}_2$ can be identified as the modulo $2$ reduction of the center of 
$SU(n)$, $\mathbb{Z}_n/2\mathbb{Z}_n\cong
\mathbb{Z}_{n}/\mathbb{Z}_{n/2}\cong \mathbb{Z}_2$ for $n$ even.
By comparison, the fundamental group of the space of all configurations with the spatial A b.c.,
discussed above, also gives $\mathbb{Z}_2$ in the present case, and these correspond as they did for the case of fixed endpoints. 
Note that all of these homotopy results hold in particular for $n=2$, for which they were seen in the main text.
 
The next step in the semiclassical analysis is the calculation of the holonomy (Berry phase) for a loop
in the space of minimal geodesics. The calculation is a special case of the more general one in the following subsection,
so the discussion continues there.

\subsubsection{Example: flag manifold of 
\texorpdfstring{$SU(n)$}{SU(n)}}

The flag manifold for $SU(n)$ is the space of sequences of nested subspaces of $\mathbb{C}^n$ of the form (a filtration)
\begin{equation}
\{{\bf 0}\}\subset V_1\subset V_2\subset\cdots\subset V_{n-1}\subset V_n=\mathbb{C}^n,
\end{equation}
where each $V_l$ has dimension $l$. A point in the space can be represented by a choice of an orthonormal basis 
$v_l$ ($l=1$, \ldots, $n$), viewed as a sequence or indexed set $(v_l)_l$, where each $V_l$ is spanned by $v_1$, \ldots, $v_l$. 
The subspaces are unchanged by multiplication of the basis vectors by a phase, so this space is indeed the homogeneous
(or coset) space $\mathbb{F}_n=U(n)/[U(1)\times\cdots \times U(1)]\cong SU(n)/S[U(1)\times\cdots\times U(1)]$, the complex flag manifold,
with (complex) dimension $n(n-1)$. Thus it is of the form ${\cal G}/(B\cap{\cal G})$, where ${\cal G}=SU(n)$, and 
$B\cap{\cal G}=S[U(1)\times\cdots\times U(1)]$.
A convenient non-redundant general notation for a point in $\mathbb{F}_n$ uses a fixed indexed set of $n$ distinct real eigenvalues,
and then a point in $\mathbb{F}_n$ is represented by a self-adjoint matrix $Q$ with the given eigenvalues; the $v_l$ are normalized 
eigenvectors. This generalizes the case
for the complex Grassmannian. We can choose for example the eigenvalues $1-2(l-1)/(n-1)$, $l=1$, \ldots, $n$, and we will see that
for this choice inversion acts as $Q\to -Q$. Usually, we will describe a point in the space using the standard choice of 
orthonormal basis $(v_l^{(0)})_l$ for $\mathbb{C}^n$,
together with a unitary map that takes the basis to a representative of the equivalence class that is a point in $\mathbb{F}_n$. 
The standard basis in $\mathbb{C}^n$ will be viewed as the origin in $\mathbb{F}_n$ (or we can refer to the origin as $Q=Q_0$, 
which is the diagonal matrix with the eigenvalues as stated), and the isotropy subgroup of that point 
is $S[U(1)\times\cdots\times U(1)]$. We note again that the flag manifolds arise as the spaces ${\cal G}_\mathbb{C}/B$ of coherent states 
for a spin that transforms as an irreducible representation $V$ of $\cal G$ in the case when the highest weight of $V$ lies in the 
interior of the Weyl chamber, or equivalently when the parameters $a_l$ are strictly positive for all $l=1$, \ldots, $r$, 
which are the most generic cases; here we consider only the cases ${\cal G}=SU(n)$, where $r=n-1$.

We claim that the inversion or antipode of a point represented by
the sequence $(v_1,\ldots, v_n)$ is the same sequence in reverse order, $(v_n,\ldots,v_1)$ [modulo (as always) the action of the isotropy subgroup
of that point]. That is because this map takes each subspace in the sequence to its orthogonal complement, and reverses the inclusions. 
For the standard basis, for $l=1$, \ldots, $n-1$, each standard subspace $V_l^{(0)}$ can be viewed as the highest weight vector 
in the $l$th fundamental representation of $SU(n)$. Hence the map takes each such highest weight to the opposite or lowest weight 
in the dual representation, and clearly this corresponds to what we need. 

A tangent vector at the origin can be viewed as a choice of an element of the Lie algebra of $SU(n)$, modulo the Lie algebra of the isotropy
subgroup of the origin. Again, there is a continuum of orbits of tangent vectors. We will move directly to the tangent vectors
that generate minimal geodesics to the inversion or antipode of the origin. Starting from the origin, one map to
its antipode is given in $U(n)$ by the unitary $n\times n$ matrix
\begin{equation}
\widetilde{\rm id}_n=\left(\begin{array}{cccccc} &&&&&1 \\&&&&1&\\&&&\cdot&&\\&&\cdot&&&\\&\cdot&&&&\\1&&&&& \end{array} \right), 
\end{equation}
the antidiagonal analog of the identity (elements omitted are zeroes). Note this matrix has determinant $(-1)^{\lfloor n/2\rfloor}$, 
where $\lfloor x\rfloor$ is the floor
function, that is, the largest integer less than or equal to $x$. Consider the tangent vector at the origin represented by the
Lie algebra element (a traceless self-adjoint $n\times n$ matrix) 
\begin{equation}
\Sigma_x=\left(\begin{array}{ccccccc} &&&&&&1 \\&&&&&1&\\&&&&\cdot&&\\&&& * &&&\\&&\cdot&&&&\\&1&&&&&\\1&&&&&& \end{array} \right),
\label{Sigma_x}
\end{equation}
where the $*$ is the element at the center, which is present and $=0$ if $n$ is odd, but absent if $n$ is even
(the remainder of the antidiagonal is $1$s). Other tangent vectors at the origin in the same orbit can be obtained by the conjugation action
of the isotropy subgroup on this one, so that the $1$s in the upper right quadrant can be replaced by phase factors, with their
complex conjugates at the transpose positions at the bottom left. Then the path of unitary matrices $e^{-i\theta\Sigma_x/2}$,
where $\theta$ is a real parameter, begins at the identity at $\theta=0$, and at $\theta=\pi$ it is $-i\Sigma_x$, plus an elementary matrix with
a $1$ at the center and zeroes elsewhere if $n$ is odd; thus $e^{-i\pi\Sigma_x/2}$ is a unitary with determinant $1$, which is almost the same, 
and has similar effect (i.e.\ within multiplying basis vectors by phase factors), as $-i\,\widetilde{\rm id}_n$. Thus the path is 
a minimal geodesic from the origin to its antipode.

We mentioned already that, for any $\cal G$ and any self-dual irreducible representation $V$ of $\cal G$, there is a projection map
from the flag manifold for $\cal G$ onto the homogeneous (or target) space for $V$. Then for ${\cal G}=SU(n)$,
the description of the inversion map and of the path
from the origin to its antipode must be preserved by the projection map, due to the symmetry they all possess.
In particular, for the complex Grassmannian $\mathbb{G}_{n,n/2}$ (for which the parabolic subgroup is maximal),
the tangent vector should agree, up to conjugation by the isotropy subgroup, with the one we had
for that space. Indeed, the tangent vector $\delta Q$ that we had there corresponds to a Lie algebra element
of the same form as $\delta Q$ itself, and for $a={\rm id}_{n/2}$ it is essentially the same as the tangent vector here,
up to a permutation that reverses the order of the final $n/2$ basis vectors. Thus the following calculations for $\mathbb{F}_n$
also transfer without change to the target spaces for $SU(n)$ and any self-dual $V$, or compatible values of $\Theta_l$.

Next, we turn to the homotopy groups for the $SU(n)$ flag manifolds. $\mathbb{F}_n$ is simply connected. Its second homotopy
group is $\pi_2(\mathbb{F}_n)\cong \pi_1(S[U(1)\times\cdots\times U(1)])\cong \mathbb{Z}^{n-1}$, in agreement with
the earlier remark, because $r=n-1$. The space of minimal geodesics with fixed endpoints can be viewed as the quotient
of the isotropy subgroup of (say) the origin, modulo the isotropy subgroup ${\cal G}_0$ of one minimal geodesic through that point, such as that 
described above. For the latter, ${\cal G}_0$ consists of elements $(u_1,\ldots,u_n)$, where $u_l$ are phase factors
with $\prod_l u_l=1$, and further $u_l=u_{n+1-l}$. Then the quotient space is a product of $\lfloor n/2 \rfloor$ circles 
$(\mathbb{S}^1)^{\lfloor n/2\rfloor}$, corresponding to the choice of phase factors in the upper right block of the tangent vector. 
$\pi_1$ of this space is $\cong \mathbb{Z}^{\lfloor n/2\rfloor}$, which differs from $\pi_2(\mathbb{F}_n)\cong\mathbb{Z}^{n-1}$ if $n>2$. 
The space of minimal geodesics with a variable endpoint (together with its antipode)
is ${\cal G}/{\cal G}_0$, where ${\cal G}=SU(n)$ and ${\cal G}_0$ is as described. Its fundamental group 
$\pi_1({\cal G}/{\cal G}_0)$ is $\cong \pi_0({\cal G}_0)$. If $n$ is even, ${\cal G}_0$ consists of indexed sets $(u_l)_l$
of phase factors $u_l$ with $\prod_{l=1}^{n/2} u_l^2=1$, or $\prod_{l=1}^{n/2} u_l=\pm 1$, and clearly the space of solutions 
has two connected components, so $\pi_1({\cal G}/{\cal G}_0)=\mathbb{Z}_2$. If $n$ is odd, we have 
$u_{(n+1)/2}\prod_{l=1}^{(n-1)/2} u_l^2=1$, or $\prod_{l=1}^{(n-1)/2} u_l=\pm u_{(n+1)/2}^{-1/2}$, and this time the space of solutions 
has a single connected component, giving trivial $\pi_1$. In both cases, this fundamental group of the space of minimal geodesics 
can be identified as $C/2C$, the reduction modulo $2$ of the center $C$ of $SU(n)$. Notice how allowing the endpoint to be variable changes 
the fundamental group to a small finite group, which is a quotient of the former one. A projection of the flag manifold to the homogeneous space 
for some self-dual irreducible representation $V$ induces a homomorphism of the fundamental group of the space of minimal geodesics, so
that, for $n$ even, $\pi_1$ in fact remains $\pi_1=\mathbb{Z}_2$, which in particular agrees with the case of $\mathbb{G}_{n,n/2}$ above,
while $\pi_1$ is always trivial for $n$ odd. 

Now we turn to the calculation of holonomy (Berry phase) for a loop in the space of minimal geodesics. 
We wish to calculate our invariant $\cal I$ for such a loop.

As a warm-up exercise, we calculate again the effect of translation of the texture (minimal geodesic) in $x$ by $2L$. In the first part 
of the following, $n$ can be either even or odd. As in the main text
(towards the end of Subsec.\ \ref{subsec:pathintmodtrans}), for this case the bulk contribution to $\pi{\cal N}'$ is zero,
and the full holonomy comes from the boundary contribution at the branch cut. The $Q$ field at the cut traverses the same curve 
in the target space as the minimal geodesic itself, passing through $-Q_0$, and making a full circuit back to its starting point, say at $Q_0$.
(The holonomy for this loop is a particular case of the definition of the invariant $\cal C$, and was calculated earlier.)
Beginning from the highest weight state in $V$, the loop is obtained by applying the group element $g=e^{-i\theta\Sigma_x/2}\in SU(n)$
as $\theta$ increases from $0$ to $2\pi$. 
The Berry connection in such a case is given at the origin by the expectation of $-i\Sigma_x$ in the highest weight state. That is zero, because
$\Sigma_x$ is off-diagonal in the basis in which the highest weight vector is an eigenvector of the Cartan subalgebra that consists of traceless 
diagonal matrices. The connection is also zero at all other points on the loop, for symmetry reasons. Then the holonomy
results solely from the final $U(1)$ gauge transformation of the adiabatically-transported state at $\theta=2\pi$ back to 
the original gauge. 

At $\theta=2\pi$, for $n$ even we have $g=-{\rm id}_n$, which belongs to the center, so the holonomy 
is simply the element $-{\rm id}_n\in SU(n)$ applied to the highest weight vector.
By Weyl's construction \cite{fulton1991}, for $n$ either even or odd, the highest weight vector can be viewed as the tensor product of $\widetilde{a}_l$
copies of $v_l^{(0)}$ for $l=1$, \ldots, $n$, where $\widetilde{a}_n=0$ and $a_l=\widetilde{a}_l-\widetilde{a}_{l+1}\geq 0$ ($l=1$, \ldots, $n-1$)
in the earlier notation in which the $a_l$s are the coefficients of the fundamental weights,
and the tensor product is mapped into a certain space, related to the Young diagram in which the $l$th row has length $\widetilde{a}_l\geq0$,
$l=1$, \ldots, $n-1$. In this language, the non-negative integer $a_l$ is the number of columns with $l$ boxes.
We discussed the action of the center of $SU(n)$ in Sec.\ \ref{sec:centerexpl} above. In the present case, for which $n$ is even, 
we obtain $(-1)^{\sum_{l=1}^n \widetilde{a}_l}=(-1)^{\sum_{l=1}^{n-1} la_l}$. For a self-dual irreducible
representation $V$, $a_l=a_{n-l}$, and this reduces to $(-1)^{na_{n/2}/2}=(-1)^{n\Theta_{n/2}/(2\pi)}$.
Hence if $n/2$ is even, this holonomy is always $1$, while if $n/2$ is odd, we obtain $(-1)^{\Theta_{n/2}/\pi}$
as in the main text case of $SU(2)$ (note it does not depend on $a_l$ for $l\neq n/2$).
This agrees exactly with the earlier evaluation as ${\cal C}={\cal T}'^2$ in the cases $SU(n)$, $n$ even,
and is $-1$ precisely when $V$ is quaternionic as well as self dual.

A similar calculation can be done for $n$ odd. In this case, at $\theta=2\pi$, $e^{-i\pi \Sigma_x}$ 
is not an element of the center, but it is an element of a Cartan subgroup of $SU(n)$, which for the flag manifold
is the isotropy subgroup of the origin. Acting on the highest weight vector of a self-dual irreducible representation
$V$, it can be evaluated as $(-1)^{\sum_{l=1}^{(n-1)/2} 2la_l}=+1$;
this gives the net holonomy. [For this calculation, note that $e^{-i\pi \Sigma_x}$ acts as $-1$ on all $v_l^{(0)}$ 
except for $v_{(n+1)/2}^{(0)}$, on which it acts as $+1$. Then form the product
of these phase factors, each one $\widetilde{a}_l$ times, and use $a_l=a_{n-l}$.] 
Note that the modulo $2$ reduction of the center is trivial for $n$ odd, so the result had to be $+1$;
all self-dual irreducible representations are real.

Finally, we consider the holonomy for a more general loop in the space of minimal geodesics. In general, we are interested
in the case in which the endpoints (which are antipodes of one another) of the minimal geodesics can vary 
along the loop, and the homotopy invariant $\cal I$, which we wish to calculate, depends only on the 
homotopy equivalence class of such a loop. But each such homotopy class contains many homotopy classes
of loops with the endpoints fixed, which can be deformed to one another when the endpoints are allowed to vary
(like the ``belt trick'' again). So it is sufficient to calculate $\cal I$ for loops on for which the endpoints are fixed, 
and the result should depend at most on the element of the modulo $2$ reduction $C/2C$ of the center $C$ of $SU(n)$ 
that corresponds to the homotopy class of the loop of minimal geodesics with variable endpoints. 
(The same strategy was used in the main text.) When the endpoints of the minimal geodesic
at $x=0$, $L$ are fixed at $Q_0$, $-Q_0$, respectively, we can place the branch cut at $x=0^+$, and then
the evaluation of $\cal I$ reduces to that of $\pi \cal N$ for the map of $\mathbb{S}^2$ into $\mathbb{F}_n$
obtained by a loop of such minimal geodesics. If we now consider the loop in $\mathbb{F}_n$ swept out by the {\em midpoint} 
of those minimal geodesics, we see that, by symmetry between the two halves of any minimal geodesic, $e^{i\pi\cal N}$ 
is equal to the holonomy of the irreducible representation $V$ as it moves around that loop in $\mathbb{F}_n$, 
and that is what we now calculate. The approach is general, but will be carried out here only for the $SU(n)$ flag manifold.

A loop of minimal geodesics with fixed endpoints is equivalent to a loop in the isotropy subgroup
$H=S[U(1)\times\cdots\times U(1)]$ [again, with elements $(u_1,\ldots, u_n)$] of the origin in $\mathbb{F}_n$. We will consider such loops
that are obtained by exponentiating a generator in the Lie algebra of $H$,
times a parameter $\phi$, say, chosen so that at $\phi=2\pi$ we obtain an element of the isotropy
subgroup of the initial minimal geodesic, which means the path gives a closed loop in the space
of minimal geodesics with endpoints fixed. We consider a generator of the isotropy
subgroup, which in $SU(n)$ is given by a matrix of the form $\widetilde{h}={\rm diag}(\widetilde{b}_1,\ldots,\widetilde{b}_n)$, 
such that ${\rm tr}\,\widetilde{h}=\sum_{l=1}^n \widetilde{b}_l=0$.
Then the path in $H$ is given by $u_l(\phi)=e^{-i\phi \widetilde{b}_l}$ for $l=1$, \ldots, $n$.
We may wish to choose the generator so that at $\phi=2\pi$ we in fact obtain an element of the center
of $SU(n)$, which means for some integer $k$, we must have $\widetilde{b}_l=k/n$ (mod $1$) for all $l$, as well as 
$\sum_l \widetilde{b}_l=0$ as real numbers. For example, we can choose $\widetilde{b}_1=\cdots=\widetilde{b}_{l-1}=k/n$, 
$\widetilde{b}_n=k/n - k$. Such a choice is not in fact necessary; more generally, we only have to ensure
that an element of the isotropy subgroup of the initial minimal geodesic is reached at $\phi=2\pi$, which implies
the conditions $\widetilde{b}_l=\widetilde{b}_{n+1-l}$ (mod $1$), as well as $\sum_l \widetilde{b}_l=0$ as real numbers.
We will see that then the same holonomy is obtained, provided
the loop lies in the same homotopy class.

For the initial minimal geodesic, we again consider the path given by $e^{-i\theta\Sigma_x/2}$ acting 
on the origin (or highest weight vector), with $\theta$ running from $0$ to $\pi$. The unitary map
from the origin to the midpoint at $\theta=\pi/2$ is given for $n$ even by $U_{\rm mid}=({\rm id}_n-i\Sigma_x)/\sqrt{2}$; 
for $n$ odd, $e^{-i\theta\Sigma_x/2}$ has a similar form in the rows and columns
with index $\neq (n+1)/2$, and an element $1$ at the center of the matrix.
We want to apply the loop ${\rm diag}(u_1(\phi),\ldots,u_n(\phi))$ to $U_{\rm mid}$ applied to
the highest weight vector, and calculate the holonomy for that loop. This can be simplified by transforming
the loop back so that it starts at the highest weight vector, resulting in the loop in $SU(n)$ generated by
the Lie algebra element $U_{\rm mid}^{-1}\widetilde{h}U_{\rm mid}$. For $n$ even, the latter is given by
(the reason for writing it in this particular way will become apparent when we consider $n$ odd)
\begin{equation}
\left(\begin{array}{cccccccc} 
(\widetilde{b}_1+\widetilde{b}_n)/2 &&&&&&&-i(\widetilde{b}_1-\widetilde{b}_n)/2 \\
&\cdot&&&&&\cdot&\\
&&\cdot&&&\cdot&&\\
&&&(\widetilde{b}_{n/2}+\widetilde{b}_{n+1-n/2})/2 &-i(\widetilde{b}_{n/2}-\widetilde{b}_{n+1-n/2})/2&&&\\
&&&i(\widetilde{b}_{n/2}-\widetilde{b}_{n+1-n/2})/2&(\widetilde{b}_{n/2}+\widetilde{b}_{n+1-n/2})/2&&&\\
&&\cdot&&&\cdot&&\\
 &\cdot&&&&&\cdot&\\
i(\widetilde{b}_1-\widetilde{b}_n)/2&&&&&&&(\widetilde{b}_1+\widetilde{b}_n)/2
\end{array} \right).
\end{equation}
For $n$ odd, we have only to replace $n/2$ by $(n-1)/2$ 
at every occurrence in this expression, and there is an additional element $\widetilde{b}_{(n+1)/2}$ at the center of the matrix.
(Note that the trace is always zero, as it must be.)
We can now calculate the holonomy for the loop similarly as in the earlier example. For the Berry connection,
we require the expectation of this Lie algebra element in the highest weight state in the self-dual irreducible representation $V$. 
That is given by the sum over $l=1$, \ldots, $n-1$ of the $l$th diagonal element times $\widetilde{a}_l$ for each $l$. 
Recall that, for a self-dual irreducible representation $V$, $a_l=a_{n-l}$ for all $l$. 
If we group together terms for the columns in the Young diagram that correspond by that relation, we see that for each $l\neq n/2$ we obtain
a multiple of $\sum_{l'=1}^n\widetilde{b}_{l'}=0$. For $n$ even, there remain also $a_{n/2}$ columns of $n/2$ boxes in the Young diagram, 
and the sum $a_{n/2}\sum_{l=1}^{n/2}(b_l+b_{n+1-l})/2$ of those terms is again zero.
Hence the net holonomy is given by the action on $V$ of the chosen element ${\rm diag}(u_1(2\pi),\ldots)={\rm diag}(u_1,\ldots)$ 
of the isotropy subgroup of the initial minimal gepdesic (the element should be conjugated by $U_{\rm mid}^{-1}$ as we did for $\widetilde{h}$, but 
by construction of the isotropy subgroup of the initial minimal geodesic, the two unitary matrices commute, so it reduces to the action 
as stated). That action gives for the holonomy $\prod_{l=1}^n u_l^{\widetilde{a}_l}$, where $\prod_{l=1}^n u_l=1$ and $u_l=u_{n+1-l}$.
Grouping together terms for columns that corresponding by duality as we did for the expectation of $U_{\rm mid}^{-1}\widetilde{h}U_{\rm mid}$, 
we find for each pair $a_l$, $a_{n-l}$ with $l\neq n/2$ a power $(\prod_{l=1}^n u_l)^{a_l}=1$. Hence if $n$ is odd, the net holonomy is $+1$.
For $n$ even, there is an additional factor $(u_1 u_2\cdots u_{n/2})^{a_{n/2}}$. In this case, the product 
$u_1 u_2\cdots u_{n/2}=\pm 1$, depending on which of the two connected components of the isotropy subgroup of the minimal geodesic 
the element belongs to. All of this holds in particular if ${\rm diag}(u_1,\ldots, u_n)$ is an element of the center $C\cong \mathbb{Z}_n$ 
of $SU(n)$, and so also gives the value obtained as the action of the nontrivial element of $C/2C\cong \mathbb{Z}_2$ for $SU(n)$, $n$ even,
namely $(-1)^{a_{n/2}}=(-1)^{\Theta_{n/2}/\pi}$. In both cases $n$ odd and even, the holonomy is independent of $a_l=\Theta_l/\pi$ 
for $l\neq n/2$.

The conclusion in this example, the flag manifold of $SU(n)$, is that, whether $n$ is even or odd, the holonomy for
a nontrivial loop in the space of minimal geodesics (the collective coordinate space for the spatial A b.c.\ in the sigma model) 
is given by the action of the nontrivial element of the modulo $2$ reduction of the center on the representation $V$, when the nontrivial
cases occur (i.e.\ for $n$ even), and by $+1$ otherwise. Hence the states in the collective-coordinate Hilbert space
for the A b.c.\ must lie in the same conjugacy class (in $C^*_{{\rm ord}\;2}$) of irreducible representations as $V$ in all cases. 
These statements also descend to the other target spaces,
with $SU(n)$ and inversion symmetry, that we have considered, including the complex Grassmannian; they can be obtained 
as a quotient of the flag manifold (in particular, the same $SU(n)$ group elements can be used in the calculations). 
We expect that a similar conclusion holds for all 
target spaces in the models we consider, but we will not pursue a more general argument.

\subsubsection{General \texorpdfstring{$SU(n)$}{SU(n)} case}

Although the holonomy for the collective-coordinate spaces for the group $SU(n)$, the spatial A b.c., and a general self-dual irreducible
irreducible representation $V$ has already been determined, it will be useful later to understand the isotropy subgroup, ${\cal G}_0\subset{\cal G}$,
under which the texture (minimal geodesic) is invariant in a little more detail.

First, we find the subgroup in $SU(n)$ under which $\Sigma_x$ in eq.\ (\ref{Sigma_x}) is invariant, where $SU(n)$ acts on the Lie algebra 
element $\Sigma_x$ by conjugation. The subgroup of $U(n)$ that preserves $\Sigma_x$ must map its eigenspaces into themselves.
The eigenvalues are $1$, $-1$, each with multiplicity $\lfloor (n/2)\rfloor$, and a further eigenvalue $0$ if and only if $n$ is odd.
Hence the subgroup of $SU(n)$ that preserves $\Sigma_x$ is isomorphic to $S[U(n/2)\times U(n/2)]$ if $n$ is even, and $S[U((n-1)/2)\times U(1)\times 
U((n-1)/2)]$ if $n$ is odd; we emphasize that a change of basis was used, so the matrices are not block diagonal in the original basis.

For the isotropy subgroup of the texture (minimal geodesic), the preceding subgroup must be intersected with the isotropy
subgroup ${\cal P}\cap {\cal G}$ of a point in the sigma model target space ${\cal G}/({\cal P}\cap{\cal G})$. For 
self-dual $V$, the latter has the form 
\begin{equation}
S[U(n_1)\times U(n_2)\times\cdots \times U(n_k)\times\cdots \times U(n_2)\times U(n_1)]
\end{equation}
for some integer $k$
Here we listed the subgroups in order, using the same basis as earlier (in which the matrices are block diagonal), 
and the index on each $n_l$ increases until the middle
is reached at $n_k$, then decreases; the unpaired factor $U(n_k)$, $n_k\geq 0$, appears only (or at most) once, and the $n_l$, $l<k$, are positive. 
Hence $n_k= n$ modulo $2$. We note that $n_k=0$ only if $n$ is even and the self-dual fundamental weight $\omega_{n/2}$ appears (i.e.\ $a_{n/2}>0$)
in the highest weight for $V$.

For the intersection, similar to the earlier special cases, and first disregarding the restriction $S$ to determinant $1$
(it will be reinstated at the end), for the subgroups which are copies of $U(n_l)$ ($l<k$)
that correspond, the intersection consists of a single copy of $U(n_l)$, embedded diagonally in the subgroup $U(n_l)\times U(n_l)$. 
For the unpaired subgroup $U(n_k)$, its intersection with the subgroup of $U(n)$ leaving $\Sigma_x$ invariant has a similar form as the latter. 
So in the intersection this factor splits to become isomorphic to $U(n_k/2)\times U(n_k/2)$ if $n_k$ is even, and to $U((n_k-1)/2)\times U(1)\times 
U((n_k-1)/2)$ if $n_k$ is odd; again, in the original basis, the $U(\lfloor (n_k/2)\rfloor)\times U(\lfloor(n_k/2)\rfloor)$ 
groups do not consist of block diagonal matrices, while the $U(1)$ factor, if present, appears as the only nonzero entry
in the central row and column. 

For the restriction to determinant $1$, the product of the determinants of the diagonal blocks must equal $1$. The effect of this can be somewhat 
complicated. For later use, we give only the result for the complex Grassmannian with $m=n/2$ ($n$ even), refining the result of 
Sec.\ \ref{App:grass} slightly. The isotropy subgroup of the texture
is $S[U(n/2)_{\rm diag}]$, where the subscript $\rm diag$ denotes the diagonal embedding into $U(n/2)\times U(n/2)$ as before. Thus
the condition is that, if $u\in U(n/2)$, we impose $(\det u)^2=1$, or $\det u=\pm 1$. We can express any such $u$ as a power of $e^{2\pi i/n}$
times a matrix $u'\in SU(n/2)$; the even powers of $e^{2\pi i/n}$ belong to $SU(n/2)$. Then if $n/2$ is odd, we have 
${\cal G}_0\cong SU(n/2)\times \mathbb{Z}_2$, but if $n/2$ is even, ${\cal G}_0$ is a nontrivial extension of $SU(n/2)$ by $\mathbb{Z}_2$ 
(not a semidirect product), with two connected components in both cases. [The case $n=2$ agrees with the earlier result that the space of textures is 
${\cal G}/{\cal G}_0\cong SU(2)/\mathbb{Z}_2\cong SO(3)$.] This completes what we will say about the isotropy subgroup ${\cal G}_0$ 
for a minimal geodesic for all $SU(n)$ cases.

\subsubsection{Representations for functions on a homogeneous space}

We conclude this Appendix by discussing the decomposition of the Hilbert spaces of
collective coordinate states (in semiclassical quantization) for both P and A spatial boundary conditions
into irreducible representations of the global symmetry group $\cal G$, in particular
the conjugacy classes of irreducible representations of
$\cal G$ to which they belong and, where possible, the multiplicities of each irreducible representation also.
We anticipate that the states in the full Hilbert space of the sigma model lie in the
same conjugacy class as the collective-coordinate states in each case, but will not
investigate that aspect. We will use cases of $SU(n)$ as examples, and while we will not obtain general
formulas for the multiplicities, we will discuss them in particular cases.

We have seen that all cases involve a homogeneous space, which for the P b.c.\
is just the sigma model target space. In the A b.c.\ case, a homogeneous
space different from the target space is involved, and in addition there may be a
nontrivial $\cal G$-invariant connection (vector potential) on that space, which
has the effect of shifting the conjugacy class from the trivial class (the identity
in $C^*$) to a nontrivial class. For these reasons, we begin with a general
discussion of harmonic analysis for a homogeneous space. (The following is an abstract approach.
Another approach that may be useful in practice would be to use some
set of coordinates on the space that transform in a representation, and
then consider polynomials in those coordinates. For example, for $\mathbb{S}^2$
we discussed cubic harmonics, and in the examples in the preceding subsections 
one could use the matrix $Q$ defined there.)

For a compact homogeneous (or coset, or quotient) space ${\cal G}/{\cal H}$, where $\cal G$ is a compact group
and $\cal H$ is a (fixed) subgroup of $\cal G$, for each point in the space there is an isotropy subgroup
in $\cal G$ defined as the set of elements that leave the point fixed, and for any point the
isotropy subgroup is isomorphic to $\cal H$. First we consider ordinary
functions on ${\cal G}/{\cal H}$ (as required in particular for all cases of the P spatial b.c.). 
Here we recall that $\cal G$ can be identified
as the product ${\cal G}_L\times {\cal G}_R$ of two copies of $\cal G$, modulo
the diagonal subgroup $\cal G$. As a homogeneous space, $\cal G$ has a left action
of ${\cal G}_L$ by left multiplication and a left action of ${\cal G}_R$ by right multiplication; 
see Appendix \ref{PetWeyl}. The coset space ${\cal G}/{\cal H}$ can be obtained as the quotient of $\cal G$
by the action of the (fixed) subgroup ${\cal H}\subset {\cal G}_R$, and then the left action 
of ${\cal G}_L$ remains as a group of nontrivial symmetries of the quotient space.
Any function on the coset space can be lifted to a function on ${\cal G}$, which must
be invariant under the right-multiplication left action by $\cal H$. We discussed in Appendix
\ref{PetWeyl} how the functions on $\cal G$ decompose into finite vector spaces (``irreducible
spaces'') of functions that transform as an irreducible representation of ${\cal G}_L$
and as the dual representation of ${\cal G}_R$. Then the space of functions on $\cal G$
that arise as lifts of functions on ${\cal G}/{\cal H}$ can be decomposed into
subspaces of the irreducible spaces such that they are invariant under the left action
of $\cal H$ by right multiplication. The problem of finding the $\cal G$-representation content 
of the functions on ${\cal G}/{\cal H}$ has now been reduced to a representation theory problem: 
find the $\cal H$-invariant subspaces for each irreducible representation of $\cal G$. This can be recognized
as a part of the problem of finding the ``branching rules'' which, for any subgroup
${\cal H}\subset{\cal G}$ and an irreducible representation, say  $V_\nu$ (where $\nu$ stands for a hishest weight), 
of $\cal G$, tell us the multiplicity of each irreducible representation of $\cal H$ in the decomposition of $V_\nu$;
this is outlined for some cases involving $U(n)$ on page 80 of Ref.\ \cite{fulton1991}, which will be useful
in the following. For example, for $G=SO(3)$, and the homogeneous space $\mathbb{S}^2=SO(3)/SO(2)$, 
where ${\cal H}=SO(2)$ is viewed as generated by $J_3$ ($\cal H$ consists of rotations about the 
$3$ axis in $\mathbb{R}^3$), the invariant subspaces  are spanned by the matrix elements of the irreducible
representations of $SO(3)$ (the Wigner matrices), the invariant subspaces
have $m_R=0$, and the functions are combinations of spherical harmonics.
The multiplicity of each total $SO(3)_L$ angular momentum $j$ is then $1$,
in agreement with the direct analysis of functions on $\mathbb{S}^2$.

For the case of a spatial P b.c., the 
collective coordinate space is the target space ${\cal G}/({\cal P}\cap{\cal G})$
of the sigma model, where now we again assume $\cal G$ is simply connected,
and where $\cal P$ is again the parabolic subgroup in ${\cal G}_\mathbb{C}$ that
leaves the highest weight $\lambda$ of the corresponding irreducible representation $V$ invariant
up to multiplication by a complex scalar. Our first results hold even if $V$ is not self dual.

First, consider the flag manifold for any such $\cal G$, for which
${\cal P}=B$, the Borel subgroup; this target space arises when $V$ is a generic irreducible representation,
that is, one whose highest weight is in the interior of the Weyl chamber. Then $B\cap{\cal G}$ is a 
Cartan subgroup of $\cal G$. For any irreducible representation of $\cal G$, say $V_\nu$ with 
highest weight $\nu$, there is an irreducible space of functions on 
$\cal G$ that transforms as $V_\nu\otimes V_\nu^*$
under ${\cal G}_L\times{\cal G}_R$, and it contains a subspace  of functions invariant
under $B\cap{\cal G}\subset {\cal G}_R$, which is spanned by the functions in the weight zero subspace
in $V_\nu^*$ under ${\cal G}_R$. Of course, each of these transforms as $V_\nu$
under ${\cal G}_L$. Hence the multiplicity of the irreducible representation $V_\nu$ in the Hilbert
space of functions on ${\cal G}/(B\cap{\cal G})$ is the dimension of the weight-zero subspace
in $V_\nu^*$ (which is the same dimension as in $V_{\lambda'}$); this dimension is zero if $V_\nu$ 
is not in the identity conjugacy class. 

For an irreducible representation $V$ with highest weight $\lambda$ that is on a face, not the interior, of the Weyl chamber,
the subgroup of the highest weight vector is a larger parabolic subgroup $\cal P$, $B\subset{\cal P}$
in ${\cal G}_\mathbb{C}$. Then the conditions on the functions lifted to $\cal G$ are more stringent, 
and only a subspace of those found for the flag manifold are still admissible. Thus the preceding result gives an upper
bound on the multiplicity of $V_\nu$ for each $\nu$, for general $\cal P$. For example, the simplest 
nontrivial irreducible representation in the identity conjugacy class is the adjoint. For the flag manifold,
it occurs with multiplicity $r$, the rank of $\cal G$; for ${\cal G}=SU(n)$, $r=n-1$. On the other hand, for the
complex Grassmannians for $SU(n)$, which arise when the highest weight in $V$ is a multiple of a fundamental weight $\omega_m$, 
we have ${\cal H}={\cal P}\cap {\cal G}\cong S[U(m)\times U(n-m)]$ ($1\leq m\leq n-1$), 
and it is not difficult to see that the subspace of the adjoint invariant under $\cal H$ has dimension $1$, 
so the multiplicity is $1$.  
More generally, for $SU(n)$, if the highest weight 
in $V$ is a sum of fundamental weights with $p$ nonzero coefficients, the multiplicity of the adjoint is $p$. 

For the case of the complex Grassmannians, these multiplicities can also be found, and easily extended, using the branching rule approach,
in terms of the Littlewood-Richardson (LR) coefficients \cite{fulton1991} by direct use of the formula in Exercise 6.11 a) on p.\ 80 of 
Ref.\ \cite{fulton1991}, with attention to the difference between $U(n)$ and $SU(n)$. [For more general $SU(n)$ target spaces, 
the use of the formula has to be iterated.] As this is the most efficient approach (and also can be modified for the A b.c. case), 
we go into more detail. We recall that the fusion rule, or Clebsch-Gordan, multiplicities arise when the tensor product of irreducible
representations, say on vector spaces $V_\rho$, $V_\mu$ (where in general $\rho$, $\mu$, $\nu$, \dots
refer to highest weights; we continue to reserve the symbol $\lambda$ for the highest weight of the chosen self-dual representation $V$, associated 
with the spin chain or sigma model) is decomposed as a direct sum,
\begin{equation}
V_\rho\otimes V_\mu\cong\bigoplus_\nu N_{\rho\mu}^\nu V_\nu,
\end{equation}
where the coefficient $N_{\rho\mu}^\nu$ is the number of copies of $V_\nu$ in the direct sum. (Clearly, $N_{\rho\mu}^\nu=N_{\mu\rho}^\nu$.)
For $U(n)$, many of the irreducible representations can be obtained by Weyl's construction, by decomposing iterated tensor products
of the defining $n$-dimensional representation into irreducible components. For such irreducible representations, 
$\rho$, $\mu$, $\nu$, \dots can be also be viewed as partitions (or Young diagrams) 
with at most $n$ parts (or rows in the diagram), and the coefficients $N_{\rho\mu}^\nu$ are the LR coefficients, 
which can be calculated by using the LR rule \cite{fulton1991}.
Then if for each partition, say $\rho$, $|\rho|$ denotes the number being partitioned,
or the number of boxes in the diagram, $N_{\rho\mu}^\nu=0$ unless $|\nu|=|\rho|+|\mu|$.
[The most general irreducible representations of $U(n)$
can be obtained from the preceding ones; the action of $u\in U(n)$ on a general irreducible representation is given by multiplying
the action on one of the preceding representations by an integer power of $\det u$ \cite{Dieck}. (The case when the power is $1$ 
is the same as the tensor product with the one-dimensional irreducible representation with diagram consisting of a single column 
of $n$ boxes.) For these, the integer powers add under tensor product.]
Further, for the branching of such an irreducible representation $V_\nu$ of $U(n)$ on passing to the subgroup 
$U(m)\times U(n-m)\subset U(n)$, the branching formula is
\begin{equation}
V_\nu\cong\bigoplus_{\rho,\mu}N_{\rho\mu}^\nu V_\rho\otimes V_\mu,
\end{equation}
where $V_\rho\otimes V_\mu$ is viewed as an ``external'' tensor product of representations of representations $V_\rho$, $V_\nu$
of $U(m)$ and $U(n-m)$, respectively. (The fact that the same coefficients $N_{\rho\mu}^\nu$ appear in both formulas is a consequence
of Frobenius reciprocity for the symmetric group, which is related to $U(n)$ through Schur-Weyl duality \cite{fulton1991}.)  

When we apply the preceding formulas to our problem, which involves irreducible representations of $SU(n)$, we should first remember \cite{fulton1991}
that irreducible representations of $SU(n)$ are in one-one correspondence with Ferrers-Young diagrams with at most $n-1$ [$=r$, the rank of $SU(n)$] 
rows; as we have mentioned already, the number of columns of length $l$, $l=1$, \ldots, $r$ is equal to $a_l$ for that irreducible
representation of $SU(n)$. More generally, an irreducible representation of $U(n)$ represented by a diagram with at most $n$ rows, 
and with an integer power of additional $\det u$ factors of the action of $U(n)$,
restricts to an irreducible representation of $SU(n)$ that is isomorphic to the irreducible representation which corresponds to the 
diagram with the columns of length $n$ (which if present, appear first) removed, and any powers of $\det u$ ignored, as now $\det u=1$. 
In view of this, we can first consider our problem in terms 
of representations of $U(n)$, and it will be sufficient to consider the irreducible representations that arise from a diagram, without additional 
integer powers; for these, we can directly apply the branching rule formula above. For the complex Grassmannians $U(n)/[U(m)\times U(n-m)]$
[also $\cong SU(n)/S[U(n)\times U(n-m)]$], for each irreducible representation of $U(n)$ [or $SU(n)$] we must find the multiplicity, 
given by the branching formula, into a representation of $U(m)\times U(n-m)$ that is trivial modulo some integer power of $\det u$
[resp., into a trivial representation of $S[U(n)\times U(n-m)]$].
In particular, the external tensor product $V_\rho\otimes V_\mu$ must be invariant under the subgroup $SU(m)\times SU(n-m)$. 
This implies that the diagrams $\rho$, $\mu$ must be rectangular with $m$, $n-m$ rows, respectively. Further, for a nonzero multiplicity
we must have $|\nu|=|\rho|+|\mu|$. There is one more condition, which when written in the form with prefix $S$ is as follows. There is 
a $U(1)$ subgroup in $S[U(n)\times U(n-m)]$ whose elements commute with those of $SU(m)\times SU(n-m)$. If we consider the block diagonal
matrices of the form $(u_1,u_2)$, where $u_1$, $u_2$ are elements of $U(1)$, each embedded as multiples of the identity matrix 
in $U(m)$, $U(n-m)$, respectively, then the prefix $S$ imposes the condition, 
in terms of the scalars $u_1$, $u_2$, that $u_1^mu_2^{n-m}=1$. Then invariance under this subgroup imposes the condition that, 
for $\rho$ and $\mu$ of the rectangular form already specified, the number of columns must be equal.

It is a striking fact that, for two diagrams that are rectangular with equal numbers of columns, the LR multiplicities are at most one; 
this can be readily proved from the LR rule. (Hence in particular, this is the case for the adjoint representation.) 
In more detail, from the LR rule, the only diagrams, or irreducible representations $V_\nu$ of $SU(n)$, 
that arise with nonzero multiplicity in the Hilbert space of functions on a complex Grassmannian are those in the conjugacy class of the identity
in $C^*$, with $a_l$ columns of length $l$ ($1\leq l\leq n-1$), where $a_l>0$ only if either $l\leq\min(m,n-m)$ or $l\geq \max(m,n-m)$ 
(the others must be zero), with the condition $a_l=a_{n-l}$ for all $l$, and so $V_\nu$ is self dual; the multiplicity is $1$ for each of these. 
This completely determines the decomposition of the Hilbert space of functions on a complex Grassmannian $\mathbb{G}_{n,m}$.

Another approach to this problem (for any group $\cal G$) arises from the semiclassical approach to the spin chain, at present for even length $N$.
If we consider all spins on one sublattice parallel, then they transform in the irreducible
representation with highest weight $N\lambda/2$ (where again $\lambda$ is the highest weight of $V=V_\lambda$),
and in the dual for the other sublattice. This two-spin problem itself maps to the sigma model
in $0+1$ dimensions, that is with no $x$ dependence of $\vec{n}$, and this is the same as the collective
coordinate space we are considering. Then the irreducible representations in the Hilbert space can be found
by decomposing the tensor product $V_{N\lambda/2}\otimes V_{N\lambda/2}^*$ into irreducible components
(again using the fusion rules or Clebsch-Gordan decomposition). When $N\lambda/2$ is large, the multiplicities
of the irreducible for each fixed highest weight $\nu$ stabilize (i.e.\ tend to a limit); they are the same as the numbers we seek,
and so depend only on the parabolic subgroup of the highest weight vector in $V_\lambda$ or $V_{N\lambda/2}$.
For ${\cal G}=SU(n)$, the fusion rule multiplicities are given by the LR coefficients, as above
(see Ref.\ \cite{fulton1991}, pp.\ 225, 424, 456). For the complex Grassmannians, the diagrams $V_{N\lambda/2}$ and its dual
are rectangular, and now it is clearly visible that the results for the multiplicity of each $V_\nu$ are the same as in the first approach.
For example, if the highest weight $\lambda$ is a multiple of $\omega_1$ (so $a_1>0$, all others are zero), then for $\nu$ we can have 
only $a_1=a_{n-1}>0$, all other $a_l$ zero; these representations are easily seen to be what occur if a Schwinger boson construction is used 
for $V_{N\lambda/2}$ and its dual, or if the representations $V_\nu$ are found in terms of singlet valence bonds.

We comment here that it turns out that some irreducible representations do not occur in the spectrum of functions on a complex Grassmannian. 
Not all self-dual irreducible representations occur, except in the cases $m=\lfloor n/2\rfloor$. The complex representations, discussed earlier, 
do not occur, not even those in the correct conjugacy class. 

For the spatial A b.c., we must consider cases in which the representation $V$ is self dual; the simplest cases
are then those in which $n$ is even and the sigma model target space is $\mathbb{G}_{n,n/2}$.
For these, the collective-coordinate space of textures (or minimal geodesics) has been identified as $SU(n)/S[U(n/2)_{\rm diag}]$, 
where $S[U(n/2)_{\rm diag}]$ was discussed in the preceding subsection. This space is doubly connected, that is its 
fundamental group $\pi_1=\mathbb{Z}_2$. Accordingly, it has a simply-connected double cover, which is just $SU(n)/SU(n/2)_{\rm diag}$, 
where $SU(n/2)_{\rm diag}$ is the copy of $SU(n/2)$ embedded as the diagonal in $SU(n/2)\times SU(n/2)\subset SU(n)$. 
There is a $\mathbb{Z}_2$ symmetry of the double cover, which commutes with the action of $SU(n)$, and arises from the quotient group 
$C/2C\cong\mathbb{Z}_2$ of the center $C\cong \mathbb{Z}_n$ of $SU(n)$; taking the quotient by that symmetry recovers the 
collective coordinate space. In the special case $n=2$, discussed in the main text, 
we note that $SU(1)$ is the trivial group, and $SU(2)/\mathbb{Z}_2\cong SO(3)$. As in that case, the connection that arises 
from the invariant $\cal I$ gives a holonomy which is $(-1)^{\Theta_{n/2}/\pi}$ raised 
to the power of the number of times that the path (in time) wraps around the nontrivial cycle in the collective coordinate space.
The sum over this number gives a projection in the Hilbert space, and hence the states in the collective-coordinate Hilbert space can be obtained 
by first finding the irreducible representations for functions on the double cover, and then for $\Theta_{n/2}=0$ (mod $2\pi$) 
[resp., $\pi$ (mod $2\pi$)] selecting (i.e.\ projecting onto) only those that are even (resp., odd) under the action of the nontrivial 
element of $C/2C$. This selection rule has been seen earlier in this Appendix, but now we identify the irreducible representations 
and in some cases the multiplicities also.

For this case, we can again obtain some first results by hand from the representations of ${\cal G}_R$. For $\Theta_{n/2}=0$ (mod $2\pi$),
the singlet has multiplicity $1$. For the adjoint, we can use the explicit matrix representation of the Lie algebra itself as traceless
self-adjoint $n\times n$ matrices, which transform as the adjoint representation, and we find that the subspace that commutes with 
$SU(n/2)_{\rm diag}$ is spanned by the matrices $\Sigma_x$, $\Sigma_y$, $\Sigma_z$, where $\Sigma_x$ appeared earlier, 
and in general these matrices are $\Sigma_\mu=\sigma_\mu\otimes {\bf 1}_{n/2}$, where $\mu=x$, $y$, or $z$, and $\sigma_\mu$ are the $2\times 2$ 
Pauli matrices, so each $\Sigma_\mu$ consists of four blocks of size $n/2\times n/2$, each block a multiple of the identity.
(We return to further implications of the structure used here in a moment.) Hence for the adjoint, the multiplicity is $3$,
the same for all (even) $n$, and the same as in the main text.

For further cases, the branching formula approach is more efficient. In the present case, if the $SU(n)_R$ irreducible representation 
is $V_\nu$, and we view this as a representation of $U(n)$ corresponding to a Young diagram $\nu$ with at most $n-1$ rows, 
then we can decompose it into irreducible representations of the $U(n/2)\times U(n/2)$ subgroup as before, which are external
tensor products $V_\rho\otimes V_\mu$. But now the final condition we must impose is that the external tensor product $V_\rho\otimes V_\mu$, 
viewed as a representation of the diagonal $SU(n/2)_{\rm diag}\subset SU(n/2)\times SU(n/2)$, must produce an $SU(n/2)_{\rm diag}$
invariant (or singlet) representation. This means that the representations $V_\rho$, $V_\mu$ must be dual to each other 
when restricted to give representations of $SU(n/2)$. So we have a general formula for the multiplicity for the (dual of the) given $V_\nu$, namely
\begin{equation}
\sharp(V_\nu^*)=\sum_\rho N_{\rho\overline{\rho}}^\nu,
\end{equation}
where $N_{\rho\mu}^\nu$ are the LR coefficients as before, $\nu$ has at most $n-1$ rows, the sum is over diagrams $\rho$ with 
at most $n/2$ rows (so the dimension of $V_\rho$ is nonzero), and $\overline{\rho}$ means the diagrams that correspond to the 
$SU(n/2)$ representation dual to that corresponding to $\rho$. The simplest example of interest is the case when $\nu$ 
is a single column of $n/2$ boxes (a self-dual representation), which can arise when $\Theta_{n/2}=\pi$ (mod $2\pi$); 
we expect the lowest-energy states in the collective-coordinate space Hilbert space to transform in this representation. 
From the LR rule, $N_{\rho\overline{\rho}}^\nu$ is nonzero, and equal to $1$, when and only when $\rho$ is a single column 
of $|\rho|=0$, $1$, \ldots, $n/2$ boxes, and $\overline{\rho}$ is a single column of $(n/2)-|\rho|$ boxes, and these.
Hence the multiplicity of this (self-dual) representation is $(n/2)+1$. For example, for $n=4$, this representation is the six-dimensional
$SO(6)$ vector representation of $SU(4)\cong SO(6)$, and the multiplicity is $3$, which can also be easily obtained by hand. 
We note that the general case agrees with the case $n=2$ in the main text, for which the invariance under $SU(1)_{\rm diag}$ has no effect. 
In general, we do not expect all these states to be degenerate in energy, but the evident 
symmetry under $|\rho|\to (n/2)-|\rho|$ should lead to degeneracy, except for the case $|\rho|=n/4$ when $n=0$ (mod $4$). 
[Indeed, the multiplicity is even if and only if $n/2$ is odd, and this agrees with the fact that for these representations ${\cal T}'^2=(-1)^{n/2}$,
so that for $n/2$ odd an even multiplicity is expected, due to our whole picture of the A sector.]
Other examples become more complicated than these.

For this problem, there is also another useful approach, which we now describe. The diagonal subgroup $U(n/2)_{\rm diag}$ in $U(n)$
commutes with a $U(2)$ subgroup, as we first saw in terms of the Lie algebra. [The $SU(2)$ subgroup of this $U(2)$ can be viewed 
as the double cover of the $SO(3)$ group that contains the $O(2)_{\rm sp}$ subgroup, which represents translations and reflections 
of the texture in space. Thus something like it should arise in every case.] 
These subgroups constitute a Howe pair, and this structure
can be useful in analyzing our problem. The defining representation of $U(n)$ can be viewed as a tensor product of the $n/2$-dimensional
and $2$-dimensional defining representations of $U(n/2)$ and $U(2)$, respectively. Then using Weyl's construction,
the irreducible representation of $U(n)$ corresponding to the Young diagram $\nu$ can be decomposed into irreducible
representations of $U(n/2)\times U(2)$ in the form
\begin{equation}
V_\nu\cong \bigoplus_{\rho,\mu} C_{\rho\mu}^\nu V_\rho\otimes V_\mu,
\end{equation}
where all diagrams have the same size, that is, $|\nu|=|\rho|=|\mu|=d$, $V_\rho$ is the representation of $U(n/2)$ corresponding to the 
diagram $\rho$, and $V_\mu$ is the representation of $U(2)$ corresponding to the diagram $\mu$. Finally, the multiplicities $C_{\rho\mu}^\nu$ 
in the direct sum in this case are the same as those in the decomposition of a product of irreducible characters (or tensor product 
of irreducible representations) of the symmetric group $S_d$, namely
\begin{equation}
\chi_\rho\chi_\mu=\sum_\nu C_{\rho\mu}^\nu \chi_\nu,
\end{equation}
where $\chi_\rho$ and so on are irreducible characters. (Hence $C_{\rho\mu}^\nu\geq 0$ and, as all characters of $S_d$ are real, 
$C_{\rho\mu}^\nu$ is invariant under any permutation of $\rho$, $\mu$, and $\nu$.) [See Exercise 6.11 b) in Ref.\ \cite{fulton1991}, p.\ 80.]
For our problem, we can assume $\nu$ has at most $n-1$ rows, and impose that $V_\rho$ is a singlet when viewed as a representation 
of $SU(n/2)$, which means it must be a rectangle with $n/2$ rows (and $d$ must be a multiple of $n/2$). Then the multiplicity of $V_\nu^*$ 
that we seek is also given by
\begin{equation}
\sharp(V_\nu^*)=\sum_\mu C_{\rho\mu}^\nu \dim V_\mu,
\end{equation}
where $\rho$ is a rectangle as specified. The coefficients $C_{\rho\mu}^\nu$ can be obtained one by one 
by the elementary orthogonality of characters of $S_d$ for each $d$, but are not easy to calculate efficiently; there does not seem to be a 
simple combinatorial rule as there is for the LR coefficients. 
However, for the example we considered before, in which $\nu$ is a single column of $n/2$ boxes, we have $\rho=\nu$ and it is easy to see that then 
$C_{\nu\mu}^\nu=0$ unless $\mu$ is a single row of $n/2$ boxes, in which case $C_{\nu\mu}^\nu=1$ [or see Ref.\ \cite{fulton1991}, using either of the 
special cases in Exercise 4.51 b) on page 61], and so $\sharp(V_\nu^*)=\dim V_\mu=(n/2)+1$, which reproduces our previous result. 
Other examples include $n=2$, when the results of the main text are recovered: $\sharp(V_\nu^*)=\dim V_\nu$, 
where $\nu=\rho=\mu$ has one row, and $C_{\nu\nu}^\nu=1$. For general $n>2$ and when $V_\nu$ is the adjoint representation of $SU(n)$, 
the multiplicity is already known to be $3$; it arises when $\mu$ has rows of lengths $(n/2)+1$, $(n/2)-1$, corresponding 
to the adjoint of $SU(2)$, with $\dim V_\mu=3$, while $C_{\rho\mu}^\nu$ should equal $1$, and $C_{\rho\mu}^\nu \dim V_\mu$ 
should be $0$ for other $\mu$. The verification of both points is straightforward for $n=4$ by using the character table of $S_4$, 
and for $n=6$ and $n=8$ the result can be confirmed using published
tables \cite{james_kerber_1981} of $C_{\rho\mu}^\nu$. 

Next, we address the question of complex irreducible representations $V_\nu^*$ of ${\cal G}_L$. Suppose, for example, that $|\nu|=n/2$, 
and $\nu$ has two columns of lengths $l$, $l'$, both nonzero. Then either $l=l'=n/4$, so $a_{n/4}=2$, or $l\neq l'$, and $a_l=a_{l'}=1$. 
These representations $V_\nu$ (and their duals) are complex. Again, $\rho$ must be 
a single column of $n/2$ boxes, and then $\mu$ must be the partition conjugate to $\nu$ (see Ref.\ \cite{fulton1991}, p. 61), 
meaning the one with rows and columns interchanged (i.e.\ reflected across the $45^\circ$ leading diagonal \cite{fulton1991,james_kerber_1981}). 
Hence the multiplicity for one of these representations (or its dual) must be $\sharp(V_\nu^*)=\sharp(V_\nu)=\dim V_\mu=|l-l'|+1$. 
We note that $|l-l'|$ has the same parity as $n/2$, so $V_\mu$ is quaternionic when $n/2$ is odd. For $\nu$ with $|\nu|=n/2$ and more than 
two columns, the multiplicity is zero by a similar argument. We conclude that complex irreducible representations are plentiful in the general case 
$n>2$, $n$ even, though not every one occurs.

Finally, we recall our results, going back to the Introduction, and extended in this Appendix, which showed that
if the self-dual irreducible representation $V$ is real, then ${\cal C}=+1$, while if $V$ is quaternionic, ${\cal C}=-1$.
We know that there is an 't Hooft anomaly for gauging of the inversion symmetry, and the group of spatial translations and reflections
$O(2)_{\rm sp}$ must be lifted to $Pin_-(2)_{\rm sp}$, so that translation by $2L$ of any state in the A sector has eigenvalue $-1$
(and hence there is an anomaly in modular $T^2$ transformations), precisely when ${\cal C}=-1$. The following simple remark, using the preceding 
decomposition with $C_{\rho\mu}^\nu$, shows that the latter point holds for the states in the collective-coordinate Hilbert space in the cases 
we have been considering. The size of the diagram $\mu$ must be $|\mu|=|\nu|$ (and $=|\rho|$), with $|\nu|/(n/2)=\Theta_{n/2}/\pi$ (mod $2$). 
An irreducible representation $V_\mu$ of $SU(2)$ is real when $|\mu|$ is even, and quaternionic when $|\mu|$ is odd.
This is exactly the expected result; it holds when $V_\nu$ is complex as well as when it is self dual. We also point out 
that if $V_\nu$ is self dual, $V_\nu$ and $V_\nu^*$ are isomorphic and have the same index 
(value of ${\cal T}'^2$), while $V_\rho$ is a singlet, and so real. Hence $V_\mu$ must have the same index as $V_\nu$, as we have seen. 
(We can consider in particular $V_\nu=V$, which implies $\nu=\rho$. The multiplicity of $V_\nu^*$ for $\nu=\rho$ is at least as large as $|\nu|+1$.) 
We expect a similar result for other target spaces and other ${\cal G}$.

\end{appendix}

\bibliography{992.bib}
\end{document}